%% file: SOFARI.tex
\documentclass[12pt]{article}
\usepackage{amsmath}
\usepackage{graphicx}
\usepackage{enumerate}
\usepackage{natbib}
\usepackage{url} 
\usepackage{bibunits}
\usepackage{algorithm}
\usepackage{algorithmic}

\usepackage{graphicx}
\usepackage{multirow}
\usepackage{hhline}
\usepackage{enumitem}
\usepackage{hyperref}
\usepackage{bm}
\usepackage{multirow,multicol}
\usepackage{verbatim}
\usepackage{amsfonts,amssymb,amsthm,color,authblk}
\usepackage{xcolor}
\definecolor{myblue}{RGB}{0, 0, 0}
\newcommand{\blue}[1]{{\color{myblue}#1}}
\input{./command.tex}

\newcommand{\blind}{1}

\usepackage{setspace}   
\usepackage{etoolbox}  

\AtBeginEnvironment{thebibliography}{\small\setstretch{0.93}}

\begin{document}

\def\spacingset#1{\renewcommand{\baselinestretch}%
{#1}\small\normalsize} \spacingset{1}


\if1\blind
{
  \title{\bf SOFARI: High-Dimensional Manifold-Based Inference\thanks{
This work was supported by  National Key R\&D Program of China (Grant 2022YFA1008000), Natural Science Foundation of China (Grants 72071187, 11671374, 71731010, and 71921001), Fundamental Research Funds for the Central Universities (Grants  WK3470000017 and WK2040000027), and NSF Grant DMS-2324490. Zemin Zheng and Xin Zhou are the co-corresponding authors.
}
  }
  \author{Zemin Zheng$^1$, Xin Zhou$^1$, Yingying Fan$^2$ and Jinchi Lv$^2$
\medskip\\
University of Science and Technology of China$^1$ and University of Southern California$^2$
\\
}
\date{June 26, 2025}
  \maketitle
} \fi

\if0\blind
{
  \bigskip
  \bigskip
  \bigskip
  \begin{center}
    {\LARGE\bf SOFARI: High-Dimensional Manifold-Based Inference}
\end{center}
  \medskip
} \fi

\bigskip
\begin{abstract}
Multi-task learning is a widely used technique for harnessing information from various tasks. Recently, the sparse orthogonal factor regression (SOFAR) framework, based on the sparse singular value decomposition (SVD) within the coefficient matrix, was introduced for interpretable multi-task learning, enabling the discovery of meaningful latent feature-response association networks across different layers. However, conducting precise inference on the latent factor matrices has remained challenging due to the orthogonality constraints inherited from the sparse SVD constraints. In this paper, we suggest a novel approach called the high-dimensional manifold-based SOFAR inference (SOFARI), drawing on the Neyman near-orthogonality inference while incorporating the Stiefel manifold structure imposed by the SVD constraints. By leveraging the underlying Stiefel manifold structure that is crucial to enabling inference, SOFARI provides easy-to-use bias-corrected estimators for both latent left factor vectors and singular values, for which we show to enjoy the asymptotic mean-zero normal distributions with estimable variances. We introduce two SOFARI variants to handle strongly and weakly orthogonal latent factors, where the latter covers a broader range of applications. We illustrate the effectiveness of SOFARI and justify our theoretical results through simulation examples and a real data application in economic forecasting.
\end{abstract}

\noindent%
\vfill

\spacingset{1.9} 

\section{Introduction} \label{Sec.new1}

Multi-task learning has gained significant popularity in modern big data applications, particularly in scenarios where the same set of covariates is employed to predict multiple target responses, as seen in applications like autonomous driving. A widely used model for multi-task learning is the multi-response regression model given by
\begin{align}\label{model}
\Y = \X\C^* + \E,
\end{align}
where $\Y\in\R^{n\times q}$ represents the response matrix, $\X\in\R^{n\times p}$ is the fixed design matrix, $\C^* \in \R^{p\times q}$ is the true, unknown regression coefficient matrix, and $\E\in\R^{n\times q}$ stands for the mean-zero random noise matrix. Here, $n$ denotes the sample size, $p$ represents the dimensionality of the covariate vector, and $q$ is the dimensionality of the response vector.
When both $p$ and $q$ are substantially larger than the sample size $n$, accurately estimating $\C^*$ becomes a formidable challenge due to high dimensionality. To address this challenge, structural assumptions are often imposed on $\C^*$ through techniques such as matrix factorization, such as the singular value decomposition (SVD), which facilitates dimensionality reduction. Among these structural assumptions, two commonly adopted ones are the low-rankness and sparsity. These assumptions form the basis for various regularization techniques developed to simultaneously reduce dimensionality and select relevant features.

Methods for simultaneously achieving sparse recovery and low-rank estimation of $\C^*$ can be broadly categorized into two classes. The first class involves the direct estimation of $\C^*$ using various regularization techniques, as discussed in previous works \citep{Bunea2012,chen2012sparse,chendong2013}. The second class focuses on reconstructing the parameter matrix by initially estimating its sparse SVD components and then combining them, as demonstrated in related research \citep{mishra2017,uematsu2017sofar,zheng2019scalable,chen2022stagewise}. An advantage of the latter approach is that the sparsity assumption applied to different SVD components naturally leads to the interpretation of a sparse latent factor model. In such model, each latent factor represents a sparse linear combination of the original predictors, and different responses can be associated with distinct sets of latent factors. The importance of each latent factor can be measured by examining the magnitude of the corresponding singular value. For example, in the analysis of yeast eQTL data discussed in \cite{uematsu2017sofar} and \cite{chen2022stagewise}, it revealed the existence of three latent pathways (i.e., latent factors). These pathways predominantly consisted of some original predictors which are certain upstream genes, downstream genes, and a combination of both, respectively.

While the sparse SVD structural assumption on $\C^*$ offers an enticing level of interpretability, estimating this sparse SVD structure poses challenges. This complexity arises from the simultaneous presence of orthogonality constraints and the need for sparsity across different SVD components. In essence, achieving both sparsity and orthogonality can be two inherently conflicting objectives within a single statistical  inference framework. For example, the sparsity pattern of a matrix may no longer hold after applying a conventional orthogonalization process such as the QR factorization. To tackle this dilemma, \cite{uematsu2017sofar} introduced the sparse orthogonal factor regression (SOFAR) method. SOFAR allows for the simultaneous attainment of sparse and orthogonal estimates of the factor matrices by formulating them within an orthogonality-constrained regularization framework. Nonetheless, the development of a valid statistical inference procedure for quantifying the uncertainty associated with these estimation results remains a challenging task under the SVD constraints. This paper focuses on addressing such  a challenge. 

Our inference method builds upon the SOFAR framework \citep{uematsu2017sofar}. It is well recognized that estimates obtained through regularization methods can be susceptible to bias issues, primarily due to the use of penalty functions. Consequently, these estimators are not directly applicable in statistical inference problems, such as hypothesis testing. To correct such bias and calculate valid p-values, several statistical inference methods have been introduced. One line of research suggests debiasing the regularized estimators by inverting the Karush--Kuhn--Tucker condition associated with the corresponding optimization problem 
\citep{javanmard2014confidence, van2014asymptotically} or, equivalently, using the relaxed projection approach \citep{zhang2014confidence}. 
In situations involving high-dimensional nuisance parameters, \cite{doubleML2018} devised a double machine learning framework. This framework introduces a score function vector that is locally insensitive to nuisance parameters of high dimensionality, allowing for bias correction in the existence of nuisance parameters. In the context of statistical inference for high-dimensional principal component analysis (PCA), \cite{jankova2021biased} proposed a debiased sparse PCA estimator. They also constructed confidence intervals and hypothesis tests for inferring the first eigenvector and the corresponding largest eigenvalue.
However, extending this inference procedure to the remaining principal components poses a nontrivial challenge due to the accumulation of noise from previous eigenvectors and eigenvalues.

In our model setting, both the left and right singular vectors of $\C^*$ are unknown, and during our statistical inference for one singular vector, the remaining ones together become high-dimensional nuisance parameters. Furthermore, these nuisance parameters must adhere to the orthogonality and unit-length constraints imposed by the SVD structure, restricting them to some underlying manifold. 
If one ignores such manifold structure and directly uses the SVD constraints to calculate the associated statistics for the inference task,
the deficiency in the degrees of freedom will render the construction of the inference procedure invalid or ineffective; this will be made clear in Section \ref{new.Sec.3.1}. 
In fact, incorporating the manifold structure is crucial in optimization problems that are subject to orthogonality or SVD constraints. For instance, in nonnegative independent component analysis (ICA), \cite{PLUMBLEY2005} utilized the manifold structure to calculate gradients and developed various algorithms under orthogonality constraints. \cite{Derenskifemba} applied the ICA framework to functional data analysis, achieving nonparametric estimation accuracy. Additionally, \cite{chen2012sparse} investigated the manifold structure of the coefficient matrix under SVD constraints in reduced rank regression. They demonstrated the consistency of the coefficient matrix and its SVD components when the dimensionality is fixed. These studies collectively offer valuable insights into leveraging the manifold structure for algorithmic and estimation improvements. Despite these advancements, exploring the integration of the manifold structure into the inference procedures for high-dimensional multi-task learning remains an uncharted area.

To overcome these aforementioned challenges, we suggest in this paper a new manifold-based inference procedure named high-dimensional manifold-based SOFAR inference (SOFARI). Here, the manifold is defined as the Riemannian space where the nuisance parameters lie in under the sparse SVD constraints. We adapt the idea proposed in \cite{doubleML2018} to construct the Neyman near-orthogonal score function on the manifold. More specifically, instead of constructing a score function which is locally insensitive to the nuisance parameters on the full Euclidean space, we need only the local insensitiveness to hold on the manifold induced by the SVD constraint. Such approach allows us to construct bias-corrected estimators under the sparse SVD constraint.  Depending on the correlation level between the latent factors, 
our suggested inference procedure takes two different forms. The first form, named SOFARI$_s$, deals with strongly orthogonal latent factors where different latent factors have correlations vanishing quickly as sample size increases;  examples in modern data science applications where such assumption makes sense include biclustering with sparse SVD \citep{lee2010biclustering}, sparse principal component analysis \citep{shen2008sparse}, and sparse factor analysis \citep{bai2008large}. The second form, named SOFARI, applies to a wider range of multi-response applications where latent factors can have stronger correlations among each other (which we name as weak orthogonality condition). We derive the asymptotic distribution for each bias-corrected estimator under the regularity conditions, allowing us to construct valid confidence intervals for latent factors. Using similar idea, we also construct the debiased leading singular values, and establish the asymptotic normality for each of them.

We conduct simulation studies to verify that our bias-corrected estimators for latent left factors and leading singular values indeed enjoy the asymptotic mean-zero normal distributions as established in our theory. We also show by numerical examples that the confidence intervals constructed from our asymptotic distributions have valid coverage under both the strong and weak orthogonality conditions. In addition, we examine the robustness of SOFARI when the correlations among latent factors violate our technical assumption using a simulation study; our results demonstrate that SOFARI is still applicable even beyond the scenario described by our technical assumptions.  Finally, we apply our inference methods to the economic forecasting data set and obtain some highly interpretable results revealing interesting dependence structure among economic variables.   

The rest of the paper is organized as follows. In Section \ref{new.Sec.3}, we present the SOFARI framework under both the strong and weak orthogonality constraints. In Section \ref{new.Sec.4}, we establish the asymptotic normalities of the debiased SOFARI estimates. Section \ref{new.Sec.5} presents several simulation examples to demonstrate the finite-sample performance of the newly suggested method. Section \ref{new.Sec.7} discusses some implications and extensions of our work.  \blue{In the Supplementary Material, we provide additional theoretical results, simulation studies, and real data details,
as well as all the proofs and secondary technical details.}

\section{High-dimensional manifold-based inference} \label{new.Sec.3} 

Denote by
$\C^*=\L^*\D^*\V\strans$
the singular value decomposition (SVD) of the true coefficient matrix $\C^*$ in model \eqref{model}, where $\L^*= \left(\l_1^*,\ldots,\l_{r^*}^*\right) \in\R^{p\times r^*}$ and $\V^* = \left(\v_1^*, \ldots, \v_{r^*}^*\right) \in\R^{q\times r^*}$ are the orthonormal matrices consisting of left and right singular vectors, respectively, $\D^* = \diag{d_1^*,\ldots, d_{r^*}^*}\in\R^{r^*\times r^*}$ is the diagonal matrix of nonzero singular values, and $r^*$ is the rank of $\C^*$. We define $\U^*=\L^*\D^*=\left(\u_1^*, \ldots, \u_{r^*}^*\right) \in\R^{p\times r^*}$ as the left factor matrix. 

We are interested in estimating and inferring $\u_k^*$'s, for which purpose  $\boldeta_k^* = \big(\v_1\strans, \ldots, \v_{r^*}\strans$, $\u_1\strans,\ldots, \u_{k-1}\strans, \u_{k+1}\strans, \ldots, \u_{r^*}\strans\big)\trans  \in \mathbb{R}^{ q r^* + p (r^*-1) }$ is the high dimensional unknown nuisance parameter vector. For technical simplicity, we assume the true rank $r^*$ is given and satisfies that $r^*\geq 2$; the case of $r^*=1$ is much simpler and not considered here because the orthogonality constraint no longer exists.   In practice, we can identify the rank of the latent SVD structure in advance by some self-tuning selection method such as the one developed in \cite{Xin2019Adaptive} that enjoys the rank selection consistency.

We introduce the new manifold-based SOFARI inference procedure in the next subsections.
Depending on the correlation level among the latent factors, the suggested SOFARI inference procedure takes two different forms: the basic form SOFARI$_s$ for strongly orthogonal factors and the general form SOFARI for weakly orthogonal factors, where the notions of strongly and weakly orthogonal factors will be made clear later.
SOFARI is more broadly applicable than SOFARI$_s$, thanks to its relaxed constraint on correlation level.  


\subsection{
SOFARI\texorpdfstring{$_s$}{s} under strongly orthogonal factors} \label{new.Sec.3.1}

Given an initial biased estimate $\u_k$ of $\u_k^*$ for a given $k\in \{1,\cdots, r^*\}$, we describe the construction of a debiased estimate with ${ \boldeta_k^* }$ the unknown nuisance parameter vector. 
To alleviate the impacts of nuisance parameters, 
we will make use of the Neyman orthogonality scores \citep{Neyman59,doubleML2018} and find a vector $\wt{\psi}_k(\u_k,\boldeta_k) { \in \mathbb{R}^{p  }}$ of score functions for $\u_k$ with nuisance parameter { $\boldeta_k = ( \v_1\trans, \ldots,  \v_{r^*}\trans, \u_1\trans,\ldots, \u_{k-1}\trans, \u_{k+1}\trans, \ldots, $ $\u_{r^*}\trans)\trans \in \mathbb{R}^{q r^* + p (r^*-1)   }$ }that satisfies two properties: 
first, the expectation of $\wt{\psi}_k$ at the true parameter values $(\u_k^*, \boldeta^*_k)$ is asymptotically vanishing; second, $\wt{\psi}_k$ satisfies the Neyman near-orthogonality condition with respect to the nuisance parameters in the sense that $\wt{\psi}_k$ is approximately insensitive to $\boldeta_k $ when evaluated at $(\u_k^*, \boldeta^*_k)$ locally. 

Moreover, in this section, we consider the case of strongly orthogonal factors such that $\sum_{j \neq k} |\u_j\strans\wh{\bSigma}\u_k^*|  = o(n^{-1/2})$, to deal with the intrinsic bias issue induced by correlations between the latent factors, which is specified in Section \ref{Biasipssue} of the Supplementary Material. It is worth pointing out that for many learning problems such as biclustering with sparse SVD \citep{lee2010biclustering}, sparse principal component analysis \citep{shen2008sparse}, and sparse factor analysis \citep{bai2008large}, the design matrix can be regarded as the identity matrix and thus the strong orthogonality condition holds naturally for the latent factors when any different $\u_j^*$ and $\u_k^*$ are orthogonal.

We start with the following constrained least-squares loss function 
\begin{align}
	&L(\u_k,\boldeta_k) = (2n)^{-1}\norm{\Y - \sum_{i=1}^{r^*} \X\u_i\v_i\trans}_F^2, \label{constraint00}\\
	&\text{subject to } ~\u_i\trans\u_j = 0 ~ \text{for} ~ 1\leq i \neq j\leq r^* ~ \text{and} ~ \V\trans\V = \I_{r^*}, \label{SVDc}
\end{align}
where $\V = \left(\v_1, \ldots, \v_{r^*}\right)$ is the matrix of right singular vectors. 

Denote by $\wh{\bSigma}=n^{-1}\X\trans\X$. A natural starting point for the score function for $\u_k$ is the partial derivative of loss function $L$ with respect to $\u_k$, which can be simplified as
\begin{align}\label{lossf1}
	\der{L}{\u_k} = \wh{\bSigma}\u_k - n^{-1}\X\trans\Y\v_k
\end{align}
under constraint \eqref{SVDc}. However, the above partial derivative is sensitive to the nuisance parameter vector $\boldeta_k$ even if it is within a shrinking neighborhood of $\boldeta_k^*$ since its derivative with respect to $\v_k$ does not vanish. 
To correct this, we define a modified score function vector for $\u_k$ as 
\begin{align}
	\wt{\psi}_k(\u_k,\boldeta_k) = \der{L}{\u_k} - \M\der{L}{\boldeta_k}, \nonumber
\end{align}
where matrix $\mathbf{M}=\left[ \mathbf{M}_{1}^{v}, \ldots, \mathbf{M}_{r^*}^{v}, \mathbf{M}_{1}^{u}, \ldots, \mathbf{M}_{k-1}^{u}, \mathbf{M}_{k+1}^{u}, \ldots,  \mathbf{M}_{r^*}^{u}\right]$  will be chosen such that $\wt{\psi}_k(\u_k,\boldeta_k)$ is approximately insensitive to $\boldeta_k$ under the SVD constraint \eqref{SVDc}. Note that the construction of $\M$ will depend on $k$, but we make such dependence implicit whenever no confusion. When confusion arises, we write the corresponding matrix as $\M^{(k)}$. 
Here, submatrices $\mathbf{M}_{i}^{u} \in\R^{p\times p}$ and $\mathbf{M}_{j}^{v} \in\R^{p\times q}$ correspond to $\u_i$ and $\v_j$  for $1 \leq i \leq r^*$ with $i \neq k$ and $1 \leq j  \leq r^*$, respectively. \blue{We will show that the most important construction of $\mathbf{M}$ lies in $\M^v_k$, to be detailed in Proposition \ref{prop:rankr2}, while other submatrices can be set as zero. }

Based on the modified score function vector $\wt{\psi}_k$, 
we can exploit the bias correction idea \citep{javanmard2014confidence, van2014asymptotically} and define a debias function for the initial estimate $\u_k$ as
\begin{align*}
	\psi_k(\u_k,\boldeta_k) = \u_k - \W\wt{\psi}_k(\u_k,\boldeta_k),
\end{align*}
where matrix $\W \in \R^{p \times p}$ will be constructed to correct the bias in the initial estimator. In this paper, we consider the initial estimator as the SOFAR estimator  $(\wt\u_i,\wt\v_i)_{i=1}^{r^*}$ { formally defined in Definition \ref{lemmsofar}} and propose valid constructions of $\M$ and $\W$ so that the bias-corrected estimator $\psi_k(\wt\u_k,\wt\boldeta_k)$ enjoys asymptotic normality with mean $\u^*_k$ and estimable variance for statistical inference, where $\wt\boldeta_k$ is the nuisance parameter constructed using the initial SOFAR estimator.
Similarly to $\M$, we make the dependence of $\W$ on $k$ implicit whenever no confusion, and write it as $\W_k$ when confusion arises.

We stress that constructions of $\M$ and $\W$ should not be considered separately because the former can affect effectiveness of the latter. For example, it may be tempting to directly leverage the SVD constraints in $\der{\wt{\psi}_k}{\boldeta_k}$ and calculate the derivatives in Euclidean space to construct $\M$. However, such an $\M$ leads to a deficiency in the degrees of freedom and hence results in the nonexistence of a valid $\W$ matrix for our ultimate goal of bias correction for $\u_k$. See Section \ref{appsec:issue} of the Supplementary Material for the detailed derivations. 

We next provide details on a construction of matrix $\mathbf{M}$ that can lead to a valid construction of $\W$. A natural way of making $\wt{\psi}_k$ insensitive to the nuisance parameter vector $\boldeta_k$ is requiring that $\der{\wt{\psi}_k}{\boldeta_k}$ be asymptotically vanishing. Such simple requirement, however, does not take the SVD constraints on $\boldeta_k$ into account.
To address this issue, we suggest a new manifold-based inference framework.
Specifically, instead of requiring that the score function vector $\wt{\psi}_k$ be locally insensitive to the nuisance parameters on the full Euclidean space, we need only the local insensitiveness to hold on the manifolds induced by the SVD constraints. Such distinction will relax the restriction and save the degrees of freedom, thereby providing more flexible choices for matrix $\M$. To this end, we first provide the gradient of $\wt{\psi}_k$ on the corresponding manifolds in the proposition below.

\begin{proposition}\label{prop:deri2}
	Under the SVD constraint \eqref{SVDc}, the  
 orthonormal vectors ${\v}_i$ with $1 \leq i \leq r^*$ belong to the Stiefel manifold $\operatorname{St}(1,q) = \{\v \in \mathbb{R}^q : \v\trans\v = 1   \}$.
 The gradient of $\wt{\psi}_k$ on the manifold is $\Q\big(\der{\wt{\psi}_k}{\boldeta_k}\big)$, where $\Q = \diag{\I_q - \v_1\v_1\trans, \dots, \I_q - \v_{r^*}\v_{r^*}\trans, \I_{p(r^* - 1)}}$  and $\der{\wt{\psi}_k}{\boldeta_k}$ is the regular derivative vector on the Euclidean space.
\end{proposition}

In light of Proposition \ref{prop:deri2} above, under the SVD constraint we can make $\wt{\psi}_k$ approximately insensitive to $\v_i$ by requiring that $\Q\big(\der{\wt{\psi}_k}{\boldeta_k}\big)$ be asymptotically vanishing. 
Based on this result, the proposition below provides a convenient choice of matrix $\M$ for SOFARI$_s$.

\begin{proposition}\label{prop:rankr2}
	When the construction of $\M$ is given by 
	\begin{align*}
		&\M^v_k =  -z_{kk}^{-1}\wh{\bSigma}\C_{-k}, \ \ \M_i^v = \0, \ \ \M_i^u = \0 \ \   \text{ for } 1 \leq i \leq r^* \ \text{and} \ i \neq k 
	\end{align*}
	with $\C_{-k} = \sum_{\substack{i \neq k}}\u_i\v_i\trans$ and $z_{kk} = \u_k\trans\wh{\bSigma}\u_k$, it holds that
	\begin{align}
		\left(\der{\wt{\psi}_k}{\boldeta_k\trans}\right)\Q = \left(\derr{L}{\u_k}{\boldeta_k\trans} - \M\derr{L}{\boldeta_k}{\boldeta_k\trans}\right)\Q
		= [\0_{p \times q(k-1) }, \bDelta, \0_{p \times [q(r^* - k) + p(r^*-1) ]}], \nonumber
	\end{align}
	where $\bDelta = \left\{\wh{\bSigma}(\C-\C^*) - n^{-1}\X\trans\E\right\} (\I_q - \v_k\v_k\trans)$.
	
\end{proposition}

\blue{Based on the construction of $\M$, after plugging in consistent SOFAR initial estimates $\wh{\C}$, we can show that $\bDelta$ and thus $\left(\der{\wt{\psi}_k}{\boldeta_k\trans}\right)\Q$ will be asymptotically vanishing. Then the modified score function $\wt{\psi}_k$ is locally insensitive to the nuisance parameters.} We next discuss the corresponding construction of matrix $\W$. The following definition is necessary to facilitate our theoretical presentation.

\begin{definition}[Approximate Inverse]\label{defi2:acceptable}
A $p\times p$ matrix $\wh{\bTheta} = (\wh{\btheta}_1,\ldots,\wh{\btheta}_p)\trans$ is called an approximate inverse matrix of $\wh{\bSigma}$ if there exists some positive constant $C$ such that 1) $\norm{\I - \wh{\bTheta}\wh{\bSigma}}_{\max}\leq C\sqrt{(\log p)/n}$ and 
	2) $\max_{1\leq i\leq p}\norm{\wh{\btheta}_i}_{0} \leq s_{\max}$ and $\max_{1\leq i\leq p}\norm{\wh{\btheta}_i}_{2} \leq C$. 
\end{definition}
Definition \ref{defi2:acceptable} above requires mainly that the approximate inverse matrix $\wh{\bTheta}$ satisfies an entrywise approximation error bound of rate $\sqrt{(\log p)/n}$ and a rowwise sparsity level $s_{\max}$ with the length of each row bounded from above. These are typical properties for high-dimensional precision matrix estimation and can be achieved by many existing approaches, such as the nodewise Lasso estimate \citep{meinshausen2006high} and ISEE \citep{fan2016innovated}. Proposition \ref{prop:rankr3} specifies our construction of matrix $\W$ for the second step.

\begin{proposition}\label{prop:rankr3}
	When $\I_{r^*-1} -z_{kk}^{-1}{\U}_{-k}\trans\wh{\bSigma} {\U}_{-k}$ is nonsingular and	
	\[\W = \widehat{\bTheta} \left\{  \I_p +   z_{kk}^{-1}\wh{\bSigma}{\U}_{-k}(\I_{r^*-1} -z_{kk}^{-1}{\U}_{-k}\trans\wh{\bSigma} {\U}_{-k})^{-1}{\U}_{-k}\trans\right\}  \]
	with $\U_{-k} = [\u_1, \ldots, \u_{k-1}, \u_{k+1}, \ldots, \u_{r^*}]$ and $\widehat{\bTheta}$ an approximate inverse of $\wh{\bSigma}$, for $\M$ constructed in Proposition \ref{prop:rankr2} it holds that
	$$\W( \I_p - \M_k^v\v_k\u_k\trans + \M_k^v\C_{-k}\trans )\wh{\bSigma} = \widehat{\bTheta}\wh{\bSigma}.$$
\end{proposition}

Now let us gain some insight into the constructions of matrices $\M$ and $\W$  in Propositions \ref{prop:rankr2} and \ref{prop:rankr3}, respectively. First, based on Proposition \ref{prop:rankr2}, as long as an estimate $\widetilde{\C}$ for $\C^*$ with consistent SVD components  $(\widetilde{\u}_i, \widetilde{\v}_i)_{i = 1}^{r^*}$ is available, we can plug them into matrix $\M$ to form $\wt{\mathbf{M}}^{(k)}$ 
so that $\wt{\psi}_k$ with $\wt{\mathbf{M}}^{(k)}$ is approximately insensitive to the nuisance parameter vector $\boldeta_k$ when $\v_1, \ldots, \v_{r^*}$ are constrained to be on the corresponding manifolds.

Next, it can be obtained from  Lemma \ref{prop:rankr1} in Section \ref{sec:proof:l2} of the Supplementary Material that the main term in the debias estimator 
takes the form of
\begin{align}\label{ubias-plug-in}
 \u_k - \W \wt{\psi}_k(\u_k,{\boldeta_k}^*)  =  & \ \u_k^* + \left[\I_p - \W (\I_p - \M_k^v\v_k\u_k\trans + \M_k^v\C_{-k}\trans )\wh{\bSigma}\right]({\u}_k - \u_k^*) \\
 & ~ -  \W \Big[ \sum_{\substack{j}}\M_j^v \C_{-j}\strans\wh{\bSigma} \u_j^* +\bdelta_k +  \bepsilon_k \Big].  \nonumber
\end{align}
\blue{The third term above consists of three components each multiplied by matrix $\W$, where $\sum_{\substack{j}}\M_j^v \C_{-j}\strans\wh{\bSigma} \u_j^*$ is the intrinsic bias term, $\bdelta_k$ is an error term that vanishes asymptotically when plugging in consistent SOFAR estimates, and $\bepsilon_k$ is the distribution term.}
In addition, a valid matrix $\W$ should be an approximate inverse (cf. Definition \ref{defi2:acceptable}) of the matrix $( \I_p - \M_k^v\v_k\u_k\trans + \M_k^v\C_{-k}\trans )\wh{\bSigma}$, so that the bias term above can be smaller than the root-$n$ order.
Proposition \ref{prop:rankr3} gives an explicit construction of such $\W$ for SOFARI$_s$,
where its component $ \I_p +   z_{kk}^{-1}\wh{\bSigma}{\U}_{-k}(\I_{r^*-1} -z_{kk}^{-1}{\U}_{-k}\trans\wh{\bSigma} {\U}_{-k})^{-1}{\U}_{-k}\trans$ is indeed the inverse of $\I_p - \M_k^v\v_k\u_k\trans + \M_k^v\C_{-k}\trans $.

By calculating the initial SOFAR estimate with SVD components $(\widetilde{\u}_i, \widetilde{\v}_i)_{i = 1}^{r^*}$, our debiased estimate for $\u_k^*$ is defined as
\begin{align}\label{debes}
	\wh{\u}_k & = \widetilde{\u}_k - \wt{\W}_k\wt{\psi}_k(\widetilde{\u}_k,\widetilde{\boldeta}_k) 
= \widetilde{\u}_k - \wt{\W}_k\Big(\der{L}{\u_k} - \wt{\mathbf{M}}^{(k)}\der{L}{\boldeta_k}\Big)\Big|_{(\widetilde{\u}_k,\widetilde{\boldeta}_k)},
\end{align}
where $\wt{\mathbf{M}}^{(k)}=\big[\0_{p \times q(k-1) }, \wt{\mathbf{M}}_{k}, \0_{p \times [q(r^* - k) + p(r^*-1) ]}  \big]$ with $\wt{\M}_{k} = -\wt{z}_{kk}^{-1}\wh{\bSigma}\wt{\C}_{-k}$ and $\widetilde{z}_{kk} = \widetilde{\u}_k\trans\wh{\bSigma}\widetilde{\u}_k$, and $\wt{\W}_k$ are as given in Propositions \ref{prop:rankr2} and \ref{prop:rankr3} after plugging in $(\widetilde{\u}_i, \widetilde{\v}_i)_{i = 1}^{r^*}$. Since the constructions of $\{\wh{\u}_k\}_{k = 1}^{r^*}$
do not rely on each other, we can calculate the SOFARI$_s$ statistics for different $\u_k^*$'s simultaneously through parallel computing for large-scale applications. 


\subsection{
SOFARI under weakly orthogonal factors} \label{new.Sec.3.2}

The SOFARI$_s$ procedure in Section \ref{new.Sec.3.1} deals with the setting of strongly orthogonal latent factors satisfying { $\sum_{j \neq k} |\u_j\strans\wh{\bSigma}\u_k^*| = o(n^{-1/2})$.} 
When the latent factors correlations do not vanish faster than a root-$n$ rate, SOFARI$_s$ may not work since the intrinsic bias caused on the stronger correlations can invalidate the asymptotic distribution. To address this, we next propose the general SOFARI inference procedure which is applicable to a broader range of multi-response applications under a weaker assumption on the latent factor correlation. 

Different from SOFARI$_s$ that considers all unknown parameters in the constrained least-squares loss function \eqref{SVDc}, 
the general SOFARI works by removing the top $k - 1$ layers from the response matrix via subtracting their estimates when making inference on $\u_k^*$. After removing the previous layers, the intrinsic bias can be controlled when the magnitude of the singular value corresponding to the current layer dominates the remaining ones even when the latent factors are \textit{not} strongly orthogonal to each other. 

Let us now describe the construction of the debiased estimate of $\u_k^*$ in SOFARI for each given $k$ with $1 \leq k \leq r^*$. 
Based on the SOFAR estimates $\wt{\u}_i$ of $\u_i^*$ and $\wt{\v}_i$ of $\v_i^*$, we have the surrogate for $\C^{*(1)} = \sum_{i = 1}^{k-1}\u_i^*\v_i\strans$ as
$\wh{\C}^{(1)} = \sum_{i=1}^{k-1} \wt{\u}_i \wt{\v}_i\trans.$
When $k = 1$, we define $\wh{\C}^{(1)} = \0$ since there is no previous layer to be removed. By subtracting the surrogate $\X\wh{\C}^{(1)}$ of the previous $k - 1$ layers from response matrix $\Y$, the constrained least-squares loss function associated with our inference problem takes the form 
\begin{align}
  &  L(\u_k,\boldeta_k) = (2n)^{-1}\norm{\Y - \X\wh{\C}^{(1)} - \sum_{i = k}^{r^*} \X\u_i\v_i\trans}_F^2, \nonumber \\
&\text{subject to } ~\u_i\trans\u_j = 0 ~ \text{for} ~ k\leq i \neq j\leq r^*  ~ \text{and} ~ (\widetilde{\V}^{(k)})\trans\widetilde{\V}^{(k)} = \I_{r^*}, \label{lossr}
\end{align}
where $\boldeta_k = \left(\u_{k+1}\trans, \ldots, \u_{r^*}\trans, \v_{k}\trans,  \ldots, \v_{r^*}\trans\right)\trans$ is the remaining nuisance parameter vector and  $\widetilde{\V}^{(k)} = \left[\wt{\v}_1, \ldots, \wt{\v}_{k-1}, \v_k, \ldots, \v_{r^*}\right]$ is the matrix of right singular vectors with the first $k - 1$ columns replaced by SOFAR estimates.

Then the vector of modified score functions for $\u_k$ can be defined similarly as
\begin{align*}
    \wt{\psi}_k(\u_k,\boldeta_k) = \der{L}{\u_k} - \M\der{L}{\boldeta_k},
\end{align*}
where $\mathbf{M}=\left[\mathbf{M}_{k}^{v}, \ldots, \mathbf{M}_{r^*}^{v}, \mathbf{M}_{k+1}^{u}, \ldots,  \mathbf{M}_{r^*}^{u}\right]$ with $\mathbf{M}_{i}^{u} \in\R^{p\times p}$ for each $k+1 \leq i \leq r^*$ and $\mathbf{M}_{j}^{v} \in\R^{p\times q}$ for each $k \leq j  \leq r^*$. \blue{Similarly, we will show that the most important part of construction of matrix $\mathbf{M}$ lies in $\M^v_k$.} 
Based on $\wt{\psi}_k$, we also suggest the debiased function 
\begin{align*}
    \psi_k(\u_k,\boldeta_k) = \u_k - \W\wt{\psi}_k(\u_k,\boldeta_k).
\end{align*}

Denote by $\C^{*(2)} = \sum_{i = k+1}^{r^*}\u_i^*\v_i\strans$, $\C^{*(2)}_{-j} = \sum_{i = k+1, i \neq j}^{r^*}\u_i^*\v_i\strans$, ${\C}^{(2)} = \sum_{i=k+1}^{r^*} {\u}_i {\v}_i\trans$, $\U^{(2)} = [\u_{k+1}, \ldots, \u_{r^*}]$, and $ \Q = \diag{  \mathbf{I}_{q}-\boldsymbol{v}_{k} \boldsymbol{v}_{k}^{T}, \ldots,  \mathbf{I}_{q}-\boldsymbol{v}_{r^*} \boldsymbol{v}_{r^*}^{T}, \I_{p(r^*-k)}  }$. The two propositions below play similar roles as Propositions \ref{prop:rankr2} and \ref{prop:rankr3} for our SOFARI inference procedure. 


%
%

\begin{proposition}\label{prop:rankapo2}
When the construction of $\M$ is given by 
\begin{align*}
	&\M^v_k =  -z_{kk}^{-1}\wh{\bSigma}\C^{(2)}, \ \ \M_i^v = \0, \ \ \M_i^u = \0 \  \text{ for } k+1 \leq i \leq r^*, 
\end{align*}
it holds that
\begin{align*}
	\left(\der{\wt{\psi}_k}{\boldeta_k\trans}\right)\Q = \left(\derr{L}{\u_k}{\boldeta_k\trans} - \M \derr{L}{\boldeta_k}{\boldeta_k\trans} \right)\Q
	= [\bDelta, \0_{p \times ( p+ q)(r^*-k)}],
\end{align*}
where $\bDelta = \left\{ \wh{\bSigma}\big(\wh{\C}^{(1)}-\C^{*(1)} + \sum_{i= k}^{r^*} (\u_i\v_i\trans - \u_i^*\v_i\strans) \big) - n^{-1}\X\trans\E\right\} (\I_q - \v_k\v_k\trans)$.

\end{proposition}

\begin{proposition}\label{prop:rankapo3}
	When $\I_{r^*-k} -z_{kk}^{-1}(\U^{(2)})\trans\wh{\bSigma} \U^{(2)}$ is nonsingular and
\[   \W = \widehat{\bTheta} \left\{  \I_p +   z_{kk}^{-1}\wh{\bSigma}\U^{(2)}\left(\I_{r^*-k} -z_{kk}^{-1}(\U^{(2)})\trans\wh{\bSigma} \U^{(2)}\right)^{-1}(\U^{(2)})\trans\right\} \]
with $\widehat{\bTheta}$ an approximate inverse of $\wh{\bSigma}$, for $\M$  constructed in Proposition \ref{prop:rankapo2} it holds that
\[  \W  \left( \I_p - \M_k^v\v_k\u_k\trans + \M_k^v (\C^{(2)})\trans \right)\wh{\bSigma}  = \widehat{\bTheta}\wh{\bSigma}.\]
\end{proposition}

 Constructions of matrices $\M$ and $\W$ for SOFARI are provided in Propositions \ref{prop:rankapo2} and \ref{prop:rankapo3}, respectively. Compared to those for SOFARI$_s$ in Propositions \ref{prop:rankr2} and \ref{prop:rankr3}, matrices $\M$ and 
$\W$ here are similar but no longer take the first $k - 1$ layers into account since they have been removed from the response matrix by subtracting their surrogates. \blue{Similar to Proposition 2, $\bDelta$ in Proposition 4 is asymptotically vanishing after plugging in consistent SOFAR initial estimates.}
Then the new debiased SOFARI estimate for $\u_k^*$ is constructed as 
\begin{align*}
    \wh{\u}_k = \psi_k(\widetilde{\u}_k,\widetilde{\boldeta}_k) = \widetilde{\u}_k - {\color{black}\wt{\W}_k}\wt{\psi}_k(\widetilde{\u}_k,\widetilde{\boldeta}_k) = \widetilde{\u}_k - \wt{\W}_k\Big(\der{L}{\u_k} - \wt{\M}^{(k)}\der{L}{\boldeta_k}\Big)\Big|_{(\widetilde{\u}_k,\widetilde{\boldeta}_k)},
\end{align*}
where $\wt{\mathbf{M}}^{(k)}=\big[ \wt{\mathbf{M}}_{k}, \mathbf{0}_{p \times  (p+ q) (r^*-k)}  \big]$ with $\wt{\M}_{k} = -\wt{z}_{kk}^{-1}\wh{\bSigma}\wt{\C}^{(2)}$, and $\wt{\W}_k$ are defined as in Propositions \ref{prop:rankapo2} and \ref{prop:rankapo3} after plugging in the SOFAR estimated SVD components $(\widetilde{\u}_i, \widetilde{\v}_i)_{i = k}^{r^*}$.

Given that $\M_i^v = \0$ for $k + 1 \leq i \leq r^*$ by Proposition \ref{prop:rankapo2}, the corresponding intrinsic bias term in $\wt{\psi}_k(\u_k^*,\boldeta^*_{k})$ would become $\M_k^v (\C^{*(2)})\trans \wh{\bSigma} \u_k^*$ in view of Lemma \ref{prop:rankapo1} in Section \ref{sec:proof:l1} of the Supplementary Material. 
After plugging consistent SOFAR estimates into $\M_k^v$, we can show that
\begin{align}\label{laybias2}
\|\M_k^v (\C^{*(2)})\trans \wh{\bSigma} \u_k^*\|_{\infty}  \asymp \sum_{j = k+1}^{r^*} (d_j^{*2}/d_k^*) |\l_j\strans\wh{\bSigma}\l_k^*|,
\end{align}
where the right-hand side will be assumed to take the order of $o(n^{-1/2})$ so that the intrinsic bias is under control.
Moreover, it is clear that both the gaps between singular values and the correlations between latent factors (as measured by $\l_j\strans\wh{\bSigma}\l_k^*$) will play key roles in determining the magnitude of the intrinsic bias.
When the gaps between nonzero singular values are large enough or the singular values decay dramatically such as in the spiked eigen-structure, 
we do not necessarily require the strong orthogonal latent factor constraint that $\sum_{j \neq k} |\l_j\strans\wh{\bSigma}\l_k^*| = o(n^{-1/2})$, but allow the latent factors to be somewhat correlated. See Conditions \ref{con:nearlyorth} and \ref{con:orth:rankr} formally summarizing these two scenarios in the next section. 
In this regard, SOFARI is indeed applicable to a wider range of applications compared to SOFARI$_s$.

\section{Asymptotic properties of SOFARI} \label{new.Sec.4}

In this section, we will establish the asymptotic distributions for both  SOFARI\texorpdfstring{$_s$}{s} (the basic form) and SOFARI (the general form)  suggested in Section \ref{new.Sec.3} that correspond to the settings of strongly orthogonal factors and  weakly orthogonal factors, respectively. 

\subsection{Technical conditions} \label{new.Sec.4.1}


%

To facilitate the technical analysis, we will need to introduce some regularity conditions. To do so, we first provide the following definition to characterize the properties of SOFAR SVD estimates. Denote by $s_u = \norm{\U^*}_0$ and $s_v = \norm{\V^*}_0$.

\begin{definition}[SOFAR SVD estimates]\label{lemmsofar}
	A $p \times q$ matrix  $\wt{\C}$ with SVD components $ (\widetilde{\L}, \wt{\D},  \widetilde{\V})$ is called an acceptable estimator of matrix $\C^*$ if it satisfies that with probability at least $1 - \theta_{n,p,q}^{\prime}$ for some asymptotically vanishing $\theta_{n,p,q}^{\prime}$, we have the estimation error bounds 
	\begin{align*}
		& (a) ~ \norm{\wt{\D}-\D^*}_F
		+ \norm{\wt{\U} - \U^*}_F
		+ \norm{\wt{\V}\wt{\D} - \V^*\D^*}_F \leq  c \gamma_n, \\
		& (b)~
		 \norm{\wt{\D}-\D^*}_0
		 + \norm{\wt{\U} - \U^*}_0
		 + \norm{\wt{\V}\wt{\D} - \V^*\D^*}_0
		\leq (r^*+s_u +s_v)[1+o(1)], 
 	\end{align*}
	where $\wt \U$ is defined analogously to $\U^*$ as $\wt\U = \wt\L\wt\D$,   $ \gamma_n = ({r^*}+s_u+s_v)^{1/2}\eta_n^2\{n^{-1}\log(pq)\}^{1/2}$, $\eta_{n}=1+\delta^{-1 / 2} \left(\sum_{j=1}^{r^*}(d_{1}^{*} / d_{j}^{*})^{2} \right)^{1 / 2}$, and $c$ and $\delta$ are some positive constants. 
\end{definition}

Definition \ref{lemmsofar} above lists the properties of the SOFAR SVD estimates, which can be ensured by Theorem 2 of \cite{uematsu2017sofar} under some regularity conditions. Since rank $r^*$ is assumed to be given, the SVD components $ (\widetilde{\L}, \wt{\D},  \widetilde{\V})$ are of the same dimensions as their population counterparts. The tail probability $\theta_{n,p,q}^{\prime}$ has been shown to decay polynomially in feature dimensionality $p$. Although some other factor regression methods \citep{mishra2017,chen2022stagewise} can also accurately recover the signals in each layer, they may not be suitable as the initial estimates for the suggested SOFARI inference procedure since the exact orthogonality is generally not enforced precisely.
Specifically, we make the technical assumptions below. 


\begin{condition}\label{cone}
The 
error matrix $\E\sim\N(\0,\I_n\otimes\bSigma_e)$ 
    with the maximum eigenvalue of $\bSigma_e$ bounded from above.
\end{condition}

\begin{condition}\label{con3}
	There exist some sparsity level $s \geq  \max\{s_{\max}, 3(r^* + s_u + s_v)\}$  with $s_{\max}$ defined in Definition \ref{defi2:acceptable}, and positive constants $\rho_l$ and $\rho_u$  such that
 \begin{align*}
 \rho_l < \min_{\bdelta\in\R^p}\left\{ \frac{\norm{\wh{\bSigma}\bdelta}_2}{\norm{\bdelta}_2} : \norm{\bdelta}_0\leq s \right\} \leq \max_{\bdelta\in\R^p}\left\{ \frac{\norm{\wh{\bSigma}\bdelta}_2}{\norm{\bdelta}_2} : \norm{\bdelta}_0\leq s \right\} < \rho_u.
\end{align*}
\end{condition}

\begin{condition}\label{con4}
	The nonzero eigenvalues $d^{*2}_{i}$ of matrix $\C\strans\C^*$ satisfy that $d^{*2}_{i}  - d^{*2}_{i+1} \geq  \delta_1 d^{*2}_{i}$
	for some positive constant $\delta_1 > 1 - \rho_l/\rho_u$ with $1 \leq i \leq r^*$ and $r^* \gamma_n = o(d^{*}_{r^*})$.
\end{condition}



\begin{condition}[Strong orthogonality]\label{con:nearlyorth}
	The nonzero squared singular values $d^{*2}_{i}$ are at the constant level and $\sum_{j \neq k} |\l_j\strans\wh{\bSigma}\l_k^*| = o(n^{-1/2})$ for each given $k$ with $1 \leq k \leq r^*$. 
\end{condition}

\begin{condition}[Weak orthogonality]\label{con:orth:rankr}
	The nonzero squared singular values $d^{*2}_{i}$ and the latent factors jointly satisfy that $\sum_{j = k+1}^{r^*} (d_j^{*2}/d_k^*) |\l_j\strans \wh{\bSigma} \l_k^*| = o(n^{-1/2})$ for each $k$, $1 \leq k < r^*$.  
\end{condition}

Similar to \cite{javanmard2014confidence, zhang2014confidence}, the Gaussian assumption in Condition \ref{cone} above is imposed to simplify the technical analysis. Our theoretical results can be extended to non-Gaussian errors using a similar central limit theorem argument to that in \cite{van2014asymptotically}. Condition \ref{con3} assumes that the $s$-sparse eigenvalues of $\wh{\bSigma}$ are bounded, which is also imposed in \cite{uematsu2017sofar,zheng2019scalable} to ensure the identifiability of significant features. The corresponding sparsity level $s$ has been shown to be at least of order $O\{n /(\log p)\}$ with asymptotic probability one based on the concept of the robust spark in \cite{lv2013impacts} when the rows of design matrix $\X$ are sampled independently from the multivariate 
elliptical distributions.

Condition \ref{con4} requires distinction between the nonzero singular values so that different latent factors are separable. It also assumes $d^{*}_{r^*}$ to be larger than the asymptotically vanishing rate $r^*\gamma_n$ with  convergence rate $\gamma_n$ given in Definition \ref{lemmsofar}. Such assumptions are standard in factor-based regressions \citep{uematsu2017sofar,zheng2019scalable}. The lower bound on $\delta_1$ is imposed such that $\rho_l d^{*2}_{i} > \rho_u d^{*2}_{i + 1}$ to avoid possible equivalence in ${z}_{ii}^{*} = \u_i\strans\wh{\bSigma}\u_i^*$ for different $i$'s, where $z_{ii}^*$ can be understood as the strengths of latent factors from different layers. Moreover, it can guarantee the existence of matrix $\W$ suggested in Propositions \ref{prop:rankr3} and \ref{prop:rankapo3} by meeting the requirement on matrix nonsingularity with a strictly diagonal dominance argument.

In particular, Conditions \ref{cone}--\ref{con4} will be exploited for the theoretical analyses of both SOFARI$_s$ and SOFARI. Further, Conditions \ref{con:nearlyorth} and \ref{con:orth:rankr} correspond to the settings of strong orthogonality and weak orthogonality between latent factors, respectively. \blue{The singular values are assumed to be bounded in Condition \ref{con:nearlyorth} mainly for technical simplicity, and they can indeed be diverging as long as the singular values and latent factors jointly satisfy a similar bound to that in Condition \ref{con:orth:rankr}.} It is clear that if Condition \ref{con:nearlyorth} holds, Condition \ref{con:orth:rankr} will be satisfied automatically. Thus, the requirement on correlations between latent factors for SOFARI is indeed weaker than that for SOFARI$_s$. 
We present the theoretical guarantees for these two versions of the SOFARI inference procedure in Sections \ref{new.Sec.4.2} and \ref{new.Sec.4.3}, respectively.

\subsection{Asymptotic theory of SOFARI\texorpdfstring{$_s$}{s}  } \label{new.Sec.4.2}

Let us denote by $\wt{\U}  = [ \wt{\u}_{1}, \ldots, \wt{\u}_{r^*}]$, $\wt{\D}  = \diag{ \wt{d}_{1}, \ldots, \wt{d}_{r^*}}$, and
$\wt{\V}  = [ \wt{\v}_{1},  \ldots, \wt{\v}_{r^*}]$ the SOFAR SVD components given in Definition \ref{lemmsofar}.
For each given $k$ with $1 \leq k \leq r^*$, recall that $z_{kk}^{*} = {\u}_k\strans\wh{\bSigma}{\u}_k^{*}$ and its estimate $\wt{z}_{kk}^{} = \wt{\u}_k\trans\wh{\bSigma}\wt{\u}_k$. 
Corresponding to $\M$ and $\W$ suggested for SOFARI$_s$ in Section \ref{new.Sec.3.1}, we define $\M_{k}^{*} = -z_{kk}^{*-1}\wh{\bSigma}\C_{-k}^*$, $\wt{\M}_{k} = -\wt{z}_{kk}^{-1}\wh{\bSigma}\wt{\C}_{-k}$, and 
\begin{align*}
		&\W_k^{*} = \widehat{\bTheta} \left\{\I_p + z_{kk}^{*-1}\wh{\bSigma}\U_{-k}^*(\I_{r^*-1} -z_{kk}^{*-1}\U_{-k}\strans\wh{\bSigma} \U_{-k}^*)^{-1}\U_{-k}\strans\right\}, \\
        &\wt{\W}_k  =   \widehat{\bTheta} \left\{  \I_p +   \wt{z}_{kk}^{-1}\wh{\bSigma}\wt{\U}_{-k}(\I_{r^*-1} -\wt{z}_{kk}^{-1}\wt{\U}_{-k}\trans\wh{\bSigma} \wt{\U}_{-k})^{-1}\wt{\U}_{-k}\trans\right\}.
\end{align*}
Observe that here, $\C^*_{-k} = \U_{-k}^*\V_{-k}\strans$ and $\wt{\C}_{-k} = \widetilde{\U}_{-k}\widetilde{\V}_{-k}\trans$ with $\U_{-k}^*, \V_{-k}^*, \widetilde{\U}_{-k}$, and $\widetilde{\V}_{-k}$ the corresponding submatrices after taking off the $k$th columns. 
Furthermore, let us define 
\begin{align} \label{kappa}
\kappa_n = \max\{s_{\max}^{1/2} , (r^*+s_u+s_v)^{1/2}, \eta_n^2\} (r^*+s_u+s_v)\eta_n^2\log(pq)/\sqrt{n},
\end{align}
which will be the key order for the error term. With $\wt\M_k$ and $\wt\W_k$, the theorem below provides the asymptotic distribution of the proposed estimator $\wh{\u}_k$ in \eqref{debes} for SOFARI$_s$.

\begin{theorem}[Inference on $\u_k^*$]\label{theorkr}
	Assume that Conditions \ref{cone}--\ref{con:nearlyorth} hold and $\wh{\bTheta}$ and $\wt{\C}$ satisfy Definitions \ref{defi2:acceptable} and \ref{lemmsofar}, respectively. Then for each given $k$ with $1 \leq k \leq r^*$ and an arbitrary vector $\a\in\mathcal{A}=\{\a\in\R^p:\norm{\a}_0\leq m,\norm{\a}_2=1\}$ satisfying $m^{1/2}\kappa_n = o(1)$, we have 
	\begin{align*}
		\sqrt{n}\a\trans(\wh{\u}_k-\u_k^*) = h_k + t_k, 
	\end{align*}
	where the distribution term $h_k = \a\trans \W^{*}_k(\X\trans\E\v_k^* - \M_{k}^{*} \E\trans\X {\u}_k^{*})/\sqrt{n} \sim \N(0,\nu_k^2)$ with 
	\begin{align*}
    \nu_k^2 = \a\trans\W_k^*(z_{kk}^{*}\M_{k}^{*}\bSigma_e\M_{k}\strans + \v_k\strans\bSigma_e\v_k^* \wh{\bSigma} - 2\wh{\bSigma}\u_k^*  \v_k\strans\bSigma_e\M_{k}^{* T})\W_k\strans\a.
	\end{align*}
    Moreover, the bias term $t_k = O_p(m^{1/2}\kappa_n)$ holds with probability at least $1 - \theta_{n,p,q}$,
	where 
		\begin{equation}\label{thetapro}
	    \theta_{n,p,q} = \theta_{n,p,q}^{\prime} +  2(pq)^{1-c_0^2/2}
	\end{equation}
    with $\theta_{n,p,q}^{\prime}$ given in Definition \ref{lemmsofar} and some constant $c_0 > \sqrt{2}$.
\end{theorem}

Theorem \ref{theorkr} above establishes the inference results for each important latent left factor vector $\u_k^*$ 
so that the tools of hypothesis testing and confidence interval on the composition of features in each latent factor are now available. Under the sparse and low-rank settings, $\max\{s_{\max}^{1/2} , (r^*+s_u+s_v)^{1/2}, \eta_n^2\}$ is relatively small so that the main requirement for the validity of the asymptotic normal distribution is $\sqrt{n} \gg \log(pq)$ in view of $m^{1/2}\kappa_n = o(1)$, which is similar to the standard constraint $\sqrt{n} \gg \log p$ for the inference of univariate response regressions \citep{javanmard2014confidence, van2014asymptotically, zhang2014confidence}. However, the inference of $\u_k^*$ here is much more challenging than that of the univariate response regression coefficient vector since we need to deal with not only the intertwined nuisance parameter vector $\v_k^*$ in the same layer, but also those unknown singular vectors from the other important layers. Under the structural constraints including the orthogonality and unit lengths on the singular vectors, our manifold-based technical arguments exploit the geodesic and the Taylor expansion on the tangent space to control the error term; 
the approximation errors caused by the estimates $\wt{\u}_{i}$ and $\wt{\v}_{i}$ of the nuisance parameter vectors  yield the secondary term $\max\{s_{\max}^{1/2} , (r^*+s_u+s_v)^{1/2}, \eta_n^2\}$ in $\kappa_n$ for the overall errors.


Besides inference for the latent left factors,  inference on the singular values is also meaningful as they measure signal strengths of latent factors on the response vector. 
Since $d_k^{*2}$ corresponds to the $k$th eigenvalue of matrix $\C\strans\C^*$, we aim to make statistical inference on these nonzero squared singular values. To this end, we will make use of the relationship that $d_k^{*2} = \|\u_k^*\|_2^2$. Nevertheless, the inference of $d_k^{*2}$ is not straightforward from that of $\u_k^*$ since $d_k^{*2}$ is a quadratic sum of components of $\u_k^*$, and the corresponding components of the debiased estimate $\widehat{\u}_k$ are correlated in an unknown and complicated fashion. 

We will address this issue by constructing a debiased estimate directly from $\|\widetilde{\u}_k\|_2^2$ and derive its asymptotic distribution. 
In a similar spirit to the construction of $\widehat{\u}_k$, by utilizing the modified score function $\wt{\psi}_k$, we define the debiased estimate for $d_k^{*2}$ as
\begin{align} \label{eigenestimate}
\wh{d}_k^2 =  \norm{\wt{\u}_k}_2^2 - 2 \wt{\u}_k\trans \wt{\W}_k\wt{\psi}_k(\wt{\u}_k,\wt{\boldeta}_k).
\end{align}
The theorem below reveals that $\wh{d}_k^2$ introduced in (\ref{eigenestimate}) is valid for the inference of $d_k^{*2}$.
\begin{theorem}[Inference on $d_k^{*2}$]\label{coro:ortho:dk}
Assume that all the conditions of Theorem \ref{theorkr} are satisfied and $(r^*+s_u+s_v)^{1/2}\kappa_n = o(1)$. 
	Then for each $k$ with $1 \leq k \leq r^*$, we have 
	\begin{align*}
		\sqrt{n}( \wh{d}_k^2 - d_k^{*2} )
		 = h_{d_k} + t_{d_k},  
	\end{align*}
    where the distribution term $h_{d_k} = 2\u_k\strans\W_k^*(\X\trans\E\v_k^* - \M_{k}^{*} \E\trans\X {\u}_k^{*})/ \sqrt{n} \sim \N(0,\nu_{d_k}^2)$ with 
	\begin{align*}
    \nu_{d_k}^2 = 4\u_k\strans\W_k^*(z_{kk}^{*}\M_{k}^* \bSigma_e \M_{k}^{* T} + \v_k\strans\bSigma_e\v_k^* \wh{\bSigma} - 2\wh{\bSigma} \u_k^*\v_k\strans\bSigma_e\M_{k}^{* T}) \W_k\strans\u_k^*.
	\end{align*}
	Moreover, the error term $t_{d_k} = O_p\{(r^*+s_u+s_v)^{1/2}\kappa_n\}$ holds with probability at least $1 - \theta_{n,p,q}$ with $\theta_{n,p,q}$ given in \eqref{thetapro}.
\end{theorem}
Theorem \ref{coro:ortho:dk} above provides the asymptotic distributions for the nonzero squared singular values $d_k^{*2}$ with $1 \leq k \leq r^*$ so that the significance levels of the latent factors can be inferred. Our proof indeed shows that $\wh{d}_k^2 - d_k^{*2}$ and $2\u_k\strans (\wh{\u}_k - \u_k^*)$ have the same asymptotic distribution.
Then applying a similar argument to that for proving Theorem \ref{theorkr}, the distribution term $h_{d_k}$ here can be obtained by replacing $\a$ with $2\u_k^*$ in $h_{k}$. Consequently, the validity requirement needs to be strengthened to $(r^*+s_u+s_v)^{1/2}\kappa_n = o(1)$ in comparison to $m^{1/2}\kappa_n = o(1)$ (cf. Theorem \ref{theorkr}). Under such assumptions, the error term $t_{d_k}$ will converge to zero with the same probability bound $1 - \theta_{n,p,q}$ as in Theorem \ref{theorkr}. 

{
Since the population variances $\nu_{k}^2$ and $\nu_{d_k}^2$ presented in Theorems \ref{theorkr} and \ref{coro:ortho:dk} are unknown in practice, we can use some consistent estimate of the error covariance matrix $\bSigma_e$ along with the initial SOFAR estimates to obtain their surrogates. 
The following definition characterizes the desired property for the estimate $\wt{\bSigma}_e$ of $\bSigma_e$.
\begin{definition}\label{defi:error}
    A $q \times q$ matrix $\wt{\bSigma}_e$ is an acceptable estimator of ${\bSigma}_e$ if $\norm{\wt{\bSigma}_e - {\bSigma}_e}_2 = o_p(1). $
\end{definition}

The definition above only requires the estimation consistency of $\wt{\bSigma}_e$.
In practice, the error covariance matrix $\bSigma_e$ can be estimated by first obtaining the residual matrix estimator $\widetilde{\mathbf{E}}$ of $\mathbf{E}$ from the SOFAR regression and then recovering the error covariance matrix via some existing covariance estimation techniques such as the hard-thresholding \citep{bickel2008covariance} or adaptive thresholding \citep{cai2011adaptive}.
Based on $\wt{\bSigma}_e$ and the initial SOFAR estimates, we can define 
}
\begin{align}
    \wt{\nu}_k^2 &= \a\trans \wt{\W}_k (\wt{z}_{kk} \wt{\M}_{k} \wt{\bSigma}_e \wt{\M}_{k}\trans + \wt{\v}_k\trans \wt{\bSigma}_e\wt{\v}_k \wh{\bSigma} - 2\wh{\bSigma} \wt{\u}_k \wt{\v}_k\trans \wt{\bSigma}_e \wt{\M}_{k}^{T})\wt{\W}_k\trans\a, \label{nuukcase1}\\
    \wt{\nu}_{d_k}^2 &= 4\wt{\u}_k\trans \wt{\W}_{k}(\wt{z}_{kk} \wt{\M}_{k} \wt{\bSigma}_e \wt{\M}_{k}^{T} + \wt{\v}_k\trans \wt{\bSigma}_e\wt{\v}_k \wh{\bSigma} - 2\wh{\bSigma} \wt{\u}_k \wt{\v}_k\trans \wt{\bSigma}_e \wt{\M}_{k}^{T}) \wt{\W}_k\trans \wt{\u}_k. \label{nudkcase2}
\end{align} 
The theorem below gives the estimation accuracy of variance estimates $\wt{\nu}_k^2$ and $\wt{\nu}_{d_k}^2$ introduced in (\ref{nuukcase1}) and (\ref{nudkcase2}) above, respectively. 
\begin{theorem}[Variance estimation]\label{coro:var:rank2uk}
Assume that all the conditions of Theorem \ref{coro:ortho:dk} are satisfied and $ \wt{\bSigma}_e$ is an acceptable estimator. Then for each $k$ with $1 \leq k \leq r^*$, we have 
	\begin{align*}
		|\wt{\nu}_{k}^2 - {\nu}_{k}^2 | \leq  \widetilde{C} m \gamma_n \ \text{ and } \  
		|\wt{\nu}_{d_k}^2 - {\nu}_{d_k}^2 | \leq  \widetilde{C}(r^*+s_u+s_v) \gamma_n 
	\end{align*}
	hold with probability at least $1- \theta_{n,p,q}$, where $\gamma_n = ({r^*}+s_u+s_v)^{1/2}\eta_n^2\{n^{-1}\log(pq)\}^{1/2}$, $\theta_{n,p,q}$ is given in \eqref{thetapro}, and $\widetilde{C} > 0$ is some constant.

\end{theorem}

\subsection{Asymptotic theory of SOFARI} \label{new.Sec.4.3}

For each given $k$ with $1 \leq k \leq r^*$, related to $\M$ and $\W$ for SOFARI in Section \ref{new.Sec.3.2}, we slightly abuse the notation and redefine $\M_{k}^{*} = -z_{kk}^{*-1}\wh{\bSigma}\C^{*(2)}$, $\wt{\M}_{k} = -\wt{z}_{kk}^{-1}\wh{\bSigma}\wt{\C}^{(2)}$, and
\begin{align*}
		&\W_k^{*} = \widehat{\bTheta} \left\{  \I_p +   z_{kk}^{*-1}\wh{\bSigma}\U^{*(2)}(\I_{r^*-k} -z_{kk}^{-1}(\U^{*(2)})\trans\wh{\bSigma} \U^{*(2)})^{-1}(\U^{*(2)})\trans\right\}, \\
		&\wt{\W}_k  =   \widehat{\bTheta} \left\{  \I_p +   \wt{z}_{kk}^{-1}\wh{\bSigma}\wt{\U}^{(2)}(\I_{r^*-k} -\wt{z}_{kk}^{-1}(\wt{\U}^{(2) })\trans\wh{\bSigma} \wt{\U}^{(2)})^{-1}(\wt{\U}^{(2)})\trans\right\}.
\end{align*}
Recall that here, $\C^{*(2)} = \sum_{i = k+1}^{r^*}\u_i^*\v_i\strans$, $\wt{\C}^{(2)} = \sum_{i = k+1}^{r^*}\widetilde{\u}_i\widetilde{\v}_i\trans$, $\U^{*(2)} = [\u_{k+1}^*, \ldots, \u_{r^*}]$, and $\wt{\U}^{(2)} = [\wt{\u}_{k+1}, \ldots, \wt{\u}_{r^*}]$ contain only the last $r^* - k$ layers since the previous $k - 1$ layers have been removed in the constrained least-squares problem \eqref{lossr}. 
Furthermore, denote by 
\begin{align*}
    \kappa_n^{(k)} =  \kappa_n \max\big\{1, d_k^{*-1}, d_k^{*-2}\big\}  + \gamma_n d_k^{*-3} d_{k+1}^{*} \big(\sum_{i=1}^{k-1}d_i^*\big)
\end{align*}
with $\kappa_n$ given in \eqref{kappa}
and $d_{r^*+1}^* = 0$.
The theorem below guarantees the asymptotic distribution of the proposed estimator $\wh{\u}_k$ in Section \ref{new.Sec.3.2} with the above defined $\wt\W_k$ and $\wt\M_k$, where $\kappa_n^{(k)}$ becomes the key order of the corresponding error term.

\begin{theorem}[Inference on $\u_k^*$]\label{theorkapor}
	Assume that Conditions \ref{cone}--\ref{con4} and \ref{con:orth:rankr}  hold, and $\wh{\bTheta}$, $\wt{\C}$ satisfy Definitions \ref{defi2:acceptable} and \ref{lemmsofar}, respectively. Then for each given $k$ with $1 \leq k \leq r^*$ and an arbitrary vector
	$\a\in\mathcal{A}=\{\a\in\R^p:\norm{\a}_0\leq m,\norm{\a}_2=1\}$ satisfying $m^{1/2}\kappa_n^{(k)} = o(1) $,
	we have
	\begin{align*}
		\sqrt{n}\a\trans(\wh{\u}_k-\u_k^*) = h_{k} + t_{k}, 
	\end{align*}
	where the distribution term $h_{k} = \a\trans \W^{*}_k(\X\trans\E\v_k^* - \M_{k}^{*} \E\trans\X {\u}_k^{*})/\sqrt{n} \sim \N(0,\nu_k^2)$ with
	\begin{align*}
    \nu_k^2 = \a\trans\W_k^*(z_{kk}^{*}\M_{k}^{*}\bSigma_e\M_{k}\strans + \v_k\strans\bSigma_e\v_k^* \wh{\bSigma} - 2\wh{\bSigma}\u_k^*  \v_k\strans\bSigma_e\M_{k}^{* T})\W_k\strans\a.
	\end{align*}
	Moreover, the error term $ t_{k} = O_p(m^{1/2}\kappa_n^{(k)} )$
 holds with probability at least $1 - \theta_{n,p,q}$ with $\theta_{n,p,q}$ given in \eqref{thetapro}.
\end{theorem}

Similar to Theorem \ref{theorkr} for SOFARI$_s$, Theorem \ref{theorkapor} above provides the asymptotic normality of the SOFARI debiased estimator $\wh{\u}_k$ for each $k$ with $1 \leq k \leq r^*$. Besides similar but slightly different definitions for $\M_{k}^{*}$ and $\W_{k}^{*}$, the main distinction between these two theorems lies in the order $\kappa_n^{(k)}$ of the error term. While its first part is similar to $\kappa_n$ since $d_k^{*}$ is generally no smaller than one, $\kappa_n^{(k)}$ above contains an extra part $\gamma_n d_k^{*-3} d_{k+1}^{*} (\sum_{i=1}^{k-1}d_i^*)$. 
This is induced by the additional approximation error when we replace the top $k - 1$ layers with their SOFAR estimates, 
reflected by term $\wt{\M}_k  (\C^{*(1)} - \wh{\C}^{(1)})\trans  \wh{\bSigma} \widetilde{\u}_k$ in $\bdelta$ of Lemma \ref{prop:rankapo1} in Section \ref{sec:proof:l1} of the Supplementary Material. Since the convergence rate $\gamma_n$ of the initial SOFAR estimates can decay polynomially with sample size $n$, assuming $\gamma_n d_k^{*-3} d_{k+1}^{*} (\sum_{i=1}^{k-1}d_i^*) = o(1)$ should be a mild condition on the singular values. In contrast, Theorem \ref{theorkapor} broadens substantially the range of applications for our suggested SOFARI procedure since it relies on the weaker Condition \ref{con:orth:rankr} instead of Condition \ref{con:nearlyorth}.

For the inference on the eigenvalues of matrix $\C\strans\C^*$, the debiased estimates $\wh{d}_k^2$ for $d_k^{*2}$ can be defined similarly as in \eqref{eigenestimate} except for plugging in the corresponding matrices $\wt{\M}_k$ and $\wt{\W}_k$ defined in this section. The theorem below validates the hypothesis testing on $d_k^{*2}$.

\begin{theorem}[Inference on $d_k^{*2}$]\label{theorkukcorok}
Assume that all the conditions of Theorem \ref{theorkapor} are satisfied and  $(r^*+s_u+s_v)^{1/2} \kappa_n^{(k)} = o(1)$.  
Then for each $k$ with $1 \leq k \leq r^*$, we have
	\begin{align*}
		\sqrt{n}( \wh{d}_k^2 -   d_k^{*2} )
		 = h_{d_k} + t_{d_k}, 
	\end{align*}
where the distribution term $h_{d_k} = 2\u_k\strans \W^{*}_k(\X\trans\E\v_k^* - \M_{k}^{*} \E\trans\X {\u}_k^{*})/\sqrt{n} \sim \N(0,\nu_{d_k}^2)$ with
\begin{align*}
    \nu_{d_k}^2 = 4\u_k\strans\W_k^*(z_{kk}^{*}\M_{k}^* \bSigma_e \M_{k}^{* T} + \v_k\strans\bSigma_e\v_k^* \wh{\bSigma} - 2\wh{\bSigma} \u_k^*\v_k\strans\bSigma_e\M_{k}^{* T}) \W_k\strans\u_k^*.
\end{align*}
Moreover, the bias term $t_{d_k} =  O_p\{(r^*+s_u+s_v)^{1/2}  d_k^*\kappa_n^{(k)} \}$ holds with probability at least $1 - \theta_{n,p,q}$ with $\theta_{n,p,q}$ given in \eqref{thetapro}.
\end{theorem}

Finally, similar to \eqref{nuukcase1} and \eqref{nudkcase2},
denote by $ \wt{\nu}_{k}^2$ and $\wt{\nu}_{d_k}^2$ the variance estimates obtained by plugging the initial SOFAR estimates into ${\nu}_{k}^2$ and ${\nu}_{d_k}^2$ in Theorems \ref{theorkapor} and \ref{theorkukcorok}, respectively.  The theorem below provides the estimation accuracy of these two variance estimates. 
\begin{theorem}[Variance estimation]\label{coro:var:rank22uk}
	Assume that all the conditions of Theorem \ref{theorkukcorok} hold and $ \wt{\bSigma}_e$ is an acceptable estimator.
    Then for each $k$ with $1 \leq k \leq r^*$, we have 
\begin{align*}
		&|\wt{\nu}_{k}^2 - {\nu}_{k}^2| \leq \widetilde{C}^{\prime} m  (r^*+s_u+s_v)^{1/2} \eta_n^2\{n^{-1}\log(pq)\}^{1/2} d_k^{*-1}, \\
		&|\wt{\nu}_{d_k}^2 - {\nu}_{d_k}^2| \leq \widetilde{C}^{\prime}  (r^*+s_u+s_v)^{3/2} \eta_n^2\{n^{-1}\log(pq)\}^{1/2}d_k^{*}
\end{align*}
    hold with probability at least $1- \theta_{n,p,q}$, where $\theta_{n,p,q}$ is given in \eqref{thetapro} and $\widetilde{C}^{\prime} > 0$ is a constant.
\end{theorem}

\section{Simulation studies}\label{new.Sec.5}

In this section, we investigate the finite-sample performance of the suggested SOFARI inference procedure relative to the asymptotic theory established in Section \ref{new.Sec.4}.
The simulation setup is presented in Section \ref{new.Sec.5.1} of the Supplementary Material. In addition, we consider two settings of different dimensions. In setting $1$, we choose $(n, p, q) = (200, 25, 15)$, while setting $(n, p, q) = (200, 50, 30)$ in
setting $2$.
We would like to emphasize that both settings give rise to the high-dimensional regime since the total dimensionality due to both features and responses is $p*q$, exceeding greatly the available sample size $n$. 

\begin{figure}[tp]\small
	\centering
	\includegraphics[width=0.75\linewidth]{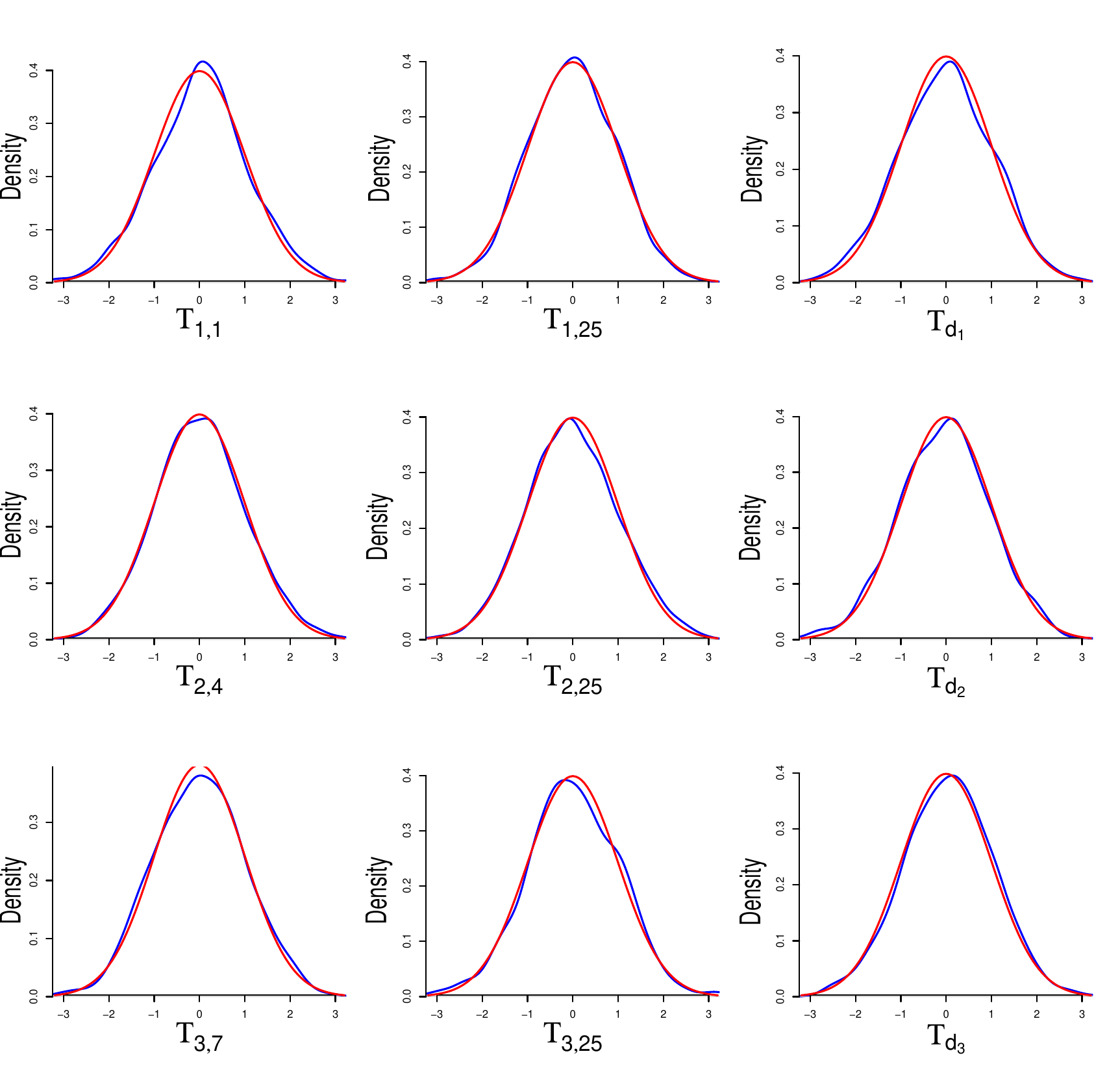}
	\caption{ \scriptsize{The kernel density estimates (KDEs) for the distributions of the SOFARI estimators on the latent left factor vectors (i.e., the left singular vectors weighted by the corresponding singular values) in different sparse SVD layers, and the squared singular values against the target standard normal density based on $1000$ replications for setting 1. Left panel: the KDEs of  $\mathrm{T}_{1,1}, \mathrm{T}_{2,4}$, and $\mathrm{T}_{3,7}$; middle panel:  the KDEs of $\mathrm{T}_{1,25}, \mathrm{T}_{2,25}$, and $\mathrm{T}_{3,25}$; right panel: the KDEs of $\mathrm{T}_{d_1}$, $\mathrm{T}_{d_2}$, and $\mathrm{T}_{d_3}$, all viewed from top to bottom. The blue curves represent the KDEs for SOFARI estimators, whereas the red curves stand for the target standard normal density.}}\label{figure1} 
\end{figure}

For implementation of SOFARI, the rank of multi-response regression model \eqref{model} is identified 
beforehand using the self-tuning selection method developed in \cite{Xin2019Adaptive}. The initial estimate $\wt{\C} = (\wt{\L}, \wt{\D}, \wt{\V})$ is obtained from the SOFAR procedure \citep{uematsu2017sofar} with the entrywise $L_1$-norm penalty (SOFAR-L) and the precision matrix of the covariates is estimated with the nodewise Lasso \cite{meinshausen2006high} as suggested in \cite{van2014asymptotically}. Moreover, we exploit the adaptive thresholding method \citep{cai2011adaptive} in the covariance estimation for the random errors. 

We choose the significance level $\alpha = 0.05$ for statistical inference and repeat the simulation $1000$ times for each setting. 
We employ two performance measures to evaluate the inference results: the average coverage probability (CP) and the average length (Len) of the $(1 - \alpha) 100\%$ (i.e., $95\%$) confidence intervals for the unknown population parameters over the $1000$ replications.
Specifically, for each individual unknown parameter $u^*$, denote by  $\mathrm{CI}$ the corresponding $95\%$ confidence interval of $u^*$ constructed using SOFARI. Then the two performance measures are defined as
$
		{\operatorname{CP}} =  \wh{\mathbb{P}}\left[u^* \in \mathrm{CI}\right] \ \text{ and } \ 
		{\operatorname{Len}} =  \operatorname{
			length}\left(\mathrm{CI}\right),$
respectively, where $\wh{\mathbb{P}}$ denotes the  empirical probability measure. Here, CP is the empirical version of the expectation for the conditional coverage probability given both parameters and the covariate matrix. To verify the asymptotic normalities of the SOFARI estimators, we 
further define the standardized quantities for each $k = 1, \ldots, r^*$ and $j = 1, \ldots, p$, 
$	\mathrm{T}_{{k,j}} = \sqrt{n}(\wh{u}_{k,j} - {u}_{k,j}^*)/ \wt{\nu}_{{k,j}} \ \text{ and } \ \mathrm{T}_{d_k} = \sqrt{n}(\wh{d}_{k}^2 - d_k^{*2})/ \wt{\nu}_{d_k},$
where $\wt{\nu}_{{k,j}}^2$ and $\wt{\nu}_{d_k}^2$ are the corresponding variance estimates given in Theorem \ref{coro:var:rank22uk}. 

\begin{table}\small
	\centering
	\caption{\label{table:uj1}
	\scriptsize{The average performance measures of SOFARI on the individual components of the latent left factor vectors (i.e., the left singular vectors weighted by the corresponding singular values) in different sparse SVD layers, and the squared singular values $(d_1^{*2}, d_2^{*2}, d_3^{*2} ) = (100^2, 15^2, 5^2)$ over $1000$ replications.}}
	\smallskip
	\resizebox{0.62\textwidth}{!}{
		\begin{tabular}{c|ccccccccc}
			\hline
			Setting& & ${\operatorname{CP}}$  & ${\operatorname{Len}}$ &    & ${\operatorname{CP}}$  & ${\operatorname{Len}}$ & & ${\operatorname{CP}}$  & ${\operatorname{Len}}$    \\
			\hline
			1&$u_{1,1}^*$
			& 0.939  &0.384  & $u_{2,4}^*$   & 0.949  & 0.406 & $u_{3,7}^*$    & 0.953  & 0.405 \\
			&$u_{1,2}^*$
			& 0.957   &0.402  & $u_{2,5}^*$   & 0.947  & 0.406 & $u_{3,8}^*$    & 0.942  & 0.404 \\
			&$u_{1,3}^*$
			& 0.946   &0.403  & $u_{2,6}^*$   & 0.952  & 0.407 & $u_{3,9}^*$    & 0.948  & 0.403 \\
			&$u_{1,p-2}^*$
			& 0.939   &0.409  & $u_{2,p-2}^*$   & 0.938  & 0.409 & $u_{3,p-2}^*$    & 0.956  & 0.411 \\
			&$u_{1,p-1}^*$
			& 0.952   &0.410  & $u_{2,p-1}^*$   & 0.941  & 0.410 & $u_{3,p-1}^*$    & 0.958  & 0.411 \\
			&$u_{1,p}^*$
			& 0.958   &0.393  & $u_{2,p}^*$   & 0.948  & 0.393 & $u_{3,p}^*$    & 0.951  & 0.395 \\[1pt]
			&$d_1^{*2}$
			& 0.944   &77.225  & $d_2^{*2}$    & 0.949  & 11.760 & $d_3^{*2}$     & 0.952  &  3.924 \\
			\hline
			2&$u_{1,1}^*$
			& 0.944   &0.277  & $u_{2,4}^*$   & 0.934  & 0.287 & $u_{3,7}^*$    & 0.948  & 0.286 \\
			&$u_{1,2}^*$
			& 0.939   &0.288  & $u_{2,5}^*$   & 0.946  & 0.288 & $u_{3,8}^*$    & 0.947  & 0.287 \\
			&$u_{1,3}^*$
			& 0.954   &0.289  & $u_{2,6}^*$   & 0.943  & 0.288 & $u_{3,9}^*$    & 0.941  & 0.286 \\
			&$u_{1,p-2}^*$
			& 0.950   &0.292  & $u_{2,p-2}^*$   & 0.956  & 0.293 & $u_{3,p-2}^*$    & 0.947  & 0.291 \\
			&$u_{1,p-1}^*$
			& 0.948   &0.291  & $u_{2,p-1}^*$   & 0.947  & 0.291 & $u_{3,p-1}^*$    & 0.940  & 0.289 \\
			&$u_{1,p}^*$
			& 0.958   &0.281  & $u_{2,p}^*$   & 0.940  & 0.282 & $u_{3,p}^*$    & 0.948  & 0.280 \\[1pt]
			&$d_1^{*2}$
			& 0.949   &55.795  & $d_2^{*2}$    & 0.943  & 8.457 & $d_3^{*2}$     & 0.943  &  2.798 \\
			\hline
		\end{tabular} }
\end{table}
The rank of the latent sparse SVD structure is  identified consistently as $r = 3$ over both two settings. Let us examine the asymptotic normalities of the different SOFARI estimators. 
We calculate the kernel density estimates (KDEs) for the standardized quantities $\mathrm{T}_{{k,j}}$  for both the nonzero and zero components of the vector $\u^*_k = d_k^* \l_k^*$, as well as the KDEs for $\mathrm{T}_{d_k}$ corresponding to the nonzero  $d_k^{*2}$'s. These KDEs are similar across the  two model settings, and
thus we only present in Figure \ref{figure1} the kernel density plots for setting 1 corresponding to the first nonzero component $u^*_{{k,3(k-1)+1}}$,  the last zero component $u^*_{{k,p}}$, and $d_k^{*2}$ in each latent sparse SVD layer, with $1 \leq k \leq 3$. 
By comparing the KDEs for the SOFARI estimators to the standard normal density, we see from Figure \ref{figure1} that the empirical distributions of the standardized SOFARI estimates all mimic closely the standard normal distribution, justifying our asymptotic normality theory established in Section \ref{new.Sec.4}. 

To ease the presentation, we report the performance measures of the SOFARI estimates for the three nonzero components and the last three zero components of $\u_k^*$ as well as the squared singular value $d_k^{*2}$ in each latent sparse SVD layer with $1 \leq k \leq 3$ over the two settings and summarize the results in Table \ref{table:uj1}. 
It is clear to see from Table \ref{table:uj1} that the average coverage probabilities of the corresponding $95\%$ confidence intervals constructed by SOFARI are all very close to the target level of $95\%$. 
Moreover, we can observe that the average lengths of the $95\%$ confidence intervals for different $u_{k,j}^*$ in each latent sparse SVD layer are relatively stable over $j$. 
For the squared singular value $d_k^{*2}$, there is a decreasing trend in the average length of the $95\%$ confidence interval  as $k$ increases, which is in line with our asymptotic theory in Section \ref{new.Sec.4} that the asymptotic variance of the SOFARI estimate for $d_k^{*2}$ depends on the magnitude of the nonzero singular value (cf. Theorem \ref{coro:var:rank22uk}). 


\blue{
Besides this simulation example, we have also examined the robustness and effectiveness of SOFARI when some technical assumptions are violated. Due to the space limit, these numerical results are presented in Section \ref{sec:app:simu:B} of the Supplementary Material.}



\section{Application to the federal reserve economic data} \label{new.Sec.6} 

We now showcase the practical utility of the suggested SOFARI inference procedure on an economic forecasting application. In particular, we will focus on analyzing the monthly macroeconomic data set from the federal reserve economic database FRED-MD in \cite{mccracken2016fred}. 
This data set is comprised of $660$ monthly observations for $134$ macroeconomic variables from January 1960 to December 2014.
Among those variables, we are interested in the interpretable multi-task learning problem of forecasting some typical macroeconomic indicators such as the consumer price index (CPI), interest rates, the unemployment rate, and the stock market price index simultaneously. Besides them, we also select several important variables considered in \cite{carriero2019large}, including the personal income, money supply, housing, and exchange rates, giving rise to a total of $q = 20$ response variables. See Section \ref{appsec:D} of the Supplementary Material for the list of the $20$ selected responses along with their descriptions.

\begin{table}[tp]
	\centering
	\caption{\scriptsize{Estimated squared singular values and the lengths of the corresponding 95\% confidence intervals for the real data application in Section \ref{new.Sec.6}.}}\label{tablerealdd}
	\resizebox{0.3\textwidth}{!}{
	\begin{tabular}{lccc}
		\hline	
		& $\wh{d}_1^{2}$ & $\wh{d}_2^{2}$ & $\wh{d}_3^{2}$\\
		\hline
		Value  &  2545.431 & 7.546 & 1.215 \\
		Len & 13.576 & 1.127 & 0.591\\
		\hline
	\end{tabular} }
\end{table}

\begin{table}[tp]
	\centering
	\caption{\scriptsize{The numbers of significant features with the target FDR level at $5\%$ for the latent left factor vectors (i.e., the left singular vectors weighted by the corresponding singular values) in different sparse SVD layers for the real data application in Section \ref{new.Sec.6}.}
	}\label{tablerealuu}   
	\resizebox{0.2\textwidth}{!}{
		\begin{tabular}{lccc}
			\hline	
			& $\wh{\u}_1$ & $\wh{\u}_2$ & $\wh{\u}_3$ \\
			\hline
			Num  &  20 & 41  & 18  \\
			\hline
		\end{tabular} }

\end{table}

We treat the remaining macroeconomic variables except for the four with missing values as covariates for predicting the multiple responses. 
To alleviate the issue of high correlations among the covariates, following \cite{zheng2021nonsparse} we choose randomly one representative covariate from those highly correlated economic variables whose correlations are above $0.9$ in magnitude, resulting in $94$ representative covariates. Further, to adapt to times series data, we transform both responses and covariates through differencing and logarithmic transformation as in \cite{mccracken2016fred}. We also incorporate the first to fourth lags of the responses and covariates into the design matrix of features and standardize each column of the feature and response matrices to have mean zero and standard deviation one, similarly as in \cite{tsknockoff2021}. 
Finally, the preprocessed data set contains $p = 456$  features and $q = 20$ responses with a total sample size of $n = 654$. 

\begin{figure}[tp]
	\centering
	\includegraphics[width=0.8\linewidth]{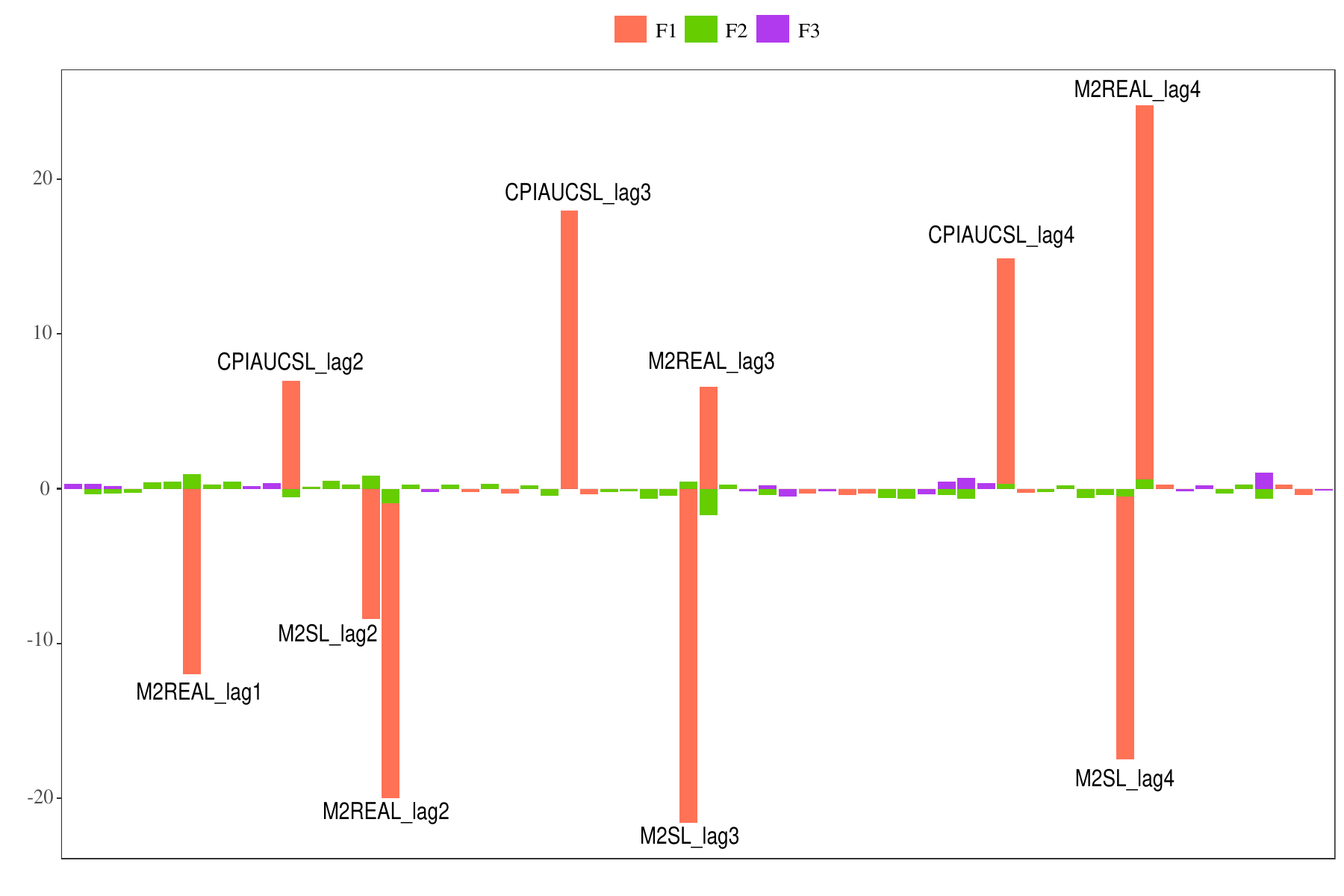}
	\caption{\scriptsize{Bar charts of the significant features in the top three latent left factor vectors (i.e., the left singular vectors weighted by the corresponding singular values). The significant features correspond to ones presented in Table \ref{tablerealuu}. Different colors F1, F2, and F3 correspond to the three factors ranked by the estimated singular values, respectively. The $y$-axis indicates the magnitude and signs of the corresponding coefficients in each factor.}}\label{figure2} 
\end{figure}

\blue{To analyze this data set, we fit multi-response regression model \eqref{model} using the SOFAR estimator \citep{uematsu2017sofar} with the entrywise $L_1$-norm penalty (SOFAR-L) due to its nice prediction performance as shown in Section \ref{appsec:D} of the Supplementary Material.}
The estimated rank is $r = 3$, which means that there are three important latent left factor vectors. 
Then we apply the SOFARI procedure to the full data set 
at significance level $\alpha = 0.05$. 
The initial estimate is obtained by SOFAR-L and the estimation for the feature precision matrix and the error covariance matrix is done similarly as that in Section \ref{new.Sec.5}. 
We summarize the estimated squared singular values as well as the lengths of the corresponding 95\% confidence intervals 
in Table \ref{tablerealdd}. It is clear that the three latent sparse SVD layers are well separated and the first latent left factor vector is the most important one since its contribution toward the total variation is over $99\%$ in terms of the squared singular values.

In addition, we are interested in studying the composition of the latent left factor vectors $\u_k^*$'s by testing which original covariates have significant latent weights in different sparse SVD layers $k=1,2,3$.  
Note that here we are performing the large-scale multiple testing in light of the total feature dimensionality $p = 456$. To account for the multiple testing problem, we first apply the SOFARI procedure to obtain the p-values for individual features, and then use the BHq procedure \citep{benjamini1995controlling} to control the false discovery rate (FDR) at the level $5\%$ for each layer. The results summarized in Table \ref{tablerealuu} reveal that the top three latent left factors are highly sparse in their dependency on the original features whose dimensionality is $p = 456$.

To gain insights into the composition of the three latent left factor vectors, we plot the corresponding weights of the significant original features in each latent left factor vector as a bar chart in Figure \ref{figure2}, where the significant original features are the same as the ones presented in Table \ref{tablerealuu}. 
It can be seen from Figure \ref{figure2} that there are ten features whose coefficients are much larger than the remaining ones so we mark them with their abbreviated names. In fact, all those ten features are included simultaneously in both the first and second latent left factor vectors, except for the third lag of CPIAUCSL (the overall CPI) which appears only in the first latent left factor vector. This reveals that some important features can contribute to more than one latent left factor vectors. 

Furthermore, the ten most important features are essentially three macroeconomic variables with different lags, namely CPIAUCSL (the overall CPI), M2SL (M2 money stock), and M2REAL (real M2 money stock). \blue{To help interpret the relationships between macroeconomic features and response variables, we also look at estimation results of the right singular matrix \(\V^*\) from initial SOFAR estimates.}  Specifically, the overall CPI measures inflation based on a basket of consumer goods and services. Given the positive factor coefficients and the negative factor loadings, CPI or inflation tends to have negatively correlate with several responses such as interest rates. Such phenomenon is sensible since when the central bank lowers interest rates to stimulate the economy, the inflation level tends to increase. The other two variables relate to money supply. In particular, M2 money stock is a measure of money supply that includes cash, checking deposits, and non-cash assets, while real M2 money stock is the value of M2 money stock deflated by CPI. 
The negative factor coefficients along with the negative factor loading suggests a positive relationship between M2 money stock and the response S\&P 500 stock price index. 
Overall, the SOFARI procedure can be exploited to assess feature significance in the latent left factor vectors across different sparse SVD layers for real applications involving high-dimensional multi-task learning inference with interpretability and flexibility.

\section{Discussions}\label{new.Sec.7}

We have investigated the problem of high-dimensional inference on the latent sparse SVD structure under the model of multi-response regression. Our technical analysis has revealed that the use of the underlying Stiefel manifold structure is key to the success of such inferential task on the latent factors. 
The resulting SOFARI estimators for the latent left factor vectors and singular values have been shown to enjoy asymptotic normalities with justified asymptotic variance estimates.
\blue{Moreover, our proposed method can be combined with a sample splitting technique to substantially mitigate the requirement on the sparsity level, which is presented in Section \ref{app:sec:split} of the Supplementary Material.} 

\blue{It is noteworthy that inference for latent right factor vectors is not straightforward from that of the left factor vectors and can be even more challenging. The key difficulty lies in that when we target at the right factor vectors, the manifold induced by the SVD constraints on the left singular vectors would not help in constructing the modified score function due to asymmetry of the left and the right singular vectors in the response matrix. This prevents the manifold of left singular vectors from saving degrees of freedom under the SVD constraints. New techniques will be needed to deal with this interesting yet challenging topic.}




\bibliographystyle{apalike}
\bibliography{references}



\newpage
\appendix
\setcounter{page}{1}
\setcounter{section}{0}
\renewcommand{\theequation}{A.\arabic{equation}}
\setcounter{equation}{0}

\begin{center}{\bf \Large Supplementary Material to ``SOFARI: High-Dimensional Manifold-Based Inference''}

\bigskip

Zemin Zheng, Xin Zhou, Yingying Fan and Jinchi Lv
\end{center}


\noindent This Supplementary Material contains some additional theoretical results, simulation studies, and real data details, as well as the proofs of all main results, lemmas, and additional technical details. All the notation is the same as defined in the main body of the paper. We will also use $c$ to denote a generic positive constant whose value may vary from line to line throughout the proofs.

\blue{
\section{The intrinsic bias issue and algorithms}

\subsection{The intrinsic bias issue}\label{Biasipssue}

Let us first introduce a lemma below to gain some insights into the intrinsic bias issue in the simple case of $r^* = 2$ and $k = 1$. 

\begin{lemma}\label{3-7:prop:2}
    Under the SVD constraint \eqref{SVDc}, for an arbitrary { $\M = \left[\M_1,\M_2,\M_3\right]$ with $\M_1, \M_2 \in\R^{p \times q}$ and $\M_3 \in\R^{p \times p}$}, it holds that
	\begin{align*}
		\wt{\psi}_1(\u_1,\boldeta^*_1)
		= (\I_p - \M_1\v_1\u_1\trans + \M_1\v_2\u_2\trans)\wh{\bSigma}(\u_1 - \u_1^*) + (\M_1\v_2^* + \M_2\v_1^*)\u_2\strans\wh{\bSigma}\u_1^* + \bdelta + \bepsilon,
	\end{align*}
	where $\bdelta = \M_1\left\{(\v_1 - \v_1^*)\u_1\trans - (\v_2\u_2\trans - \v_2^*\u_2\strans)\right\}\wh{\bSigma}(\u_1 - \u_1^*)$ and
	\begin{align*}
		\bepsilon = n^{-1} \left\{ \M_1 \E\trans \X \u_1 + \M_2 \E\trans \X \u_2^* + \M_3 \X\trans \E \v_2^*  \right\} - n^{-1} \X\trans \E \v_1^*.
	\end{align*}
\end{lemma}

In view of Lemma \ref{3-7:prop:2} above, we see immediately that the score function vector $\wt{\psi}_1$ at the true parameter  values $(\u_1^*, \boldeta^*_1)$ is
\begin{align*}
	\wt{\psi}_1(\u_1^*,\boldeta^*_1)
	= (\M_1\v_2^* + \M_2\v_1^*)\u_2\strans\wh{\bSigma}\u_1^* + \bepsilon^*,
\end{align*}
where $\bepsilon^* = n^{-1} \left\{ \M_1 \E\trans \X \u_1^* + \M_2 \E\trans \X \u_2^* + \M_3 \X\trans \E \v_2^*  \right\} - n^{-1} \X\trans \E \v_1^*$. Hence, the expectation of $\wt{\psi}_1$ at $(\u_1^*, \boldeta^*_1)$ will be determined by $(\M_1\v_2^* + \M_2\v_1^*)\u_2\strans\wh{\bSigma}\u_1^*$. It is an intrinsic bias term associated with this inference problem, induced by the correlation between the two latent factors $\X \u_1^*$ and $\X \u_2^*$. In order to design a valid inference procedure, we need some orthogonality between the latent factors so that
\begin{align}\label{laybias}
	\|(\M_1\v_2^* + \M_2\v_1^*)\u_2\strans\wh{\bSigma}\u_1^*\|_{\infty}
	= o(n^{-1/2}).
\end{align}
Then the intrinsic bias term can become secondary and does not affect the asymptotic distribution. In the case of strongly orthogonal factors where  $|\u_2\strans\wh{\bSigma}\u_1^*| = o(n^{-1/2})$, the nuisance parameters from different layers can be separable,  and \eqref{laybias} will hold under suitable choices of matrix $\M$.


For a general rank $r^*$, Lemma \ref{prop:rankr1} in Section \ref{sec:proof:l2} 
shows that the corresponding intrinsic bias term would be $\M_k^v\C_{-k}\strans \wh{\bSigma} \u_k^*$, where $\C_{-k}^* = \sum_{\substack{i \neq k}}\u_i^*\v_i\strans$. Then the assumption of strongly orthogonal latent factors naturally generalizes as
\begin{align*}
	\sum_{j \neq k} |\u_j\strans\wh{\bSigma}\u_k^*|  = o(n^{-1/2}),
\end{align*}
which together with suitably chosen matrix $\M$ can control the intrinsic bias. 

\subsection{Algorithms}\label{Algo}

In this section, we present Algorithms \ref{alg1} and \ref{alg12} for our SOFARI$_s$ and SOFARI procedures, respectively. 
Here, we denote by \(\mathbf{e}_k \in \mathbb{R}^p \) the \(p\)-dimensional unit vector with $1$ at the $k$th component and $0$ elsewhere. In addition, the code for implementing our methods is publicly available on GitHub (\url{https://github.com/xinaut/SOFARI}).}

\begin{algorithm}
	\caption{ \blue{ SOFARI$_s$ } }
	\label{alg1}
	\resizebox{\textwidth}{!}{%
	\begin{minipage}{1.2\textwidth}
	\blue{
	\begin{algorithmic}[1]
	\STATE \textbf{Input:}  Data $\mathbf{X} \in \R^{n \times p}, \mathbf{Y} \in \R^{n \times q}$
	\STATE \textbf{Initial Step:} Determine the rank $\wh{r}$ and compute initial SOFAR estimates $\big\{\wt{d}_i, \wt{\boldsymbol{\ell}}_i, \wt{\boldsymbol{r}}_i \big\}_{i=1}^{\wh{r}}$
	\FOR{$k = 1, \ldots, \wh{r}$}
		\STATE $\mathbf{M}$-step: Compute $\wt{\mathbf{M}}^{(k)}=\big[\0_{p \times q(k-1) }, \wt{\mathbf{M}}_{k}, \0_{p \times [q(\hat{r} - k) + p(\hat{r}-1) ]}  \big]$ with $\wt{\M}_{k} = -\wt{z}_{kk}^{-1}\wh{\bSigma}\wt{\C}_{-k}$ \\[0.5em]
		\STATE $\mathbf{W}$-step: Compute $\wt{\W}_k  =   \widehat{\bTheta} \left\{  \I_p +   \wt{z}_{kk}^{-1}\wh{\bSigma}\wt{\U}_{-k}(\I_{\hat{r}-1} -\wt{z}_{kk}^{-1}\wt{\U}_{-k}\trans\wh{\bSigma} \wt{\U}_{-k})^{-1}\wt{\U}_{-k}\trans\right\} $ 
		\STATE Debiased estimate: For  $\wt{\boldeta}_k = \left(\wt{\u}_1\trans,\ldots,  \wt{\u}_{\hat{r}}\trans, \wt{\v}_1\trans, \ldots, \wt{\v}_{k-1}\trans, \wt{\v}_{k+1}\trans, \ldots, \wt{\v}_{\hat{r}}\trans\right)\trans$, compute
		\begin{align}
		&\wh{\u}_k  = \widetilde{\u}_k - \wt{\W}_k\Big(\der{L}{\u_k} - \wt{\mathbf{M}}^{(k)}\der{L}{\boldeta_k}\Big)\Big|_{(\widetilde{\u}_k,\widetilde{\boldeta}_k)}, \nonumber \\[5pt]
	 &\wh{d}_k^2 =  \norm{\wt{\u}_k}_2^2 - 2 \wt{\u}_k\trans \wt{\W}_k\wt{\psi}_k(\wt{\u}_k,\wt{\boldeta}_k). \nonumber
	\end{align}
		\STATE Variance estimate: 
	\begin{align*}
		\wt{\nu}_{u_k}^2 &= \e_k\trans \wt{\W}_k (\wt{z}_{kk} \wt{\M}_{k} \wt{\bSigma}_e \wt{\M}_{k}\trans + \wt{\v}_k\trans \wt{\bSigma}_e\wt{\v}_k \wh{\bSigma} - 2\wh{\bSigma} \wt{\u}_k \wt{\v}_k\trans \wt{\bSigma}_e \wt{\M}_{k}^{T})\wt{\W}_k\trans\e_k, \\[5pt]
		\wt{\nu}_{d_k}^2 &= 4\wt{\u}_k\trans \wt{\W}_{k}(\wt{z}_{kk} \wt{\M}_{k} \wt{\bSigma}_e \wt{\M}_{k}^{T} + \wt{\v}_k\trans \wt{\bSigma}_e\wt{\v}_k \wh{\bSigma} - 2\wh{\bSigma} \wt{\u}_k \wt{\v}_k\trans \wt{\bSigma}_e \wt{\M}_{k}^{T}) \wt{\W}_k\trans \wt{\u}_k. 
	\end{align*} 
	\ENDFOR
	\STATE \textbf{Output:} Debiased estimate and variance estimate $\{ \wh{\u}_k, \wh{d}_k^2, \wt{\nu}^2_{u_k}, \wt{\nu}_{d_k}^2 \}_{k=1}^{\wh{r}}$
	\end{algorithmic} }
	\end{minipage}%
	}
	\end{algorithm}

	\begin{algorithm}
	\caption{\blue{SOFARI}}
	\label{alg12}
	\resizebox{\textwidth}{!}{%
	\begin{minipage}{1.2\textwidth}
	\blue{
	\begin{algorithmic}[1]
	\STATE \textbf{Input:} Data $\mathbf{X} \in \R^{n \times p}, \mathbf{Y} \in \R^{n \times q}$
	\STATE \textbf{Initial Step:} Determine the rank $\wh{r}$ and compute initial SOFAR estimates $\big\{\wt{d}_i, \wt{\boldsymbol{\ell}}_i, \wt{\boldsymbol{r}}_i \big\}_{i=1}^{\wh{r}}$
	\FOR{$k = 1, \ldots, \wh{r}$}
		\STATE $\mathbf{M}$-step:  $\wt{\mathbf{M}}^{(k)}=\big[ \wt{\mathbf{M}}_{k}, \mathbf{0}_{p \times  (p+ q) (\hat{r}-k)}  \big]$ with $\wt{\M}_{k} = -\wt{z}_{kk}^{-1}\wh{\bSigma}\wt{\C}^{(2)}$ \\[0.5em]
		\STATE $\mathbf{W}$-step: Compute
		$
		   \wt{\W}_k  =   \widehat{\bTheta} \left\{  \I_p +   \wt{z}_{kk}^{-1}\wh{\bSigma}\wt{\U}^{(2)}(\I_{\hat{r}-k} -\wt{z}_{kk}^{-1}\wt{\U}^{(2) T}\wh{\bSigma} \wt{\U}^{(2)})^{-1}(\wt{\U}^{(2) })\trans\right\}
		$ \\[0.5em]
		\STATE Debiased estimate:  For   $\wt{\boldeta}_k = \left(\wt{\u}_{k+1}\trans, \ldots, \wt{\u}_{\hat{r}}\trans, \wt{\v}_{k}\trans,  \ldots, \wt{\v}_{\hat{r}}\trans\right)\trans$, compute
		\begin{align*}
			   &\wh{\u}_k = \psi_k(\widetilde{\u}_k,\widetilde{\boldeta}_k) = \widetilde{\u}_k - \wt{\W}_k\wt{\psi}_k(\widetilde{\u}_k,\widetilde{\boldeta}_k) = \widetilde{\u}_k - \wt{\W}_k\Big(\der{L}{\u_k} - \wt{\M}^{(k)}\der{L}{\boldeta_k}\Big)\Big|_{(\widetilde{\u}_k,\widetilde{\boldeta}_k)}, \\
			&\wh{d}_k^2 =  \norm{\wt{\u}_k}_2^2 - 2 \wt{\u}_k\trans \wt{\W}_k\wt{\psi}_k(\wt{\u}_k,\wt{\boldeta}_k).
		\end{align*}
		   \STATE Variance estimate: 
	\begin{align*}
		\wt{\nu}_{u_k}^2 &= \e_k\trans \wt{\W}_k (\wt{z}_{kk} \wt{\M}_{k} \wt{\bSigma}_e \wt{\M}_{k}\trans + \wt{\v}_k\trans \wt{\bSigma}_e\wt{\v}_k \wh{\bSigma} - 2\wh{\bSigma} \wt{\u}_k \wt{\v}_k\trans \wt{\bSigma}_e \wt{\M}_{k}^{T})\wt{\W}_k\trans\e_k, \\
		\wt{\nu}_{d_k}^2 &= 4\wt{\u}_k\trans \wt{\W}_{k}(\wt{z}_{kk} \wt{\M}_{k} \wt{\bSigma}_e \wt{\M}_{k}^{T} + \wt{\v}_k\trans \wt{\bSigma}_e\wt{\v}_k \wh{\bSigma} - 2\wh{\bSigma} \wt{\u}_k \wt{\v}_k\trans \wt{\bSigma}_e \wt{\M}_{k}^{T}) \wt{\W}_k\trans \wt{\u}_k. 
	\end{align*} 
	\ENDFOR
	\STATE \textbf{Output:} Debiased estimate and variance estimate $\{ \wh{\u}_k, \wh{d}_k^2, \wt{\nu}^2_{u_k}, \wt{\nu}_{d_k}^2 \}_{k=1}^{\wh{r}}$
	\end{algorithmic}}
	\end{minipage}%
	}
	\end{algorithm}

\section{Additional simulation results}\label{sec:app:simu:B}

\subsection{Simulation setup}\label{new.Sec.5.1}

For the simulation example in Section \ref{new.Sec.5},
we consider a similar setup of the multi-response regression model \eqref{model} to that in \cite{mishra2017} so that the latent factors are weakly orthogonal to each other allowing for correlations among the latent factors. 
Specifically, we assume that the true regression coefficient matrix $\mathbf{C}^{*}=\sum_{k=1}^{r^*} d_{k}^{*} \l_{k}^{*} \mathbf{v}_{k}\strans$ satisfies that $r^* = 3$, $d_{1}^{*}=100, d_{2}^{*}=15, d_{3}^{*}=5$, and
\begin{align*}
	\l_{k}^{*}=\check{\l}_{k} /\|\check{\l}_{k}\|_{2}  & \ \text{ with }
	\check{\l}_{k}=\big(\operatorname{rep}(0,s_1(k-1)), \operatorname{unif}\left(S_{1}, s_1\right), \operatorname{rep}(0, p - ks_1)\big)\trans, \\
	\mathbf{v}_{k}^{*}=\check{\mathbf{v}}_{k} /\left\|\check{\mathbf{v}}_{k}\right\|_{2} & \ \text{ with }  \check{\mathbf{v}}_{k}=\big(\operatorname{rep}(0,s_2(k-1)), \operatorname{unif}\left(S_{2}, s_2\right), \operatorname{rep}(0, q - ks_2)\big)\trans.
\end{align*}
Here, $\operatorname{unif}\left({S}, s\right)$ denotes an $s$-dimensional random vector with i.i.d. components from the uniform distribution on set ${S}$, $\operatorname{rep}(a,s)$ represents an $s$-dimensional vector with identical components $a$, $S_{1} = \{-1, 1\}$, $S_2 = [-1,-0.3] \cup[0.3,1]$, $s_1 = 3$, and $s_2 = 3$. 

Given the matrix of left singular vectors $\L^* = (\l_1^*, \ldots, \l_{r^*}^*)$, we can find a matrix $\L_{\perp}^{*} \in \mathbb{R}^{p \times\left(p-r^{*}\right)}$ such that $\mathbf{P}=\left[\L^{*}, \L_{\perp}^{*}\right] \in \mathbb{R}^{p \times p}$ is nonsingular. The covariate matrix $\mathbf{X}$ is generated following the three steps specified below. First, we create matrix $\mathbf{X}_{1} \in \mathbf{R}^{n \times r^{*}}$ by drawing a random sample of size $n$ from $N\left(\mathbf{0}, \mathbf{I}_{r *}\right)$.
Second, denote by $\wt{\x} \sim {N}(\mathbf{0}, \bSigma_X)$,  $\wt{\x}_1 = \L^{* {T}} \wt{\x}$,  and $\wt{\x}_{2}=\L_{\perp}^{* {T}} \wt{\x}$, where the population covariance matrix is given by $\bSigma_X=\left(0.3^{|i-j|}\right)_{p \times p}$. We then generate matrix $\mathbf{X}_{2} \in \mathbb{R}^{n \times\left(p-r^{*}\right)}$ by drawing a random sample of size $n$ from the conditional distribution of $\wt{\x}_{2}$ given $\wt{\x}_{1}$. 
Third, the covariate matrix $\mathbf{X}$ is finally defined as $\mathbf{X}=\left[\mathbf{X}_{1}, \mathbf{X}_{2}\right] \mathbf{P}^{-1}$ 
so that the latent factors $n^{-1/2}\X\l_i^*$ are weakly orthogonal to each other.

We further assume that the rows of the error matrix $\E$ are i.i.d. copies from $N\left(\mathbf{0}, \sigma^{2} \boldsymbol{\Sigma}_E\right)$ with $\boldsymbol{\Sigma}_E=\left(0.3^{|i-j|}\right)_{q \times q}$ that is independent of covariate matrix $\mathbf{X}$, where the noise level $\sigma^{2}$ is set such that the signal-to-noise ratio (SNR) $\left\| \mathbf{X} (d_{r^*}^{*}\l_{r^*}^{*} \mathbf{v}_{r^*}\strans)\right\|_{F} /\|\mathbf{E}\|_{F}$ is equal to $1$. 

\subsection{Simulation example 2}\label{new.Sec.5.2}

The setup of the second simulation example is similar to that in \cite{uematsu2017sofar}. The major difference with the simulation setup in Section \ref{new.Sec.5.1} is that we now do \textit{not} assume any particular form of the orthogonality constraint on the latent factors, allowing for \textit{stronger} correlations among the latent factors. This means the technical assumptions in Conditions \ref{con:nearlyorth} and \ref{con:orth:rankr} may be violated. The challenging setup here is designed to test the robustness of the SOFARI inference procedure when some of the orthogonality conditions are not satisfied. 
Specifically, we assume that the rows of covariate matrix $\X$ are i.i.d. and drawn directly from $N(\mathbf{0}, \bSigma_X)$ with covariance matrix  $\bSigma_X = (0.3^{|i-j|})_{p \times p}$. The true underlying coefficient matrix $\mathbf{C}^{*}$ 
follows the same latent sparse SVD structure as that in simulation example 1, except that $d_{1}^{*}$ increases from $100$ to $200$ and both $s_1$ and $s_2$ increase from $3$ to $5$. 
Similarly, we consider two settings for the second simulation example, and the remaining setups for settings 3 and 4 are the same as those for settings 1 and 2 in Section \ref{new.Sec.5.1}, respectively. 

\clearpage

\begin{table}[t]
	\centering
	\caption{\label{table:uj2sofar} \small
		The average performance measures of SOFARI on the individual components of the latent left factor vectors (i.e., the left singular vectors weighted by the corresponding singular values) in different sparse SVD layers, and the squared singular values $ (d_1^{*2}, d_2^{*2}, d_3^{*2} ) = (200^2, 15^2, 5^2)$ over $1000$ replications for simulation example 2 in Section \ref{new.Sec.5.2}.}
	\smallskip
	\scalebox{0.95}{
		\begin{tabular}{c|ccccccccc}
			\hline	
			Setting& & ${\operatorname{CP}}$  & ${\operatorname{Len}}$ &    & ${\operatorname{CP}}$  & ${\operatorname{Len}}$ & & ${\operatorname{CP}}$  & ${\operatorname{Len}}$    \\
			\hline
			3&$u_{1,1}^*$
			& 0.937   &0.400  & $u_{2,6}^*$   & 0.945  & 0.417 & $u_{3,11}^*$    & 0.948  & 0.418 \\
			&$u_{1,2}^*$
			& 0.945  &0.417  & $u_{2,7}^*$   & 0.945  & 0.416 & $u_{3,12}^*$    & 0.935  & 0.418 \\
			&$u_{1,3}^*$
			& 0.957   &0.416  & $u_{2,8}^*$   & 0.948  & 0.407 & $u_{3,13}^*$    & 0.947  & 0.418 \\
			&$u_{1,4}^*$
			& 0.943   &0.417  & $u_{2,9}^*$   & 0.949  & 0.417 & $u_{3,14}^*$    & 0.933  & 0.418 \\
			&$u_{1,5}^*$
			& 0.941   &0.417  & $u_{2,10}^*$   & 0.937  & 0.416 & $u_{3,15}^*$    & 0.947  & 0.417 \\
			&$u_{1,p-4}^*$
			& 0.952   &0.418  & $u_{2,p-4}^*$   & 0.948  & 0.417 & $u_{3,p-4}^*$    & 0.965  & 0.418 \\
			&$u_{1,p-3}^*$
			& 0.954   &0.417  & $u_{2,p-3}^*$   & 0.947  & 0.416 & $u_{3,p-3}^*$    & 0.950  & 0.417 \\
			&$u_{1,p-2}^*$
			& 0.949   &0.417  & $u_{2,p-2}^*$   & 0.943  & 0.416 & $u_{3,p-2}^*$    & 0.927  & 0.418 \\
			&$u_{1,p-1}^*$
			& 0.941   &0.416  & $u_{2,p-1}^*$   & 0.942  & 0.415 & $u_{3,p-1}^*$    & 0.952  & 0.417 \\
			&$u_{1,p}^*$
			& 0.945   &0.400  & $u_{2,p}^*$   & 0.950  & 0.399 & $u_{3,p}^*$    & 0.946  & 0.400 \\[1pt]
			&$d_1^{*2}$
			& 0.949   &165.637  & $d_2^{*2}$    & 0.955  & 11.015 & $d_3^{*2}$     & 0.944  &  4.135 \\
			\hline
			4
			&$u_{1,1}^*$
			& 0.955   &0.287  & $u_{2,6}^*$   & 0.944  & 0.295 & $u_{3,11}^*$    & 0.949  & 0.295 \\
			&$u_{1,2}^*$
			& 0.947   &0.296  & $u_{2,7}^*$   & 0.942  & 0.295 & $u_{3,12}^*$    & 0.922  & 0.295 \\
			&$u_{1,3}^*$
			& 0.938   &0.296  & $u_{2,8}^*$   & 0.945  & 0.295 & $u_{3,13}^*$    & 0.940  & 0.296 \\
			&$u_{1,4}^*$
			& 0.945  &0.297  & $u_{2,9}^*$   & 0.939  & 0.295 & $u_{3,14}^*$    & 0.942  & 0.295 \\
			&$u_{1,5}^*$
			& 0.946   &0.296  & $u_{2,10}^*$   & 0.945  & 0.295 & $u_{3,15}^*$    & 0.948  & 0.295 \\
			&$u_{1,p-4}^*$
			& 0.952   &0.296  & $u_{2,p-4}^*$   & 0.952  & 0.295 & $u_{3,p-4}^*$    & 0.950  & 0.295 \\
			&$u_{1,p-3}^*$
			& 0.944   &0.297  & $u_{2,p-3}^*$   & 0.953  & 0.295 & $u_{3,p-3}^*$    & 0.941  & 0.295 \\
			&$u_{1,p-2}^*$
			& 0.948   &0.297  & $u_{2,p-2}^*$   & 0.948  & 0.296 & $u_{3,p-2}^*$    & 0.942  & 0.296 \\
			&$u_{1,p-1}^*$
			& 0.947   &0.296  & $u_{2,p-1}^*$   & 0.947  & 0.295 & $u_{3,p-1}^*$    & 0.945  & 0.295 \\
			&$u_{1,p}^*$
			& 0.948   &0.287  & $u_{2,p}^*$   & 0.926  & 0.286 & $u_{3,p}^*$    & 0.956  & 0.285 \\[1pt]
			&$d_1^{*2}$
			& 0.949   &117.649  & $d_2^{*2}$    & 0.952  & 8.028 & $d_3^{*2}$     & 0.941  &  2.927 \\
			\hline
		\end{tabular} }
\end{table}

\clearpage

Table \ref{table:uj2sofar} summarizes the average performance measures of different SOFARI estimates under simulation example 2. Similar to simulation example 1, the rank of the latent sparse SVD structure is identified consistently as $r = 3$. From Table \ref{table:uj2sofar}, we can see that the average coverage probabilities of the $95\%$ confidence intervals constructed by SOFARI for the representative parameters are still very close to the target level of $95\%$. Furthermore, it can be seen that the average lengths of the $95\%$ confidence intervals for different components of the latent left factors across different settings are also stable over both $j$ and $k$. This demonstrates that the suggested SOFARI inference procedure can still apply and perform well even when the correlations among the latent factors may no longer be weak, provided that the eigengap among the nonzero singular values are sufficiently large.

\blue{
\subsection{Simulation example 3}\label{new.Sec.5.3}
We consider the setup similar to that of simulation example 2 in Section \ref{new.Sec.5.2} but containing weakly sparse signals. 
To be specific, we set $(n, p, q, r^*) = (200, 50, 30, 3)$ and in each layer, the left singular vector contains $8$ strong signals and $12$ weak signals, while the right singular vector contains $8$ strong signals and $3$ weak signals. For the left and right singular vectors, they are generated as follows:
\begin{align*}
	&\check{\l}_1=(\text{unif}(S_1,8),\text{rep}(0,30),\text{unif}(S_2,12))\trans,\\
	&\check{\l}_2=(\text{rep}(0,4),\text{unif}(S_2,12),\text{unif}(S_1,8),\text{rep}(0,26))\trans,\\
	&\check{\l}_3=(\text{rep}(0,20),\text{unif}(S_2,12),\text{unif}(S_1,8),\text{rep}(0,10))\trans,\\
	&\check{\mathbf{v}}_1=(\text{unif}(S_3,8),\text{rep}(0,19),\text{unif}(S_4,3))\trans,\\
	&\check{\mathbf{v}}_2=(\text{rep}(0,6),\text{unif}(S_4,3),\text{unif}(S_3,8),\text{rep}(0,13))\trans,\\
	&\check{\mathbf{v}}_3=(\text{rep}(0,19),\text{unif}(S_3,8),\text{unif}(S_4,3))\trans, \\
	&\l_{k}^{*}=\check{\l}_{k} /\|\check{\l}_{k}\|_{2}, \mathbf{v}_{k}^{*}=\check{\mathbf{v}}_{k} /\left\|\check{\mathbf{v}}_{k}\right\|_{2}, k = 1, 2, 3.
	\end{align*}
We set $S_{1} = \{-1, 1\}$ for relatively strong signals,  $S_2 = S_4 = [-0.1,-0.01] \cup[0.01,0.1]$ for weak signals, and $S_3 = [-1,-0.6] \cup[0.6,1]$ for moderate signals. The other setup is similar to that of simulation example 2 in Section \ref{new.Sec.5.2}.

The simulation results are summarized in Table \ref{table:3supp}. In view of the results, we can see that the average coverage probabilities of the $95\%$ confidence intervals constructed by SOFARI for the representative parameters are still close to the target level of $95\%$. Moreover, the average lengths of the $95\%$ confidence intervals for different components of the latent left factors across different settings are stable over both $j$ and $k$. This shows that the suggested SOFARI inference procedure can still apply and perform well under some weakly sparse settings.
}

\begin{table}[t]
	\centering
	\caption{\label{table:3supp} \small
		\blue{The average performance measures of SOFARI on the individual components of the latent left factor vectors (i.e., the left singular vectors weighted by the corresponding singular values) in different sparse SVD layers, and the squared singular values $ (d_1^{*2}, d_2^{*2}, d_3^{*2} ) = (200^2, 15^2, 5^2)$ over $1000$ replications for simulation example 3 in Section \ref{new.Sec.5.3}.}}
	\smallskip
	\scalebox{0.95}{
		\begin{tabular}{c|lcclcclcc}
			\hline
			Setting &  & ${\operatorname{CP}}$ & ${\operatorname{Len}}$ &  & ${\operatorname{CP}}$ & ${\operatorname{Len}}$ &  & ${\operatorname{CP}}$ & ${\operatorname{Len}}$ \\
			\hline
			5& $u_{1,1}^*$ & 0.937 & 0.279 & $u_{2,17}^*$ & 0.940 & 0.298 & $u_{3,33}^*$ & 0.944 & 0.297 \\
			& $u_{1,2}^*$ & 0.928 & 0.289 & $u_{2,18}^*$ & 0.937 & 0.297 & $u_{3,34}^*$ & 0.941 & 0.296 \\
			& $u_{1,3}^*$ & 0.928 & 0.288 & $u_{2,19}^*$ & 0.945 & 0.297 & $u_{3,35}^*$ & 0.931 & 0.296 \\
			& $u_{1,4}^*$ & 0.948 & 0.290 & $u_{2,20}^*$ & 0.945 & 0.298 & $u_{3,36}^*$ & 0.943 & 0.297 \\
			& $u_{1,5}^*$ & 0.932 & 0.289 & $u_{2,21}^*$ & 0.944 & 0.298 & $u_{3,37}^*$ & 0.926 & 0.296 \\
			& $u_{1,6}^*$ & 0.931 & 0.289 & $u_{2,22}^*$ & 0.940 & 0.298 & $u_{3,38}^*$ & 0.929 & 0.296 \\
			& $u_{1,7}^*$ & 0.945 & 0.289 & $u_{2,23}^*$ & 0.945 & 0.298 & $u_{3,39}^*$ & 0.946 & 0.297 \\
			& $u_{1,8}^*$ & 0.941 & 0.288 & $u_{2,24}^*$ & 0.946 & 0.298 & $u_{3,40}^*$ & 0.942 & 0.296 \\
			\cline{2-10}
			& $u_{1,39}^*$ & 0.935 & 0.289 & $u_{2,5}^*$ & 0.948 & 0.299 & $u_{3,21}^*$ & 0.942 & 0.296 \\
			& $u_{1,40}^*$ & 0.937 & 0.288 & $u_{2,6}^*$ & 0.940 & 0.298 & $u_{3,22}^*$ & 0.939 & 0.296 \\
			& $u_{1,41}^*$ & 0.944 & 0.289 & $u_{2,7}^*$ & 0.927 & 0.298 & $u_{3,23}^*$ & 0.941 & 0.297 \\
			& $u_{1,42}^*$ & 0.935 & 0.288 & $u_{2,8}^*$ & 0.926 & 0.298 & $u_{3,24}^*$ & 0.943 & 0.297 \\
			& $u_{1,43}^*$ & 0.932 & 0.289 & $u_{2,9}^*$ & 0.943 & 0.298 & $u_{3,25}^*$ & 0.928 & 0.296 \\
			& $u_{1,44}^*$ & 0.944 & 0.288 & $u_{2,10}^*$ & 0.934 & 0.298 & $u_{3,26}^*$ & 0.939 & 0.297 \\
			& $u_{1,45}^*$ & 0.940 & 0.289 & $u_{2,11}^*$ & 0.943 & 0.299 & $u_{3,27}^*$ & 0.934 & 0.297 \\
			& $u_{1,46}^*$ & 0.934 & 0.288 & $u_{2,12}^*$ & 0.933 & 0.298 & $u_{3,28}^*$ & 0.947 & 0.297 \\
			& $u_{1,47}^*$ & 0.926 & 0.289 & $u_{2,13}^*$ & 0.932 & 0.298 & $u_{3,29}^*$ & 0.939 & 0.297 \\
			& $u_{1,48}^*$ & 0.935 & 0.288 & $u_{2,14}^*$ & 0.942 & 0.298 & $u_{3,30}^*$ & 0.950 & 0.297 \\
			& $u_{1,49}^*$ & 0.931 & 0.288 & $u_{2,15}^*$ & 0.935 & 0.298 & $u_{3,31}^*$ & 0.942 & 0.296 \\
			& $u_{1,50}^*$ & 0.926 & 0.279 & $u_{2,16}^*$ & 0.947 & 0.298 & $u_{3,32}^*$ & 0.944 & 0.296 \\
			&$d_1^{*2}$
			& 0.935   &114.765  & $d_2^{*2}$    & 0.939  & 8.883 & $d_3^{*2}$     & 0.938  &  2.892 \\
			\hline
			\end{tabular}
			 }
\end{table}

\begin{table}[t]\small
	\centering
	\caption{The list of 20 selected responses for the real data application in Section \ref{new.Sec.6}.}\label{table4}
	\smallskip
	\begin{tabular}{ll}
		\hline	
		Variable&  Description  \\
		\hline
		RPI &  Real personal income    \\
		INDPRO & Total industrial production 
		\\
		CUMFNS & Capacity utilization: manufacturing    \\
		UNRATE &  Civilian unemployment rate    \\
		PAYEMS & Total number of employees on non-agricultural payrolls  \\
		CES0600000007&Average weekly hours: goods-producing \\
		HOUST & Total housing starts \\
		DPCERA3M086SBEA& Real personal consumption expenditures \\
		NAPMNOI & ISM manufacturing: new orders index\\
		CMRMTSPLx& Real manufacturing and trade industries sales\\
		FEDFUNDS & Effective federal funds rate  \\
		T1YFFM & 1-Year treasury constant maturity minus FEDFUNDS \\
		T10YFFM & 10-Year treasury constant maturity minus FEDFUNDS \\
		BAAFFM & Moody's baa corporate bond minus FEDFUNDS \\
		EXUSUKx &U.S.-U.K. exchange rate \\
		WPSFD49207
		& Producer price index for finished goods \\
		PPICMM &  Producer price index for commodities \\
		CPIAUCSL & Consumer price index for all items \\
		PCEPI & Personal consumption expenditure implicit price deflator \\
		S\&P 500 &  S\&P's common stock price index: composite \\
		\hline
	\end{tabular} 
\end{table}
\clearpage

\section{Additional real data results for the federal reserve economic data} \label{appsec:D}

\begin{table}[tp]
	\centering
	\caption{Prediction errors of different methods for the real data application in Section \ref{new.Sec.6}.}\label{tablereal01}
	\smallskip
	\begin{tabular}{lccccc}
		\hline	
		& SOFAR-L &  SOFAR-GL &RRR& RRSVD & SRRR \\
		\hline
		Prediction error& 0.921 &  0.935 &1.775& 1.002 & 0.968 \\
		\hline
	\end{tabular} 
\end{table}

We provide in Table \ref{table4} above the list of 20 selected responses along with their descriptions for the real data application in Section \ref{new.Sec.6}. In addition, we show the prediction performance of different methods on this data set based on the multi-response regression model \eqref{model}. Specifically, we consider the SOFAR estimator \citep{uematsu2017sofar} with the entrywise $L_1$-norm penalty (SOFAR-L) or the rowwise $(2,1)$-norm penalty (SOFAR-GL), reduced rank regression (RRR), reduced rank regression with sparse SVD (RSSVD) \cite{chen2012reduced}, and sparse reduced rank regression (SRRR) \citep{chen2012sparse}. Specifically, we treat the first $474$ observations as the training sample, identify the rank of the multi-response regression model in the same fashion as in Section \ref{new.Sec.5}, and fit the model using each of those five methods. The prediction error $\norm{\Y - \X\widehat{\C}}_F^2/(n_1 q)$  is calculated based on the test sample consisting of the remaining $n_1 = 180$ observations. 
Table \ref{tablereal01} reports the prediction errors for all the methods. We see from Table \ref{tablereal01} that the sparse learning methods tend to have much better prediction performance than the nonsparse learning approach of reduced rank regression. In particular, SOFAR-L enjoys the highest prediction accuracy followed closely by SOFAR-GL, which indicates that the latent sparse SVD structure assumed in SOFAR provides a better approximation to the true underlying data structure.

\section{Application to the yeast eQTL data}\label{appsec:yeast}
We also demonstrate the effectiveness of our SOFARI method by analyzing a yeast expression quantitative trait loci (eQTL) data set described by \cite{brem2005landscape}, previously studied in \cite{uematsu2017sofar}. In this eQTL data analysis, the primary goal is to investigate the associations between eQTLs, i.e., genomic regions harboring DNA sequence variants, and the expression levels of genes within specific signaling pathways. This data set originally contains $n = 112$ samples with $2957$ genetic markers and $6216$ genes. Based on the SOFAR estimator, our SOFARI procedure can further evaluate the feature importance in the latent SVD structure. So we preprocess the data following the same procedure as that in \cite{uematsu2017sofar}. Specifically, a marginal screening is performed to obtain $p = 605$ genetic markers and we focus on $q = 54$ genes belonging to the yeast mitogen-activated protein kinases signaling pathway.

Since SOFAR-L estimate demonstrated nice performance on this data set in \cite{uematsu2017sofar}, we first implement SOFAR-L to fit the multi-response regression model \eqref{model} and then apply our SOFARI procedure similarly as in Section \ref{new.Sec.6} to conduct inference at  significance level $\alpha = 0.05$. First, the rank of model \eqref{model} is estimated as $3$.  We then summarize the estimated squared singular values and the lengths of the corresponding 95\% confidence intervals in Table \ref{table:app:d2}. We can see from Table \ref{table:app:d2} that the first three latent sparse SVD layers are significant and the first squared singular value is substantially larger than the remaining ones, indicating that the leading component captures the majority of the overall variation in the data. Furthermore, we apply our SOFARI procedure to obtain individual p-values of the compositions of the three latent left factor vectors and then use the BHq procedure to control the FDR at the $5\%$ level for each layer. Consequently, the numbers of significant features in the three layers are 74, 18, and 15, respectively, resulting in a total of 107 nonzeros in the left factor matrix and 101 distinct markers. It reveals certain sparsity patterns across these layers.




To further investigate the prediction aspect of SOFARI, we randomly split the data set into 80\% for training (418 samples) and the remaining 20\% for testing (105 samples). This splitting process is repeated 100 times. For each split, we first apply the SOFARI method and integrate the BHq procedure to identify nonzero elements of each left singular vector with the target FDR level of 5\%. Then we refit the training data using SOFAR-L by constraining on the set of identified nonzero elements and evaluate predictive accuracy through the corresponding test set. For comparison, we also directly estimate the coefficient matrix via SOFAR-L on the training set and compute the prediction loss on the test set. The resulting average prediction errors, with standard errors in parentheses, are 0.734 (0.005) and 0.742 (0.006), respectively, indicating that refitting the data based on our SOFARI method (i.e., the former one) can enhance the predictive performance.

\begin{table}[tp]
	\centering
	\caption{Estimated squared singular values and the lengths of the corresponding 95\% confidence intervals for the yeast eQTL data application in Section \ref{appsec:yeast}.}\label{table:app:d2}
	\smallskip
	\begin{tabular}{lccc}
		\hline	
		& $\wh{d}_1^{2}$ & $\wh{d}_2^{2}$ & $\wh{d}_3^{2}$\\
		\hline
		Value  &  25.869 & 3.946 &2.304 \\
		Len & 2.400 & 0.841 & 0.555\\
		\hline
	\end{tabular} 
\end{table}


\section{Sample splitting techniques}\label{app:sec:split}
In this section, we gain some insights into how the sample splitting techniques can be incorporated into the debiased procedure to weaken the sparsity constraints. 
Although the sample splitting technique can help weaken the sparsity constraints, this approach also possesses certain limitations. In particular, random partitioning can introduce variability, and larger sample sizes are typically needed to ensure efficiency.
Thus, for the sake of simplicity, we only present the theoretical results of the inference on $\u^*_k$ in strongly orthogonal factors cases. Nevertheless, it is noteworthy that the generalization of inference on $\u^*_k$ in weakly orthogonal factors cases and inference on $d^{*2}_k$ under both strongly and weakly orthogonal factors cases follows a similar analysis.

For simplicity, we assume that sample size is $2n$ and the sample is randomly split into two of equal size. The second fold of data is used to obtain the initial SOFAR estimates $ \widetilde{\C}$ and the corresponding SVD components $\{\wt{\u}_i, \wt{\v}_i\}_{i=1}^{r^*}$. Meanwhile, the first fold of data, still denoted as $(\X, \Y)$, is employed to compute the approximate inverse $\wh{\bTheta}$ and the debiased estimator $\wh{\u}_k^{\operatorname{split}}$ for each layer.
Then in the strongly orthogonal factors case, similar to \eqref{debes}, our debiased estimate for $\u_k^*$ can be defined as 
\begin{align}
	\wh{\u}_k^{\operatorname{split}} &  = \widetilde{\u}_k - \wt{\W}_k\wt{\psi}_k(\widetilde{\u}_k,\widetilde{\boldeta}_k) \nonumber\\
	&= \widetilde{\u}_k - \wt{\W}_k\Big(\der{L}{\u_k} - \wt{\mathbf{M}}^{(k)}\der{L}{\boldeta_k}\Big)\Big|_{(\widetilde{\u}_k,\widetilde{\boldeta}_k)}, \nonumber
\end{align}
where $\wt{\mathbf{M}}^{(k)}=\big[\0_{p \times q(k-1) }, \wt{\mathbf{M}}_{k}, \0_{p \times [q(r^* - k) + p(r^*-1) ]}  \big]$ with $\wt{\M}_{k} = -\wt{z}_{kk}^{-1}\wh{\bSigma}\wt{\C}_{-k}$, and $\wt{\W}_k$ are as given in Propositions \ref{prop:rankr2} and \ref{prop:rankr3} after plugging in the SOFAR estimates for the SVD components derived from the second fold of data.

Denote by 
\begin{align}
	\kappa_n^{\prime}  &= \max\{s_{\max}^{1/2}, (r^*+s_u+s_v)^{1/2}\eta_n^2\} (r^*+s_u+s_v)^{1/2}\eta_n^2 \log(pq)/\sqrt{n}. \label{kappa:ds} 
\end{align}
The following theorem shows that the debiased estimator $\wh{\u}_k^{\operatorname{split}}$ based on the sample splitting technique enjoys the asymptotic normality.
\begin{theorem}\label{corollary1}
	Assume that Conditions \ref{cone}--\ref{con:nearlyorth} hold, and $\wh{\bTheta}$ and $\wt{\C}$ satisfy Definitions \ref{defi2:acceptable} and \ref{lemmsofar}, respectively. Then for each given $k$ with $1 \leq k \leq r^*$ and an arbitrary vector $\a\in\mathcal{A}=\{\a\in\R^p:\norm{\a}_0\leq m,\norm{\a}_2=1\}$ satisfying $m^{1/2}\kappa_n^{\prime} = o(1)$, we have 
	\begin{align*}
		\sqrt{n}\a\trans(\wh{\u}_k^{\operatorname{split}} -\u_k^*) = h_k + t_k, 
	\end{align*}
	where the distribution term $h_k = \a\trans \W^{*}_k(\X\trans\E\v_k^* - \M_{k}^{*} \E\trans\X {\u}_k^{*})/\sqrt{n} \sim \N(0,\nu_k^2)$ with 
	\begin{align*}
    \nu_k^2 = \a\trans\W_k^*(z_{kk}^{*}\M_{k}^{*}\bSigma_e\M_{k}\strans + \v_k\strans\bSigma_e\v_k^* \wh{\bSigma} - 2\wh{\bSigma}\u_k^*  \v_k\strans\bSigma_e\M_{k}^{* T})\W_k\strans\a.
	\end{align*}
    Moreover, the bias term 
	$t_k =  O_p\big(m^{1/2}\kappa_n^{\prime} \big)$ holds with probability at least $1 - \theta_{n,p,q}$,
	where 
		\begin{equation}\label{thetapro77}
	    \theta_{n,p,q}^{\prime \prime} = \theta_{n,p,q}^{\prime} +  2 (pq)^{-cs_0}
	\end{equation}
    with $s_0 = \max\{s_{\max}, r^* + s_u + s_v\}$, $\theta_{n,p,q}^{\prime}$ given in Definition \ref{lemmsofar}, and some positive constant $c$.
\end{theorem}
Theorem \ref{corollary1} above establishes the asymptotic normality results of each $\u_k^*$ in the strongly orthogonal factors case. By exploiting the sample splitting technique, we can see that the requirement for the bias term $t_k$ to be asymptotic vanishing is weaker than that in Theorem \ref{theorkr}. To have an explicit view of this, recall that $ \gamma_n = ({r^*}+s_u+s_v)^{1/2}\eta_n^2\{n^{-1}\log(pq)\}^{1/2}$ is the SOFAR estimation rate. Then the error rate $\kappa_n$ in Theorem \ref{theorkr} can be written as 
\begin{align} 
	\kappa_n = \max\{s_{\max}^{1/2}\eta_n^{-2} , (r^*+s_u+s_v)^{1/2}\eta_n^{-2}, 1\} \gamma_n^2/\sqrt{n}, \nonumber
\end{align}
whereas the error rate $\kappa_n^{\prime}$ given in \eqref{kappa:ds} can be formulated as 
\begin{align}
	\kappa_n^{\prime} = \max\{s_{\max}^{1/2} (r^*+s_u+s_v)^{-1/2}\eta_n^{-2}, \eta_n^{-2}, 1 \} \gamma_n^2/\sqrt{n}. \nonumber 
\end{align}
Since $\eta_n^2$ given in Definition \ref{lemmsofar} is larger than $1$ and can even diverge with $n$, the first two term in $\kappa_n$ are reduced by a factor $(r^*+s_u+s_v)^{-1/2}$ in the corresponding terms of $\kappa_n^{\prime}$. In addition, note that the third term in both $\kappa_n$ and $\kappa_n^{\prime}$ is $\gamma_n^2/\sqrt{n}$, which is the inherent bias induced by the SOFAR initial estimates. Thus, if the first two terms in $\kappa_n$ dominate, the sample splitting technique can help weaken the sparsity constraints.

\section{Proofs of Theorems \ref{theorkr}--\ref{corollary1} and Propositions \ref{prop:deri2}--\ref{prop:rankapo3}} \label{new.Sec.A}

\subsection{Proof of Theorem \ref{theorkr}} \label{new.Sec.A.1}

The proof of Theorem \ref{theorkr} consists of two parts. The first part establishes the theoretical results under the rank-2 case, while the second part further extends the results to the general rank case. Let us denote by $\mathcal{E}_0$ the event on which the inequalities in Definition \ref{lemmsofar} hold. By Definition \ref{lemmsofar}, its probability is at least $1 - \theta_{n,p,q}^{\prime}$. Moreover,
we define $\mathcal{E}_1 = \{ n^{-1}\norm{\X\trans\E}_{\max} \leq c_1[n^{-1}\log(pq)]^{1/2} \}$, where $c_1$ is some positive constant.
Since $\E\sim\N(\0,\I_n\otimes\bSigma_e)$ under Condition \ref{cone}, by the same argument as in Step 2 of the proof of Theorem 1 in \cite{uematsu2017sofar}, we know that event $\mathcal{E}_1$ holds  with probability at least $1- 2(pq)^{1-c_0^2/2}$, where $c_0 > \sqrt{2}$ is some positive constant.  Then we have that event $\mathcal{E} = \mathcal{E}_0 \cap \mathcal{E}_1$ holds with probability at least $1- \theta_{n,p,q}$ with $\theta_{n,p,q} = \theta_{n,p,q}^{\prime} + 2(pq)^{1-c_0^2/2}$. To ease the technical presentation, we will condition on event $\mathcal{E}$ throughout the proof. 


\bigskip

\noindent\textbf{Part 1: Proof for the rank-2 case.} Under the strongly orthogonal factors, since the technical arguments for the theoretical results of $\u_1^*$ and $\u_2^*$ are basically the same, we present the proof only for $\u_1^*$ here for simplicity.
In the case that $r^* = 2$ and $k = 1$, for $\M$ and $\W$ constructed in \eqref{m2cons} and \eqref{w2const} in the proofs of Propositions \ref{prop:rankr2} and \ref{prop:rankr3}, after plugging the initial SOFAR estimates satisfying Definition \ref{lemmsofar} we can obtain that 
\begin{align*}
&\wt{\M}_1 = -\wt{z}_{11}^{-1}\wh{\bSigma}\widetilde{\u}_2\widetilde{\v}_2\trans \ \text{ and } \ \wt{\W}_1 =  \widehat{\bTheta}\{\I_p + (\wt{z}_{11} - \wt{z}_{22})^{-1}\wh{\bSigma}\wt{\u}_2\wt{\u}_2\trans\},
\end{align*}
where $\wt{z}_{11} = \wt{\u}_1\trans\wh{\bSigma}\wt{\u}_1$ and $\wt{z}_{22} = \wt{\u}_2\trans\wh{\bSigma}\wt{\u}_2.$
By Lemma \ref{lemmzzz} in Section \ref{new.Sec.B.4}, it holds that $| \wt{z}_{11} - \wt{z}_{22} | \geq c$, which entails that 
$\wt{z}_{11} \neq \wt{z}_{22}$. 
Then we see that $\wt{\W}_1$ is well-defined.

Let us recall that
\begin{align*}
\wh{\u}_1 & ={\psi}_1(\wt{\u}_1,\wt{\boldeta}_1) = \wt{\u}_1  - \wt{\W}_1\wt{\psi}_1(\wt{\u}_1,\wt{\boldeta}_1) \\
& = \wt{\u}_1  - \wt{\W}_1\wt{\psi}_1(\wt{\u}_1,\boldeta_1^*) + \wt{\W}_1(\wt{\psi}_1(\wt{\u}_1,\boldeta_1^*) -\wt{\psi}_1(\wt{\u}_1,\wt{\boldeta}_1)).
\end{align*}
It follows from Lemma \ref{3-7:prop:2} and the fact that the initial estimates satisfy Definition \ref{lemmsofar} that 
\begin{align*}
\wt{\u}_1 - \wt{\W}_1\wt{\psi}_1(\wt{\u}_1,\boldeta^*_1) & = \u_1^* - \wt{\W}_1\wt{\bepsilon}_1 - \wt{\W}_1\wt{\bdelta}_1  - \wt{\W}_1 \wt{\M}_1\v_2^* \u_2\strans\wh{\bSigma}\u_1^* \\
&\quad + \left[\I_p - \wt{\W}_1(\I_p - \wt{\M}_1\wt{\v}_1\wt{\u}_1\trans+\wt{\M}_1\wt{\v}_2\wt{\u}_2\trans)\wh{\bSigma}\right] (\wt{\u}_1-\u_1^*),
\end{align*}
where
\begin{align}
&\wt{\bepsilon}_1 
=  n^{-1}\wt{\M}_1 \E\trans\X\wt{\u}_1  - n^{-1}\X\trans\E\v_1^*,   \label{epeq1}\\[5pt]
&\wt{\bdelta}_1 = \wt{\M}_1\left\{(\wt{\v}_1 - \v_1^*)\wt{\u}_1\trans - (\wt{\v}_2\wt{\u}_2\trans - \v_2^*\u_2\strans)\right\}\wh{\bSigma}(\wt{\u}_1 - \u_1^*). \label{deltaeq1}
\end{align}
In view of Definition \ref{lemmsofar},  $\wt{\v}_2\trans \wt{\v}_1 = 0$, and $\wt{\v}_2\trans \wt{\v}_2 = 1$, for each given $\a\in\mathcal{A}=\{\a\in\R^p:\norm{\a}_0\leq m,\norm{\a}_2=1\}$, we can represent $\sqrt{n}\a\trans(\wh{\u}_1-\u_1^*) $ as 
\begin{align}\label{eqafinal}
\sqrt{n}\a\trans(\wh{\u}_1-\u_1^*)  =  &-\sqrt{n} \a\trans\wt{\W}_1\wt{\bepsilon}_1 -\sqrt{n} \a\trans\wt{\W}_1\wt{\bdelta}_1 + \sqrt{n}\a\trans\big(\I_p - \wt{\W}_1\T_1\big)(\wt{\u}_1-\u_1^*) \nonumber\\
&- \sqrt{n}\a\trans\wt{\W}_1\wt{\M}_1\v_2^*\u_2\strans\wh{\bSigma}\u_1^* - \sqrt{n}\a\trans\wt{\W}_1( \wt{\psi}_1(\wt{\u}_1,\wt{\boldeta}_1) - \wt{\psi}_1(\wt{\u}_1,\boldeta^*_1)),
\end{align}
where
$\T_1 =(\I_p - \wt{z}_{11}^{-1}\wh{\bSigma}\wt{\u}_2\wt{\u}_2\trans )\wh{\bSigma}.$
We will show that the last four terms on the right-hand side of \eqref{eqafinal} above are asymptotically vanishing for $\wh{\bTheta}$ satisfying Definition \ref{defi2:acceptable} and $\wt{\C}$ satisfying Definition \ref{lemmsofar}.

First, under Conditions \ref{con3}--\ref{con:nearlyorth},   by Lemma \ref{lemma:1rk2} in Section \ref{new.Sec.B.6} we have that 
\begin{align}\label{eq:th:1rk2}
|\a\trans\wt{\W}_1 \wt{\bdelta}_1 | \leq c m^{1/2}(r^*+s_u+s_v)  \eta^4_n \{n^{-1}\log(pq)\}.
\end{align}
Second, for term $\a\trans(\I_p -\wt{\W}_1\T_1)(\wt{\u}_1-\u_1^*)$,
it follows from \eqref{w3const} in the proof of Proposition \ref{prop:rankr3} that $\I_p - \wt{\W}_1\T_1 = \I_p - \wh{\bTheta}\wh{\bSigma}.$ Since  $\wh{\bTheta}$ is an acceptable estimator satisfying Definition \ref{defi2:acceptable}, it holds that
\begin{align}
	\norm{\I_p - \wh{\bTheta}\wh{\bSigma}}_{\max}\leq c\{n^{-1}\log(pq)\}^{1/2}. \nonumber
\end{align}
In addition, by Definition \ref{lemmsofar} it follows that 
\begin{align*}
	&\norm{\wt{\u}_1-\u_1^*}_0 \leq \norm{\wt{\U}-\U^*}_0 \leq c(r^*+ s_u+s_v), \\
	&\norm{\wt{\u}_1-\u_1^*}_2 \leq \norm{\wt{\U}-\U^*}_F \leq c(r^*+ s_u+s_v)^{1/2}\eta_n^2\{n^{-1}\log(pq)\}^{1/2}.
\end{align*}
Then we can deduce that 
\begin{align}\label{eq:th:4rk2}
\abs{\a\trans(\I_p-\wt{\W}_1\T_1)(\wt{\u}_1-\u_1^*)}
& \leq \norm{\a}_1\norm{(\I_p - \wh{\bTheta}\wh{\bSigma})(\wt{\u}_1-\u_1^*)}_{\max} \nonumber     \\
& \leq c m^{1/2}\norm{\I_p - \wh{\bTheta}\wh{\bSigma}}_{\max}\norm{\wt{\u}_1-\u_1^*}_1 \nonumber \\
&\leq c m^{1/2}\norm{\I_p - \wh{\bTheta}\wh{\bSigma}}_{\max}\norm{\wt{\u}_1-\u_1^*}_0^{1/2}\norm{\wt{\u}_1-\u_1^*}_2 \nonumber \\ 
&\leq cm^{1/2}(r^*+s_u+s_v)\eta_n^2\{n^{-1}\log(pq)\}.
\end{align}	

Third, under Conditions \ref{con3}--\ref{con:nearlyorth}, an application of Lemma \ref{lemma:gap} in Section \ref{new.Sec.B.7} yields that
\begin{align}\label{fdsfsad}
	|\a\trans\wt{\W}_1\wt{\M}_1\v_2^*\u_2\strans\wh{\bSigma}\u_1^*| =  o( m^{1/2} n^{-1/2}).
\end{align}
Furthermore, for the last term $\a\trans\wt{\W}_1(\wt{\psi}_1(\wt{\u}_1,\boldeta^*_1) -\wt{\psi}_1(\wt{\u}_1,\wt{\boldeta}_1))$, under Conditions \ref{con3}--\ref{con:nearlyorth}, by Lemma \ref{prop:taylor12} in Section \ref{new.Sec.B.2} it holds that 
\begin{align}\label{eq:th:1rk2tay}
|& \a\trans\wt{\W}_1(\wt{\psi}_1(\wt{\u}_1,\boldeta^*_1) -\wt{\psi}_1(\wt{\u}_1,\wt{\boldeta}_1)) | \nonumber \\
& \leq  c\max\{s_{\max}^{1/2}, (r^*+s_u+s_v)^{1/2}, \eta_n^2 \}(r^*+s_u+s_v)\eta_n^2\{n^{-1}\log(pq)\}.
\end{align}
Thus, combining \eqref{eqafinal}--\eqref{eq:th:1rk2tay} leads to
\begin{align}
\sqrt{n}\a\trans(\wh{\u}_1-\u_1^*)  =  - \sqrt{n}\a\trans\wt{\W}_1\wt{\bepsilon}_1 + t^{\prime}, \nonumber
\end{align}
where $t^{\prime} = O\left[m^{1/2}\{s_{\max}^{1/2}, (r^*+s_u+s_v)^{1/2}, \eta_n^2 \}(r^*+s_u+s_v)\eta_n^2\log(pq)/\sqrt{n}\right]$.

We now proceed with analyzing term $- \sqrt{n}\a\trans\wt{\W}_1\wt{\bepsilon}_1$. Let us define
\begin{align}\label{eqh1}
h_1 =
-\a\trans \W^{*}_1 \M_{1}^* \E\trans\X {\u}_1^{*} / \sqrt{n}
+ \a\trans\W^*_1\X\trans\E\v_1^*/ \sqrt{n},
\end{align}
where
$\W^{*}_1 =   \widehat{\bTheta}\{\I_p + (z_{11}^{*} - z_{22}^{*})^{-1}\wh{\bSigma}{\u}_2^{*}{\u}_2\strans\}$, $\M_{1}^* = -z_{11}^{*-1}\wh{\bSigma}{\u}_2^{*}{\v}_2\strans$,  $z_{11}^{*} = {\u}_1\strans\wh{\bSigma}{\u}_1^{*}$, and $z_{22}^{*} = {\u}_2\strans\wh{\bSigma}{\u}_2^{*}$.
Under Conditions \ref{con3}--\ref{con:nearlyorth}, it follows from Lemma \ref{lemmzzz} in Section \ref{new.Sec.B.4} that $|{z}_{11}^* - {z}_{22}^* | \geq c$, which implies that ${z}_{11}^* \neq {z}_{22}^*$. Then we can see that $\W_1^*$ is well-defined. Moreover, under Conditions \ref{con3}--\ref{con:nearlyorth}, by Lemma \ref{lemma:1rk3} in Section \ref{new.Sec.B.8}, we have that 
\begin{align*}
\abs{-\sqrt{n}\a\trans\wt{\W}_1\wt{\bepsilon}_1 - h_1} \leq c m^{1/2}(r^*+s_u + s_v)^{3/2}\eta_n^2\log(pq)/\sqrt{n} .
\end{align*}
Hence, $\sqrt{n}\a\trans(\wh{\u}_1-\u_1^*)$ can be rewritten  as
\begin{align*}
\sqrt{n}\a\trans(\wh{\u}_1-\u_1^*) = h_1 + t_1,
\end{align*}
where  $t_1=O\left[m^{1/2}\{s_{\max}^{1/2}, (r^*+s_u+s_v)^{1/2}, \eta_n^2 \}(r^*+s_u+s_v)\eta_n^2\log(pq)/\sqrt{n}\right]$.

Finally, we will investigate the distribution of $h_1$.
For the sake of clarity, denote by 
\begin{align*}
& \balpha_1 = \X\u_1^*, ~ \bbeta_1  = -\M_{1}\strans \W_1\strans\a / \sqrt{n}, \\[5pt]
& \balpha_2 = \X\W_1\strans\a / \sqrt{n}, ~ \bbeta_2 = \v_1^*.
\end{align*}
Observe that all of them are independent of $\E$.
Then we can rewrite $h_1$ as
\begin{align}\label{eqh2}
h_1 & = \balpha_1\trans\E\bbeta_1 + \balpha_2\trans\E\bbeta_2 \nonumber\\
&= (\balpha_1\otimes\bbeta_1)\trans\vect{\E} + (\balpha_2\otimes\bbeta_2)\trans\vect{\E},
\end{align}
where $\vect{\E}\in\R^{nq}$ denotes the vectorization of $\E$.

By Condition \ref{cone} that $\E\sim\N(\0,\I_n\otimes\bSigma_e)$, it holds that $h_1$ is normally distributed.
Furthermore, we have that $\mathbb{E}(h_1|\X) = 0$ and variance
\begin{align}\label{eqh3}
\text{var}(h_1|\X)
& = (\balpha_1\otimes\bbeta_1)\trans(\I_n\otimes\bSigma_e)(\balpha_1\otimes\bbeta_1)
+ (\balpha_2\otimes\bbeta_2)\trans(\I_n\otimes\bSigma_e)(\balpha_2\otimes\bbeta_2)   \nonumber          \\
& \quad\quad +2(\balpha_1\otimes\bbeta_1)\trans(\I_n\otimes\bSigma_e)(\balpha_2\otimes\bbeta_2).
\end{align}
After some simplification, we can obtain that 
\begin{align}\label{eqh4}
\text{var}(h_1|\X)
& = \balpha_1\trans\I_n\balpha_1\bbeta_1\trans\bSigma_e\bbeta_1
+ \balpha_2\trans\I_n\balpha_2\bbeta_2\trans\bSigma_e\bbeta_2
+ 2\balpha_1\trans\I_n\balpha_2\bbeta_1\trans\bSigma_e\bbeta_2                    \nonumber      \\
& = \u_1\strans\wh{\bSigma}\u_1^* \cdot \a\trans\W_1^*\M_{1}^*\bSigma_e\M_{1}\strans\W_1\strans\a +\v_1\strans\bSigma_e\v_1^* \cdot \a\trans\W_1^*\wh{\bSigma}\W_1\strans\a  \nonumber\\
& \quad - 2\a\trans\W_1^*\wh{\bSigma} \u_1^* \v_1\strans\bSigma_e\M_{1}\strans\W_1\strans\a,
\end{align}
which completes the proof for the rank-2 case.

\bigskip

\noindent\textbf{Part 2: Extension to the general rank case.} We now extend the results using similar arguments to those in the first part to the inference of $\u_k^*$ for each given $k$ with $1 \leq k \leq r^*$. Note that
\begin{align*}
\wh{\u}_k & ={\psi}_k(\wt{\u}_k,\wt{\boldeta}_k) = \wt{\u}_k  - \wt{\W}_k\wt{\psi}_k(\wt{\u}_k,\wt{\boldeta}_k)\\
&= \wt{\u}_k  - \wt{\W}_k\wt{\psi}_k(\wt{\u}_k,\boldeta^*_k) + \wt{\W}_k(\wt{\psi}_k(\wt{\u}_k,\boldeta^*_k) -\wt{\psi}_k(\wt{\u}_k,\wt{\boldeta}_k)).
\end{align*}
Then by Propositions \ref{prop:rankr2}--\ref{prop:rankr3}, Lemma \ref{prop:rankr1}, and the initial estimates satisfying Definition \ref{lemmsofar}, we can deduce that 
\begin{align}
\sqrt{n}\a\trans(\wh{\u}_k-\u_k^*)  &=  - \sqrt{n}\a\trans\wt{\W}_k\wt{\bepsilon}_k - \sqrt{n}\a\trans\wt{\W}_k\wt{\bdelta}_k -\sqrt{n} \a\trans\wt{\W}_k(\wt{\psi}_k(\wt{\u}_k,\wt{\boldeta}_k) - \wt{\psi}_k(\wt{\u}_k,\boldeta^*_k))  \nonumber\\[5pt]
& \quad + \sqrt{n}\a\trans(\I_p - \wt{\W}_k\mathbf{T}_k)(\wt{\u}_k-\u_k^*)  - \sqrt{n}\a\trans\wt{\W}_k\wt{\M}_k\C_{-k}\strans \wh{\bSigma} \u_k^*, \label{ukrankeq}
\end{align}
where
\begin{align}
&\wt{\bepsilon}_k =  n^{-1}\wt{\M}_k\E\trans\X\wt{\u}_k - n^{-1}\X\trans\E\v_k^*, \label{eprankr} \\[5pt]
&\wt{\bdelta }_k =   \left\{\wt{\M}_k(\wt{\v}_k - \v_k^*)\wt{\u}_k\trans- \wt{\M}_k( \wt{\C}_{-k}\trans - \C_{-k}\strans)\right\}\wh{\bSigma} (\wt{\u}_k - \u_k^*), \label{delrankr}\\[5pt]
&\wt{\W}_k = \widehat{\bTheta} \left\{  \I_p +   \wt{z}_{kk}^{-1}\wh{\bSigma}\wt{\U}_{-k}(\I_{r^*-1} -\wt{z}_{kk}^{-1}\wt{\U}_{-k}\trans\wh{\bSigma} \wt{\U}_{-k})^{-1}\wt{\U}_{-k}\trans\right\}, \label{eqwknear1} \\[5pt]
& \mathbf{T}_k = ( \I_p - \wt{\M}_k\wt{\v}_k\wt{\u}_k\trans + \wt{\M}_k\wt{\C}_{-k}\trans )\wh{\bSigma}, \ \ \wt{\M}_k =  -\wt{z}_{kk}^{-1}\wh{\bSigma}\wt{\C}_{-k}. \nonumber
\end{align}

Let us further define
\begin{align}\label{eqhknear}
h_k = -\a\trans \W^{*}_k \M_{k}^{*} \E\trans\X {\u}_k^{*}/\sqrt{n}
+  \a\trans\W^*_k\X\trans\E\v_k^*/\sqrt{n},
\end{align}
where $ \M_{k}^{*} = -z_{kk}^{*-1}\wh{\bSigma}\C_{-k}^* , ~z_{kk}^{*} = {\u}_k\strans\wh{\bSigma}{\u}_k^{*}$, and 
\begin{align}
& \W_k^{*}=   \widehat{\bTheta} \left\{  \I_p +   z_{kk}^{*-1}\wh{\bSigma}\U_{-k}^*(\I_{r^*-1} -z_{kk}^{*-1}\U_{-k}\strans\wh{\bSigma} \U_{-k}^*)^{-1}\U_{-k}\strans\right\}. \label{eqwknear2}
\end{align}
By Lemma \ref{lemm:wexist} in Section \ref{new.Sec.B.9}, we see that $\wt{\W}_k $ and $\W_k^{*}$  are both well-defined. Then we will bound the terms on the right-hand side of \eqref{ukrankeq} above, which will be conditional on $\wh{\bTheta}$ satisfying Definition \ref{defi2:acceptable} and $\wt{\C}$ satisfying Definition \ref{lemmsofar}.

Observe that $\norm{\a}_0\leq m$ and $\norm{\a}_2=1$.
First, by Proposition \ref{prop:rankr3} it holds that $\I_p - \wt{\W}_k\mathbf{T}_k = \I_p - \wh{\bTheta}\wh{\bSigma}$. Then by the same argument as for \eqref{eq:th:4rk2}, it follows that
\begin{align}\label{h1}
\abs{\a\trans(\I_p -\wt{\W}_k\mathbf{T}_k)(\wt{\u}_k-\u_k^*)} \leq cm^{1/2}(r^*+s_u+s_v)\eta_n^2\{n^{-1}\log(pq)\}.
\end{align}
Under Conditions \ref{con3}--\ref{con:nearlyorth}, an application of Lemma \ref{prop:taylor12} in Section \ref{new.Sec.B.2} gives that 
\begin{align*}
& |\a\trans\wt{\W}_k (\wt{\psi}_k(\wt{\u}_k,\wt{\boldeta}_k) - \wt{\psi}_k(\wt{\u}_k,\boldeta_k^*) )| \\[5pt]
& \leq  cm^{1/2}\{s_{\max}^{1/2}, (r^*+s_u+s_v)^{1/2}, \eta_n^2 \}(r^*+s_u+s_v)\eta_n^2\{n^{-1}\log(pq)\}.
\end{align*}
Moreover,  under Conditions \ref{con3}--\ref{con:nearlyorth}, using Lemmas  \ref{lemma:1rkr}--\ref{lemma:1rk3r} in Sections \ref{new.Sec.B.12}--\ref{new.Sec.B.14}, respectively, we can deduce that 
\begin{align}
&|\a\trans\wt{\W}_k \wt{\bdelta}_k | \leq cm^{1/2}    (r^*+s_u+s_v)\eta_n^4 \{n^{-1}\log(pq)\}, \label{h2}\\[5pt]
&| \a\trans\wt{\W}_k\wt{\M}_k\C_{-k}\strans \wh{\bSigma} \u_k^*| = o(m^{1/2} n^{-1/2}), \label{h3}\\[5pt]
&\abs{-\a\trans\wt{\W}_k\wt{\bepsilon}_k - h_k/ \sqrt{n}} \leq c m^{1/2}   (r^*+s_u + s_v)^{3/2}\eta_n^2\{n^{-1}\log(pq)\}.
\end{align}

Therefore, combining the above results yields that 
\begin{align*}
\sqrt{n}\a\trans(\wh{\u}_k-\u_k^*) = h_k + t_k, 
\end{align*}
where
$ t_k = O\big(m^{1/2}\{s_{\max}^{1/2}, (r^*+s_u+s_v)^{1/2}, \eta_n^2 \}(r^*+s_u+s_v)\eta_n^2\log(pq)/\sqrt{n} \big).$
The distribution of $ h_k$ can be derived using similar arguments as for \eqref{eqh1}--\eqref{eqh4}. Consequently, we can obtain that $h_k$ is normally distributed by Condition \ref{cone}, $\mathbb{E}( h_k|\X) = 0$, and variance
\begin{align*}
\nu_k^2 = \text{var}( h_k|\X) = \a\trans\W_k^*(z_{kk}^{*}\M_{k}^{*}\bSigma_e\M_{k}\strans + \v_k\strans\bSigma_e\v_k^* \wh{\bSigma} - 2\wh{\bSigma}\u_k^*  \v_k\strans\bSigma_e\M_{k}^{* T})\W_k\strans\a.
\end{align*}
This completes the proof of Theorem \ref{theorkr}. 

\subsection{Proof of Theorem \ref{coro:ortho:dk}} \label{d1rkrproof}

Similar to the proof of Theorem \ref{theorkr} in Section \ref{new.Sec.A.1}, the proof of Theorem \ref{coro:ortho:dk} also contains two parts. In particular, the first part establishes the desired results under the rank-2 case, while the second part extends further the results to the general rank case.

\bigskip

\noindent\textbf{Part 1: Proof for the rank-2 case.} With the presence of strongly orthogonal factors, since the technical arguments for the results of $d_1^{*2}$ and $d_2^{*2}$ are rather similar, we will mainly present the proof for $d_1^{*2}$ here for brevity. Recall that
$	\wh{\u}_1 ={\psi}_1(\wt{\u}_1,\wt{\boldeta}_1) = \wt{\u}_1  - \wt{\W}_1\wt{\psi}_1(\wt{\u}_1,\wt{\boldeta}_1),$ where  $\wt{\W}_1 =  \widehat{\bTheta}\{\I_p + (\wt{z}_{11} - \wt{z}_{22})^{-1}\wh{\bSigma}\wt{\u}_2\wt{\u}_2\trans\}.$
By some calculations, we can show that 
\begin{align}\label{singd}
& \norm{\wt{\u}_1}_2^2 - \norm{{\u}_1^*}_2^2 = 2 \u_1\strans (\wt{\u}_1 - \u_1^*) +  (\wt{\u}_1 - \u_1^*)\trans (\wt{\u}_1 - \u_1^*) \nonumber \\[5pt]
& = 2 \u_1\strans (\wt{\u}_1 - \u_1^* -   \wt{\W}_1\wt{\psi}_1(\wt{\u}_1,\wt{\boldeta}_1)) + 2 \u_1\strans \wt{\W}_1\wt{\psi}_1(\wt{\u}_1,\wt{\boldeta}_1)+ \|\wt{\u}_1 - \u_1^*\|_2^2\nonumber \\[5pt]
&=  2 \u_1\strans (\wh{\u}_1 - \u_1^*)   + 2 \wt{\u}_1\trans \wt{\W}_1\wt{\psi}_1(\wt{\u}_1,\wt{\boldeta}_1)  + 2( \u_1^* - \wt{\u}_1)\trans \wt{\W}_1\wt{\psi}_1(\wt{\u}_1,\wt{\boldeta}_1) + \|\wt{\u}_1 - \u_1^*\|_2^2.
\end{align}
To show that $\wh{d}_1^2 =  \norm{\wt{\u}_1}_2^2 - 2 \wt{\u}_1\trans \wt{\W}_1\wt{\psi}_1(\wt{\u}_1,\wt{\boldeta}_1)$ is a valid debiased estimate of $d_1^{*2} = \norm{{\u}_1^*}_2^2$, we will first prove that $ 2( \u_1^* - \wt{\u}_1)\trans \wt{\W}_1\wt{\psi}_1(\wt{\u}_1,\wt{\boldeta}_1) + \|\wt{\u}_1 - \u_1^*\|_2^2$ is asymptotically negligible and then establish that $ 2 \u_1\strans (\wh{\u}_1 - \u_1^*)$ is asymptotically normal.

For simplicity,
denote by $\brho =  \u_1^* - \wt{\u}_1$.
Observe that
\begin{align*}
	\brho\trans \wt{\W}_1\wt{\psi}_1(\wt{\u}_1,\wt{\boldeta}_1) &= \brho\trans \wt{\W}_1\wt{\psi}_1(\wt{\u}_1,\boldeta^*_1)
	+ \brho\trans \wt{\W}_1( \wt{\psi}_1(\wt{\u}_1,\wt{\boldeta}_1) - \wt{\psi}_1(\wt{\u}_1,\boldeta^*_1)).
\end{align*}
In light of Lemma \ref{3-7:prop:2}, after plugging in the initial estimates we can obtain that 
\begin{align}
	\brho\trans&\wt{\W}_1\wt{\psi}_1(\wt{\u}_1,\boldeta^*_1) =  \brho\trans\wt{\W}_1\wt{\bepsilon}_1 + \brho\trans\wt{\W}_1\wt{\bdelta}_1 - \brho\trans\wt{\W}_1 \mathbf{T}_1(\u_1^* - \wt{\u}_1) + \brho\trans\wt{\W}_1 \wt{\M}_1 \v_2^* \u_2\strans\wh{\bSigma}\u_1^* \nonumber \\[5pt]
	&=  \brho\trans\wt{\W}_1\wt{\bepsilon}_1 +  \brho\trans\wt{\W}_1\wt{\bdelta}_1 + \brho\trans(\I_p -\wt{\W}_1 \mathbf{T}_1)\brho - \brho\trans\brho + \brho\trans\wt{\W}_1 \wt{\M}_1 \v_2^* \u_2\strans\wh{\bSigma}\u_1^*, \label{cxxzd}
\end{align}
where $\wt{\M}_1 =  -\wt{z}_{11}^{-1}\wh{\bSigma}\wt{\u}_2\wt{\v}_2\trans$,  $\T_1 =(\I_p - \wt{z}_{11}^{-1}\wh{\bSigma}\wt{\u}_2\wt{\u}_2\trans )\wh{\bSigma}$, and
\begin{align*}
	&\wt{\bepsilon}_1 =  n^{-1}\wt{\M}_1 \E\trans\X\wt{\u}_1  - n^{-1}\X\trans\E\v_1^*, \\[5pt]
	&\wt{\bdelta}_1 = \wt{\M}_1 \left\{(\wt{\v}_1 - \v_1^*)\wt{\u}_1\trans - (\wt{\v}_2\wt{\u}_2\trans - \v_2^*\u_2\strans)\right\}  \wh{\bSigma}(\wt{\u}_1 - \u_1^*).
\end{align*}
We will then show that $\brho\trans \wt{\W}_1( \wt{\psi}_1(\wt{\u}_1,\wt{\boldeta}_1) - \wt{\psi}_1(\wt{\u}_1,\boldeta^*_1))$ and all the terms on the right-hand side of \eqref{cxxzd} above are asymptotically vanishing, which will be conditional on $\wh{\bTheta}$ satisfying Definition \ref{defi2:acceptable} and $\wt{\C}$ satisfying Definition \ref{lemmsofar}.

First, by Definition \ref{lemmsofar} it holds that 
\begin{align}\label{brho}
	\norm{\brho}_0 \leq \norm{\wt{\U}-\U^*}_0 \leq c(r^* + s_u + s_v) \ \text{ and } \   \norm{\brho}_2 \leq \norm{\wt{\U}-\U^*}_F \leq c \gamma_n,
\end{align}
where $ \gamma_n = (r^*+s_u+s_v)^{1/2}\eta_{n}^2\{n^{-1}\log(pq)\}^{1/2}$.
Then it is immediate to see that
\begin{align}
	\|\brho\|_2^2 \leq c(r^*+s_u+s_v) \eta_{n}^4 \{n^{-1}\log(pq)\}. \label{cxxzd6}
\end{align}
Under Conditions \ref{con3}--\ref{con:nearlyorth}, from \eqref{brho} and Lemmas \ref{prop:taylor12} and \ref{lemma:1rk2}--\ref{lemma:gap} in Sections \ref{new.Sec.B.2} and \ref{new.Sec.B.6}--\ref{new.Sec.B.7}, respectively, we can deduce that 
\begin{align}
	&|\brho\trans\wt{\W}_1(\wt{\psi}_1(\wt{\u}_1,\boldeta^*_1) -\wt{\psi}_1(\wt{\u}_1,\wt{\boldeta}_1)) | \nonumber  \\[5pt]
	& \quad \leq  c\max\{s_{\max}^{1/2}, (r^*+s_u+s_v)^{1/2}, \eta_n^2\}(r^*+s_u+s_v)^2\eta_n^4\{n^{-1}\log(pq)\}^{3/2}, \label{cxxzd2}\\[5pt]
		&	|\brho\trans{\wt{\W}_1} \wt{\bdelta}_1 |  \leq c (r^*+s_u+s_v)^2  \eta^6_n \{n^{-1}\log(pq)\}^{3/2}, \label{cxxzd4}\\[5pt]
	&  	| \brho\trans\wt{\W}_1 \wt{\M}_1 \v_2^* \u_2\strans\wh{\bSigma}\u_1^*|
	=  o((r^*+s_u+s_v)\eta_{n}^2\{\log(pq)\}^{1/2}n^{-1}). \label{cxxzd3}
\end{align}
Moreover, similar to \eqref{eq:th:4rk2}, it can be shown that
\begin{align}\label{cxxzd5}
	\abs{\brho\trans(\I_p -\wt{\W}_1\T_1)\brho}
	& \leq \norm{\brho}_1\norm{(\I_p - \wh{\bTheta}\wh{\bSigma})\brho}_{\max} \nonumber     \\
	& \leq \norm{\brho}_0^{1/2} \norm{\brho}_2 \norm{\I_p - \wh{\bTheta}\wh{\bSigma}}_{\max}\norm{\brho}_0^{1/2} \norm{\brho}_2  \nonumber \\
	&\leq c(r^*+s_u+s_v)^2\eta_n^4\{n^{-1}\log(pq)\}^{3/2}.
\end{align}	

It remains to bound term $\brho\trans\wt{\W}_1\wt{\bepsilon}_1$.
We see that 
\begin{equation}
	|\brho\trans\wt{\W}_1\wt{\bepsilon}_1|  \leq | n^{-1}\brho\trans\wt{\W}_1 \wt{\M}_1 \E\trans\X\wt{\u}_1 | +  | n^{-1}\brho\trans\wt{\W}_1\X\trans\E\v_1^*|.  \label{czxcaw}
\end{equation}
Let us bound the first term $| n^{-1}\brho\trans\wt{\W}_1 \wt{\M}_1 \E\trans\X\wt{\u}_1 |$ on the right-hand side of (\ref{czxcaw}) above. Note that Condition \ref{con:nearlyorth} entails that the nonzero eigenvalues $d^{*2}_{i}$ are at the constant level. Then under Conditions \ref{con3}--\ref{con:nearlyorth}, by parts (b) and (c) of Lemma \ref{lemmauv} in Section \ref{new.Sec.B.3} and Definition \ref{lemmsofar} with $\norm{\widetilde{\v}_2 }_2 =1$,  we have that 
\begin{align}
	&\norm{\wt{\M}_1}_2 \leq \norm{\wt{z}_{11}^{-1}\wh{\bSigma}\widetilde{\u}_2\widetilde{\v}_2\trans}_2 \leq |\wt{z}_{11}^{-1}| \norm{\wh{\bSigma}\widetilde{\u}_2 }_2 \norm{\widetilde{\v}_2 }_2  \leq c. \label{zxcdqaz}
\end{align}
Further, under Conditions \ref{con3}--\ref{con:nearlyorth}, it follows from \eqref{brho} and part (c) of Lemma \ref{lemma:1rk4} in Section \ref{new.Sec.B.5} that 
\begin{align}
	\norm{\brho\trans\wt{\W}_1}_2 \leq  c(r^*+s_u+s_v)\eta_{n}^2\{n^{-1}\log(pq)\}^{1/2}.  \label{zxcdqaz2}
\end{align}

Denote by $s = c(r^* + s_u + s_v)$.
It is easy to see that
\begin{align*}
	\norm{\brho\trans \wt{\W}_1 \wt{\M}_1}_0 & = \norm{ (\wt{z}_{11}^{-1}\brho\trans\wt{\W}_1\wh{\bSigma}\wt{\l}_2) \cdot \wt{d}_2\wt{\v}_2\trans   }_0 \leq  \norm{ \wt{d}_2\wt{\v}_2   }_0 \\
 & \leq  \norm{ \v_2^* }_0  + \norm{ \wt{d}_2(\wt{\v}_2 - \v_2^*)   }_0 \leq  s,
\end{align*}
where the last step is due to $\norm{ \v_2^* }_0 \leq s_v$ and part (a) of Lemma \ref{lemmauv} in Section \ref{new.Sec.B.3}. Moreover, from Definition \ref{lemmsofar} and $\norm{{\u}_1^*}_0 \leq s_u$, it holds for sufficiently large $n$ that 
\begin{align*}
	&\norm{\wt{\u}_1}_0 \leq \norm{\wt{\u}_1 - {\u}_1^*}_0 + \norm{{\u}_1^*}_0  \leq c(r^* + s_u + s_v), \\ 
&\norm{\wt{\u}_1}_2 \leq \norm{\wt{\u}_1 - {\u}_1^*}_2 + \norm{{\u}_1^*}_2  \leq c.
\end{align*}
With the aid of  $n^{-1}\norm{\X\trans\E}_{\max} \leq c\{n^{-1}\log(pq)\}^{1/2} $, it follows that
\begin{align}
	n^{-1}\norm{\E\trans\X\wt{\u}_1}_{2,s} &\leq s^{1/2} n^{-1}\norm{\E\trans\X\wt{\u}_1 }_{\max}  \leq  s^{1/2} n^{-1}\norm{\E\trans\X }_{\max}\norm{\wt{\u}_1}_0^{1/2}\norm{\wt{\u}_1}_2 \nonumber \\ &
	\leq c (r^*+s_u+s_v)\{n^{-1}\log(pq)\}^{1/2}. \label{sdaqessss}
\end{align}
 Note that here, for an arbitrary vector $\x$, $\norm{\x}_{2,s}^2=\max_{\abs{S}\leq s}\sum_{i\in S}x_i^2$ with $S$ denoting an index set. Hence, combining the above results gives that
\begin{align}
	| n^{-1}\brho\trans\wt{\W}_1 \wt{\M}_1 \E\trans\X\wt{\u}_1 |
	&\leq \|\brho\trans\wt{\W}_1\wt{\M}_1\|_2  \|  n^{-1} \E\trans\X\wt{\u}_1 \|_{2,s} \nonumber \\[5pt]
	&\leq \|\brho\trans\wt{\W}_1\|_2 \|\wt{\M}_1\|_2 \|  n^{-1} \E\trans\X\wt{\u}_1 \|_{2,s} \nonumber \\[5pt]
	&\leq  c   (r^*+s_u+s_v)^{2}\eta_{n}^2\{n^{-1}\log(pq)\}.  \label{zcasdq}
\end{align}

We next bound the second term $| n^{-1}\brho\trans\wt{\W}_1\X\trans\E\v_1^*|$ on the right-hand side of  \eqref{czxcaw} above.
Denote by
$\phi_i = \wt{\w}_i\trans n^{-1}\X\trans\E\v^*_1$, where $\wt{\w}_i\trans$ represents the $i$th row of ${\wt{\W}_1}$.
Then we see that ${ n^{-1}\brho\trans\wt{\W}_1\X\trans\E\v_1^*} = \brho\trans \boldsymbol{\phi}$, where $ \boldsymbol{\phi}=(\phi_i) \in\R^p$.
Under Conditions \ref{con3}--\ref{con:nearlyorth}, by Lemma \ref{lemma:1rk4} in Section \ref{new.Sec.B.5} it holds that
\begin{align*}
	\max_{1 \leq i \leq p} \abs{\phi_i}
	&\leq \max_{1 \leq i \leq p} \norm{ \wt{\w}_i}_1 n^{-1}\norm{\X\trans\E\v^*_1}_{\max}
	\\
 &\leq \max_{1 \leq i \leq p} \norm{\wt{\w}_i}_0^{1/2} \norm{\wt{\w}_i}_2 \norm{n^{-1}\X\trans\E}_{\max}\norm{\v^*_1}_0^{1/2}\norm{\v^*_1}_2 \\[5pt]
	&
	\leq  c s_v^{1/2}  \max\{s_{\max}, (r^*+s_u+s_v)\}^{1/2} \{n^{-1}\log(pq)\}^{1/2}.
\end{align*}
This together with \eqref{brho} entails that 
\begin{align}\label{zcasdq2}
    |n^{-1}\brho\trans&\wt{\W}_1\X\trans\E\v_1^*| = \abs{ \brho\trans\boldsymbol{\phi}} \leq \norm{ \brho}_1\norm{\boldsymbol{\phi}}_{\max} \leq \norm{ \brho}_0^{1/2}\norm{\brho}_2 \norm{\boldsymbol{\phi}}_{\max} \nonumber\\[5pt]
	& \leq c \max\{s_{\max}, (r^*+s_u+s_v)\}^{1/2} (r^*+s_u+s_v)^{3/2} \eta_n^2 \{n^{-1}\log(pq)\}.
\end{align}
Then it follows from \eqref{czxcaw}, \eqref{zcasdq}, and \eqref{zcasdq2} that 
\begin{align}
	|\brho\trans\wt{\W}_1\wt{\bepsilon}_1| \leq  c  \max\{s_{\max}, (r^*+s_u+s_v)\}^{1/2} (r^*+s_u+s_v)^{3/2} \eta_n^2 \{n^{-1}\log(pq)\}. \label{bbgd}
\end{align}
Thus, combining \eqref{cxxzd6}--\eqref{cxxzd5} and \eqref{bbgd} leads to 
\begin{align}
	&\|\wt{\u}_1 - \u_1^*\|_2^2 \leq c(r^*+s_u+s_v) \eta_{n}^4 \{n^{-1}\log(pq)\}, \label{opopwq2} \\[5pt]
	& \big|( \u_1^* - \wt{\u}_1)\trans \wt{\W}_1\wt{\psi}_1(\wt{\u}_1,\wt{\boldeta}_1) \big| \nonumber \\[5pt]
	&\quad \leq c  \max\{s_{\max}^{1/2}, (r^*+s_u+s_v)^{1/2},\eta_{n}^2\} (r^*+s_u+s_v)^{3/2}\eta_{n}^2\{n^{-1}\log(pq)\}. \label{opopwq}
\end{align}

Finally, for \eqref{singd} we can rewrite it as 
\begin{align}\label{distu1}
	\norm{\wt{\u}_1}_2^2 - \norm{{\u}_1^*}_2^2 -  2 \wt{\u}_1\trans &\wt{\W}_1\wt{\psi}_1(\wt{\u}_1,\wt{\boldeta}_1)
	=  2 \u_1\strans (\wh{\u}_1 - \u_1^*)  + t^{\prime},
\end{align}
where
$t^{\prime} = O( \max\{s_{\max}^{1/2}, (r^*+s_u+s_v)^{1/2},\eta_{n}^2\} (r^*+s_u+s_v)^{3/2}\eta_{n}^2\{n^{-1}\log(pq)\} )$. In addition, replacing $\a$ with $ 2\u_1^*$ and using similar arguments as in the first part of the proof of Theorem \ref{theorkr}, we can obtain that 
\begin{align}\label{distu2}
	\sqrt{n} (2 \u_1\strans)(\wh{\u}_1-\u_1^*) = h_{d_1} + t^{\prime \prime}, \   h_{d_1}     \sim \N(0,\nu_{d_1}^2),
\end{align}
where the error term $ t^{\prime \prime} $ satisfies that 
$$  t^{\prime \prime} = O( \max\{s_{\max}^{1/2}, (r^*+s_u+s_v)^{1/2},\eta_{n}^2\}(r^*+s_u+s_v)^{3/2}\eta_n^2\log(pq)/\sqrt{n}), $$
and the distribution term $  h_{d_1}$ and its variance are given by
\begin{align*}
	& h_{d_1} =
	2\u_1\strans {\W}_1^{*}\M_{1}^{*} \E\trans\X {\u}_1^{*} / \sqrt{n}
	+2\u_1\strans{\W}_1^*\X\trans\E\v_1^*/ \sqrt{n}, \\[5pt]
	&\nu_{d_1}^2
	=4 \u_1\strans\wh{\bSigma}\u_1^* \cdot \u_1\strans{\W}_1^*\M_{1}^*\bSigma_e\M_{1}\strans{\W}_1\strans \u_1^* +4\v_1\strans\bSigma_e\v_1^* \cdot\u_1\strans{\W}_1^*\wh{\bSigma}{\W}_1\strans \u_1^* \\
	& \qquad  \quad
	- 8\u_1\strans{\W}_1^*\wh{\bSigma} \u_1^* \cdot \u_1\strans{\W}_1^*\M_{1}^*\bSigma_e\v_1^*,                      
\end{align*}
respectively. Therefore, in view of \eqref{distu1} and \eqref{distu2} we can deduce that 
\begin{align*}
	\sqrt{n}(\norm{\wt{\u}_1}_2^2 - \norm{{\u}_1^*}_2^2 -  2 \wt{\u}_1\trans \wt{\W}_1\wt{\psi}_1(\wt{\u}_1,\wt{\boldeta}_1))
	= h_{d_1} + t_{d_1}
\end{align*}
with $h_{d_1}     \sim \N(0,\nu_{d_1}^2)$ and 
$$  t_{d_1} = t^{\prime } +  t^{\prime \prime}  = O(  \max\{s_{\max}^{1/2}, (r^*+s_u+s_v)^{1/2},\eta_{n}^2\}(r^*+s_u+s_v)^{3/2}\eta_n^2\log(pq)/\sqrt{n}), $$
which completes the proof for the rank-2 case.

\bigskip

\noindent\textbf{Part 2: Extension to the general rank case.} Now we extend the results to deal with the general rank case using similar arguments as in the first part and establish the results on $d_k^{*2}$ for each given $k$ with $1 \leq k \leq r^*$. Observe that
$	\wh{\u}_k ={\psi}_k(\wt{\u}_k,\wt{\boldeta}_k) = \wt{\u}_k  - \wt{\W}_k\wt{\psi}_k(\wt{\u}_k,\wt{\boldeta}_k).$
Similar to \eqref{singd}, it holds that 
\begin{align}
\widehat{d}_k^2 - d_k^{*2} &=  \norm{\wt{\u}_k}_2^2 - 2 \wt{\u}_k\trans \wt{\W}_k\wt{\psi}_k(\wt{\u}_k,\wt{\boldeta}_k) - \norm{{\u}_k^*}_2^2 \nonumber \\[5pt]
&=  2 \u_k\strans (\wh{\u}_k - \u_k^*)
+ 2( \u_k^* - \wt{\u}_k)\trans \wt{\W}_k\wt{\psi}_k(\wt{\u}_k,\wt{\boldeta}_k) + \|\wt{\u}_k - \u_k^*\|_2^2. \nonumber
\end{align}
We will first prove that $ 2( \u_k^* - \wt{\u}_k)\trans \W_k\wt{\psi}(\wt{\u}_k,\wt{\boldeta}_k) + \|\wt{\u}_k - \u_k^*\|_2^2$ is asymptotically negligible and then show that $ 2 \u_k\strans (\wh{\u}_k - \u_k^*)$ is asymptotically normal.

Let us define $\brho_k =  \u_k^* - \wt{\u}_k$. It is easy to see that 
\begin{align*}
\brho_k\trans \wt{\W}_k\wt{\psi}_k(\wt{\u}_k,\wt{\boldeta}_k) &= \brho_k\trans \wt{\W}_k\wt{\psi}_k(\wt{\u}_k,\boldeta^*_k)
+ \brho_k\trans \wt{\W}_k( \wt{\psi}_k(\wt{\u}_k,\wt{\boldeta}_k) - \wt{\psi}_k(\wt{\u}_k,\boldeta^*_k)).
\end{align*}
By Lemma \ref{prop:rankr1}, with the initial estimates we can deduce that 
\begin{align}\label{sdczca}
&\brho_k\trans\wt{\W}_k\wt{\psi}_k(\wt{\u}_k,\boldeta^*_k) =  \brho_k\trans\wt{\W}_k\wt{\bepsilon}_k + \brho_k\trans\wt{\W}_k\wt{\bdelta}_k - \brho_k\trans\wt{\W}_k \mathbf{T}_k(\u_k^* - \wt{\u}_k) + \brho_k\trans\wt{\W}_k \wt{\M}_k\C_{-k}\strans \wh{\bSigma} \u_k^* \nonumber\\[5pt]
&~=  \brho_k\trans\wt{\W}_k\wt{\bepsilon}_k +  \brho_k\trans\wt{\W}_k\wt{\bdelta}_k + \brho_k\trans(\I_p -\wt{\W}_k \mathbf{T}_k)\brho_k - \brho_k\trans\brho_k + \brho_k\trans\wt{\W}_k\wt{\M}_k\C_{-k}\strans \wh{\bSigma} \u_k^*,
\end{align}
where $\wt{\bepsilon}_k, \wt{\bdelta }_k, \wt{\W}_k$ are given in \eqref{eprankr}--\eqref{eqwknear1}, respectively, $\wt{\M}_k =  -\wt{z}_{kk}^{-1}\wh{\bSigma}\wt{\C}_{-k} $, and $\mathbf{T}_k = ( \I_p - \wt{\M}_k\wt{\v}_k\wt{\u}_k\trans + \wt{\M}_k\wt{\C}_{-k}\trans )\wh{\bSigma}$.

Next we aim to show that all the terms on the right-hand side of \eqref{sdczca} above and $\brho_k\trans \wt{\W}_k( \wt{\psi}_k(\wt{\u}_k,\wt{\boldeta}_k) - \wt{\psi}_k(\wt{\u}_k,\boldeta^*_k))$ are asymptotically vanishing. First, similar to \eqref{brho} it follows from Definition \ref{lemmsofar} that 
\begin{align}\label{eqwdrhio}
\norm{\brho_k}_0 \leq c(r^* + s_u + s_v) \ \text{ and } \   \norm{\brho_k}_2 \leq c(r^*+s_u+s_v)^{1/2}\eta_{n}^2\{n^{-1}\log(pq)\}^{1/2}.
\end{align}
Under Conditions \ref{con3}--\ref{con:nearlyorth}, from this and Lemmas \ref{prop:taylor12} and \ref{lemma:1rkr}--\ref{lemma:gapr} in Sections \ref{new.Sec.B.2} and \ref{new.Sec.B.12}--\ref{new.Sec.B.13}, respectively, we can obtain that 
\begin{align*}
&|\brho_k\trans \wt{\W}_k( \wt{\psi}_k(\wt{\u}_k,\wt{\boldeta}_k) - \wt{\psi}_k(\wt{\u}_k,\boldeta^*_k)) | \nonumber \\
&\quad \leq   c\max\{s_{\max}^{1/2}, (r^*+s_u+s_v)^{1/2}, \eta_n^2\}(r^*+s_u+s_v)^2\eta_n^4\{n^{-1}\log(pq)\}^{3/2}, \\[5pt]
&|\brho_k\trans{\wt{\W}_k} \wt{\bdelta}_k | \leq c   (r^*+s_u+s_v)^2  \eta^6_n \{n^{-1}\log(pq)\}^{3/2}, \\[5pt]
&  	|  \brho_k\trans\wt{\W}_k\wt{\M}_k\C_{-k}\strans \wh{\bSigma} \u_k^*|
=  o( (r^*+s_u+s_v)\eta_{n}^2\{\log(pq)\}^{1/2}n^{-1}).
\end{align*}
Further, an application of similar arguments as for \eqref{cxxzd5} yields that 
\begin{align*}
\abs{\brho_k\trans(\I_p -\wt{\W}_k\T_k)\brho_k} \leq c(r^*+s_u+s_v)^2\eta_n^4\{n^{-1}\log(pq)\}^{3/2}.
\end{align*}
It also follows from \eqref{eqwdrhio}  that
\begin{align*}
\|\brho_k\|_2^2 \leq c(r^*+s_u+s_v) \eta_{n}^4 \{n^{-1}\log(pq)\}.
\end{align*}

It remains to examine term $\brho_k\trans\wt{\W}_k\wt{\bepsilon}_k$. Note that 
\begin{align}
|\brho_k\trans\wt{\W}_k\wt{\bepsilon}_k| \leq | n^{-1}\brho_k\trans\wt{\W}_k\wt{\M}_k\E\trans\X\wt{\u}_k| +  | n^{-1}\brho_k\trans\wt{\W}_k\X\trans\E\v_k^*| \label{sadqz}
\end{align}
and $\wt{\M}_k = \wt{z}_{kk}^{-1} \wh{\bSigma}\widetilde{\U}_{-k}\widetilde{\V}_{-k}\trans$.
Under Conditions \ref{con3}--\ref{con:nearlyorth}, by Lemma \ref{rankr:boundm0} in Section \ref{new.Sec.B.10},  we have that 
\begin{align}\label{sadqf}
	\norm{\wt{\M}_k}_2 \leq c.
\end{align}
From part (a) of Lemma \ref{lemmauv} in Section \ref{new.Sec.B.3} and $ \sum_{i =1}^{r^*}  \norm{ {\v}_i^*   }_0  = s_v$, it holds that 
\begin{align}
\norm{\brho_k\trans\wt{\W}_k\wt{\M}_k}_0 &= \norm{ \sum_{i \neq k} ( \brho\trans \wt{\W}_k \wt{z}_{ii}^{-1}\wh{\bSigma} \wt{\l}_i) \cdot \wt{d}_i \wt{\v}_i\trans }_0 \leq  \sum_{i \neq k}  \norm{ \wt{d}_i\wt{\v}_i   }_0  \nonumber\\
&\leq   \sum_{i \neq k}  \norm{ {\v}_i^*   }_0   + \sum_{i \neq k}\norm{ \wt{d}_i(\wt{\v}_i - \v_i^*)   }_0  \leq  c(r^* + s_u + s_v). \label{czkjdkq}
\end{align}
Under Conditions \ref{con3}--\ref{con:nearlyorth}, based on the above sparsity bound, \eqref{sadqf}, and part (e) of Lemma \ref{rankr:aw} in Section \ref{new.Sec.B.11}, an application of similar arguments to those for \eqref{czxcaw}--\eqref{bbgd} results in 
\begin{align}\label{rhoep3}
|\brho_k\trans\wt{\W}_k\wt{\bepsilon}_k|
\leq  c \max\{s_{\max}, (r^*+s_u+s_v)\}^{1/2}    (r^*+s_u+s_v)^{3/2}\eta_{n}^2\{n^{-1}\log(pq)\}.
\end{align}
Thus, combining the above results yields that
\begin{align*}
&\|\wt{\u}_k - \u_k^*\|_2^2 \leq c(r^*+s_u+s_v) \eta_{n}^4 \{n^{-1}\log(pq)\}, \nonumber\\[5pt]
& \big|( \u_k^* - \wt{\u}_k)\trans \wt{\W}_k\wt{\psi}_k(\wt{\u}_k,\wt{\boldeta}_k)\big| \nonumber\\[5pt]
&\quad \leq c \max\{s_{\max}^{1/2} , (r^*+s_u+s_v)^{1/2}, \eta_n^2\}   (r^*+s_u+s_v)^{3/2}\eta_{n}^2\{n^{-1}\log(pq)\}.
\end{align*}

For \eqref{sdczca}, it can be rewritten as
\begin{align*}
\norm{\wt{\u}_k}_2^2 - \norm{{\u}_k^*}_2^2 -  2 \wt{\u}_k\trans &\wt{\W}_k\wt{\psi}_k(\wt{\u}_k,\wt{\boldeta}_k)
=  2 \u_k\strans (\wh{\u}_k - \u_k^*)  + t^{\prime}_{d_k},
\end{align*}
where
$t^{\prime}_{d_k} = O( \max\{s_{\max}^{1/2} , (r^*+s_u+s_v)^{1/2}, \eta_n^2\}   (r^*+s_u+s_v)^{3/2}\eta_{n}^2\{n^{-1}\log(pq)\} )$. Moreover, replacing $\a$ with $ 2\u_k^*$ and using similar arguments as in the second part of the proof of Theorem \ref{theorkr},  we can show that 
\begin{align*}
\sqrt{n} (2 \u_k\strans)(\wh{\u}_k-\u_k^*)  = h_{d_k} + t^{\prime \prime}_{d_k}, \ \  h_{d_k}     \sim \N(0,\nu_{d_k}^2),
\end{align*}
where the error term $ t^{\prime \prime}_{d_k} $ satisfies that 
$$t^{\prime \prime}_{d_k} = O\big(\max\{s_{\max}^{1/2} , (r^*+s_u+s_v)^{1/2}, \eta_n^2\}   (r^*+s_u+s_v)^{3/2}\eta_{n}^2\log(pq)/\sqrt{n}\big), $$
and the distribution term $h_{d_k}$ and its variance are represented as 
\begin{align*}
& h_{d_k} =
-2\u_k\strans \W_k^{*} \M_{k}^{*} \E\trans\X {\u}_k^{*} / \sqrt{n}
+ 2\u_k\strans\W_k^*\X\trans\E\v_k^*/ \sqrt{n}, \\[5pt]
&\nu_{d_k}^2
= 4\u_k\strans\wh{\bSigma}\u_k^* \cdot \u_k\strans\W_k^*\M_{k}^*\bSigma_e\M_{k}^{* T}\W_k\strans\u_k^* +4\v_k\strans\bSigma_e\v_k^* \cdot  \u_k\strans\W_k^*\wh{\bSigma}\W_k\strans\u_k^*  \\[5pt]
& \qquad  \quad
- 8 \u_k\strans\W_k^*\wh{\bSigma} \u_k^*\v_k\strans\bSigma_e\M_{k}^{* T}\W_k\strans\u_k^*,
\end{align*}
respectively. Therefore, we can obtain that
\begin{align*}
	\norm{\wt{\u}_k}_2^2 - \norm{{\u}_k^*}_2^2 -  2 \wt{\u}_k\trans &\wt{\W}_k\wt{\psi}_k(\wt{\u}_k,\wt{\boldeta}_k)
	=   h_{d_k}  + t_{d_k}
\end{align*}
with $\ h_{d_k}     \sim \N(0,\nu_{d_k}^2)$ and
$$  t_{d_k} = t^{\prime }_{d_k} +  t^{\prime \prime}_{d_k}  = O( \max\{s_{\max}^{1/2} , (r^*+s_u+s_v)^{1/2}, \eta_n^2\}   (r^*+s_u+s_v)^{3/2}\eta_{n}^2\log(pq)/\sqrt{n}). $$
This concludes the proof of Theorem \ref{coro:ortho:dk}.

\subsection{Proof of  Theorem \ref{coro:var:rank2uk} } \label{nearr:sec14}

From Condition \ref{con:nearlyorth}, we see that the nonzero eigenvalues $d^{*2}_{i}$ are at the constant level. It follows from  Definition \ref{lemmsofar} that 
\begin{align*}
	& \norm{{\u}_k^*}_0 \leq s_u, \ \norm{\wt{\u}_k -  {\u}_k^*}_0 \leq c(r^* + s_u + s_v), \ \norm{\wt{\u}_k}_0 \leq c(r^* + s_u + s_v),  \\
	&\norm{{\u}_k^*}_2 \leq c, \  \norm{\wt{\u}_k -  {\u}_k^*}_2 \leq c (r^*+s_u+s_v)^{1/2} \eta_n^2\{n^{-1}\log(pq)\}^{1/2}, \ \norm{\wt{\u}_k}_2 \leq c.
\end{align*}
For some $s_0$ and $\tau_n$, let us define $p$-dimensional vectors $\wt{\a}$ and $\a^*$ that satisfy
\begin{align*}
	&\norm{{\a}^*}_0 \leq s_0, \ \norm{\wt{\a} - \a^*}_0 \leq s_0, \ \norm{\wt{\a}}_0 \leq s_0, \\
	& \norm{\a^*}_2 \leq c, \ \ \norm{\wt{\a} - \a^*}_2 \leq \tau_n, \ \norm{\wt{\a}}_2 \leq c.
\end{align*}
We also define
$ \wt{\nu}^2_e =\varphi_1 + \varphi_2  - 2\varphi_3$ and $ {\nu}^2 =\varphi_1^* + \varphi_2^*  - 2\varphi_3^*$, where
\begin{align*}
	&\varphi_1 = \wt{\u}_k\trans\wh{\bSigma}\wt{\u}_k \wt{\a}\trans\wt{\W}_{k}\wt{\M}_{k}\bSigma_e\wt{\M}_{k}\trans\wt{\W}_k\trans\wt{\a},
	\ \ \varphi_1^* = \u_k\strans\wh{\bSigma}\u_k^*  \a\strans\W_k^*\M_{k}^{*}\bSigma_e\M_{k}\strans\W_k\strans\a^*, \\
	&\varphi_2 =   \wt{\v}_k\trans\bSigma_e\wt{\v}_k \wt{\a}\trans\wt{\W}_k\wh{\bSigma}\wt{\W}_k\trans\wt{\a}, \ ~ ~ \varphi_2^* = \v_k\strans\bSigma_e\v_k^* \a\strans\W_k^*\wh{\bSigma}\W_k\strans\a^*, \\
	&\varphi_3 =  \wt{\a}\trans\wt{\W}_k\wh{\bSigma}\wt{\u}_k \wt{\v}_k\trans\bSigma_e\wt{\M}_{k}\trans\wt{\W}_k\trans\wt{\a}, \ \
	\varphi_3^* =  \a\strans\W_k^*\wh{\bSigma}\u_k^*  \v_k\strans\bSigma_e\M_{k}^{* T}\W_k\strans\a^*.
\end{align*}
It is easy to see that
\begin{align}\label{phiij20}
	|  \wt{\nu}^2_e -  {\nu}^2 | &\leq  | \varphi_1 - \varphi_1^*| +  | \varphi_2 - \varphi_2^*| +  2| \varphi_3 - \varphi_3^*|  \nonumber \\
	&: = A_1 + A_2 + 2 A_3.
\end{align}
Then we will bound the three terms on the right-hand side of (\ref{phiij20}) above separately, which will be conditional on $\wh{\bTheta}$ satisfying Definition \ref{defi2:acceptable} and $\wt{\C}$ satisfying Definition \ref{lemmsofar}.

\medskip

\noindent \textbf{(1). The upper bound on $A_1$}.
Recall that  $\wt{\M}_k = -\wt{z}_{kk}^{-1}  \wh{\bSigma}\widetilde{\U}_{-k} \widetilde{\V}_{-k}\trans,
{\M}_k^* = - {z}_{kk}^{*-1}  \wh{\bSigma}{\U}_{-k}^*{\V}_{-k}\strans$. By Lemma \ref{rankr:boundm0} in Section \ref{new.Sec.B.10}, it holds that 
\begin{align}
	&\norm{\wt{\M}_k}_2 \leq c, \ \
	\norm{{\M}_k^*}_2   \leq c,
	\label{b11az} \\[5pt]
	&\norm{ \wt{\M}_k  - \M_{k}^{*}}_2  \leq  c (r^*+s_u+s_v)^{1/2} \eta_n^2\{n^{-1}\log(pq)\}^{1/2}. \label{b11az2}
\end{align}
Observe that $A_1 = |\wt{z}_{kk} \wt{\a}\trans\wt{\W}_{k}\wt{\M}_{k}\bSigma_e\wt{\M}_{k}\trans\wt{\W}_k\trans\wt{\a}	- {z}_{kk}^* \a\strans\W_k^*\M_{k}^{*}\bSigma_e\M_{k}\strans\W_k\strans\a^*|.$
We denote by
\begin{align*}
	& A_{11} =  | \wt{\a}\trans\wt{\W}_{k}\wt{\M}_{k}\bSigma_e\wt{\M}_{k}\trans\wt{\W}_k\trans\wt{\a}-  \a\strans\W_k^*\M_{k}^{*}\bSigma_e\M_{k}\strans\W_k\strans\a^*|, \\[5pt]
	&A_{12} =  | \a\strans\W_k^*\M_{k}^{*}\bSigma_e\M_{k}\strans\W_k\strans\a^*|.
\end{align*}
Then under Conditions \ref{con3}--\ref{con:nearlyorth}, by parts (b) and (c) of Lemma \ref{lemmauv} in Section \ref{new.Sec.B.3}, we have that 
\begin{align}
	A_1
	&  \leq  |\wt{z}_{kk}| A_{11}  +   |\wt{z}_{kk}  - {z}_{kk}^*| A_{12} \nonumber\\
	&\leq \norm{\wt{\u}_k}_2 \norm{\wh{\bSigma} \wt{\u}_k}_2A_{11}  +   |\wt{z}_{kk}  - {z}_{kk}^*| A_{12} \nonumber\\
	& \leq c  A_{11} + c  (r^*+s_u+s_v)^{1/2} \eta_n^2 \{n^{-1}\log(pq)\}^{1/2} A_{12}. \label{a1lem80}
\end{align}

For term $A_{12}$ introduced above, by Condition \ref{cone} that $\norm{\bSigma_e }_2 \leq c$, part (c) of Lemma \ref{rankr:aw} in Section \ref{new.Sec.B.11}, and \eqref{b11az}, we can show that 
\begin{align}
	A_{12}  \leq \norm{\a\strans\W_k^* }_2  \norm{\M_{k}^* }_2  \norm{\bSigma_e }_2  \norm{\M_{k}\strans }_2 \norm{\W_k\strans\a^* }_2 \leq  c s_0. \label{a22lem80}
\end{align}
Further,  for term $A_{11}$ introduced above, it follows that 
\begin{align*}
	A_{11} & \leq  | \wt{\a}\trans\wt{\W}_{k}\wt{\M}_{k}\bSigma_e(\wt{\M}_{k}\trans\wt{\W}_k\trans\wt{\a} - \M_{k}\strans\W_k\strans\a^*)| \\
 &\quad+  |( \wt{\a}\trans\wt{\W}_{k}\wt{\M}_{k} - \a\strans\W_k^*\M_{k}^{*})\bSigma_e\M_{k}\strans\W_k\strans\a^*|  \\
	&\leq  \|\wt{\a}\trans\wt{\W}_{k}\wt{\M}_{k}\bSigma_e\|_2  \norm{\wt{\M}_{k}\trans\wt{\W}_k\trans\wt{\a} - \M_{k}\strans\W_k\strans\a^*}_2 \\
 &\quad+    \norm{ \wt{\a}\trans\wt{\W}_{k}\wt{\M}_{k} - \a\strans\W_k^*\M_{k}^{*}}_2  \|\bSigma_e\M_{k}\strans\W_k\strans\a^*\|_2.
\end{align*}
In light of Condition \ref{cone}, Lemma \ref{rankr:aw} in Section \ref{new.Sec.B.11}, \eqref{b11az}, and \eqref{b11az2}, it can be easily seen that
\begin{align*}
	\|\wt{\a}\trans\wt{\W}_{k}\wt{\M}_{k}\bSigma_e\|_2  \leq  c s_0^{1/2} \ \text{ and } \
	\|\bSigma_e\M_{k}\strans\W_k\strans\a^*\|_2  \leq  c s_0^{1/2}.
\end{align*}
Moreover, it follows from parts (c) and (d) of Lemma \ref{rankr:aw} in Section \ref{new.Sec.B.11} that 
\begin{align}
	\|\wt{\a}\trans\wt{\W}_k -   \a\strans\W_k^* \|_2 & \leq \|\wt{\a}\trans(\wt{\W}_k - \W_k^* )\|_2 + \|(\wt{\a} -   \a^*)\trans\W_k^* \|_2 \nonumber\\
	&\leq      c s_0^{1/2}  (r^*+s_u+s_v)^{1/2} \eta_n^2\{n^{-1}\log(pq)\}^{1/2} + cs_0^{1/2} \tau_n. \label{aw-awk}
\end{align}

Then from part (e) of Lemma \ref{rankr:aw} in Section \ref{new.Sec.B.11}, \eqref{b11az}, \eqref{b11az2}, and \eqref{aw-awk}, we can deduce that 
\begin{align}\label{mwamwa0}
	\norm{ \wt{\a}\trans\wt{\W}_{k}\wt{\M}_{k} - \a\strans\W_{k}^*\M_{k}^*}_2  & \leq \norm{\wt{\a}\trans\wt{\W}_{k}}_2 \norm{ \wt{\M}_{k} - \M_{k}^*   }_2 + \norm{ \wt{\a}\trans\wt{\W}_{k} - \a\strans\W_{k}^*}_2 \norm{ \M_{k}^*  }_2 \nonumber\\[5pt]
	&\leq  c  s_0^{1/2}  (r^*+s_u+s_v)^{1/2} \eta_n^2\{n^{-1}\log(pq)\}^{1/2} + c  s_0^{1/2}\tau_n.
\end{align}
Thus, combining the above terms leads to 
\begin{align}
	A_{11}
	\leq  c s_0 (r^*+s_u+s_v)^{1/2} \eta_n^2\{n^{-1}\log(pq)\}^{1/2}  +  cs_0\tau_n. \label{a11lem80}
\end{align}
Finally, using \eqref{a1lem80}, \eqref{a22lem80}, and \eqref{a11lem80}, we can obtain that 
\begin{align}\label{phia10}
	A_1
	\leq  c  s_0    (r^*+s_u+s_v)^{1/2} \eta_n^2\{n^{-1}\log(pq)\}^{1/2} + cs_0  \tau_n.
\end{align}

\smallskip

\noindent \textbf{(2). The upper bound on $A_2$}.
Notice that $$	A_2 =  | \wt{\v}_k\trans\bSigma_e\wt{\v}_k \wt{\a}\trans\wt{\W}_k\wh{\bSigma}\wt{\W}_k\trans\wt{\a} - \v_k\strans\bSigma_e\v_k^* \a\strans\W_k^*\wh{\bSigma}\W_k\strans\a^*|.$$
In view of Condition \ref{con:nearlyorth}, the nonzero eigenvalues $d^{*2}_{i}$ are at the constant level. It follows from Condition \ref{cone} that $\norm{\bSigma_e }_2 \leq c$, Definition \ref{lemmsofar} that $\norm{\wt{\v}_k}_2 = \|\v_k^* \|_2 = 1 $, and part (a) of Lemma \ref{lemmauv} in Section \ref{new.Sec.B.3} that $|  \wt{\v}_k\trans\bSigma_e\wt{\v}_k | \leq  \|  \wt{\v}_k\trans\|_2 \|\bSigma_e\|_2 \|\wt{\v}_k \|_2  \leq c$ and
\begin{align*}
	| \wt{\v}_k\trans\bSigma_e\wt{\v}_k - \v_k\strans\bSigma_e\v_k^* | &\leq   | \wt{\v}_k\trans\bSigma_e(\wt{\v}_k - \v_k^*) | +  | ( \wt{\v}_k\trans - \v_k\strans)\bSigma_e\v_k^* | \\
	&\leq  \| \wt{\v}_k\|_2 \|\bSigma_e\|_2 \|\wt{\v}_k - \v_k^* \|_2 +  \|  \wt{\v}_k - \v_k^*\|_2 \|\bSigma_e\|_2 \|\v_k^* \|_2 \\
	& \leq   c (r^*+s_u+s_v)^{1/2} \eta_n^2\{n^{-1}\log(pq)\}^{1/2}.
\end{align*}
With the aid of the triangle inequality, we can show that 
\begin{align}
	A_2 &\leq     | \wt{\v}_k\trans\bSigma_e\wt{\v}_k | |\wt{\a}\trans\wt{\W}_k\wh{\bSigma}\wt{\W}_k\trans\wt{\a} -   \a\strans\W_k^*\wh{\bSigma}\W_k\strans\a^*| \nonumber \\[5pt]
	&\quad + | \wt{\v}_k\trans\bSigma_e\wt{\v}_k - \v_k\strans\bSigma_e\v_k^* | |  \a\strans\W_k^*\wh{\bSigma}\W_k\strans\a^*| \nonumber\\[5pt]
	&\leq c  |\wt{\a}\trans\wt{\W}_k\wh{\bSigma}\wt{\W}_k\trans\wt{\a} -   \a\strans\W_k^*\wh{\bSigma}\W_k\strans\a^*| \nonumber\\[5pt]
	&\quad +    |  \a\strans\W_k^*\wh{\bSigma}\W_k\strans\a^*| c (r^*+s_u+s_v)^{1/2} \eta_n^2\{n^{-1}\log(pq)\}^{1/2}. \label{eqa22va}
\end{align}

It is easy to see that 
\begin{align*}
	&|\wt{\a}\trans\wt{\W}_k\wh{\bSigma}\wt{\W}_k\trans\wt{\a} -   \a\strans\W_k^*\wh{\bSigma}\W_k\strans\a^*| \\[5pt]
	& \leq 	|\wt{\a}\trans\wt{\W}_k\wh{\bSigma}(\wt{\W}_k\trans\wt{\a} -  \W_k\strans\a^*)|
	+ |(\wt{\a}\trans\wt{\W}_k -    \a\strans\W_k^* ) \wh{\bSigma}\W_k\strans\a^* | \\
	& =: A_{21} + A_{22}.
\end{align*}
Let us bound the two terms $A_{21}$ and $A_{22}$ introduced above separately.
Denote by $\w_{i}\strans$ and $\wt{\w}_{i}\trans$ the $i$th rows of $\W_{k}^*$ and $\wt{\W}_k$. Recall that 
\begin{align*}
	&\W^*_k = \widehat{\bTheta} \left\{  \I_p +   z_{kk}^{*-1}\wh{\bSigma} \U_{-k}^*(\I_{r^*-1} -z_{kk}^{*-1}\U_{-k}\strans\wh{\bSigma} \U_{-k}^*)^{-1}\U_{-k}\strans\right\}.
\end{align*}
Denote by  $\A^* = (z_{kk}^{*}\I_{r^*-1} -\U_{-k}\strans\wh{\bSigma} \U_{-k}^*)^{-1}$.
Then we have 
\begin{align*}
\w_i\strans & = \widehat{\btheta}_i\trans  +   \widehat{\btheta}_i\trans\wh{\bSigma}\U^*_{-k}\A^*\U_{-k}\strans \\
&=: \widehat{\btheta}_i\trans + \w_{0,i}\strans,
\end{align*}
where  $\wh{\btheta}_i\trans$ is the $i$th row of $\wh{\bTheta}$.
Since $\norm{\U_{-k}^*}_0 \leq s_u$ and $\widehat{\btheta}_i\trans   \wh{\bSigma}\U^*_{-k}\A^*$ is a vector, it follows that 
\[\| \w_{0,i}^*\|_0 = \| \U_{-k}^* \cdot (\widehat{\btheta}_i\trans   \wh{\bSigma}\U^*_{-k}\A^*)\trans \|_0 \leq s_u. \]
Together with Definition \ref{defi2:acceptable} that $ \max_{1 \leq i \leq p} \|\widehat{\btheta}_i\|_0 \leq s_{\max}$ and $ \max_{1 \leq i \leq p} \|\widehat{\btheta}_i\|_2 \leq c$, under Condition \ref{con3} it holds that
\begin{align}\label{daszcghns0}
	\max_{1 \leq i \leq p} \|\wh{\bSigma}\w_{i}^*\|_2 &\leq \max_{1 \leq i \leq p} \|\wh{\bSigma}\widehat{\btheta}_i\|_2 + \max_{1 \leq i \leq p} \|\wh{\bSigma}\w_{0,i}^*\|_2 \nonumber \\
	& \leq \max_{1 \leq i \leq p} \|\widehat{\btheta}_i\|_2 + \max_{1 \leq i \leq p} \|\w_{0,i}^*\|_2   \leq c,
\end{align}
where the last step has utilized \eqref{czmla} in the proof of Lemma \ref{rankr:aw} in Section \ref{new.Sec.B.11}.
Further, based on parts (a) and (b) of Lemma \ref{rankr:aw} and similar arguments, we can show that 
\begin{align}\label{dsaczq}
	\max_{1 \leq i \leq p} \|\wh{\bSigma}\wt{\w}_{i}\|_2  \leq c, \ \	\max_{1 \leq i \leq p}\|\wh{\bSigma}(\wt{\w}_{i} - \w_{i}^*)\|_2 \leq   c  (r^*+s_u+s_v)^{1/2} \eta_n^2\{n^{-1}\log(pq)\}^{1/2}.
\end{align}

Then using \eqref{daszcghns0}, \eqref{dsaczq}, and part (c) of Lemma \ref{rankr:aw}, we can deduce that 
\begin{align}
	&\norm{\wh{\bSigma}\W_k\strans\a^* }_2
	\leq \| \sum_{i=1}^p a_{i}^* \wh{\bSigma}\w_{i}^* \|_2 \leq  \| \a^* \|_1  \max_{1 \leq i \leq p} \|\wh{\bSigma}\w_{i}^*\|_2 \leq c \norm{\a^*}_0^{1/2}\norm{\a^*}_2 \leq  c s_0^{1/2}, \label{daszcghns} \\[5pt]
	& |  \a\strans\W_k^*\wh{\bSigma}\W_k\strans\a^*| \leq \|  \a\strans\W_k^*\|_2 \norm{\wh{\bSigma}\W_k\strans\a^* }_2 \leq c s_0, \label{daszcghns2}\\[5pt]
	&\|\wh{\bSigma}\wt{\W}_{k}\trans\wt{\a} -  \wh{\bSigma}{\W}_{k}\strans\a^*\|_2 \leq
	\|\wh{\bSigma}\wt{\W}_{k}\trans(\wt{\a} - \a^*)\|_2 + 	\|\wh{\bSigma}(\wt{\W}_{k}\trans - {\W}_{k}\strans)\a^*\|_2 \nonumber \\[5pt]
	&\qquad \qquad \qquad \leq  \| \wt{\a} - \a^* \|_1  \max_{1 \leq i \leq p} \|\wh{\bSigma}\wt{\w}_{i}\|_2 +  \| \a^*\|_1  \max_{1 \leq i \leq p} \|\wh{\bSigma}(\wt{\w}_{i} - \w_{i}^*)\|_2 \nonumber\\[5pt]
	&\qquad \qquad \qquad \leq  cs_0^{1/2} \tau_n +  c s_0^{1/2}  (r^*+s_u+s_v)^{1/2} \eta_n^2\{n^{-1}\log(pq)\}^{1/2}.  \label{daszcghns3}
\end{align}
Along with $\|\wt{\a}\trans\wt{\W}_k \|_2 \leq c s_0^{1/2}$  by part (e) of Lemma \ref{rankr:aw}, it follows that 
\begin{align}
	A_{21} &\leq \|\wt{\a}\trans\wt{\W}_k \|_2 \|  \wh{\bSigma}(\wt{\W}_{k}\trans\wt{\a} -  {\W}_{k}\strans\a^*)\|_2  \nonumber\\
	&\leq  cs_0 \tau_n +  c s_0 (r^*+s_u+s_v)^{1/2} \eta_n^2\{n^{-1}\log(pq)\}^{1/2}. \label{a222lem8}
\end{align}

For term $A_{22}$ introduced above,  
combining \eqref{aw-awk} and \eqref{daszcghns} leads to 
\begin{align}
	A_{22} &\leq \|\wt{\a}\trans\wt{\W}_k -   \a\strans\W_k^* \|_2  \| \wh{\bSigma}\W_{k}\strans\a^* \|_2 \nonumber \\
	& \leq   cs_0 \tau_n +  c s_0  (r^*+s_u+s_v)^{1/2} \eta_n^2\{n^{-1}\log(pq)\}^{1/2}. \label{a234lem8}
\end{align}
Hence, by \eqref{a222lem8} and \eqref{a234lem8} it holds that
\begin{align*}
	|\wt{\a}\trans\wt{\W}_k&\wh{\bSigma}\wt{\W}_k\trans\wt{\a} -   \a\strans\W_k^*\wh{\bSigma}\W_k\strans\a^*| 
	\leq cs_0 ( \tau_n +  (r^*+s_u+s_v)^{1/2} \eta_n^2\{n^{-1}\log(pq)\}^{1/2}).
\end{align*}
This together with \eqref{eqa22va} and \eqref{daszcghns2} yields that
\begin{align}\label{phia20}
	A_{2}
	\leq cs_0( \tau_n +  (r^*+s_u+s_v)^{1/2} \eta_n^2\{n^{-1}\log(pq)\}^{1/2}).
\end{align}

\smallskip

\noindent \textbf{(3). The upper bound on $A_3$}.
Note that
$$A_3 =  | \wt{\a}\trans\wt{\W}_k\wh{\bSigma}\wt{\u}_k \wt{\v}_k\trans\bSigma_e\wt{\M}_{k}\trans\wt{\W}_k\trans\wt{\a}-
\a\strans\W_k^*\wh{\bSigma}\u_k^*  \v_k\strans\bSigma_e\M_{k}^{* T}\W_k\strans\a^*|.$$
From Condition \ref{cone}, part (b) of Lemma \ref{lemmauv}, parts (c) and (e) of Lemma \ref{rankr:aw}, and \eqref{b11az}, we see that 
\begin{align*}
	&|\wt{\a}\trans\wt{\W}_k\wh{\bSigma}\wt{\u}_k |
	\leq \|\wt{\a}\trans\wt{\W}_k\|_2 \|\wh{\bSigma}\wt{\u}_k \|_2
	\leq  c s_0^{1/2}, \\[5pt]
	&| \v_k\strans\bSigma_e\M_{k}^{* T}\W_k\strans\a^* | \leq
	\| \v_k\strans\|_2 \|\bSigma_e\|_2 \|\M_{k}^{* T}\|_2 \|\W_k\strans\a^* \|_2
	\leq  c s_0^{1/2}.
\end{align*}
Let us define $ A_{31} = | \wt{\v}_k\trans\bSigma_e\wt{\M}_{k}\trans\wt{\W}_k\trans\wt{\a}-
\v_k\strans\bSigma_e\M_{k}^{* T}\W_k\strans\a^*  |$.
Then for term $A_3$, it holds that 
\begin{align*}
	A_3 &\leq   | \wt{\a}\trans\wt{\W}_k\wh{\bSigma}\wt{\u}_k | |\wt{\v}_k\trans\bSigma_e\wt{\M}_{k}\trans\wt{\W}_k\trans\wt{\a} - \v_k\strans\bSigma_e\M_{k}^{* T}\W_k\strans\a^*| \\[5pt]
	& \quad +   | \wt{\a}\trans\wt{\W}_k\wh{\bSigma}\wt{\u}_k -
	\a\strans\W_k^*\wh{\bSigma}\u_k^*| | \v_k\strans\bSigma_e\M_{k}^{* T}\W_k\strans\a^*|  \\[5pt]
	& \leq c s_0^{1/2}  A_{31}  + cs_0^{1/2} | \wt{\a}\trans\wt{\W}_k\wh{\bSigma}\wt{\u}_k -
	\a\strans\W_k^*\wh{\bSigma}\u_k^*|.
\end{align*}

For term $A_{31}$ introduced above, it follows from Definition \ref{lemmsofar} that $\|  \wt{\v}_k\trans \|_2 = 1$, Condition \ref{cone}, \eqref{b11az}, \eqref{mwamwa0}, part (a) of Lemma \ref{lemmauv}, and part (c) of Lemma \ref{rankr:aw} that
\begin{align*}
	A_{31} &\leq \|  \wt{\v}_k\trans \|_2 \|\bSigma_e \|_2 \| \wt{\M}_{k}\trans\wt{\W}_k\trans\wt{\a}-
	\M_{k}^{* T}\W_k\strans\a^*  \|_2 + \|  \wt{\v}_k -  \v_k^* \|_2 \|\bSigma_e \|_2 \| \M_{k}^{* T}\|_2 \|\W_k\strans\a^*  \|_2 \\[5pt]
	&\leq   c  s_0^{1/2} (r^*+s_u+s_v)^{1/2} \eta_n^2\{n^{-1}\log(pq)\}^{1/2} + c  s_0^{1/2}\tau_n.
\end{align*}
Moreover, using \eqref{aw-awk}, part (b) of Lemma \ref{lemmauv}, and part (e) of Lemma \ref{rankr:aw}, we can deduce that 
\begin{align*}
	| \wt{\a}\trans\wt{\W}_k\wh{\bSigma}\wt{\u}_k -	\a\strans\W_k^*\wh{\bSigma}\u_k^*|&\leq
	\|\wt{\a}\trans\wt{\W}_k \|_2  \|\wh{\bSigma}(\wt{\u}_k -  \u_k^*)\|_2
	+ \|\wt{\a}\trans\wt{\W}_k  -   \a\strans\W_k^* \|_2  \|\wh{\bSigma} \u_k^*\|_2 \\[5pt]
	&\leq  cs_0^{1/2} \tau_n +  c s_0^{1/2}  (r^*+s_u+s_v)^{1/2} \eta_n^2\{n^{-1}\log(pq)\}^{1/2}.
\end{align*}
It follows that
\begin{align}\label{phia30}
	A_3  \leq  cs_0 \tau_n +  c s_0 (r^*+s_u+s_v)^{1/2} \eta_n^2\{n^{-1}\log(pq)\}^{1/2}.
\end{align}

Then combining \eqref{phiij20}, \eqref{phia10}, \eqref{phia20}, and \eqref{phia30} gives that 
\begin{align}\label{phi-phik}
	|  \wt{\nu}^2_e -  {\nu}^2 | \leq  cs_0  ( \tau_n +  (r^*+s_u+s_v)^{1/2} \eta_n^2\{n^{-1}\log(pq)\}^{1/2}).
\end{align}

Let us define $\wt{\nu}^2$ with $\bSigma_e$ in $\wt{\nu}^2_e$ replaced by the 
acceptable estimator  $\wt{\bSigma}_e$ satisfying Definition \ref{defi:error}. Then by Condition \ref{cone} that  $\norm{{\bSigma}_e}_2 \leq c$ and Definition \ref{defi:error}, for sufficiently large $n$, it holds that 
\begin{align}\label{eq:error}
    \norm{\wt{\bSigma}_e }_2 \leq  \norm{{\bSigma}_e}_2 + \norm{\wt{\bSigma}_e - \bSigma_e}_2 \leq c.
\end{align}
Note that in the above three part proofs, we have only used the property of ${\bSigma}_e$ such that $\norm{{\bSigma}_e}_2 \leq c$.
Based on the observation \eqref{eq:error}, replacing ${\bSigma}_e$ with $\wt{\bSigma}_e$ in the above proofs will lead to the same result as in \eqref{phi-phik} that
\begin{align}\label{phi-phik222}
	|  \wt{\nu}^2 -  {\nu}^2 | \leq  cs_0  ( \tau_n +  (r^*+s_u+s_v)^{1/2} \eta_n^2\{n^{-1}\log(pq)\}^{1/2}).
\end{align}

When $\wt{\a} = \wt{\u}_k$ and $\a^* = \u_k^*$, we have that $s_0 = c(r^*+s_u + s_v)$ and 
$$\tau_n = c(r^*+s_u+s_v)^{1/2} \eta_n^2\{n^{-1}\log(pq)\}^{1/2}.$$ Also, we have $\wt{\nu}^2 = \wt{\nu}_{d_k}^2 $ and ${\nu}^2 = {\nu}_{d_k}^2$.
Then in light of \eqref{phi-phik222}, we see that 
\begin{align*}
	&|\wt{\nu}_{d_k}^2 - {\nu}_{d_k}^2| \leq c (r^*+s_u+s_v)^{3/2} \eta_n^2\{n^{-1}\log(pq)\}^{1/2}.
\end{align*}
When $\wt{\a} = \a^* \in\mathcal{A}(m)=\{\a\in\R^p:\norm{\a}_0\leq m, \norm{\a}_2 = 1\}$, we have $s_0 = m$ and $\tau_n=0$. Furthermore, we have that $\wt{\nu}^2 = \wt{\nu}_{k}^2 $ and ${\nu}^2 = {\nu}_{k}^2$. Similarly, from \eqref{phi-phik222} it holds that 
\begin{align*}
	&|\wt{\nu}_{k}^2 - {\nu}_{k}^2| \leq c m  (r^*+s_u+s_v)^{1/2} \eta_n^2\{n^{-1}\log(pq)\}^{1/2},
\end{align*}
which completes the proof of Theorem \ref{coro:var:rank2uk}.

\subsection{Proof of Theorem \ref{theorkapor}} \label{new.Sec.A.4}

Let us recall that
\begin{align*}
\wh{\u}_k & ={\psi}_k(\wt{\u}_k,\wt{\boldeta}_k) = \wt{\u}_k  - \wt{\W}_k\wt{\psi}_k(\wt{\u}_k,\wt{\boldeta}_k) \\
&= \wt{\u}_k  - \wt{\W}_k\wt{\psi}_k(\wt{\u}_k,\boldeta^*_k) + \wt{\W}_k(\wt{\psi}_k(\wt{\u}_k,\boldeta^*_k) -\wt{\psi}_k(\wt{\u}_k,\wt{\boldeta}_k)).
\end{align*}
Using Lemma \ref{prop:rankapo1} and Propositions \ref{prop:rankapo2}--\ref{prop:rankapo3}  and the initial estimates satisfying  Definition \ref{lemmsofar},
we can deduce that
\begin{align}\label{eqafinalreeq}
\sqrt{n}\a\trans(\wh{\u}_k-\u_k^*)  & =  - \sqrt{n}\a\trans\wt{\W}_k\wt{\bepsilon}_k - \sqrt{n}\a\trans\wt{\W}_k\wt{\bdelta}_k -\sqrt{n} \a\trans\wt{\W}_k(\wt{\psi}_k(\wt{\u}_k,\wt{\boldeta}_k) - \wt{\psi}_k(\wt{\u}_k,\boldeta^*_k) )  \nonumber\\[5pt]
&\quad + \sqrt{n}\a\trans(\I_p - \wt{\W}_k\mathbf{T}_k)(\wt{\u}_k-\u_k^*)  - \sqrt{n}\a\trans\wt{\W}_k\wt{\M}_{k} (\C^{*(2)})\trans\wh{\bSigma} \u_k^*,
\end{align}
where $\wt{\M}_{k} =  -z_{kk}^{-1}\wh{\bSigma}\wt{\C}^{(2)}, \
\mathbf{T}_k = \I_p - \wt{\M}_{k}\wt{\v}_k\wt{\u}_k\trans + \wt{\M}_{k}(\wt{\C}^{(2)})\trans\wh{\bSigma},$ and
\begin{align}
&\wt{\bepsilon}_k =   - n^{-1}\X\trans\E\v_k^* + n^{-1}\wt{\M}_k\E\trans\X\wt{\u}_k, \label{eqepweak}\\[5pt]
&\wt{\bdelta}_k =   
\wt{\M}_k ((\wt{\v}_k - \v_k^*)\wt{\u}_k\trans - (\wt{\C}^{(2)} - \C^{*(2)})\trans)  \wh{\bSigma}(\wt{\u}_k - \u_k^*) - \wt{\M}_k  (\wh{\C}^{(1)} - \C^{*(1)})\trans  \wh{\bSigma} \wt{\u}_k, \label{eqdeweak}\\[5pt]
&\wt{\W}_k  =   \widehat{\bTheta} \left\{  \I_p +   \wt{z}_{kk}^{-1}\wh{\bSigma}\wt{\U}^{(2)}(\I_{r^*-k} -\wt{z}_{kk}^{-1}\wt{\U}^{(2) T}\wh{\bSigma} \wt{\U}^{(2)})^{-1}(\wt{\U}^{(2) })\trans\right\}.  \label{eqwkge1}
\end{align}

Further, denote by 
\begin{align}
& h_{k} = -\a\trans \W^{*}_k \M_{k}^{*} \E\trans\X {\u}_k^{*}/\sqrt{n} +  \a\trans\W^*_k\X\trans\E\v_k^*/\sqrt{n}, \label{eqhkweak}
\end{align}
where $\M_{k}^{*} = -z_{kk}^{*-1}\wh{\bSigma}\C^{*(2)}$, $z_{kk}^{*} = {\u}_k\strans\wh{\bSigma}{\u}_k^{*}$, and
\begin{align}
\W_k^{*} = \widehat{\bTheta} \left\{  \I_p +   z_{kk}^{*-1}\wh{\bSigma}\U^{*(2)}(\I_{r^*-k} -z_{kk}^{*-1}(\U^{*(2)})\trans\wh{\bSigma} \U^{*(2)})^{-1}(\U^{*(2)})\trans\right\}. \label{eqwkge2}
\end{align}
In view of Lemma \ref{lemm:wexist2} in Section \ref{new.Sec.B.16}, we see that both  $\wt{\W}_k $ and $\W_k^{*}$  are well-defined. We aim to bound the terms on the right-hand side of \eqref{eqafinalreeq} above, which will be conditional on $\wh{\bTheta}$ satisfying Definition \ref{defi2:acceptable} and $\wt{\C}$ satisfying Definition \ref{lemmsofar}.

It follows from Proposition \ref{prop:rankapo3} that $\I_p - \wt{\W}_k\mathbf{T}_k = \I_p - \wh{\bTheta}\wh{\bSigma}$. Using similar arguments as for \eqref{eq:th:4rk2}, this implies that
\begin{align}
\abs{\a\trans(\I_p - \wt{\W}_k\mathbf{T}_k)(\wt{\u}_k-\u_k^*)} \leq cm^{1/2}(r^*+s_u+s_v)\eta_n^2\{n^{-1}\log(pq)\}. \nonumber
\end{align}
Under Conditions \ref{con3}--\ref{con4} and \ref{con:orth:rankr}, an application of Lemmas \ref{prop:tay2lor1222r}--\ref{lemma:k21rk2} in Sections \ref{new.Sec.B.19}--\ref{new.Sec.B.22}, respectively, leads to 
\begin{align*}
&| \a\trans\wt{\W}_k(\wt{\psi}_k(\wt{\u}_k,\boldeta^*_k) -\wt{\psi}_k(\wt{\u}_k,\wt{\boldeta}_k)) | \nonumber \\[5pt]
&\  \leq  cm^{1/2} \max\{s_{\max}^{1/2} , (r^*+s_u+s_v)^{1/2}, \eta_n^2\} (r^*+s_u+s_v)\eta_n^2\{n^{-1}\log(pq)\} \max\{d_k^{*-1}, d_k^{*-2}\}, \\[5pt]
&	\abs{-\a\trans\wt{\W}_k\wt{\bepsilon}_{k} - h_{k}  / \sqrt{n}} \leq c m^{1/2}    (r^*+s_u + s_v)^{3/2}\eta_n^2\{ n^{-1}\log(pq)\}d_k^{*-1}, \\[5pt]
&	|\a\trans\wt{\W}_k \wt{\M}_k (\C^{*(2)})\trans \wh{\bSigma} \u_k^*  | = o(m^{1/2}  n^{-1/2}), \\[5pt]
&|\a\trans\wt{\W}_k\wt{\bdelta}_{k}|
 \leq c m^{1/2} (r^*+s_u+s_v)^{1/2}\eta_{n}^2\{\log(pq)\}^{1/2} d_{k+1}^{*}d_k^{*-3}(\sum_{i=1}^{k-1}d_i^*)/n \nonumber  \\
&~ \qquad \qquad\quad +c m^{1/2}(r^*+s_u+s_v)\eta_{n}^4\{n^{-1}\log(pq)\}  d_k^{*-2} d_{k+1}^{*}. \nonumber
\end{align*}
Moreover, using similar arguments to those for \eqref{eqh1}--\eqref{eqh4},
we can show that $h_{k}  \sim \N(0,\nu_{k}^{2})$ with
\begin{align*}
	\nu_k^2 = \text{var}( h_k|\X) = \a\trans\W_k^*(z_{kk}^{*}\M_{k}^{*}\bSigma_e\M_{k}\strans + \v_k\strans\bSigma_e\v_k^* \wh{\bSigma} - 2\wh{\bSigma}\u_k^*  \v_k\strans\bSigma_e\M_{k}^{* T})\W_k\strans\a,
\end{align*}

Therefore, combining the above results yields that
\begin{align}
\sqrt{n}\a\trans(\wh{\u}_k-\u_k^*) = h_{k}  + t_{k}, \nonumber
\end{align}
where $h_{k} = -\a\trans \W^{*}_k \M_{k}^{*} \E\trans\X {\u}_k^{*}/\sqrt{n} +  \a\trans\W^*_k\X\trans\E\v_k^*/\sqrt{n}  \sim \N(0,\nu_{k}^{2})$ and
\begin{align*}
 t_{k} &= O\Big\{ m^{1/2}\max\{1, d_k^{*-1}, d_k^{*-2}\} \max\{s_{\max}^{1/2} , (r^*+s_u+s_v)^{1/2}, \eta_n^2\} (r^*+s_u+s_v)\eta_n^2  \log(pq)/\sqrt{n}  \\
    &\quad +m^{1/2}  (r^*+s_u+s_v)^{1/2}\eta_{n}^2\{n^{-1}\log(pq)\}^{1/2}  d_{k+1}^{*}d_k^{*-3}(\sum_{i=1}^{k-1}d_i^*)\Big\}.
\end{align*}
This concludes the proof of Theorem  \ref{theorkapor}.

\subsection{Proof of Theorem \ref{theorkukcorok}} \label{new.Sec.A.5}


Notice that
$	\wh{\u}_k ={\psi}_k(\wt{\u}_k,\wt{\boldeta}_k) = \wt{\u}_k  - \wt{\W}_k\wt{\psi}_k(\wt{\u}_k,\wt{\boldeta}_k).$
Similar to \eqref{singd}, it holds that 
\begin{align}
	\widehat{d}_k^2 - d_k^{*2} &=  \norm{\wt{\u}_k}_2^2 - 2 \wt{\u}_k\trans \wt{\W}_k\wt{\psi}_k(\wt{\u}_k,\wt{\boldeta}_k) - \norm{{\u}_k^*}_2^2 \nonumber \\[5pt]
	&=  2 \u_k\strans (\wh{\u}_k - \u_k^*)
	+ 2( \u_k^* - \wt{\u}_k)\trans \wt{\W}_k\wt{\psi}_k(\wt{\u}_k,\wt{\boldeta}_k) + \|\wt{\u}_k - \u_k^*\|_2^2. \label{czxdqq}
\end{align}
We then show that $ 2( \u_k^* - \wt{\u}_k)\trans \W_k\wt{\psi}(\wt{\u}_k,\wt{\boldeta}) + \|\wt{\u}_k - \u_k^*\|_2^2$ is asymptotically negligible. Denote by $\brho_k =  \u_k^* - \wt{\u}_k$.
It is easy to see that 
\begin{align*}
\brho_k\trans \wt{\W}_k\wt{\psi}_k(\wt{\u}_k,\wt{\boldeta}_k) &= \brho_k\trans \wt{\W}_k\wt{\psi}_k(\wt{\u}_k,\boldeta^*_k)
+ \brho_k\trans \wt{\W}_k( \wt{\psi}_k(\wt{\u}_k,\wt{\boldeta}_k) - \wt{\psi}_k(\wt{\u}_k,\boldeta^*_k)).
\end{align*}
From Lemma \ref{prop:rankapo1}, with the initial estimates we can deduce that 
\begin{align}\label{sdczca2}
&\brho_k\trans\wt{\W}_k\wt{\psi}_k(\wt{\u}_k,\boldeta^*_k) =  \brho_k\trans\wt{\W}_k\wt{\bepsilon}_k + \brho_k\trans\wt{\W}_k\wt{\bdelta}_k + \brho_k\trans\wt{\W}_k \mathbf{T}_k(\wt{\u}_k-\u_k^*) + \brho_k\trans\wt{\W}_k \wt{\M}_k(\C^{*(2)})\trans \wh{\bSigma} \u_k^* \nonumber\\[5pt]
&~=  \brho_k\trans\wt{\W}_k\wt{\bepsilon}_k +  \brho_k\trans\wt{\W}_k\wt{\bdelta}_k + \brho_k\trans(\I_p -\wt{\W}_k \mathbf{T}_k)\brho_k - \brho_k\trans\brho_k + \brho_k\trans\wt{\W}_k\wt{\M}_k(\C^{*(2)})\trans \wh{\bSigma} \u_k^*,
\end{align}
where $\wt{\bepsilon}_k, \wt{\bdelta }_k, \wt{\W}_k$ are given in \eqref{eqepweak}--\eqref{eqwkge1}, respectively, $\wt{\M}_{k} =  -z_{kk}^{-1}\wh{\bSigma}\wt{\C}^{(2)}$, and $
\mathbf{T}_k = \I_p - \wt{\M}_{k}\wt{\v}_k\wt{\u}_k\trans + \wt{\M}_{k}(\wt{\C}^{(2)})\trans\wh{\bSigma}$.

We will show that all terms on the right-hand side of \eqref{sdczca2} above and $\brho_k\trans \wt{\W}_k( \wt{\psi}_k(\wt{\u}_k,\wt{\boldeta}_k) - \wt{\psi}_k(\wt{\u}_k,\boldeta^*_k))$ are asymptotically vanishing. Similar to \eqref{brho}, by Definition \ref{lemmsofar} it holds that
\begin{align}
\norm{\brho_k}_0 \leq c(r^* + s_u + s_v) \ \text{ and } \   \norm{\brho_k}_2 \leq c(r^*+s_u+s_v)^{1/2}\eta_{n}^2\{n^{-1}\log(pq)\}^{1/2}. \label{czzdq}
\end{align}
Under Conditions \ref{con3}--\ref{con4} and \ref{con:orth:rankr}, an application of  \eqref{czzdq} and Lemmas \ref{prop:tay2lor1222r}--\ref{lemma:k21rk2} gives that 
\begin{align*}
&|\brho_k\trans \wt{\W}_k( \wt{\psi}_k(\wt{\u}_k,\wt{\boldeta}_k) - \wt{\psi}_k(\wt{\u}_k,\boldeta^*_k)) | \nonumber \\
& ~ ~ \leq  c\max\{s_{\max}^{1/2}, (r^*+s_u+s_v)^{1/2},\eta_n^2\} (r^*+s_u+s_v)^2\eta_n^4\{n^{-1}\log(pq)\}^{3/2} \max\{d_k^{*-1}, d_k^{*-2}\}, \\[5pt]
&  	|  \brho_k\trans\wt{\W}_k\wt{\M}_k(\C^{*(2)})\trans \wh{\bSigma} \u_k^*|=  o( (r^*+s_u+s_v)\eta_{n}^2\{\log(pq)\}^{1/2}n^{-1}), \\[3pt]
 &|\brho_k\trans\wt{\W}_k\wt{\bdelta}_{k}|
	 \leq c   (r^*+s_u+s_v)^{3/2}\eta_{n}^4\{n^{-3/2}\log(pq)\}  d_{k+1}^{*}d_k^{*-3}(\sum_{i=1}^{k-1}d_i^*) \nonumber  \\
	&~ \qquad \qquad \quad +c (r^*+s_u+s_v)^2\eta_{n}^6\{n^{-1}\log(pq)\}^{3/2}  d_k^{*-2} d_{k+1}^{*}.
\end{align*}
Further, using similar arguments as for \eqref{cxxzd5}, we can show that 
\begin{align*}
\abs{\brho_k\trans(\I_p -\wt{\W}_k\T_k)\brho_k} \leq c(r^*+s_u+s_v)^2\eta_n^4\{n^{-1}\log(pq)\}^{3/2}.
\end{align*}
It also follows from \eqref{czzdq} that
\begin{align*}
\|\brho_k\|_2^2 \leq c(r^*+s_u+s_v) \eta_{n}^4 \{n^{-1}\log(pq)\}.
\end{align*}

It remains to bound term $\brho_k\trans\wt{\W}_k\wt{\bepsilon}_k$ above. Clearly, we have
\begin{align}
	|\brho_k\trans\wt{\W}_k\wt{\bepsilon}_k|& \leq | n^{-1}\brho_k\trans\wt{\W}_k\wt{\M}_k\E\trans\X\wt{\u}_k| +  | n^{-1}\brho_k\trans\wt{\W}_k\X\trans\E\v_k^*|. \nonumber
\end{align}
Based on the property of $\wt{\W}_k$ in Lemma \ref{rankr:aw}, it follows from similar arguments as for \eqref{zcasdq2} that 
\begin{align}
	| n^{-1}\brho_k\trans\wt{\W}_k\X\trans\E\v_k^*|
	 \leq c \max\{s_{\max}, (r^*+s_u+s_v)\}^{1/2} (r^*+s_u+s_v)^{3/2} \eta_n^2 \{n^{-1}\log(pq)\}. \nonumber
\end{align}
Observe that $\wt{\M}_{k} =  -z_{kk}^{-1}\wh{\bSigma}\wt{\C}^{(2)}$.
With the aid of similar arguments as for \eqref{czkjdkq}, we can obtain that 
\begin{align*}
	\norm{\brho_k\trans\wt{\W}_k\wt{\M}_k}_0   \leq  c(r^* + s_u + s_v).
\end{align*}
Denote by $s = c(r^* + s_u + s_v)$.
Similar to \eqref{sdaqessss}, we can deduce that
\begin{align}
	n^{-1}\norm{\E\trans\X\wt{\u}_k}_{2,s} & \leq  s^{1/2} n^{-1}\norm{\E\trans\X }_{\max}\norm{\wt{\u}_k}_0^{1/2}\norm{\wt{\u}_k}_2 \nonumber \\ &
	\leq c (r^*+s_u+s_v)\{n^{-1}\log(pq)\}^{1/2} d_k^*. \nonumber
\end{align}
Then similar to \eqref{zcasdq}, an application of Lemma \ref{rankr:boundm} in Section \ref{new.Sec.B.17} results in 
\begin{align}
	| n^{-1}\brho_k\trans\wt{\W}_k\wt{\M}_k\E\trans\X\wt{\u}_k |
	&\leq \|\brho\trans\wt{\W}_k\wt{\M}_k\|_2  \|  n^{-1} \E\trans\X\wt{\u}_k \|_{2,s} \nonumber \\[5pt]
	&\leq \|\brho\trans\wt{\W}_k\|_2 \|\wt{\M}_k\|_2 \|  n^{-1} \E\trans\X\wt{\u}_k \|_{2,s} \nonumber \\[5pt]
	&\leq  c   (r^*+s_u+s_v)^{2}\eta_{n}^2\{n^{-1}\log(pq)\} d_{k+1}^* d_k^{*-1}.  \nonumber
\end{align}
It follows that
\begin{align}
	|\brho_k\trans\wt{\W}_k\wt{\bepsilon}_k|
	\leq  c \max\{s_{\max}, (r^*+s_u+s_v)\}^{1/2}    (r^*+s_u+s_v)^{3/2}\eta_{n}^2\{n^{-1}\log(pq)\}.
\end{align}

Thus, combining the above results yields that
\begin{align*}
	&\|\wt{\u}_k - \u_k^*\|_2^2 \leq c(r^*+s_u+s_v) \eta_{n}^4 \{n^{-1}\log(pq)\}, \nonumber\\[5pt]
	& \big|( \u_k^* - \wt{\u}_k)\trans \wt{\W}_k\wt{\psi}_k(\wt{\u}_k,\wt{\boldeta}_k)\big| \nonumber\\[5pt]
	&\quad \leq c \max\{s_{\max}^{1/2} , (r^*+s_u+s_v)^{1/2}, \eta_n^2\}   (r^*+s_u+s_v)^{3/2}\eta_{n}^2\{n^{-1}\log(pq)\}.
\end{align*}
Then \eqref{czxdqq} can be rewritten as 
\begin{align*}
\sqrt{n}(\widehat{d}_k^2 - d_k^{*2} ) = \sqrt{n} 2 \u_k\strans (\wh{\u}_k - \u_k^*)  
+ t_{d_k}^{\prime},
\end{align*}
where $t_{d_k}^{\prime} = O(\max\{s_{\max}^{1/2}, (r^*+s_u+s_v)^{1/2},\eta_n^2\}   (r^*+s_u+s_v)^{3/2}\eta_{n}^2{\log(pq)}/\sqrt{n})$.
For term $\sqrt{n} 2 \u_k\strans (\wh{\u}_k - \u_k^*) $ above,
replacing $\a$ with $ 2\u_k^*$ and using similar arguments as in 
 the proof of Theorem \ref{theorkapor} in Section \ref{new.Sec.A.4}, it holds that
\begin{align*}
 \sqrt{n} 2 \u_k\strans (\wh{\u}_k - \u_k^*)  = h_{d_k} + t_{d_k}^{\prime \prime}, 
\end{align*}
where the distribution term $h_{d_k} =  2\u_k\strans \W^{*}_k(\X\trans\E\v_k^* - \M_{k}^{*} \E\trans\X {\u}_k^{*})/\sqrt{n} \sim \N(0,\nu_{d_k}^2)$ with variance
\begin{align*}
    \nu_{d_k}^2 = 4\u_k\strans\W_k^*(z_{kk}^{*}\M_{k}^{*}\bSigma_e\M_{k}\strans + \v_k\strans\bSigma_e\v_k^* \wh{\bSigma} - 2\wh{\bSigma}\u_k^*  \v_k\strans\bSigma_e\M_{k}^{* T})\W_k\strans\u_k^*,
\end{align*}
and the error term $t_{d_k}^{\prime \prime}$ satisfies that 
\begin{align*}
t_{d_k}^{\prime \prime} &= O\Big\{ d_k^* \max\{1, d_k^{*-1}, d_k^{*-2}\} \max\{s_{\max}^{1/2}, (r^*+s_u+s_v)^{1/2},\eta_n^2\}  (r^*+s_u+s_v)^{3/2}\eta_n^2 \log(pq)/\sqrt{n} \\
    &\quad ~ +  (r^*+s_u+s_v)\eta_{n}^2\{n^{-1}\log(pq)\}^{1/2}  d_{k+1}^{*}d_k^{*-2}(\sum_{i=1}^{k-1}d_i^*)\Big\}.
\end{align*}

Finally, we can obtain that 
\begin{align*}
\sqrt{n}(\widehat{d}_k^2 - d_k^{*2} ) = h_{d_k}+ t_{d_k}
\end{align*}
with $h_{d_k} \sim \N(0,\nu_{d_k}^2)$, where the error term $t_{d_k}$ satisfies that 
\begin{align*}
t_{d_k} &= t_{d_k}^{\prime } + t_{d_k}^{\prime \prime} = O\Big\{(r^*+s_u+s_v)\eta_{n}^2\{n^{-1}\log(pq)\}^{1/2}  d_{k+1}^{*}d_k^{*-2}(\sum_{i=1}^{k-1}d_i^*)  \\
&\quad + d_k^* \max\{1, d_k^{*-1}, d_k^{*-2}\} \max\{s_{\max}^{1/2}, (r^*+s_u+s_v)^{1/2},\eta_n^2\}  (r^*+s_u+s_v)^{3/2}\eta_n^2 \log(pq)/\sqrt{n}   \Big\}.
\end{align*}
This completes the proof of Theorem \ref{theorkukcorok}.

\subsection{Proof of Theorem \ref{coro:var:rank22uk}} \label{weakrankr:sec28}

This proof of Theorem \ref{coro:var:rank22uk} is similar to that of  Theorem \ref{coro:var:rank2uk} in Section \ref{nearr:sec14}.
For some $s_0$, $\alpha_n$, and $\tau_n$, let us define $p$-dimensional vectors $\wt{\a}$ and $\a^*$ satisfying that 
\begin{align*}
	&\norm{\wt{\a}}_0 \leq s_0, \ \norm{\a^*}_0 \leq s_0, \  \norm{\wt{\a}}_2 \leq c\alpha_n, \ \norm{\a^*}_2 \leq c\alpha_n, \\
	&\norm{\wt{\a} - \a^*}_0 \leq s_0, \  \norm{\wt{\a} - \a^*}_2 \leq \tau_n.
\end{align*}
In addition, we define
$ \wt{\nu}^2_e =\varphi_1 + \varphi_2  - 2\varphi_3$ and $ {\nu}^2 =\varphi_1^* + \varphi_2^*  - 2 \varphi_3^*$ with 
\begin{align*}
	&\varphi_1 = \wt{\u}_k\trans\wh{\bSigma}\wt{\u}_k \wt{\a}\trans\wt{\W}_{k}\wt{\M}_{k}\bSigma_e\wt{\M}_{k}\trans\wt{\W}_k\trans\wt{\a},
	\ \varphi_1^* = \u_k\strans\wh{\bSigma}\u_k^*  \a\strans\W_k^*\M_{k}^{*}\bSigma_e\M_{k}\strans\W_k\strans\a^*, \\
	&\varphi_2 =   \wt{\v}_k\trans\bSigma_e\wt{\v}_k \wt{\a}\trans\wt{\W}_k\wh{\bSigma}\wt{\W}_k\trans\wt{\a}, \ ~ ~ \varphi_2^* = \v_k\strans\bSigma_e\v_k^* \a\strans\W_k^*\wh{\bSigma}\W_k\strans\a^*, \\
	&\varphi_3 =  \wt{\a}\trans\wt{\W}_k\wh{\bSigma}\wt{\u}_k \wt{\v}_k\trans\bSigma_e\wt{\M}_{k}\trans\wt{\W}_k\trans\wt{\a}, \
	\varphi_3^* =  \a\strans\W_k^*\wh{\bSigma}\u_k^*  \v_k\strans\bSigma_e\M_{k}^{* T}\W_k\strans\a^*.
\end{align*}
Then it holds that 
\begin{align}\label{phiij2}
	|  \wt{\nu}^2_e -  {\nu}^2 | &\leq  | \varphi_1 - \varphi_1^*| +  | \varphi_2 - \varphi_2^*| + 2 | \varphi_3 - \varphi_3^*|  \nonumber \\
	&: = A_1 + A_2 + 2 A_3.
\end{align}
We will bound the three terms in (\ref{phiij2}) above following similar analysis as in the proof of Theorem \ref{coro:var:rank2uk}, which will be conditional on $\wh{\bTheta}$ satisfying Definition \ref{defi2:acceptable} and $\wt{\C}$ satisfying Definition \ref{lemmsofar}.

\medskip

\noindent \textbf{(1). The upper bound on $A_1$}.
Observe that $$A_1 = |\wt{z}_{kk} \wt{\a}\trans\wt{\W}_{k}\wt{\M}_{k}\bSigma_e\wt{\M}_{k}\trans\wt{\W}_k\trans\wt{\a}	- {z}_{kk}^* \a\strans\W_k^*\M_{k}^{*}\bSigma_e\M_{k}\strans\W_k\strans\a^*|.$$ 
Let us define 
\begin{align*}
	& A_{11} =  | \wt{\a}\trans\wt{\W}_{k}\wt{\M}_{k}\bSigma_e\wt{\M}_{k}\trans\wt{\W}_k\trans\wt{\a}-  \a\strans\W_k^*\M_{k}^{*}\bSigma_e\M_{k}\strans\W_k\strans\a^*|, \\[5pt]
	&A_{12} =  | \a\strans\W_k^*\M_{k}^{*}\bSigma_e\M_{k}\strans\W_k\strans\a^*|.
\end{align*}
Under Conditions \ref{con3}--\ref{con4}, by part (c) of Lemma \ref{lemmauv} we have that 
\begin{align}
	|A_1|
	&  \leq  |\wt{z}_{kk}| A_{11}  +   |\wt{z}_{kk}  - {z}_{kk}^*| A_{12} \nonumber\\
	& \leq c d_k^{*2}  A_{11} + c  d_k^{*}(r^*+s_u+s_v)^{1/2} \eta_n^2 \{n^{-1}\log(pq)\}^{1/2} A_{12}. \label{a1lem8}
\end{align}
For term $A_{12}$ above, from Condition \ref{cone}, Lemma \ref{rankr:boundm} in Section \ref{new.Sec.B.17}, and part (c) of Lemma \ref{rankr:aww2} in Section \ref{new.Sec.B.18}, we can show that 
\begin{align}
	A_{12}  \leq \norm{\a\strans\W_k^* }_2  \norm{\M_{k}^* }_2  \norm{\bSigma_e }_2  \norm{\M_{k}\strans }_2 \norm{\W_k\strans\a^* }_2 \leq  c s_0 \alpha_n^2 d_k^{*-4}d_{k+1}^{*2}. \label{a22lem8}
\end{align}
Further,  for term $A_{11}$ above, it holds that
\begin{align*}
	A_{11} & \leq  | \wt{\a}\trans\wt{\W}_{k}\wt{\M}_{k}\bSigma_e(\wt{\M}_{k}\trans\wt{\W}_k\trans\wt{\a} - \M_{k}\strans\W_k\strans\a^*)|  \\
 &\quad+  |( \wt{\a}\trans\wt{\W}_{k}\wt{\M}_{k} - \a\strans\W_k^*\M_{k}^{*})\bSigma_e\M_{k}\strans\W_k\strans\a^*|  \\[5pt]
	&\leq  \|\wt{\a}\trans\wt{\W}_{k}\wt{\M}_{k}\bSigma_e\|_2  \norm{\wt{\M}_{k}\trans\wt{\W}_k\trans\wt{\a} - \M_{k}\strans\W_k\strans\a^*}_2 \\
 &\quad+    \norm{ \wt{\a}\trans\wt{\W}_{k}\wt{\M}_{k} - \a\strans\W_k^*\M_{k}^{*}}_2  \|\bSigma_e\M_{k}\strans\W_k\strans\a^*\|_2.
\end{align*}

By Condition \ref{cone}, Lemma \ref{rankr:boundm}, and Lemma \ref{rankr:aww2}, it is easy to see that 
\begin{align*}
	&\|\wt{\a}\trans\wt{\W}_{k} \wt{\M}_{k}\bSigma_e\|_2 \leq  \norm{\wt{\a}\trans\wt{\W}_{k}}_2  \norm{\wt{\M}_{k} }_2 \norm{\bSigma_e }_2 \leq  c s_0^{1/2}   d_k^{*-2} d_{k+1}^* \alpha_n, \\[5pt]
	& \|\bSigma_e\M_{k}\strans\W_k\strans\a^*\|_2 \leq \norm{\bSigma_e }_2 \norm{\M_{k}\strans }_2  \|\W_k\strans\a_2^*\|_2 \leq  c s_0^{1/2}   d_k^{*-2} d_{k+1}^* \alpha_n.
\end{align*}
In light of parts (c) and (d) of Lemma \ref{rankr:aww2}, we can deduce that 
\begin{align}
	\|\wt{\a}\trans\wt{\W}_k -   \a\strans\W_k^* \|_2 & \leq \|\wt{\a}\trans(\wt{\W}_k - \W_k^* )\|_2 + \|(\wt{\a} -   \a^*)\trans\W_k^* \|_2 \nonumber\\
	&\leq      c s_0^{1/2} \alpha_n (r^*+s_u+s_v)^{1/2} \eta_n^2\{n^{-1}\log(pq)\}^{1/2}d_k^{*-2} d_{k+1}^*  \nonumber\\
 &\quad+ cs_0^{1/2}  \tau_n. \label{cda2ef}
\end{align}
Then it follows from \eqref{cda2ef} and Lemmas \ref{rankr:boundm}--\ref{rankr:aww2}  that
\begin{align}\label{mwamwa}
	\norm{ \wt{\a}\trans\wt{\W}_{k}\wt{\M}_{k} - \a\strans\W_{k}^*\M_{k}^*}_2  & \leq \norm{\wt{\a}\trans\wt{\W}_{k}}_2 \norm{ \wt{\M}_{k} - \M_{k}^*   }_2 + \norm{ \wt{\a}\trans\wt{\W}_{k} - \a\strans\W_{k}^*}_2 \norm{ \M_{k}^*  }_2 \nonumber\\[5pt]
	&\leq  c  s_0^{1/2} \alpha_n  (r^*+s_u+s_v)^{1/2} \eta_n^2\{n^{-1}\log(pq)\}^{1/2} d_k^{*-2} \nonumber\\
 &\quad+ c  s_0^{1/2}   \tau_n d_k^{*-2} d_{k+1}^{*}. 
\end{align}
Hence, combining the above results yields that 
\begin{align}
	A_{11} \leq  c  s_0 \alpha_n^2  (r^*+s_u+s_v)^{1/2} \eta_n^2\{n^{-1}\log(pq)\}^{1/2} d_k^{*-4} d_{k+1}^{*} + c  s_0  \alpha_n \tau_n d_k^{*-4} d_{k+1}^{*2}. \label{a11lem8}
\end{align}
Therefore, with the aid of \eqref{a1lem8}, \eqref{a22lem8}, and \eqref{a11lem8}, we can obtain that 
\begin{align}\label{phia1}
	A_1
	\leq 
	c  s_0 \alpha_n^2  (r^*+s_u+s_v)^{1/2} \eta_n^2\{n^{-1}\log(pq)\}^{1/2} d_k^{*-2} d_{k+1}^{*} + c  s_0  \alpha_n \tau_n d_k^{*-2} d_{k+1}^{*2}.
\end{align}

\smallskip

\noindent \textbf{(2). The upper bound on $A_2$}.
Notice that $$	A_2 =  | \wt{\v}_k\trans\bSigma_e\wt{\v}_k \wt{\a}\trans\wt{\W}_k\wh{\bSigma}\wt{\W}_k\trans\wt{\a} - \v_k\strans\bSigma_e\v_k^* \a\strans\W_k^*\wh{\bSigma}\W_k\strans\a^*|.$$
From Condition \ref{cone}, Definition \ref{lemmsofar} that $\norm{\wt{\v}_k}_2 = \|\v_k^* \|_2 = 1 $, and part (a) of Lemma \ref{lemmauv}, it holds that $|  \wt{\v}_k\trans\bSigma_e\wt{\v}_k | \leq  \|  \wt{\v}_k\trans\|_2 \|\bSigma_e\|_2 \|\wt{\v}_k \|_2  \leq c$ and
\begin{align*}
	| \wt{\v}_k\trans\bSigma_e\wt{\v}_k - \v_k\strans\bSigma_e\v_k^* | &\leq   | \wt{\v}_k\trans\bSigma_e(\wt{\v}_k - \v_k^*) | +  | ( \wt{\v}_k\trans - \v_k\strans)\bSigma_e\v_k^* | \\
	&\leq  \| \wt{\v}_k\|_2 \|\bSigma_e\|_2 \|\wt{\v}_k - \v_k^* \|_2 +  \|  \wt{\v}_k - \v_k^*\|_2 \|\bSigma_e\|_2 \|\v_k^* \|_2 \\
	& \leq   c (r^*+s_u+s_v)^{1/2} \eta_n^2\{n^{-1}\log(pq)\}^{1/2}d_k^{*-1}.
\end{align*}
An application of the triangle inequality gives that 
\begin{align}
	A_2 &\leq     | \wt{\v}_k\trans\bSigma_e\wt{\v}_k | |\wt{\a}\trans\wt{\W}_k\wh{\bSigma}\wt{\W}_k\trans\wt{\a} -   \a\strans\W_k^*\wh{\bSigma}\W_k\strans\a^*| \nonumber \\
	&\quad + | \wt{\v}_k\trans\bSigma_e\wt{\v}_k - \v_k\strans\bSigma_e\v_k^* | |  \a\strans\W_k^*\wh{\bSigma}\W_k\strans\a^*| \nonumber\\
	&\leq c  |\a\trans\wt{\W}_k\wh{\bSigma}\wt{\W}_k\trans\a  -   \a\strans\W_k^*\wh{\bSigma}\W_k\strans\a^*| \nonumber\\
	&\quad +  c  |  \a\strans\W_k^*\wh{\bSigma}\W_k\strans\a^*|  (r^*+s_u+s_v)^{1/2} \eta_n^2\{n^{-1}\log(pq)\}^{1/2}d_k^{*-1}. \nonumber
\end{align}

Using similar arguments as for \eqref{daszcghns0}--\eqref{daszcghns3}, it follows from Lemma \ref{rankr:aww2} that 
\begin{align*}
	&\norm{\wh{\bSigma}\W_k\strans\a^* }_2 \leq  c s_0^{1/2} \alpha_n, \nonumber \\[5pt]
	&|  \a\trans\W_k^*\wh{\bSigma}\W_k\strans\a^*| \leq  \norm{\a\trans\W_k^* }_2 \norm{\wh{\bSigma}\W_k\strans\a^* }_2 \leq  c s_0 \alpha_n^2, \nonumber 
	\\[5pt]
	&\|\wh{\bSigma}\wt{\W}_k\trans\a - \wh{\bSigma} \W_k\strans\a^*\|_2 \leq 
	\|\wh{\bSigma}\wt{\W}_{k}\trans(\wt{\a} - \a^*)\|_2 + 	\|\wh{\bSigma}(\wt{\W}_{k}\trans - {\W}_{k}\strans)\a^*\|_2 \\
	&\qquad \qquad \leq c s_0^{1/2} \alpha_n +  c s_0^{1/2} \alpha_n  (r^*+s_u+s_v)^{1/2} \eta_n^2\{n^{-1}\log(pq)\}^{1/2}d_{k+1}^*d_{k}^{*-2}.
\end{align*}
Along with \eqref{cda2ef}, it holds that
\begin{align*}
	|\wt{\a}\trans\wt{\W}_k&\wh{\bSigma}\wt{\W}_k\trans\wt{\a} -   \a\strans\W_k^*\wh{\bSigma}\W_k\strans\a^*| \\
	&\leq
	\|\wt{\a}\trans\wt{\W}_k \|_2 \| \wh{\bSigma}(\wt{\W}_k\trans\wt{\a} -  \W_k\strans\a^*)\|_2
	+ \|\wt{\a}\trans\wt{\W}_k -   \a\strans\W_k^* \|_2 \| \wh{\bSigma}\W_k\strans\a^* \|_2 \\
	&\leq  c s_0 \alpha_n^2  (r^*+s_u+s_v)^{1/2} \eta_n^2\{n^{-1}\log(pq)\}^{1/2} d_{k+1}^*d_{k}^{*-2} +  cs_0 \alpha_n \tau_n.
\end{align*}
Thus, combining the above results leads to 
\begin{align}\label{phia2}
	A_{2}
	&\leq  c s_0 \alpha_n^2  (r^*+s_u+s_v)^{1/2} \eta_n^2\{n^{-1}\log(pq)\}^{1/2} d_{k}^{*-1} +  cs_0 \alpha_n \tau_n.
\end{align}

\smallskip

\noindent \textbf{(3). The upper bound on $A_3$}.
Note that
$$A_3 =  | \wt{\a}\trans\wt{\W}_k\wh{\bSigma}\wt{\u}_k \wt{\v}_k\trans\bSigma_e\wt{\M}_{k}\trans\wt{\W}_k\trans\wt{\a}-
\a\strans\W_k^*\wh{\bSigma}\u_k^*  \v_k\strans\bSigma_e\M_{k}^{* T}\W_k\strans\a^*|.$$
In view of Condition \ref{cone}, Lemma \ref{rankr:boundm}, and Lemma \ref{rankr:aww2},  we have that 
\begin{align*}
	&|\wt{\a}\trans\wt{\W}_k\wh{\bSigma}\wt{\u}_k |
	\leq \|\wt{\a}\trans\wt{\W}_k\|_2 \|\wh{\bSigma}\wt{\u}_k \|_2
	\leq  c s_0^{1/2}\alpha_n  d_k^*, \\[5pt]
	&| \v_k\strans\bSigma_e\M_{k}^{* T}\W_k\strans\a^* | \leq
	\| \v_k\strans\|_2 \|\bSigma_e\|_2 \|\M_{k}^{* T}\|_2 \|\W_k\strans\a^* \|_2
	\leq  c s_0^{1/2} \alpha_n d_k^{*-2} d_{k+1}^*.
\end{align*}
Let us define $ A_{31} = | \wt{\v}_k\trans\bSigma_e\wt{\M}_{k}\trans\wt{\W}_k\trans\wt{\a}-
\v_k\strans\bSigma_e\M_{k}^{* T}\W_k\strans\a^*  |$.
Then we can obtain that 
\begin{align*}
	A_3 &\leq   | \wt{\a}\trans\wt{\W}_k\wh{\bSigma}\wt{\u}_k | |\wt{\v}_k\trans\bSigma_e\wt{\M}_{k}\trans\wt{\W}_k\trans\wt{\a} - \v_k\strans\bSigma_e\M_{k}^{* T}\W_k\strans\a^*| \\[5pt]
	& \quad+   | \wt{\a}\trans\wt{\W}_k\wh{\bSigma}\wt{\u}_k -
	\a\strans\W_k^*\wh{\bSigma}\u_k^*| | \v_k\strans\bSigma_e\M_{k}^{* T}\W_k\strans\a^*|  \\[5pt]
	& \leq c s_0^{1/2}  \alpha_n d_k^* A_{31}  + c s_0^{1/2} \alpha_n d_k^{*-2}d_{k+1}^* |\wt{\a}\trans\wt{\W}_k\wh{\bSigma}\wt{\u}_k  -   \a\strans\W_k^*\wh{\bSigma} \u_k^*|.
\end{align*}

For term $A_{31}$ above, it follows from Condition \ref{cone}, Lemma \ref{rankr:boundm}, Lemma \ref{rankr:aww2}, and \eqref{mwamwa}  that
\begin{align*}
	A_{31} &\leq \|  \wt{\v}_k\trans \|_2 \|\bSigma_e \|_2 \| \wt{\M}_{k}\trans\wt{\W}_k\trans\wt{\a}-
	\M_{k}^{* T}\W_k\strans\a^*  \|_2 + \|  \wt{\v}_k -  \v_k^* \|_2 \|\bSigma_e \|_2 \| \M_{k}^{* T}\|_2 \|\W_k\strans\a^*  \|_2  \\[5pt]
	&\leq   c  s_0^{1/2} \alpha_n  (r^*+s_u+s_v)^{1/2} \eta_n^2\{n^{-1}\log(pq)\}^{1/2} d_k^{*-2} + c  s_0^{1/2}   \tau_n d_k^{*-2} d_{k+1}^{*}. 
\end{align*}
Further, in light of Lemma \ref{lemmauv}, Lemma \ref{rankr:aww2}, and \eqref{cda2ef}, it holds that 
\begin{align*}
	| \wt{\a}\trans\wt{\W}_k\wh{\bSigma}\wt{\u}_k -	\a\strans\W_k^*\wh{\bSigma}\u_k^*|&\leq
	\|\wt{\a}\trans\wt{\W}_k \|_2  \|\wh{\bSigma}(\wt{\u}_k -  \u_k^*)\|_2
	+ \|\wt{\a}\trans\wt{\W}_k  -   \a\strans\W_k^* \|_2  \|\wh{\bSigma} \u_k^*\|_2  \\
	&\leq  c s_0^{1/2} \alpha_n (r^*+s_u+s_v)^{1/2} \eta_n^2\{n^{-1}\log(pq)\}^{1/2}  + c s_0^{1/2} \tau_n d_k^*.
\end{align*}
Combining the above results leads to 
\begin{align}\label{phia3}
	&A_3  \leq  c  s_0 \alpha_n^2  (r^*+s_u+s_v)^{1/2} \eta_n^2\{n^{-1}\log(pq)\}^{1/2} d_k^{*-1} + c  s_0 \alpha_n   \tau_n d_k^{*-1} d_{k+1}^{*}. 
\end{align}
Hence, a combination of \eqref{phia1}, \eqref{phia2}, and \eqref{phia3} yields that
\begin{align}\label{phi-phik223}
	|  \wt{\nu}^2_e -  {\nu}^2 | \leq  c  s_0 \alpha_n^2  (r^*+s_u+s_v)^{1/2} \eta_n^2\{n^{-1}\log(pq)\}^{1/2} d_k^{*-1}  + c  s_0  \alpha_n \tau_n.
\end{align}

{
Define $\wt{\nu}^2$ with $\bSigma_e$ in $\wt{\nu}^2_e$ replaced by the 
acceptable estimator  $\wt{\bSigma}_e$ satisfying Definition \ref{defi:error}. 
Since in the above three part proofs, we have only used the property of ${\bSigma}_e$ such that $\norm{{\bSigma}_e}_2 \leq c$, using similar arguments as for \eqref{eq:error} and \eqref{phi-phik222} in the proof of Theorem \ref{coro:var:rank2uk} in Section \ref{nearr:sec14},  it follows from \eqref{phi-phik223} that
\begin{align}\label{phi-phik22}
	|  \wt{\nu}^2 -  {\nu}^2 | \leq  c  s_0 \alpha_n^2  (r^*+s_u+s_v)^{1/2} \eta_n^2\{n^{-1}\log(pq)\}^{1/2} d_k^{*-1}  + c  s_0  \alpha_n \tau_n.
\end{align}}

When $\wt{\a} = 2\wt{\u}_k$ and $\a^* = 2\u_k^*$, we have that $s_0 = c(r^*+s_u + s_v), \alpha_n = d_k^*$, and $\tau_n = c(r^*+s_u+s_v)^{1/2} \eta_n^2\{n^{-1}\log(pq)\}^{1/2}$. Then with \eqref{phi-phik22}, we can show that
\begin{align*}
	&|\wt{\nu}_{d_k}^2 - {\nu}_{d_k}^2| \leq c (r^*+s_u+s_v)^{3/2} \eta_n^2\{n^{-1}\log(pq)\}^{1/2}d_k^*.
\end{align*}
Moreover, when $\wt{\a} = \a^* \in\mathcal{A}(m)=\{\a\in\R^p:\norm{\a}_0\leq m, \norm{\a}_2 = 1\}$, we have $s_0 = m, \alpha_n = 1$, and $\tau_n=0$, which yields that 
\begin{align*}
	&|\wt{\nu}_{k}^2 - {\nu}_{k}^2| \leq c m  (r^*+s_u+s_v)^{1/2} \eta_n^2\{n^{-1}\log(pq)\}^{1/2}d_k^{*-1}.
\end{align*}
This concludes the proof of Theorem \ref{coro:var:rank22uk}.

\blue{
\subsection{Proof of Theorem \ref{corollary1}}\label{sec:app:coro7}

The proof of this theorem is similar to that of Theorem \ref{theorkr}. The main difference is that we exploit the data independence to bound the remainder terms. 
Note that the construction of the debiased estimate is given by 
\begin{align*}
	\wh{\u}_k^{\operatorname{split}} 
	= \wt{\u}_k  - \wt{\W}_k\wt{\psi}_k(\wt{\u}_k,\boldeta^*_k) + \wt{\W}_k(\wt{\psi}_k(\wt{\u}_k,\boldeta^*_k) -\wt{\psi}_k(\wt{\u}_k,\wt{\boldeta}_k)).
	\end{align*}
	Then by Propositions \ref{prop:rankr2}--\ref{prop:rankr3}, Lemma \ref{prop:rankr1}, and the initial estimates satisfying Definition \ref{lemmsofar}, it holds that  
	\begin{align}
	\sqrt{n}\a\trans(\wh{\u}_k^{\operatorname{split}}-\u_k^*)  &=  - \sqrt{n}\a\trans\wt{\W}_k\wt{\bepsilon}_k -\sqrt{n} \a\trans\wt{\W}_k(\wt{\psi}_k(\wt{\u}_k,\wt{\boldeta}_k) - \wt{\psi}_k(\wt{\u}_k,\boldeta^*_k)) - \sqrt{n}\a\trans\wt{\W}_k\wt{\bdelta}_k  \nonumber\\[5pt]
	& \quad + \sqrt{n}\a\trans(\I_p - \wt{\W}_k\mathbf{T}_k)(\wt{\u}_k-\u_k^*)  - \sqrt{n}\a\trans\wt{\W}_k\wt{\M}_k\C_{-k}\strans \wh{\bSigma} \u_k^*, \label{ukrankeq22}
	\end{align}
	where
	\begin{align}
	&\wt{\bepsilon}_k =  n^{-1}\wt{\M}_k\E\trans\X\wt{\u}_k - n^{-1}\X\trans\E\v_k^*, \label{eprankr22} \\[5pt]
	&\wt{\bdelta }_k =   \left\{\wt{\M}_k(\wt{\v}_k - \v_k^*)\wt{\u}_k\trans- \wt{\M}_k( \wt{\C}_{-k}\trans - \C_{-k}\strans)\right\}\wh{\bSigma} (\wt{\u}_k - \u_k^*),  \nonumber\\[5pt]
	&\wt{\W}_k = \widehat{\bTheta} \left\{  \I_p +   \wt{z}_{kk}^{-1}\wh{\bSigma}\wt{\U}_{-k}(\I_{r^*-1} -\wt{z}_{kk}^{-1}\wt{\U}_{-k}\trans\wh{\bSigma} \wt{\U}_{-k})^{-1}\wt{\U}_{-k}\trans\right\}, \nonumber \\[5pt]
	& \mathbf{T}_k = ( \I_p - \wt{\M}_k\wt{\v}_k\wt{\u}_k\trans + \wt{\M}_k\wt{\C}_{-k}\trans )\wh{\bSigma}, \ \ \wt{\M}_k =  -\wt{z}_{kk}^{-1}\wh{\bSigma}\wt{\C}_{-k}. \nonumber
	\end{align}
The last three terms in \eqref{ukrankeq22} follow the same argument as in \eqref{h1}--\eqref{h3} of the proof of Theorem \ref{theorkr}, which implies that  
\begin{align*}
	&	\abs{\a\trans(\I_p -\wt{\W}_k\mathbf{T}_k)(\wt{\u}_k-\u_k^*)} \leq cm^{1/2}(r^*+s_u+s_v)\eta_n^2\{n^{-1}\log(pq)\}, \\[5pt]
	&|\a\trans\wt{\W}_k \wt{\bdelta}_k | \leq cm^{1/2}    (r^*+s_u+s_v)\eta_n^4 \{n^{-1}\log(pq)\}, \\[5pt]
	&| \a\trans\wt{\W}_k\wt{\M}_k\C_{-k}\strans \wh{\bSigma} \u_k^*| = o(m^{1/2} n^{-1/2}).
\end{align*}

Denote by $h_k = -\a\trans \W^{*}_k \M_{k}^{*} \E\trans\X {\u}_k^{*}/\sqrt{n}
+  \a\trans\W^*_k\X\trans\E\v_k^*/\sqrt{n}$.
By exploiting the sample splitting technique, we can bound the first two terms of \eqref{ukrankeq22} in Lemmas \ref{lemma:1rk3r1} and \ref{lemm:tayds} as 
\begin{align*}
	&\abs{-\a\trans\wt{\W}_k\wt{\bepsilon}_k - h_k  / \sqrt{n}} \leq c m^{1/2}  (r^* +s_u + s_v)^{2}\eta_n^2\{ n^{-1}\log(pq)\}, \\
	& |\a\trans\wt{\W}_k (\wt{\psi}_k(\wt{\u}_k,\wt{\boldeta}_k) - \wt{\psi}_k(\wt{\u}_k,\boldeta_k^*) )| \\[5pt]
	& \leq  c m^{1/2} \max\{s_{\max}^{1/2}, (r^*+s_u+s_v)^{1/2}\eta_n^2  \}(r^*+s_u+s_v)^{1/2}\eta_n^2\{n^{-1}\log(pq)\}.
\end{align*}
Thus, combining above terms leads to
\begin{align*}
	\sqrt{n}\a\trans(\wh{\u}_k^{\operatorname{split}} -\u_k^*) = h_k + t_k, 
	\end{align*}
	where
	$ t_k = O\big(m^{1/2}\{s_{\max}^{1/2}, (r^*+s_u+s_v)^{1/2} \eta_n^2 \}(r^*+s_u+s_v)^{1/2}\eta_n^2\log(pq)/\sqrt{n} \big).$
Furthermore, under Condition \ref{cone}, we can obtain that $h_k$ is normally distributed with $\mathbb{E}( h_k|\X) = 0$ and variance
	\begin{align*}
	\nu_k^2 = \text{var}( h_k|\X) = \a\trans\W_k^*(z_{kk}^{*}\M_{k}^{*}\bSigma_e\M_{k}\strans + \v_k\strans\bSigma_e\v_k^* \wh{\bSigma} - 2\wh{\bSigma}\u_k^*  \v_k\strans\bSigma_e\M_{k}^{* T})\W_k\strans\a.
	\end{align*}
This completes the proof of Theorem \ref{corollary1}. 
}

\subsection{Proof of Proposition \ref{prop:deri2}} \label{nearrr:sec1}
	Under the SVD constraint \eqref{SVDc} that  $\V\trans\V = \I_{r^*}$, we have that  $${\v}_i\trans\v_i = 1$$ for $1 \leq i \leq r^*$.  
 From the definition of the Stiefel manifold given in Section \ref{sec:stiefel}, we see that all vectors $\v_i$ belong to the Stiefel manifold  $\text{St}(1, q) = \{ \v \in \mathbb{R}^q: \v\trans\v = 1   \}$.  
For function $\wt{\psi}_k$, denote by $\der{\wt{\psi}_k}{\v_i}$ with $1 \leq i \leq r^*$ the regular derivative vectors on the Euclidean space. Then under the Stiefel manifold $\text{St}(1, q)$, applying Lemma \ref{lemmgradst} in Section \ref{sec:stiefel} and similar to \eqref{grad:sti:eq}, we can show that the manifold gradient of $\wt{\psi}_k$ at $\v_i \in \text{St}(1, q) $ is given by $$(\I_q - \v_i{\v_i\trans})\der{\wt{\psi}_k}{\v_i}.$$ 

Moreover, for vectors $\u_j$ with $1 \leq j \leq r^*$ and $j \neq k$, it holds that 
$$\u_j\trans\u_j = d_j^2.$$ Since $d_j$ is unknown and its estimate varies across different estimation methods, there is no unit length constraint on ${\u}_i$ and we can take the gradient of $\wt{\psi}_k$ with respect to ${\u}_i$ directly on the Euclidean space $\mathbb{R}^p$ as $\der{\wt{\psi}_k}{\u_j}$. Recall that $\boldeta_k = \left(\u_1\trans,\ldots, \u_{k-1}\trans, \u_{k+1}\trans, \ldots, \u_{r^*}\trans, \v_1\trans, \ldots, \v_{r^*}\trans\right)\trans$. Therefore, the gradient of $\wt{\psi}_k$ on the manifold can be written as $$	\Q\big(\der{\wt{\psi}_k}{\boldeta_k}\big),$$ where $\Q = \diag{\I_{p(r^* - 1)},$ $\I_q - \v_1\v_1\trans, \dots, \I_q - \v_{r^*}\v_{r^*}\trans}$. This completes the proof of Proposition \ref{prop:deri2}.

\subsection{Proof of Proposition \ref{prop:rankr2}} \label{new.Sec.A.11}

The proof of Proposition \ref{prop:rankr2} consists of two parts. Specifically, the first part establishes the theoretical results under the rank-2 case, while the second part further extends the results to the general rank case.

\bigskip

\noindent\textbf{Part 1: Proof for the rank-2 case.} Under the strongly orthogonal factors, the technical analyses for the theoretical results of $\u_1^*$ and $\u_2^*$ are basically the same, so we present the proof only for $\u_1^*$ for simplicity.
Using the derivatives \eqref{3-7:der:1}--\eqref{3-7:der:4} in the proof of Lemma \ref{3-7:prop:2}, some calculations show that
\begin{align*}
	& \derr{L}{\boldeta_1}{\boldeta_1\trans} =
	\left[
	\begin{array}{ccc}
		\u_1\trans\wh{\bSigma}\u_1\I_q  & \0                                                    & \0 \\
		&                                                                                   &                                      \\
		\0                      &  \u_2\trans\wh{\bSigma}\u_2\I_q                                                                    &  2\v_2\u_2\trans\wh{\bSigma} - n^{-1}\Y\trans\X                   \\
		&                                                                                   &                                      \\
		\0 & -n^{-1}\X\trans\Y & \wh{\bSigma} 
	\end{array}
	\right],                                                                                                 \\
	& \derr{L}{\u_1}{\boldeta_1\trans} = \left[-n^{-1}\X\trans\Y, \0, \0\right].
\end{align*}
Observe that $n^{-1}\X\trans\Y = \wh{\bSigma}\C + \wh{\bSigma}(\C^*-\C) + n^{-1}\X\trans\E$ for a given matrix $\C = \u_1\v_1\trans + \u_2\v_2\trans$. Plugging it into the derivatives above, we have that 
\[\derr{L}{\boldeta_1}{\boldeta_1\trans} = \A + \bDelta_{a}, \ \ \derr{L}{\u_1}{\boldeta_1\trans}=\B + \bDelta_{b},\]
where
\begin{align*}
	& \A = \left[
	\begin{array}{ccc}
		\u_1\trans\wh{\bSigma}\u_1\I_q  & \0 & \0                           \\
		&                               &                                                                \\
		\0                      &   \u_2\trans\wh{\bSigma}\u_2\I_q                &  \v_2\u_2\trans\wh{\bSigma} - \v_1\u_1\trans\wh{\bSigma}\\
		&                               &                                                                \\
		\0 & -\wh{\bSigma}\u_1\v_1\trans - \wh{\bSigma}\u_2\v_2\trans & \wh{\bSigma}
	\end{array}
	\right],                                                                                     \\
	& \bDelta_{a} = \left[
	\begin{array}{ccc}
		\0 & \0                                             & \0                                       \\
		\0 & \0                                             & (\C-\C^*)\trans\wh{\bSigma} - n^{-1}\E\trans\X \\
		\0 &  \wh{\bSigma}(\C-\C^*) - n^{-1}\X\trans\E& \0
	\end{array}
	\right],                                                                                     \\
	& \B = \left[-\wh{\bSigma}\u_1\v_1\trans - \wh{\bSigma}\u_2\v_2\trans,\0,\0\right], ~
	\bDelta_{b} = \left[\wh{\bSigma}(\C - \C^*) - n^{-1}\X\trans\E,\0,\0\right].
\end{align*}

Then we aim to find matrix $\M\in\R^{p\times(p+2q)}$ that satisfies $(\B - \M\A)\Q = \0$. It is equivalent to solving equations
\begin{align*}
	&(\u_1\trans\wh{\bSigma}\u_1\M_1 + \wh{\bSigma}\u_1\v_1\trans + \wh{\bSigma}\u_2\v_2\trans)(\I_q - \v_1\v_1\trans) = \0, \\
 	&(- \M_3 \wh{\bSigma}\u_1\v_1\trans - \M_3 \wh{\bSigma}\u_2\v_2\trans + \u_2\trans\wh{\bSigma}\u_2\M_2)(\I_q - \v_2\v_2\trans) = \0, \\
	&\M_3\wh{\bSigma} + \M_2(\v_2\u_2\trans - \v_1\u_1\trans)\wh{\bSigma} = \0.
\end{align*}
Recall that $z_{11}=\u_1\trans\wh{\bSigma}\u_1$, $z_{22}=\u_2\trans\wh{\bSigma}\u_2$, and $z_{12}=\u_1\trans\wh{\bSigma}\u_2$. By the orthogonality constraint of $\v_1\trans \v_2 = 0$, the equations above can be rewritten as
\begin{align*}
	&z_{11}\M_1(\I_q - \v_1\v_1\trans) = -\wh{\bSigma}\u_2\v_2\trans, \\
 	&z_{22}\M_2(\I_q - \v_2\v_2\trans) = \M_3\wh{\bSigma}\u_1\v_1\trans, \\
	&\M_3\wh{\bSigma} + \M_2(\v_2\u_2\trans - \v_1\u_1\trans)\wh{\bSigma} = \0.
\end{align*}

Therefore, it can be seen that the choice of $\M$ with 
\begin{align}\label{m2cons}
	&\M_1 = -z_{11}^{-1}\wh{\bSigma}\u_2\v_2\trans, \ \M_2 = \mathbf{0}, \ \M_3 = \0
\end{align}
satisfies the above equations. Since $(\B - \M\A)\Q = \0$, it holds that 
\begin{align*}
	(\derr{L}{\u_1}{\boldeta_1\trans} - \M\derr{L}{\boldeta_1}{\boldeta_1\trans})\Q
	& = (\B - \M\A)\Q  + (\bDelta_{b} - \M \bDelta_{a})\Q  \\
	&= (\bDelta_{b} - \M \bDelta_{a})\Q = \left[\bDelta,\0,\0\right],
\end{align*}
where $\bDelta = \left\{\wh{\bSigma}(\C-\C^*) - n^{-1}\X\trans\E\right\} (\I_q - \v_1\v_1\trans)$. 
This concludes the proof for the rank-2 case

\bigskip

\noindent\textbf{Part 2: Extension to the general rank case.} We now extend the results using similar arguments to those in the first part to the inference of $\u_k^*$ for each given $k$ with $1 \leq k \leq r^*$. 
Utilizing the derivatives \eqref{der:1}--\eqref{der:4} and after some calculations,  for each $i, j \in \{1, \ldots, r^*\}$ with   $i \neq j$ we can show that
\begin{align*}
	&\derr{L}{\u_i}{\u_j\trans} = \0_{p\times p}, \ \derr{L}{\u_i}{\u_i\trans} = \wh{\bSigma}, \\
	&\derr{L}{\u_i}{\v_j\trans} = \0_{p\times q}, \ \derr{L}{\u_i}{\v_i\trans} = -n^{-1}\X\trans\Y, \\
	&\derr{L}{\v_i}{\u_j\trans} = \0_{q\times p}, \ \derr{L}{\v_i}{\u_i\trans} = 2\v_i\u_i\trans\wh{\bSigma} - n^{-1}\Y\trans\X, \\
	&\derr{L}{\v_i}{\v_j\trans} = \0_{q\times q}, \ \derr{L}{\v_i}{\v_i\trans} = \u_i\trans\wh{\bSigma}\u_i\I_q.
\end{align*}
Note that $n^{-1}\X\trans\Y = \wh{\bSigma}\C + \wh{\bSigma}(\C^*-\C) + n^{-1}\X\trans\E$ for a given matrix $\C = \sum_{k=1}^{r^*} \u_k\v_k\trans$. Plugging it into the  derivatives above leads to 
\begin{align*}
	\derr{L}{\u_i}{\u_j\trans} = \A^{uu}_{ij}  + \bDelta^{uu}_{ij}&, \ \derr{L}{\u_i}{\u_i\trans} = \A^{uu}_{ii}  + \bDelta^{uu}_{ii}, \\
	\derr{L}{\u_i}{\v_j\trans} = \A^{uv}_{ij}  + \bDelta^{uv}_{ij}&, \ \derr{L}{\u_i}{\v_i\trans} = \A^{uv}_{ii}  + \bDelta^{uv}_{ii}, \\
	\derr{L}{\v_i}{\u_j\trans} =  \A^{vu}_{ij}  + \bDelta^{vu}_{ij}&, \ \derr{L}{\v_i}{\u_i\trans} =  \A^{vu}_{ii}  + \bDelta^{vu}_{ii}, \\
	\derr{L}{\v_i}{\v_j\trans} =  \A^{vv}_{ij}  + \bDelta^{vv}_{ij}&, \ \derr{L}{\v_i}{\v_i\trans} =  \A^{vv}_{ii}  + \bDelta^{vv}_{ii},
\end{align*}
where
\begin{align*}
	&\A^{uu}_{ij} = \0_{p\times p}, \ \bDelta^{uu}_{ij} = \0_{p\times p}, \quad
	\A^{uu}_{ii} = \wh{\bSigma}, \ \bDelta^{uu}_{ii} = \0_{p\times p}, \\
	& \A^{uv}_{ij} = \0_{p\times q}, \ \bDelta^{uv}_{ij}= \0_{p\times q},  \quad
	\A^{uv}_{ii}= -\wh{\bSigma}\C, \ \bDelta^{uv}_{ii}=   \wh{\bSigma}(\C-\C^*) - n^{-1}\X\trans\E, \\
	&  \A^{vu}_{ij} = \0_{q \times p}, \ \bDelta^{vu}_{ij}= \0_{q\times p}, \quad
	\A^{vu}_{ii} =  \v_i\u_i\trans\wh{\bSigma} - \sum_{j \neq i} \v_j\u_j\trans\wh{\bSigma}, \ \bDelta^{vu}_{ii} =  (\C-\C^*)\trans\wh{\bSigma} - n^{-1}\E\trans\X,   \\
	& \A^{vv}_{ij}  = \0_{q\times q}, \ \bDelta^{vv}_{ij}= \0_{q\times q}, \quad
	\A^{vv}_{ii} =  \u_i\trans\wh{\bSigma}\u_i\I_q, \ \bDelta^{vv}_{ii} = \0_{q\times q}.
\end{align*}

We next calculate the term
\[ 	\left(\der{\wt{\psi}_k}{\boldeta_k\trans}\right)\Q = \left(\derr{L}{\u_k}{\boldeta_k\trans} - \M\derr{L}{\boldeta_k}{\boldeta_k\trans}\right)\Q.\]
It holds that 
\begin{align*}
	\begin{aligned}
	\frac{\partial^{2} L}{\partial \boldsymbol{\eta}_k \partial \boldsymbol{\eta}_k^{T}} & =\left[\begin{array}{cc}
	\left(\mathbf{A}_{i j}^{v v}\right)_{\substack{1 \leq i \leq r^* \\
	1 \leq j \leq r^*}}  & \left(\mathbf{A}_{i j}^{v u}\right)_{\substack{1 \leq i \leq r^* \\
	1 \leq j \leq r^*, j \neq k}}\\
	 \left(\mathbf{A}_{i j}^{u v}\right)_{\substack{1 \leq i \leq r^*, i \neq k \\
	1 \leq j \leq r^*}}  & \left(\mathbf{A}_{i j}^{u u}\right)_{\substack{1 \leq i \leq r^*, i \neq k \\
	1 \leq j \leq r^*, j \neq k}}
	\end{array}\right]
	\\[5pt]
 &\quad+\left[\begin{array}{cc}
		\left(\boldsymbol{\Delta}_{i j}^{v v}\right)_{\substack{1 \leq i \leq r^* \\
		1 \leq j \leq r^*}} & \left(\boldsymbol{\Delta}_{i j}^{v u}\right)_{\substack{1 \leq i \leq r^* \\
		1 \leq j \leq r^*, j \neq k}} \\
		\left(\boldsymbol{\Delta}_{i j}^{u v}\right)_{\substack{1 \leq i \leq r^*, i \neq k \\
		1 \leq j \leq r^*}} & \left(\boldsymbol{\Delta}_{i j}^{u u}\right)_{\substack{1 \leq i \leq r^*, i \neq k \\
		1 \leq j \leq r^*, j \neq k}}
		\end{array}\right],\\
	\frac{\partial^{2} L}{\partial \boldsymbol{u}_{k} \partial \boldsymbol{\eta}_k^{T}} & =\left[\left(\mathbf{A}_{k j}^{u v}\right)_{1 \leq j \leq r},\left(\mathbf{A}_{k j}^{u u}\right)_{1 \leq j \leq r^*, j \neq k}\right]+\left[\left(\boldsymbol{\Delta}_{k j}^{u v}\right)_{1 \leq j \leq r^*}, \left(\boldsymbol{\Delta}_{k j}^{u u}\right)_{1 \leq j \leq r^*, j \neq k}\right].
	\end{aligned}
\end{align*}
Then we aim to find matrix $\M$ satisfying that 
\begin{align*}
	& \left[ \mathbf{M}^{v}, \mathbf{M}_{-k}^{u}\right] \left[\begin{array}{cc}
	\left(\mathbf{A}_{i j}^{v v}\right)_{\substack{1 \leq i \leq r^* \\
	1 \leq j \leq r^*}}  & \left(\mathbf{A}_{i j}^{v u}\right)_{\substack{1 \leq i \leq r^* \\
	1 \leq j \leq r^*, j \neq k}}\\
	 \left(\mathbf{A}_{i j}^{u v}\right)_{\substack{1 \leq i \leq r^*, i \neq k \\
	1 \leq j \leq r^*}}  & \left(\mathbf{A}_{i j}^{u u}\right)_{\substack{1 \leq i \leq r^*, i \neq k \\
	1 \leq j \leq r^*, j \neq k}}
	\end{array}\right] \Q   \\
 & = \left[\left(\mathbf{A}_{k j}^{u v}\right)_{1 \leq j \leq r},\left(\mathbf{A}_{k j}^{u u}\right)_{1 \leq j \leq r^*, j \neq k}\right],
\end{align*}
where  $\mathbf{M}^{v}=\left[\mathbf{M}_{1}^{v}, \ldots, \mathbf{M}_{r^*}^{v}\right], \mathbf{M}^{u}=\left[\mathbf{M}_{1}^{u}, \ldots, \mathbf{M}_{r^*}^{u}\right]$, $\mathbf{M}^{u}_{-k}$ is obtained by removing $\mathbf{M}_{k}^{u}$ in $\mathbf{M}^{u}$,   $\mathbf{M}_{i}^{u} \in \mathbb{R}^{p \times p}$ and $\mathbf{M}_{j}^{v} \in \mathbb{R}^{p \times q}$ with $i, j \in \{1, \ldots, r^*\}$, and $\Q = \diag{\I_q - \v_1\v_1\trans, \dots, \I_q - \v_{r^*}\v_{r^*}\trans, \I_{p(r^* - 1)}}$.

Based on the above analysis, after some calculations we can deduce that 
\begin{align*}
	& \M^u_i \wh{\bSigma} + \M^v_i ( \v_i\u_i\trans - \sum_{j \neq i} \v_j\u_j\trans )\wh{\bSigma} = \0
	\ \  \text{for} \ i= 1, \ldots, r^* \ \text{with} \ i \neq k, \\
	& (  -\M^u_i\wh{\bSigma}\C + \M^v_i\u_i\trans\wh{\bSigma}\u_i )(\I_q - \v_i\v_i\trans) = \0
	\ \text{for} \ i= 1, \ldots, r^* \ \text{with} \ i \neq k, \\
	& (\M^v_k \u_k\trans\wh{\bSigma}\u_k  + \wh{\bSigma}\C)(\I_q - \v_k\v_k\trans) = \0.
\end{align*}
Let us recall that $\C = \sum_{j =1}^{r^*} \u_j\v_j\trans$.
Using the orthogonality constraints of $\v_i\trans \v_j = 0$  for  $i, j \in \{1, \ldots, r^*\}$ with   $i \neq j$, we can rewrite the above equations as
\begin{align*}
	& \M^u_i \wh{\bSigma} + \M^v_i ( \v_i\u_i\trans - \sum_{j \neq i} \v_j\u_j\trans )\wh{\bSigma} = \0
	\ \  \text{for} \ i= 1, \ldots, r^* \ \text{with} \ i \neq k, \\
	&z_{ii}\M^v_i (\I_q - \v_i\v_i\trans) = \M^u_i\wh{\bSigma}\C_{-i}
	\ \text{for} \ i= 1, \ldots, r^* \ \text{with} \ i \neq k, \\
	& z_{kk}\M^v_k(\I_q - \v_k\v_k\trans) = -\wh{\bSigma}\C_{-k}.
\end{align*}

Therefore, it is clear that the choice of $\M$ with 
    \begin{align*}
        &\M^v_k =  -z_{kk}^{-1}\wh{\bSigma}\C_{-k}, \ \M_i^u = \0, \ \M_i^v = \0 \  \text{ for } \ i= 1, \ldots, r^* \ \text{with} \ i \neq k  
    \end{align*}
satisfies the above equations. Moreover, some further calculations reveal that 
	\begin{align*}
		(\derr{L}{\u_k}{\boldeta_k\trans} - \M\derr{L}{\boldeta_k}{\boldeta_k\trans})\Q
		&=  [\0_{p \times q(k-1) }, \bDelta^{uv}_{kk} (\I_q - \v_k\v_k\trans), \0_{p \times [q(r^* - k) + p(r^*-1) ]}]\\
		&=  [\0_{p \times q(k-1) }, \bDelta, \0_{p \times [q(r^* - k) + p(r^*-1) ]}],
	\end{align*}
where $\bDelta = \left\{\wh{\bSigma}(\C-\C^*) - n^{-1}\X\trans\E\right\} (\I_q - \v_k\v_k\trans)$. 
This concludes the proof of Proposition \ref{prop:rankr2}.

\subsection{Proof of Proposition \ref{prop:rankr3}} \label{new.Sec.A.12}

Similar to the proof of Proposition \ref{prop:rankr2} in Section \ref{new.Sec.A.11}, the proof of Proposition \ref{prop:rankr3} also includes two parts. In particular, the first part establishes the desired results under the rank-2 case, while the second part generalizes these results to the general rank case.

\bigskip

\noindent\textbf{Part 1: Proof for the rank-2 case.} With the strongly orthogonal factors, as the technical arguments for the results of $\u_1^{*}$ and $\u_2^{*}$ are quite similar, we will primarily present the proof for $\u_1^{*}$ here.
It follows from the construction of $\M$ in \eqref{m2cons} of the rank-2 case in the proof of Proposition \ref{prop:rankr2} that 
\begin{align*}
	\I_p - \M_1\v_1\u_1\trans + \M_1\v_2\u_2\trans
	& = \I_p - z_{11}^{-1}\wh{\bSigma}\u_2\v_2\trans\v_2\u_2\trans \\
	&= \I_p - z_{11}^{-1}\wh{\bSigma}\u_2\u_2\trans.
\end{align*}
By the Sherman--Morrison--Woodbury formula, for each $\a,\b \in \mathbb{R}^p$, we see that $\I_p + \a\b\trans$ is nonsingular if and only if $1 + \b\trans \a$ is nonzero. If it is nonsingular, then we have 
\begin{align*}
	(\I_p + \a\b\trans)^{-1} = \I_p - \frac{ \a\b\trans}{1 + \b\trans\a}.
\end{align*}
Let us define $\a = -  z_{11}^{-1}\wh{\bSigma}\u_2$ and $\b = \u_2$.
From the assumption that $z_{11} \neq z_{22}$, we see that $1 - z_{11}^{-1}z_{22} \neq 0$. This further shows that $\I_p - z_{11}^{-1}\wh{\bSigma}\u_2\u_2\trans$ is nonsingular and 
\begin{align*}
	(\I_p - \M_1\v_1\u_1\trans + \M_1\v_2\u_2\trans)^{-1} &= (\I_p - z_{11}^{-1}\wh{\bSigma}\u_2\u_2\trans)^{-1} \\
	&= \I_p + (z_{11} - z_{22})^{-1}\wh{\bSigma}\u_2\u_2\trans.
\end{align*}
Thus, we can set
\begin{align}\label{w2const}
	\W = \widehat{\bTheta}\{\I_p + (z_{11} - z_{22})^{-1}\wh{\bSigma}\u_2\u_2\trans\},
\end{align}
which satisfies that 
\begin{align}\label{w3const}
		\W(\I_p - \M_1{\v}_1{\u}_1\trans + \M_1{\v}_2{\u}_2\trans)\wh{\bSigma} =  \wh{\bTheta}\wh{\bSigma}.
\end{align}
This completes the proof for the rank-2 case.

\bigskip

\noindent\textbf{Part 2: Extension to the general rank case.} We proceed to extend the results by employing similar arguments to those in the first part to the inference of $\u_k^*$ for each given $k$ with $1 \leq k \leq r^*$.
With the aid of the definition of $\M_k^v$ in Proposition \ref{prop:rankr2}, we can deduce that 
\begin{align*}
    \I_p - \M_k^v\v_k\u_k\trans + \M_k^v\C_{-k}\trans
    &= \I_p -z_{kk}^{-1}\wh{\bSigma}\C_{-k} \C_{-k}\trans = \I_p -z_{kk}^{-1}\wh{\bSigma} \cdot \sum_{i \neq k}\u_i\v_i\trans \cdot  \sum_{j \neq k}\v_j\u_j\trans \\
	&= \I_p -z_{kk}^{-1}\wh{\bSigma} \cdot \sum_{i \neq k}  \u_i\u_i\trans = \I_p -z_{kk}^{-1}\wh{\bSigma} \U_{-k}\U_{-k}\trans.
\end{align*}
Moreover, by the Sherman--Morrison--Woodbury Formula, we have that for $\A, \B \in \mathbb{R}^{p \times (r^*-1)}$,  $\I_p + \A\B\trans$ is nonsingular if and only if $\I_{r^*-1} + \B\trans\A$ is nonsingular, and then
\[ (\I_p + \A\B\trans  )^{-1} = \I_p  - \A(\I_{r^*-1} + \B\trans\A )^{-1}\B\trans.  \]
Denote by $\A = - z_{kk}^{-1}\wh{\bSigma}{\U}_{-k}$ and $\B\trans = {\U}_{-k}\trans$. Notice that
\begin{align*}
	\I_{r^*-1} + \B\trans\A = \I_{r^*-1} -z_{kk}^{-1}{\U}_{-k}\trans\wh{\bSigma}{\U}_{-k},
\end{align*}
and by assumption, $\I_{r^*-1} -z_{kk}^{-1}{\U}_{-k}\trans\wh{\bSigma}{\U}_{-k}$ is nonsingular. Thus, we can set 
\[   \W = \widehat{\bTheta} \left\{  \I_p +   z_{kk}^{-1}\wh{\bSigma}{\U}_{-k}(\I_{r^*-1} -z_{kk}^{-1}{\U}_{-k}\trans\wh{\bSigma} {\U}_{-k})^{-1}{\U}_{-k}\trans\right\},   \]
which satisfies that 
$$ \W( \I_p - \M_k^v\v_k\u_k\trans + \M_k^v\C_{-k}\trans )\wh{\bSigma} = \widehat{\bTheta}\wh{\bSigma}.$$
This completes the proof of Proposition \ref{prop:rankr3}.

\subsection{Proof of Proposition \ref{prop:rankapo2}} \label{new.Sec.A.14}

Based on the derivatives \eqref{der:r1}--\eqref{der:r4}, 
through some calculations we can show that for each $i, j \in \{k, \ldots, r^*\}$ with $i \neq j$, 
\begin{align*}
	&\derr{L}{\u_i}{\u_j\trans} = \0_{p\times p}, \ \derr{L}{\u_i}{\u_i\trans} = \wh{\bSigma}, \\
	&\derr{L}{\u_i}{\v_j\trans} = \0_{p\times q}, \ \derr{L}{\u_i}{\v_i\trans} = -n^{-1}\X\trans\Y +  \wh{\bSigma} \wh{\C}^{(1)}, \\
	&\derr{L}{\v_i}{\u_j\trans} = \0_{q\times p}, \ \derr{L}{\v_i}{\u_i\trans} = 2\v_i\u_i\trans\wh{\bSigma} - n^{-1}\Y\trans\X + (\wh{\C}^{(1)})\trans\wh{\bSigma}, \\
	&\derr{L}{\v_i}{\v_j\trans} = \0_{q\times q}, \ \derr{L}{\v_i}{\v_i\trans} = \u_i\trans\wh{\bSigma}\u_i\I_q.
\end{align*}
Notice that $n^{-1}\X\trans\Y = \wh{\bSigma}\C + \wh{\bSigma}(\C^*-\C) + n^{-1}\X\trans\E$ for a given matrix $\C = \sum_{l=1}^{r^*} \u_l\v_l\trans$. Plugging it into the above derivatives, we rewrite the  derivatives as
\begin{align*}
	&\derr{L}{\u_i}{\u_j\trans} = \A^{uu}_{ij}  + \bDelta^{uu}_{ij}, \ \derr{L}{\u_i}{\u_i\trans} = \A^{uu}_{ii}  + \bDelta^{uu}_{ii}, \\
	&\derr{L}{\u_i}{\v_j\trans} = \A^{uv}_{ij}  + \bDelta^{uv}_{ij}, \ \derr{L}{\u_i}{\v_i\trans} = \A^{uv}_{ii}  + \bDelta^{uv}_{ii}, \\
	&\derr{L}{\v_i}{\u_j\trans} =  \A^{vu}_{ij}  + \bDelta^{vu}_{ij}, \ \derr{L}{\v_i}{\u_i\trans} =  \A^{vu}_{ii}  + \bDelta^{vu}_{ii}, \\
	&\derr{L}{\v_i}{\v_j\trans} =  \A^{vv}_{ij}  + \bDelta^{vv}_{ij}, \ \derr{L}{\v_i}{\v_i\trans} =  \A^{vv}_{ii}  + \bDelta^{vv}_{ii},
\end{align*}
where
\begin{align*}
	&\A^{uu}_{ij} = \0_{p\times p}, \ \bDelta^{uu}_{ij} = \0_{p\times p}, \quad
	\A^{uu}_{ii} = \wh{\bSigma}, \ \bDelta^{uu}_{ii} = \0_{p\times p}, \\
	& \A^{uv}_{ij} = \0_{p\times q}, \ \bDelta^{uv}_{ij}= \0_{p\times q},  \quad
	\A^{uv}_{ii}= -\wh{\bSigma}\C, \ \bDelta^{uv}_{ii}=   \wh{\bSigma}(\wh{\C}^{(1)} + \sum_{i= k}^{r^*} \u_i\v_i\trans  -\C^{*} ) - n^{-1}\X\trans\E, \\
	&  \A^{vu}_{ij} = \0_{q \times p}, \ \bDelta^{vu}_{ij}= \0_{q\times p}, \quad
	\A^{vu}_{ii} =  \v_i\u_i\trans\wh{\bSigma} - \sum_{\substack{ k \leq  l \leq r^* \\ l \neq i}} \v_l\u_l\trans\wh{\bSigma}, \\
 & \bDelta^{vu}_{ii} =  (\wh{\C}^{(1)} + \C^{(2)}-\C^*)\trans\wh{\bSigma} - n^{-1}\E\trans\X,   \\
	& \A^{vv}_{ij}  = \0_{q\times q}, \ \bDelta^{vv}_{ij}= \0_{q\times q}, \quad
	\A^{vv}_{ii} =  \u_i\trans\wh{\bSigma}\u_i\I_q, \ \bDelta^{vv}_{ii} = \0_{q\times q}.
\end{align*}

Let us recall that
$ \frac{\partial \widetilde{\psi}}{\partial \boldsymbol{\eta}_k^{T}} = \derr{L}{\u_k}{\boldeta\trans_k} - \M\derr{L}{\boldeta_k}{\boldeta_k\trans}.$ 
It holds that 
\begin{align*}
	\begin{aligned}
	\frac{\partial^{2} L}{\partial \boldsymbol{\eta}_k \partial \boldsymbol{\eta}_k^{T}}&=
	\left[\begin{array}{cc}
	\left(\mathbf{A}_{i j}^{v v}\right)_{\substack{k \leq i \leq r^* \\
	k \leq j \leq r^*}} & \left(\mathbf{A}_{i j}^{v u}\right)_{\substack{k \leq i \leq r^* \\
	k+1 \leq j \leq r^*}} \\
	 \left(\mathbf{A}_{i j}^{u v}\right)_{\substack{k+1 \leq i \leq r^* \\
	k \leq j \leq r^*}}&  \left(\mathbf{A}_{i j}^{u u}\right)_{\substack{k+1 \leq i \leq r^* \\
	k+1 \leq j \leq r^*}}
	\end{array}\right]\\[5pt]
 &\quad
	+	\left[\begin{array}{cc}
		\left(\bDelta_{i j}^{v v}\right)_{\substack{k \leq i \leq r^* \\
		k \leq j \leq r^*}} &  \left(\bDelta_{i j}^{v u}\right)_{\substack{k \leq i \leq r^* \\
		k+1 \leq j \leq r^*}}\\
		 \left(\bDelta_{i j}^{u v}\right)_{\substack{k+1 \leq i \leq r^* \\
		k \leq j \leq r^*}}&  \left(\bDelta_{i j}^{u u}\right)_{\substack{k+1 \leq i \leq r^* \\
		k+1 \leq j \leq r^*}}
		\end{array}\right],\\
	\frac{\partial^{2} L}{\partial \boldsymbol{u}_{k} \partial \boldsymbol{\eta}_k^{T}}&=\left[\left(\mathbf{A}_{k j}^{u v}\right)_{k \leq j \leq r^*},\left(\mathbf{A}_{k j}^{u u}\right)_{k+1 \leq j \leq r^*}\right]+\left[\left(\boldsymbol{\Delta}_{k j}^{u v}\right)_{k \leq j \leq r^*}, \left(\boldsymbol{\Delta}_{k j}^{u u}\right)_{k+1 \leq j \leq r^*}\right].
	\end{aligned}
\end{align*}
Then we aim to find matrix $\M$ that satisfies 
\begin{align*}
	&\left[\mathbf{M}^{v}, \mathbf{M}^{u}\right] \left[\begin{array}{cc}
	\left(\mathbf{A}_{i j}^{v v}\right)_{\substack{k \leq i \leq r^* \\
	k \leq j \leq r^*}} & \left(\mathbf{A}_{i j}^{v u}\right)_{\substack{k \leq i \leq r^* \\
	k+1 \leq j \leq r^*}} \\
	 \left(\mathbf{A}_{i j}^{u v}\right)_{\substack{k+1 \leq i \leq r^* \\
	k \leq j \leq r^*}}&  \left(\mathbf{A}_{i j}^{u u}\right)_{\substack{k+1 \leq i \leq r^* \\
	k+1 \leq j \leq r^*}}
	\end{array}\right]\Q   \\
  &= \left[\left(\mathbf{A}_{k j}^{u v}\right)_{k \leq j \leq r^*},\left(\mathbf{A}_{k j}^{u u}\right)_{k+1 \leq j \leq r^*}\right],
\end{align*}
where  $\mathbf{M}^{u}=\left[\mathbf{M}_{k+1}^{u}, \ldots, \mathbf{M}_{r^*}^{u}\right], \mathbf{M}^{v}=\left[\mathbf{M}_{k}^{v}, \ldots, \mathbf{M}_{r^*}^{v}\right]$, and
 $ \Q = \diag{  \mathbf{I}_{q}-\boldsymbol{v}_{k} \boldsymbol{v}_{k}^{T}, \ldots,  \\ \mathbf{I}_{q}-\boldsymbol{v}_{r^*} \boldsymbol{v}_{r^*}^{T}, \I_{p(r^*-k)}  }$.

Observe that $\left(\mathbf{A}_{i j}^{u u}\right)_{\substack{k+1 \leq i \leq r^* \\
k+1 \leq j \leq r^*}}  \left(\mathbf{A}_{i j}^{u v}\right)_{\substack{k+1 \leq i \leq r^* \\
k \leq j \leq r^*}} 
\left(\mathbf{A}_{i j}^{v u}\right)_{\substack{k \leq i \leq r^* \\
k+1 \leq j \leq r^*}}$ and  $ \left(\mathbf{A}_{i j}^{v v}\right)_{\substack{k \leq i \leq r^* \\
k \leq j \leq r^*}}$ are both diagonal matrices. After some calculations, we can deduce that 
\begin{align*}
	& \M^u_i \wh{\bSigma} + \M^v_i ( \v_i\u_i\trans - \sum_{ j= k + 1, j \neq i}^{r^*} \v_j\u_j\trans )\wh{\bSigma} = \0
	\ \text{ for } i= k+1, \ldots, {r^*}, \\
	& (  -\M^u_i\wh{\bSigma}\C + \M^v_i\u_i\trans\wh{\bSigma}\u_i )(\I_q - \v_i\v_i\trans) = \0
	\ \text{ for } i= k+1, \ldots, {r^*},  \\
	& (\M^v_k \u_k\trans\wh{\bSigma}\u_k  + \wh{\bSigma}\sum_{i= k}^{r^*} \u_i\v_i\trans)(\I_q - \v_k\v_k\trans) = \0.
\end{align*}
Recall that $\C = \sum_{l =1}^{r^*} \u_l\v_l\trans$.
Then by the orthogonality constraints  $\v_i\trans \v_j = 0$  for  any $i \neq j$, the equations above can be simplified as
\begin{align*}
	& \M^u_i \wh{\bSigma} + \M^v_i ( \v_i\u_i\trans -  \sum_{ l= k, j \neq i}^{r^*} \v_l\u_l\trans )\wh{\bSigma} = \0
	\ \text{ for }  i= k+1, \ldots, {r^*},  \\
	&z_{ii}\M^v_i (\I_q - \v_i\v_i\trans) = \M^u_i\wh{\bSigma}\C_{-i}
	\ \text{ for } i= k+1, \ldots, {r^*}, \\
	& z_{kk}\M^v_k(\I_q - \v_k\v_k\trans) = -\wh{\bSigma}\sum_{i= k+1}^{r^*} \u_i\v_i\trans.
\end{align*}

Therefore, we are ready to see that the choice of $\M$ with 
    \begin{align*}
        &\M^v_k =  -z_{kk}^{-1}\wh{\bSigma}\C^{(2)}, \ \ \M_i^u = \0, \ \ \M_i^v = \0 \ \text{ for } i= k+1, \ldots, {r^*}  
    \end{align*}
satisfies the above equations. Finally, with some calculations we can obtain that 
	\begin{align*}
		(\derr{L}{\u_k}{\boldeta\trans_k} - \M\derr{L}{\boldeta_k}{\boldeta\trans_k})\Q
		 = [\bDelta^{\prime},  \0_{p \times (p + q)(r^* - k)} ],
	\end{align*}
where $\bDelta^{\prime} =  \bDelta^{uv}_{kk} (\I_q - \v_k\v_k\trans) = \left\{ \wh{\bSigma}(\wh{\C}^{(1)}-\C^{*(1)} + \sum_{i= k}^{r^*} (\u_i\v_i\trans - \u_i^*\v_i\strans) ) - n^{-1}\X\trans\E \right\} (\I_q - \v_k\v_k\trans)$. 
This concludes the proof of Proposition \ref{prop:rankapo2}.

\subsection{Proof of Proposition \ref{prop:rankapo3}} \label{new.Sec.A.15}

Similar to the proof of Proposition \ref{prop:rankr3} in Section \ref{new.Sec.A.12},  we see that the existence of matrix $\W_k$ depends on the nonsingularity of matrix $\I_p - \M_k^v\v_k\u_k\trans + \M_k^v(\C^{(2)})\trans $. It follows from the definition of $\M_k^v$ in Proposition \ref{prop:rankapo2} that 
\begin{align*}
    & \I_p - \M_k^v\v_k\u_k\trans + \M_k^v(\C^{(2)})\trans
    = \I_p -z_{kk}^{-1}\wh{\bSigma}\C_{-k} (\C^{(2)})\trans \\
    &= \I_p -z_{kk}^{-1}\wh{\bSigma} \cdot \sum_{i \neq k}\u_i\v_i\trans \cdot  \sum_{j = k+1}^{r^*}\v_j\u_j\trans \\
	&= \I_p -z_{kk}^{-1}\wh{\bSigma} \cdot \sum_{i = k+1}^{r^*}  \u_i\u_i\trans = \I_p -z_{kk}^{-1}\wh{\bSigma} \U^{(2)}(\U^{(2)})\trans.
\end{align*}
Under the  assumption that $ \I_{r^*-k} -z_{kk}^{-1}({\U}^{(2)})\trans\wh{\bSigma}{\U}^{(2)}$ is nonsingular, an application of similar arguments as in the proof of  Proposition \ref{prop:rankr3} yields that  $\I_p -z_{kk}^{-1}\wh{\bSigma} \U^{(2)}(\U^{(2)})\trans$ is nonsingular and the choice of 
\[   \W = \widehat{\bTheta} \left\{  \I_p +   z_{kk}^{-1}\wh{\bSigma}\U^{(2)}(\I_{r^*-k} -z_{kk}^{-1}(\U^{(2)})\trans\wh{\bSigma} \U^{(2)})^{-1}(\U^{(2)})\trans\right\}  \]
satisfies that 
\[ \I_p - \W  ( \I_p - \M_k^v\v_k\u_k\trans + \M_k^v (\C^{(2)})\trans )\wh{\bSigma}  = \I_p - \widehat{\bTheta}\wh{\bSigma}.\]
This completes the proof of Proposition \ref{prop:rankapo3}.

\section{Some key lemmas and their proofs} \label{new.Sec.B}

\subsection{Proof of Lemma \ref{3-7:prop:2}} \label{near2:sec1}

When $r^* = 2$, the loss function  \eqref{constraint00} can be written as
\begin{align*}
	&L(\u_1,\boldeta_1) = (2n)^{-1}\norm{\Y - \X\u_1\v_1\trans - \X\u_2\v_2\trans}_F^2  \\
	&\text{subject to }
	~\u_1\trans\u_2 = 0~\text{and}~\left[\v_1,\v_2\right]\trans\left[\v_1,\v_2\right] = \I_2,
\end{align*}
where $\boldeta_1 = \left(\v_1\trans,\v_2\trans,\u_2\trans\right)\trans$.
Then under the orthogonality constraint $\v_1\trans\v_2=0$, it can be simplified as
\begin{align}\label{eqproloss}
	L = (2n)^{-1}& \Big\{\norm{\Y}_F^2 + \u_1\trans\X\trans\X\u_1\v_1\trans\v_1 + \u_2\trans\X\trans\X\u_2\v_2\trans\v_2 \nonumber\\
	&- 2\u_1\trans\X\trans\Y\v_1 - 2\u_2\trans\X\trans\Y\v_2\Big\}.
\end{align}
After some calculations with $\norm{\v_1}_2=\norm{\v_2}_2=1$, we can deduce that 
\begin{align}
	& \der{L}{\u_1} = \wh{\bSigma}\u_1 - n^{-1}\X\trans\Y\v_1,                 \label{3-7:der:1}                                      \\
	& \der{L}{\v_1} = \v_1\u_1\trans\wh{\bSigma}\u_1 - n^{-1}\Y\trans\X\u_1, \label{3-7:der:2} \\
	& \der{L}{\u_2} = \wh{\bSigma}\u_2 - n^{-1}\X\trans\Y\v_2,                                          \label{3-7:der:3}             \\
	& \der{L}{\v_2} = \v_2\u_2\trans\wh{\bSigma}\u_2 - n^{-1}\Y\trans\X\u_2. \label{3-7:der:4}
\end{align}

Utilizing the derivatives \eqref{3-7:der:2}--\eqref{3-7:der:4} with some calculations, it follows that 
\begin{align*}
	\M\der{L}{\boldeta_1}\Big|_{\boldeta_1^*}
	& = \M_1\left\{\v_1^*\u_1\trans\wh{\bSigma}(\u_1 - \u_1^*) -\v_2^*\u_2\strans\wh{\bSigma}\u_1 - n^{-1}\E\trans\X\u_1\right\} - n^{-1}\M_3\X\trans\E\v_2^* \\
	& \quad + \M_2\left\{-\v_1^*\u_2\strans\wh{\bSigma}\u_1^* - n^{-1}\E\trans\X\u_2^*\right\}                  \\
	& = (\M_1\v_1^*\u_1\trans - \M_1\v_2^*\u_2\strans)\wh{\bSigma}(\u_1 - \u_1^*) - (\M_1\v_2^* + \M_2\v_1^*)\u_2\strans\wh{\bSigma}\u_1^* \\
	& \quad - n^{-1}(\M_1\E\trans\X\u_1
	+ \M_3\X\trans\E\v_2^* + \M_2\E\trans\X\u_2^*)                                                                 \\
	& = (\M_1\v_1\u_1\trans - \M_1\v_2\u_2\trans)\wh{\bSigma}(\u_1 - \u_1^*) - (\M_1\v_2^* + \M_2\v_1^*)\u_2\strans\wh{\bSigma}\u_1^* + \bdelta_1^{\prime},
\end{align*}
where we set
\begin{align}\label{3-7:eq:prop:2:1}
	\bdelta_1^{\prime}
	& = - \left\{\M_1(\v_1 - \v_1^*)\u_1\trans - \M_1(\v_2\u_2\trans - \v_2^*\u_2\strans)\right\}\wh{\bSigma}(\u_1 - \u_1^*) \nonumber                                 \\
	& \quad - n^{-1}(\M_1\E\trans\X\u_1 + \M_2\E\trans\X\u_2^*+ \M_3\X\trans\E\v_2^* ).
\end{align}
Together with the derivative \eqref{3-7:der:1}, it holds that 
\begin{align}\label{3-7:eq:prop:2:2}
	& \wt{\psi}(\u_1,\boldeta_1^*)
	= \der{L}{\u_1}\Big|_{\boldeta_1^*} - \M\der{L}{\boldeta_1}\Big|_{\boldeta_1^*} \nonumber                 \\
	& = (\I_p - \M_1\v_1\u_1\trans + \M_1\v_2\u_2\trans)\wh{\bSigma}(\u_1 - \u_1^*) + (\M_1\v_2^* + \M_2\v_1^*)\u_2\strans\wh{\bSigma}\u_1^* + \bdelta_1,
\end{align}
where $\bdelta_1 = -\bdelta_1^{\prime} - n^{-1}\X\trans\E\v_1^*$.

Therefore, combining \eqref{3-7:eq:prop:2:1} and \eqref{3-7:eq:prop:2:2}, we can obtain that 
\begin{align*}
	\wt{\psi}(\u_1,\boldeta_1^*)
	& = (\I_p - \M_1\v_1\u_1\trans + \M_1\v_2\u_2\trans)\wh{\bSigma}(\u_1 - \u_1^*) + (\M_1\v_2^* + \M_2\v_1^*)\u_2\strans\wh{\bSigma}\u_1^* \\
	&\quad+ \bdelta + \bepsilon,
\end{align*}
where $\bdelta = \M_1\left\{(\v_1 - \v_1^*)\u_1\trans - (\v_2\u_2\trans - \v_2^*\u_2\strans)\right\}\wh{\bSigma}(\u_1 - \u_1^*)$ and
\begin{align*}
	\bepsilon = n^{-1}\left\{\M_1\E\trans\X\u_1 + \M_2\E\trans\X\u_2^* + \M_3\X\trans\E\v_2^* \right\} - n^{-1}\X\trans\E\v_1^*.
\end{align*}
This completes the proof of Lemma \ref{3-7:prop:2}.

\subsection{Lemma \ref{prop:rankr1} and its proof} \label{sec:proof:l2}
\begin{lemma}\label{prop:rankr1}
	Under the SVD constraint \eqref{SVDc} in Section \ref{new.Sec.3.1}, for an arbitrary $\mathbf{M}$ it holds that
	\begin{align}
		& \wt{\psi}_k(\u_k,\boldeta_k^*) = (\I_p - \M_k^v\v_k\u_k\trans + \M_k^v\C_{-k}\trans )\wh{\bSigma}(\u_k - \u_k^*) + \M_k^v\C_{-k}\strans \wh{\bSigma} \u_k^*  \nonumber   \\	 &\quad~ \quad~ \quad~ \quad~
		+ \sum_{\substack{j \neq k}}\M_j^v \C_{-j}\strans\wh{\bSigma} \u_j^* +\bdelta_k +  \bepsilon_k,  \nonumber 
	\end{align}
	where $\bdelta_k = \M_k^v\left\{(\v_k - \v_k^*)\u_k\trans - ( \C_{-k}\trans - \C_{-k}\strans)\right\}\wh{\bSigma} (\u_k - \u_k^*)$ and
	\begin{align*}
		\bepsilon_k =  n^{-1}\M_k^v\E\trans\X\u_k - n^{-1}\X\trans\E\v_k^* + n^{-1}\sum_{\substack{j \neq k}}  (\M_j^u \X\trans\E\v_j^* + \M_j^v \E\trans\X\u_j^*). 
	\end{align*}
\end{lemma}

\noindent \textit{Proof}.  
Under the orthogonality constraints $\v_i\trans\v_j=0$ for each $i, j \in \{1, \ldots, r^*\}$ with   $i \neq j$, we have that 
\begin{align*}
	\begin{aligned}
	L = (2 n)^{-1}\left\{\|\mathbf{Y}\|_{F}^{2}
	+2\left\langle\mathbf{Y},-\mathbf{X C}_{-k}\right\rangle
	+ \boldsymbol{u}_{k}^{T} \mathbf{X}^{T} \mathbf{X} \boldsymbol{u}_{k}\boldsymbol{v}_{k}^{T} \boldsymbol{v}_{k}
    +\left\|\mathbf{X C}_{-k}\right\|_{F}^{2}
	-2 \boldsymbol{u}_{k}^{T} \mathbf{X}^{T} \mathbf{Y} \boldsymbol{v}_{k}
	\right\}.
	\end{aligned}
\end{align*}
After some calculations with $\norm{\v_k}_2=1$, we can obtain that 
\begin{align}
     & \der{L}{\u_k} = \wh{\bSigma}\u_k - n^{-1}\X\trans\Y\v_k,                 \label{der:1}                                      \\
     & \der{L}{\v_k} = \v_k\u_k\trans\wh{\bSigma}\u_k - n^{-1}\Y\trans\X\u_k. \label{der:2}
\end{align}
For each $j \neq k$, similarly we also have that 
\begin{align}
	& \der{L}{\u_j} = \wh{\bSigma}\u_j - n^{-1}\X\trans\Y\v_j,                \label{der:3}             \\
	& \der{L}{\v_j} = \v_j\u_j\trans\wh{\bSigma}\u_j - n^{-1}\Y\trans\X\u_j. \label{der:4}
\end{align}

It follows from the derivatives \eqref{der:2}--\eqref{der:4} that  
\begin{align*}
    &\M\der{L}{\boldeta_k}\Big|_{\boldeta^*_k}
      = \M_k^v\left\{\v_k^*\u_k\trans\wh{\bSigma}(\u_k - \u_k^*) -\C_{-k}\strans\wh{\bSigma}\u_k - n^{-1}\E\trans\X\u_k\right\} \\
     & \quad~ \quad~ \quad~ \ \ \ \ + \sum_{j \neq k} \M_j^v\left\{-\C_{-j}\strans\wh{\bSigma}\u_j^* - n^{-1}\E\trans\X\u_j^*\right\}  - \sum_{j \neq k} \M_j^u \X\trans\E\v_j^* \\
	 & =    \M_k^v(\v_k^*\u_k\trans-\C_{-k}\strans )\wh{\bSigma}(\u_k - \u_k^*)
	 -  \M_k^v\C_{-k}\strans \wh{\bSigma} \u_k^*    \\
	 &\quad  - \sum_{j \neq k}\M_j^v \C_{-j}\strans\wh{\bSigma} \u_j^*   - n^{-1}\M_k^v\E\trans\X\u_k   - n^{-1} \sum_{j \neq k} \M_j^v \E\trans\X\u_j^*  - n^{-1}\sum_{j \neq k} \M_j^u \X\trans\E\v_j^* \\[5pt]
	 & =    \M_k^v(\v_k\u_k\trans\wh{\bSigma}-\C_{-k}\trans\wh{\bSigma} )(\u_k - \u_k^*)
	 -  \M_k^v\C_{-k}\strans \wh{\bSigma} \u_k^*  - \sum_{j \neq k}\M_j^v \C_{-j}\strans\wh{\bSigma} \u_j^*   +\bdelta_1^{\prime},
\end{align*}
where
\begin{align}\label{rprop:2:1}
    \bdelta_1^{\prime}
	& = - \M_k^v\left\{(\v_k - \v_k^*)\u_k\trans- ( \C_{-k}\trans - \C_{-k}\strans)\right\}\wh{\bSigma} (\u_k - \u_k^*) \nonumber\\
     & \quad - n^{-1}\M_k^v\E\trans\X\u_k - n^{-1}\sum_{j \neq k} \M_j^u \X\trans\E\v_j^*     -  \sum_{j \neq k} \M_j^v n^{-1}\E\trans\X\u_j^*.
\end{align}
Along with the derivative \eqref{der:1}, it holds that 
\begin{align}\label{rprop:2:2}
    \wt{\psi}(\u_k,\boldeta^*_k)
     & = \der{L}{\u_k}\Big|_{\boldeta^*_k} - \M\der{L}{\boldeta_k}\Big|_{\boldeta^*_k} \nonumber                 \\
     & =  ( \I_p - \M_k^v\v_k\u_k\trans + \M_k^v\C_{-k}\trans )\wh{\bSigma}(\u_k - \u_k^*)   + \M_k^v\C_{-k}\strans \wh{\bSigma} \u_k^*  \nonumber\\
     &\quad+ \sum_{j \neq k}\M_j^v \C_{-j}\strans\wh{\bSigma} \u_j^*   +\bdelta_1,
\end{align}
where $\bdelta_1 = -\bdelta_1^{\prime} - n^{-1}\X\trans\E\v_k^*$.

Hence, combining \eqref{rprop:2:1} and \eqref{rprop:2:2} yields that 
\begin{align}
    \wt{\psi}(\u_k,\boldeta^*_k)
     & =  ( \I_p - \M_k^v\v_k\u_k\trans\wh{\bSigma} + \M_k^v\C_{-k}\trans\wh{\bSigma} )(\u_k - \u_k^*)    \nonumber   \\
	 &\quad + \M_k^v\C_{-k}\strans \wh{\bSigma} \u_k^*  
	 + \sum_{j \neq k}\M_j^v \C_{-j}\strans\wh{\bSigma} \u_j^*   +\bdelta_k +  \bepsilon_k,  \nonumber
\end{align}
    where $\bdelta_k =   \left\{\M_k^v(\v_k - \v_k^*)\u_k\trans- \M_k^v( \C_{-k}\trans - \C_{-k}\strans)\right\}\wh{\bSigma} (\u_k - \u_k^*) $ and
    \begin{align*}
        \bepsilon_k =   n^{-1}\sum_{j \neq k} \M_j^u \X\trans\E\v_j^*  + n^{-1}\M_k^v\E\trans\X\u_k   +  \sum_{j \neq k} \M_j^v n^{-1}\E\trans\X\u_j^* - n^{-1}\X\trans\E\v_k^*.
    \end{align*}
This concludes the proof of Lemma \ref{prop:rankr1}.

\subsection{Lemma \ref{prop:rankapo1} and its proof} \label{sec:proof:l1}
\begin{lemma}\label{prop:rankapo1}
	Under the SVD constraint \eqref{lossr} in Section \ref{new.Sec.3.2},  for an arbitrary $\mathbf{M}$ it holds that
	\begin{align*}
		\wt{\psi}_k(\u_k,\boldeta_k^*)
		& =  \left( \I_p - \M_k^v\v_k\u_k\trans + \M_k^v (\C^{(2)})\trans \right)\wh{\bSigma}(\u_k - \u_k^*) + \M_k^v (\C^{*(2)})\trans \wh{\bSigma} \u_k^*    \\
		&\quad~ + \sum_{i= k+1}^{r^*} \M_i^v \left(\v_k^*\u_k\strans + (\C^{*(2)}_{-i})\trans \right)
		\wh{\bSigma} \u_i^* +  \bdelta_k + \bepsilon_k,
	\end{align*}
	where $\bdelta_k =
	\M_k^v \left\{(\v_k - \v_k^*)\u_k\trans - (\C^{(2)} - \C^{*(2)})\trans \right\}  \wh{\bSigma}(\u_k - \u_k^*) -\M_k^v  (\wh{\C}^{(1)} - \C^{*(1)})\trans  \wh{\bSigma} \u_k$ and
	\begin{align*}
		\bepsilon_k = \ & n^{-1}\M_k^v\E\trans\X\u_k - n^{-1}\X\trans\E\v_k^* +  n^{-1}\sum_{i = k+1}^{r^*} (\M_i^u \X\trans\E\v_i^*  	   +  \M_i^v \E\trans\X\u_i^*).
	\end{align*}
\end{lemma}

\noindent \textit{Proof}.  
Observe that the loss function \eqref{lossr} is equivalent to
\begin{align*}
	\begin{aligned}
	L= \ & (2 n)^{-1}\Big\{\|\mathbf{Y}\|_{F}^{2}-2 \boldsymbol{u}_{k}^{T} \mathbf{Y}^{T} \mathbf{X} \boldsymbol{v}_{k}+ \boldsymbol{u}_{k}^{T} \mathbf{X}^{T} \mathbf{X} \boldsymbol{u}_{k}\boldsymbol{v}_{k}^{T} \boldsymbol{v}_{k} \\
	&+2\langle\mathbf{Y},-\mathbf{X} \wh{\C}^{(1)}\rangle+2\langle\mathbf{Y},-\mathbf{X} \mathbf{C}^{(2)}\rangle+2 \boldsymbol{u}_{k}\trans \mathbf{X}^{T} \mathbf{X} \wh{\C}^{(1)} \boldsymbol{v}_{k} +2 \boldsymbol{u}_{k}\trans \mathbf{X}^{T} \mathbf{X} {\C}^{(2)} \boldsymbol{v}_{k} \\
	&+2\langle \mathbf{X} \wh{\C}^{(1)}, \mathbf{X} {\C}^{(2)} \rangle
	+\|\mathbf{X }\wh{\C}^{(1)} \|_{F}^{2} +\|\mathbf{X } {\C}^{(2)} \|_{F}^{2}\Big\} .
	\end{aligned}
\end{align*}
For each $j, j^{\prime} \in \{k+1, \ldots, r^*\}$ with $ j \neq j^{\prime}$,
we have $\v_k\trans \v_j = 0$ and $ \v_j\trans\v_{j^{\prime}} = 0$.
Then the loss function can be simplified further as
\begin{align*}
	\begin{aligned}
	L= \ &(2 n)^{-1}\Big\{\|\mathbf{Y}\|_{F}^{2}-2 \boldsymbol{u}_{k}^{T} \mathbf{Y}^{T} \mathbf{X} \boldsymbol{v}_{k}+ \boldsymbol{u}_{k}^{T} \mathbf{X}^{T} \mathbf{X} \boldsymbol{u}_{k}\boldsymbol{v}_{k}^{T} \boldsymbol{v}_{k} \\
	&+2\langle\mathbf{Y},-\mathbf{X} \wh{\C}^{(1)}\rangle+2\langle\mathbf{Y},-\mathbf{X} \mathbf{C}^{(2)}\rangle+2 \boldsymbol{u}_{k}\trans \mathbf{X}^{T} \mathbf{X} \wh{\C}^{(1)} \boldsymbol{v}_{k}  \\
	&+2\langle \mathbf{X} \wh{\C}^{(1)}, \mathbf{X} {\C}^{(2)} \rangle
	+\|\mathbf{X }\wh{\C}^{(1)} \|_{F}^{2} +   \sum_{j=k+1}^{r^*} \boldsymbol{u}_{j}^{T} \mathbf{X}^{T} \mathbf{X} \boldsymbol{u}_{j}\boldsymbol{v}_{j}^{T} \boldsymbol{v}_{j}   \Big\} .
	\end{aligned}
\end{align*}

After some calculations with $\norm{\v_k}_2=1$, we can show that  
\begin{align}
     & \der{L}{\u_k} = \wh{\bSigma}\u_k - n^{-1}\X\trans\Y\v_k + \wh{\bSigma}\wh{\C}^{(1)} \boldsymbol{v}_{k} ,                 \label{der:r1}                                      \\
     & \der{L}{\v_k} = \v_k\u_k\trans\wh{\bSigma}\u_k - n^{-1}\Y\trans\X\u_k + (\wh{\C}^{(1)})\trans\wh{\bSigma}\boldsymbol{u}_{k}. \label{der:r2}
\end{align}
For each $j \in \{k+1, \ldots, r^*\}$, similarly by  $\norm{\v_j}_2=1$ it follows that 
\begin{align}
	& \der{L}{\u_j} = \wh{\bSigma}\u_j - n^{-1}\X\trans\Y\v_j   + \wh{\bSigma}\wh{\C}^{(1)} \boldsymbol{v}_{j} ,                \label{der:r3}             \\
	& \der{L}{\v_j} = \v_j\u_j\trans\wh{\bSigma}\u_j - n^{-1}\Y\trans\X\u_j + (\wh{\C}^{(1)})\trans\wh{\bSigma}\boldsymbol{u}_{j}. \label{der:r4}
\end{align}
Note that  $\boldeta_k = \left[  \v_{k}\trans, \ldots, \v_{r^*}\trans, \u_{k+1}\trans, \ldots, \u_{r^*}\trans\right]\trans$ and  $\boldeta^*_k = \left[   \v_{k}\strans, \ldots, \v_{r^*}\strans, \u_{k+1}\strans, \ldots, \u_{r^*}\strans\right]\trans$.
Let us simplify  \eqref{der:r1}--\eqref{der:r4} using $\wh{\C}^{(1)} \boldsymbol{v}_{k} = 0$ and $\wh{\C}^{(1)} \boldsymbol{v}_{j} = 0$. It holds that
\begin{align*}
	\der{L}{\u_k}\Big|_{\boldeta^*_k}
	&= \wh{\bSigma}\u_k  - \sum_{l=1}^{r^*} \wh{\bSigma}  \u_l^*\v_l\strans  \v_k^* - n^{-1}  \X\trans\E\v_k^*=  \wh{\bSigma}(\u_k - \u_k^* )  - n^{-1}  \X\trans\E\v_k^*, \\
	\der{L}{\u_j}\Big|_{\boldeta^*_k}
	&=  \wh{\bSigma}\u_j^* - \sum_{l=1}^{r^*} \wh{\bSigma}  \u_l^*\v_l\strans  \v_j^* - n^{-1} \X\trans\E\v_j^* =  - n^{-1} \X\trans\E\v_j^*.
\end{align*}

Moreover, we can deduce that 
\begin{align*}
    \der{L}{\v_k}\Big|_{\boldeta^*_k}
	&=  \v_k^*\u_k\trans\wh{\bSigma}\u_k -\sum_{l=1}^{r^*} \v_l^*\u_l\strans \wh{\bSigma} \u_k + \sum_{i=1}^{k-1}\wt{\v}_i\wt{\u}_i\trans \wh{\bSigma}\u_k  - n^{-1}\E\trans\X\u_k\\
    &  = \v_k^*\u_k\trans\wh{\bSigma}(\u_k - \u_k^*) + \sum_{i=1}^{k-1} (\wt{\v}_i\wt{\u}_i\trans - \v_i^*\u_i\strans) \wh{\bSigma} \u_k  -\sum_{j= k+1}^{r^*} \v_j^*\u_j\strans \wh{\bSigma}\u_k - n^{-1}\E\trans\X\u_k, \\
	&=  (\v_k\u_k\trans -  \sum_{j= k+1}^{r^*} \v_j\u_j\trans )\wh{\bSigma}(\u_k - \u_k^*)  -  \sum_{j= k+1}^{r^*} \v_j^*\u_j\strans \wh{\bSigma}\u_k^* \\
 & \quad +  \sum_{j= k+1}^{r^*} (\v_j\u_j\trans -  \v_j^*\u_j\strans)\wh{\bSigma}(\u_k - \u_k^*) + \sum_{i=1}^{k-1} (\wt{\v}_i\wt{\u}_i\trans - \v_i^*\u_i\strans) \wh{\bSigma} \u_k  \\
	&\quad   - (\v_k - \v_k^*)\u_k\trans\wh{\bSigma}(\u_k - \u_k^*)  - n^{-1}\E\trans\X\u_k, \\
	\der{L}{\v_j}\Big|_{\boldeta^*_k} &= \v_j^*\u_j\strans\wh{\bSigma}\u_j^* - \sum_{l=1}^{r^*} \v_l^*\u_l\strans \wh{\bSigma} \u_j^* +  \sum_{i=1}^{k-1}\wt{\v}_i\wt{\u}_i\trans \wh{\bSigma}\u_j^* - n^{-1}\E\trans\X\u_j^* \\
	&= \sum_{i=1}^{k-1}(\wt{\v}_i\wt{\u}_i\trans - \v_i^*\u_i\strans ) \wh{\bSigma}\u_j^* - \sum_{k \leq l \leq r^*, l \neq j } \v_l^*\u_l\strans \wh{\bSigma}\u_j^* - n^{-1}\E\trans\X\u_j^*.
\end{align*}
Recall that $\mathbf{M}=\left[  \mathbf{M}_{k}^{v}, \ldots, \mathbf{M}_{r^*}^{v}, \mathbf{M}_{k+1}^{u}, \ldots,  \mathbf{M}_{r^*}^{u}\right]$.
Hence, combining the above results leads to 
\begin{align*}
	\wt{\psi}_k(\u_k,\boldeta_k^*)
	& =  ( \I_p - \M_k^v\v_k\u_k\trans + \M_k^v (\C^{(2)})\trans )\wh{\bSigma}(\u_k - \u_k^*)     \\
	&\quad +   \M_k^v  (\C^{*(2)})\trans \wh{\bSigma} \u_k^* + \sum_{i= k+1}^{r^*} \M_i^v (\v_k^*\u_k\strans + (\C^{*(2)}_{-i})\trans )   
	\wh{\bSigma} \u_i^* + \bepsilon_k   +  \bdelta_k,
\end{align*}
where $\bdelta_k =   
	 \M_k^v ((\v_k - \v_k^*)\u_k\trans - (\C^{(2)} - \C^{*(2)})\trans)  \wh{\bSigma}(\u_k - \u_k^*) $ and 
\begin{align*}
   \bepsilon_k &=   - n^{-1}\X\trans\E\v_k^* + n^{-1}\M_k^v\E\trans\X\u_k +  n^{-1}\sum_{i = k+1}^{r^*} (\M_i^u \X\trans\E\v_i^*  	   +  \M_i^v \E\trans\X\u_i^*) 
   \\ 
   &\quad - \M_k^v  (\wh{\C}^{(1)} - \C^{*(1)})\trans  \wh{\bSigma} \u_k.
\end{align*}
This completes the proof of Lemma \ref{prop:rankapo1}.

\subsection{Lemma \ref{prop:taylor} and its proof} \label{sec:proof:taylor}

The lemma below provides the Taylor expansion of $\wt{\psi}_1(\u_1,\boldeta_1)$ around $\boldeta^*_1$ on the Stiefel manifold for the rank-$2$ case; see Section \ref{new.Sec.C} for the technical background and relevant notation. 


\begin{lemma}\label{prop:taylor}
For arbitrary $\M_1 \in \mathbb{R}^{p \times q}, \M_2 \in \mathbb{R}^{p \times p}$, and $\M_3 \in \mathbb{R}^{p \times q}$, we have the first-order Taylor expansion of $\wt{\psi}_1(\u_1,\boldeta_1)$ with respect to $\boldeta_1$ in a neighborhood of $\boldeta_1^*$ given by 
	\begin{align*}
		\wt{\psi}_1\left( \u_1, \boldeta_1  \right) & = \wt{\psi}_1 \left( \u_1, \boldeta_1^* \right)
		+  ( - n^{-1}\X\trans\Y - \u_1\trans\wh{\bSigma}\u_1\M_1 )(\I_q - \v_1^*\v_1\strans) \exp^{-1}_{\v_1^*}(\v_1) \nonumber \\
		& \quad+  ( n^{-1}\M_2\X\trans\Y - \u_2\trans\wh{\bSigma}\u_2\M_3)(\I_q - \v_2^*\v_2\strans) \exp^{-1}_{\v_2^*}(\v_2)  \nonumber\\
		& \quad +  ( -\M_2 \wh{\bSigma} - 2\M_3{\v}^*_2\u_2\strans\wh{\bSigma} + n^{-1}\M_3\Y\trans\X)(\u_2 - \u_2^*)  + \r_{\v_1^*}+ \r_{\u_2^*}  + \r_{\v_2^*},
	\end{align*}
	where
	$ \exp^{-1}_{\v_1^*}(\v_1) \in T_{\v_1^*}\operatorname{St}(1,q)$ and $ \exp^{-1}_{\v_2^*}(\v_2) \in T_{\v_2^*}\operatorname{St}(1,q)$ are the tangent vectors on the corresponding Stiefel manifolds, and $\r_{\v_1^*} \in \mathbb{R}^p$, $\r_{\u_2^*} \in \mathbb{R}^p$, and $\r_{\v_2^*} \in \mathbb{R}^p$ are the Taylor remainder terms satisfying that
\[\|\r_{\v_1^*}\|_2 = O(\| \exp^{-1}_{\v_1^*}(\v_1) \|_2^2), \ \|\r_{\u_2^*}\|_2 = O(\|\u_2 - \u_2^*  \|_2^2), \ \|\r_{\v_2^*}\|_2 = O(\| \exp^{-1}_{\v_2^*}(\v_2) \|_2^2). \]
\end{lemma}

\noindent \textit{Proof}. Recall that
  $  \wt{\psi}_1(\u_1,\boldeta_1) = \der{L}{\u_1} - \M\der{L}{{\boldeta_1}},$
where $\boldeta_1 =[{\v}_1\trans,{\u}_2\trans,{\v}_2\trans]\trans\in\R^{p+2q}$.  We will prove the result by conducting the Taylor expansion with respect to ${\v}_1$, ${\u}_2$, and ${\v}_2$, respectively. In order to show the Taylor expansion clearly, let us write function $\wt{\psi}_1$ in the form 
\begin{align*}
    \wt{\psi}_1 (\u_1, \boldeta_1) = \wt{\psi}_1( \u_1,  {\v}_1, {\u}_2, {\v}_2).
\end{align*}
We will exploit the path below to carry out the Taylor expansion of $\wt{\psi}_1( \u_1,  {\v}_1, {\u}_2, {\v}_2  )$ 
\begin{align*}
    (\u_1, {\v}_1, {\u}_2, {\v}_2 ) \rightarrow  (\u_1, {\v}_1^*, {\u}_2, {\v}_2 )  \rightarrow  (\u_1, {\v}_1^*, {\u}_2, {\v}^*_2 )   \rightarrow  (\u_1, \v_1^*, \u_2^*, \v_2^* ) .
\end{align*}

Let us first treat $\u_1, {\u}_2$, and ${\v}_2$ as fixed and do the expansion of $\wt{\psi}_{1}$ with respect to ${\v}_1$. 
Since both ${\v}_1$ and $\v_1^*$ belong to set $\{ \v \in \mathbb{R}^q: \v\trans\v = 1   \}$, we have that ${\v}_1, \v_1^* \in \text{St}(1, q)$ by the definition of the Stiefel manifold. Then by the representation of orthonormal matrices on the Stiefel manifold given in \eqref{exponv2} in Section \ref{sec:stiefel}, we see that there exists some tangent vector $\bxi_1 \in T_{\v_1^*}\operatorname{St}(1,q)$ such that ${\v}_1$ can be represented through the exponential map as
\[{\v}_1 = \exp_{\v_1^*}(\bxi_1).\]
Meanwhile, the tangent vector $\bxi_1$ can be represented as $\bxi_1 = \exp_{\v_1^*}^{-1}{\v}_1$, where $\exp_{\v_1^*}^{-1}$ denotes the inverse of the exponential map.
Then by Lemma \ref{lemm:taylormanifold} in Section \ref{sec1.1.1}, we have the first-order Taylor expansion of $\wt{\psi}_{1}$ with respect to ${\v}_1$ given by 
\begin{align}\label{phivv1}
	\wt{\psi}_{1}\left(\u_1, {\v}_1, {\u}_2, {\v}_2 \right)=\wt{\psi}_{1}\left(\u_1, \v_1^*, {\u}_2, {\v}_2\right)+  \langle \nabla_{\v_1^*}  \wt{\psi}_1\left(\u_1, {\v}_1, {\u}_2, {\v}_2 \right), \bxi_1 \rangle
	+ \r_{\v_1^*},
\end{align}
where $\nabla_{\v_1^*}  \wt{\psi}_{1}\left(\u_1, {\v}_1, {\u}_2, {\v}_2 \right)$ is the gradient  of $\wt{\psi}_{1}\left(\u_1, {\v}_1, {\u}_2, {\v}_2 \right)$ with respect to ${\v}_1$ at $\v_1^*$ on the Stiefel manifold, $\langle \cdot, \cdot \rangle$ is the metric defined in \eqref{metric2} in Section \ref{sec:stiefel}, and $\r_{\v_1^*} \in \mathbb{R}^p$ is the corresponding Taylor remainder term satisfying that 
\[\|\r_{\v_1^*}\|_2 = O(\| \bxi_1 \|_2^2).\] 
Applying Lemma \ref{lemmgradst} in Section \ref{sec:stiefel}, the gradient on the Stiefel manifold $\operatorname{St}(1,q)$ is given by 
\begin{align*}
	\nabla_{\v_1^*}  \wt{\psi}_{1}(\u_1, {\v}_1, {\u}_2, {\v}_2) = (\I_q - \v_1^*\v_1\strans) \der{\wt{\psi}_{1}(\u_1, {\v}_1, {\u}_2, {\v}_2)}{{\v}_1}\Big|_{\v_1^*},
\end{align*}
where $ \der{\wt{\psi}_{1}(\u_1, {\v}_1, {\u}_2, {\v}_2)}{{\v}_1}\Big|_{\v_1^*}$ represents the partial derivative of $\wt{\psi}_{1}(\u_1, {\v}_1, {\u}_2, {\v}_2)$ with respect to ${\v}_1$ at $\v_1^*$ in the (usual) Euclidean space.

In view of \eqref{metric2} in Section \ref{sec:stiefel}, we further have that 
\begin{align*}
	\langle \nabla_{\v_1^*}  \wt{\psi}_{1}(\u_1, {\v}_1, {\u}_2, {\v}_2),  \bxi_1 \rangle &= \operatorname{tr}(	[\nabla_{\v_1^*}  \wt{\psi}_1(\u_1, {\v}_1, {\u}_2, {\v}_2)]\trans \bxi_1 ) = [\nabla_{\v_1^*}  \wt{\psi}_1(\u_1, {\v}_1, {\u}_2, {\v}_2)]\trans \bxi_1  \\ &=\der{\wt{\psi}_{1}(\u_1, {\v}_1, {\u}_2, {\v}_2)}{{\v}_1\trans}\Big|_{\v_1^*} (\I_q - \v_1^*\v_1\strans) \bxi_1,
\end{align*}
where the second equality above holds since $\nabla_{\v_1^*}  \wt{\psi}_{1}(\u_1, {\v}_1, {\u}_2, {\v}_2)$ and $\bxi_1  $ are $q$-dimensional vectors.
Hence, combining the above results leads to 
\begin{align}\label{phiv1}
    \wt{\psi}_1\left(  \u_1, {\v}_1, {\u}_2, {\v}_2  \right)&=\wt{\psi}_1\left( \u_1, \v_1^*, {\u}_2, {\v}_2  \right)+   \der{\wt{\psi}_1(\u_1, {\v}_1, {\u}_2, {\v}_2)}{{\v}_1\trans}\Big|_{\v_1^*} (\I_q - \v_1^*\v_1\strans)  \bxi_1
     \nonumber\\
     &\quad+ \r_{\v_1^*},
\end{align}
where $\|\r_{\v_1^*}\|_2 = O(\| \bxi_1 \|_2^2).$

Moreover, similar to ${\v}_1$, for the Taylor expansion with respect to ${\v}_2$ we can deduce that 
\begin{align}\label{phiv2}
    \wt{\psi}_1\left(  \u_1, \v_1^*, {\u}_2, {\v}_2   \right)&= \wt{\psi}_1\left( \u_1, \v_1^*, {\u}_2,  \v_2^*   \right)+  \der{\wt{\psi}_1\left(  \u_1, \v_1^*, {\u}_2, {\v}_2   \right)}{{\v}_2\trans}\Big|_{\v_2^*}(\I_q - \v_2^*\v_2\strans) \bxi_2
     \nonumber\\
     &\quad+ \r_{\v_2^*} ,
\end{align}
where $ \der{\wt{\psi}_{1}(\u_1, \v_1^*, {\u}_2, {\v}_2)}{{\v}_2}\Big|_{\v_2^*}$ is the partial derivative of $\wt{\psi}_{1}(\u_1, \v_1^*, {\u}_2, {\v}_2)$ with respect to ${\v}_2$ at $\v_2^*$, $\bxi_2 = \exp_{\v_2^*}^{-1}({\v}_2)$ is the corresponding tangent vector, and $\r_{\v_2^*} \in \mathbb{R}^p$ is the Taylor remainder term satisfying that 
\[\|\r_{\v_2^*} \|_2 = O(\|\bxi_2 \|_2^2).\]

Since there is no unit length constraint on ${\u}_2$, 
we can take the Taylor expansion of $\wt{\psi}$ with respect to ${\u}_2$ directly on the Euclidean space $\mathbb{R}^p$. It gives that 
\begin{align}\label{phiu2}
    \wt{\psi}_1\left( \u_1, \v_1^*, {\u}_2,  \v_2^*   \right)& = \wt{\psi}_1\left( \u_1, \v_1^*, \u_2^*,  \v_2^*   \right)+  \der{\wt{\psi}_1\left( \u_1, \v_1^*, {\u}_2,  \v_2^*    \right)}{{\u}_2\trans}\Big|_{\u_2^*} ({\u}_2 - \u_2^*)
    \nonumber\\
    &\quad+ \r_{\u_2^*},
\end{align}
where  $ \der{\wt{\psi}_{1}(\u_1, \v_1^*, {\u}_2,  \v_2^* )}{{\u}_2}\Big|_{\u_2^*}$ is the partial derivative of $\wt{\psi}_{1}(\u_1, \v_1^*, {\u}_2,  \v_2^* )$ with respect to ${\u}_2$ at $\u_2^*$, and $\r_{\u_2^*} \in \mathbb{R}^p$ is the corresponding Taylor remainder term satisfying that 
\[\|\r_{\u_2^*}\|_2 =  O( \| {\u}_2 - \u_2^* \|_2 ^2).\]
Combining \eqref{phiv1}--\eqref{phiv2}, we can obtain that 
\begin{align}
    \wt{\psi}_1&\left( \u_1,  {\v}_1, {\u}_2,  {\v}_2   \right) = \wt{\psi}_1\left( \u_1, \v_1^*, \u_2^*, \v_2^* \right)
    +  \der{\wt{\psi}_1( \u_1,  {\v}_1, {\u}_2,  {\v}_2   )}{{\v}_1\trans}\Big|_{\v_1^*}(\I_q - \v_1^*\v_1\strans) \bxi_1 \nonumber \\
    &\quad+  \der{\wt{\psi}_1( \u_1, \v_1^*, {\u}_2, {\v}_2   )}{{\v}_2\trans}\Big|_{\v_2^*}(\I_q - \v_2^*\v_2\strans) \bxi_2 +  \der{\wt{\psi}_1( \u_1, \v_1^*, {\u}_2,  {\v}^*_2   )}{{\u}_2\trans}\Big|_{\u_2^*} ({\u}_2 - \u_2^*)   \nonumber\\
    & \quad  + \r_{\v_1^*}+ \r_{\u_2^*}   + \r_{\v_2^*}.\label{phi1}
\end{align}

On the other hand, it follows from the definition of $\wt{\psi}_1(\u_1,{\boldeta_1}) $ with ${\boldeta_1} = \left[{\v}_1\trans,{\u}_2\trans,{\v}_2\trans\right]\trans$ 
that 
\begin{align*}
	\wt{\psi}_1(\u_1,{\boldeta_1}) = \der{L}{\u_1} - \M\der{L}{{\boldeta_1}}= \der{L}{\u_1} - \M_1\der{L}{{\v}_1}- \M_2\der{L}{{\u}_2}- \M_3\der{L}{{\v}_2}.
\end{align*}
Through some calculations with \eqref{3-7:der:1}--\eqref{3-7:der:4}, we can show that 
\begin{align*}
	&\der{\wt{\psi}_1(\u_1, {\v}_1, {\u}_2, {\v}_2   )}{{\v}_1\trans}\Big|_{\v_1^*} = - n^{-1}\X\trans\Y - \u_1\trans\wh{\bSigma}\u_1\M_1,\\
	&\der{\wt{\psi}_1(\u_1, \v_1^*, {\u}_2, {\v}_2    )}{{\v}_2\trans}\Big|_{\v_2^*} = n^{-1}\M_2\X\trans\Y - {\u}_2\trans\wh{\bSigma}{\u}_2\M_3, \\
	&\der{\wt{\psi}_1(\u_1, \v_1^*, {\u}_2, {\v}_2^*    )}{{\u}_2\trans}\Big|_{\u_2^*} = -\M_2 \wh{\bSigma} - 2\M_3{\v}_2^*\u_2\strans\wh{\bSigma}+ n^{-1}\M_3\Y\trans\X.
\end{align*}
Then plugging them into \eqref{phi1} entails that 
\begin{align*}
	\wt{\psi}_1\left( \u_1, {\boldeta_1}  \right) & = \wt{\psi}_1 \left( \u_1, \boldeta_1^* \right)
	+  ( - n^{-1}\X\trans\Y - \u_1\trans\wh{\bSigma}\u_1\M_1 )(\I_q - \v_1^*\v_1\strans) \bxi_1 \nonumber \\
	& \quad+  ( n^{-1}\M_2\X\trans\Y - {\u}_2\trans\wh{\bSigma}{\u}_2\M_3)(\I_q - \v_2^*\v_2\strans) \bxi_2  \nonumber\\
	&  \quad +  ( -\M_2 \wh{\bSigma} - 2\M_3{\v}^*_2\u_2\strans\wh{\bSigma} + n^{-1}\M_3\Y\trans\X)({\u}_2 - \u_2^*)  + \r_{\v_1^*}+ \r_{\u_2^*}   + \r_{\v_2^*}.
\end{align*}
Since $\bxi_1 = \exp_{\v_1^*}^{-1}({\v}_1)$ and $\bxi_2 = \exp_{\v_2^*}^{-1}({\v}_2)$, this concludes the proof of Lemma \ref{prop:taylor}.

\subsection{Lemma \ref{prop:taylor12} and its proof} \label{new.Sec.B.2}

\begin{lemma}\label{prop:taylor12}
Assume that all the conditions of Theorem \ref{theorkr} are satisfied. Then under the strongly orthogonal factors, for $\wt{\W}_k$ given in \eqref{eqwknear1} and an arbitrary
$\a\in\R^p $, with probability at least
$1- \theta_{n,p,q}$ we have
\begin{align*}
	& |\a\trans\wt{\W}_k (\wt{\psi}_k(\wt{\u}_k,\wt{\boldeta}_k) - \wt{\psi}_k(\wt{\u}_k,\boldeta_k^*) )| \\[5pt]
	& \leq  c \norm{\a}_0^{1/2}\norm{\a}_2 \max\{s_{\max}^{1/2}, (r^*+s_u+s_v)^{1/2}, \eta_n^2\} (r^*+s_u+s_v)\eta_n^2\{n^{-1}\log(pq)\},
\end{align*}
where $\theta_{n,p,q}$ is given in \eqref{thetapro} and $c$ is some positive constant.
\end{lemma}

\noindent \textit{Proof}. The proof of Lemma \ref{prop:taylor12} consists of two parts. Specifically, the first part establishes the desired results under the rank-2 case, while the second part further extends the results to the general rank case.

\bigskip

\noindent\textbf{Part 1: Proof for the rank-2 case.}
For the rank-2 case with strongly orthogonal factors, since the technical arguments for the inference of $\u_1^*$ and $\u_2^*$ are similar, for simplicity we will present the proof only for $\u_1^*$ here. When we use the initial estimates $(\wt{\u}_1, \wt{\u}_2, \wt{\v}_1, \wt{\v}_2)$ satisfying Definition \ref{lemmsofar}, by
Lemma \ref{prop:taylor} the first-order Taylor expansion of $\wt{\psi}_1\left( \wt{\u}_1, \wt{\boldeta}_1   \right) $ at $\boldeta_1^*$ is given by 
\begin{align*}
	\wt{\psi}_1\left( \wt{\u}_1, \wt{\boldeta}_1   \right) & = \wt{\psi}_1 \left( \wt{\u}_1, \boldeta_1^* \right)
	+  ( - n^{-1}\X\trans\Y -  \wt{\u}_1\trans\wh{\bSigma} \wt{\u}_1\M_1 )(\I_q - \v_1^*\v_1\strans) \exp^{-1}_{\v_1^*}(\wt{\v}_1) \nonumber \\
	& \quad+  ( n^{-1}\M_2\X\trans\Y - \wt{\u}_2\trans\wh{\bSigma}\wt{\u}_2\M_3)(\I_q - \v_2^*\v_2\strans) \exp^{-1}_{\v_2^*}(\wt{\v}_2)  \nonumber\\
	& \quad +  ( -\M_2 \wh{\bSigma} - 2\M_3{\v}_2^*\u_2\strans\wh{\bSigma}+ n^{-1}\M_3\Y\trans\X)(\wt{\u}_2 - \u_2^*)  + {\r}_{\v_1^*}+ {\r}_{\u_2^*}   + {\r}_{\v_2^*},
\end{align*}
where the Taylor remainder terms satisfy that
\[ \|\r_{\v_1^*}\|_2 = O(\| \exp^{-1}_{\v_1^*}(\wt{\v}_1) \|_2^2), \ \|\r_{\u_2^*}\|_2  = O(\|\wt{\u}_2 - \u_2^*  \|_2^2), \ \|\r_{\v_2^*}\|_2  = O(\| \exp^{-1}_{\v_2^*}(\wt{\v}_2) \|_2^2). \]

Moreover, when the construction of $\M = \left[\M_1,\M_2,\M_3\right]$ is given as 
\begin{align*}
	&\M_1 = -(\wt{\u}_1\trans\wh{\bSigma} \wt{\u}_1)^{-1}\wh{\bSigma}\wt{\u}_2\wt{\v}_2\trans, \ \M_2 = \mathbf{0}, \ \M_3 = \0,
\end{align*}
it is immediate to see that
\begin{align}
		\wt{\psi}_1\left( \wt{\u}_1, \wt{\boldeta}_1  \right) & = \wt{\psi}_1 \left( \wt{\u}_1, \boldeta_1^* \right)
	+ (- n^{-1}\X\trans\Y + \wh{\bSigma}\wt{\u}_2\wt{\v}_2\trans)(\I_q - \v_1^*\v_1\strans) \exp^{-1}_{\v_1^*}(\wt{\v}_1)  \nonumber\\
	 &\quad + {\r}_{\v_1^*}+ {\r}_{\u_2^*}   + {\r}_{\v_2^*}.\label{phi3}
\end{align}
We aim to bound the difference between $\wt{\psi}_1\left( \wt{\u}_1, \wt{\boldeta}_1  \right)$ and $\wt{\psi}_1\left( \wt{\u}_1, \boldeta_1^* \right)$, which will be divided into two parts.

\medskip

\noindent\textbf{(1). Upper bounds on $\norm{ \exp^{-1}_{\v_1^*}(\wt{\v}_1)}_0$, $\norm{ \exp^{-1}_{\v_1^*}(\wt{\v}_1)}_2$, $\norm{ \exp^{-1}_{\v_2^*}(\wt{\v}_2)}_0 $, and $\norm{ \exp^{-1}_{\v_2^*}(\wt{\v}_2)}_2$}.
Denote by ${\bxi}_1 = \exp^{-1}_{\v_1^*}(\wt{\v}_1)$ the tangent vector, so that $ \exp_{\v_1^*}({\bxi}_1) = \wt{\v}_1$. Since $\r_{\v_1^*} = O(\| \bxi_1 \|_2^2)$, if ${\bxi}_1 = \0$ we need only to bound term $ {\r}_{\u_2^*} + {\r}_{\v_2^*}$ in \eqref{phi3} above. Without loss of generality, let us assume that ${\bxi}_1 \neq \0$. 
Observe that the $q$-dimensional tangent vector ${\bxi}_1 \in T_{\v_1^*}\operatorname{St}(1,q)$, where $T_{\v_1^*}\operatorname{St}(1,q)$ is the tangent space of the Stiefel manifold $\operatorname{St}(1,q)$ at $\v_1^*$. Then from Lemma \ref{lemmgeodsti} in Section \ref{sec:stiefel}, we have the explicit form of the corresponding geodesic 
\begin{align*}
    \gamma(t ; \v_1^*, \bxi_1) = \v_1^* \cdot \cos( \norm{{\bxi}_1}_2 t) + \frac{{\bxi}_1}{\norm{{\bxi}_1}_2}  \cdot \sin(\norm{{\bxi}_1}_2 t).
\end{align*}
Hence, in view of the definition of the exponential map in \eqref{expon} in Section \ref{sec1.1.1}, it holds that
\begin{align}\label{eqxi}
     \wt{\v}_1 & = \exp_{\v_1^*}({\bxi}_1) = \gamma(1 ; \v_1^*, \bxi_1) \nonumber\\
     &= \v_1^* \cdot  \cos( \norm{{\bxi}_1}_2 ) + \frac{{\bxi}_1}{\norm{{\bxi}_1}_2} \cdot  \sin(\norm{{\bxi}_1}_2).
\end{align}

Moreover, we claim that $\sin(\norm{{\bxi}_1}_2) \neq 0$ when ${\bxi}_1 \neq \0$. Otherwise, if $\sin(\norm{{\bxi}_1}_2)  = 0$ it implies that $\cos( \norm{{\bxi}_1}_2 )  = \pm 1$ and then $ \wt{\v}_1 = \pm \v_1^* $ by \eqref{eqxi}. When $ \wt{\v}_1 =  \v_1^* $, we have ${\bxi}_1 = \0$, which is a contradiction. On the other hand, if $\wt{\v}_1 =  -\v_1^*$, we have $\norm{\wt{\v}_1 - \v_1^* }_2 = \norm{2\v_1^*}_2 =2$. Then $\wt{\v}_1$ is not a consistent estimator of $\v_1^*$, which is a contradiction to Definition \ref{lemmsofar}. 
Thus, we have that $\sin(\norm{{\bxi}_1}_2) \neq 0$. Then it follows from \eqref{eqxi} that
\begin{align}\label{eqxi1}
    {\bxi}_1 =  \left(  \wt{\v}_1 -  \v_1^* \cos( \norm{{\bxi}_1}_2 ) \right) \cdot \frac{\norm{{\bxi}_1}_2}{\sin(\norm{{\bxi}_1}_2)}.
\end{align}
Since $\norm{{\bxi}_1}_2/ \sin(\norm{{\bxi}_1}_2) \neq 0$,
 we can deduce that 
\begin{align}
    \norm{{\bxi}_1}_0 &= \norm{  \wt{\v}_1 -  \v_1^* \cos( \norm{{\bxi}_1}_2 )}_0 =  \norm{  (\wt{\v}_1 -  \v_1^*) +  \v_1^*(1-\cos( \norm{{\bxi}_1}_2 ) )}_0 \nonumber\\
    &\leq \norm{ \wt{\v}_1 -  \v_1^* }_0 + \norm{    \v_1^* (1-\cos( \norm{{\bxi}_1}_2 ) )}_0 \nonumber\\
    &\leq c(r^*+s_u +s_v), \label{bxi0} 
\end{align}
where the last inequality above follows from Lemma \ref{lemmauv} in Section \ref{new.Sec.B.3} and $\norm{\v_1^* }_0 = s_v$. 

We next derive the upper bound on $ \|{\bxi}_1\|_2$.
Since $\bxi_1 \in  T_{\v_1^*}\operatorname{St}(1,q)$, from \eqref{tagvec} in Section \ref{sec:stiefel} we see that
$\v_1\strans\bxi_1 = 0$. An application of Lemma 3 in \cite{chen2012sparse} leads to 
\[ \norm{\bxi_1}_2  =   O(\norm{\wt{\v}_1 - \v_1^*}_2   ).  \]
Together with Lemma \ref{lemmauv}, it yields that
\begin{align}\label{eqr11}
	\norm{\bxi_1}_2 \leq c\norm{\wt{\v}_1 - \v_1^{*}}_{2} \leq c(r^*+s_u+s_v)^{1/2}\eta_n^2\{n^{-1}\log(pq)\}^{1/2}/d_1^*. 
\end{align}
Further, applying similar arguments to $\bxi_2 = \exp^{-1}_{\v_2^*}(\wt{\v}_2)$, we can obtain that 
\begin{align}
	&\norm{ \exp^{-1}_{\v_2^*}(\wt{\v}_2)}_0  \leq c(r^*+s_u +s_v), \label{eqr21}\\
	&\|\exp^{-1}_{\v_2^*}(\wt{\v}_2)\|_2
	\leq c(r^*+s_u+s_v)^{1/2}\eta_n^2\{n^{-1}\log(pq)\}^{1/2}/d_2^*. \label{eqr22}
\end{align}

\smallskip

\noindent\textbf{(2). The upper bound on $|\a\trans \wt{\W}_1(\wt{\psi}_1\left(\wt{\u}_1, \wt{\boldeta}_1 \right) - \wt{\psi}_1\left( \wt{\u}_1, \boldeta_1^* \right)) | $}.
By the Taylor expansion of $\wt{\psi}_1\left(\wt{\u}_1, \wt{\boldeta}_1 \right)$ in \eqref{phi3}, it holds that 
\begin{align}\label{adeee}
	|\a\trans \wt{\W}_1(\wt{\psi}_1\left(\wt{\u}_1, \wt{\boldeta}_1 \right) - \wt{\psi}_1\left( \wt{\u}_1, \boldeta_1^* \right)) | &
	\leq |\a\trans\wt{\W}_1(- n^{-1}\X\trans\Y + \wh{\bSigma}\wt{\u}_2\wt{\v}_2\trans)(\I_q - \v_1^*\v_1\strans){\bxi}_1|
   \nonumber\\[5pt]
	&\quad  +	|\a\trans\wt{\W}_1( \r_{\v_1^*}+ \r_{\u_2^*}   + \r_{\v_2^*}) |.
\end{align} 
Let us first bound term $|\a\trans\wt{\W}_1(- n^{-1}\X\trans\Y + \wh{\bSigma}\wt{\u}_2\wt{\v}_2\trans)(\I_q - \v_1^*\v_1\strans){\bxi}_1|$.
Notice that  $n^{-1}\X\trans\Y = \wh{\bSigma}\u_1^*\v_1\strans + \wh{\bSigma}\u_2^*\v_2\strans + n^{-1}\X\trans\E$. Along with $\v_1\strans\v_1^* =1$, it gives that 
\begin{align*}
	& (- n^{-1}\X\trans\Y + \wh{\bSigma}\wt{\u}_2\wt{\v}_2\trans)(\I_q - \v_1^*\v_1\strans)  \\
 &= ( -\wh{\bSigma}\u_1^*\v_1\strans + \wh{\bSigma}(\wt{\u}_2\wt{\v}_2\trans - \u_2^*\v_2\strans) -n^{-1}\X\trans\E)(\I_q - \v_1^*\v_1\strans) \\[5pt]
	&= ( \wh{\bSigma}(\wt{\u}_2\wt{\v}_2\trans - \u_2^*\v_2\strans) -n^{-1}\X\trans\E)(\I_q - \v_1^*\v_1\strans).
\end{align*}
Denote by $\wh{\bDelta} = \wh{\bSigma}(\wt{\u}_2\wt{\v}_2\trans - \u_2^*\v_2\strans) -n^{-1}\X\trans\E$.
It follows that
\begin{align}\label{dassd}
	|\a\trans \wt{\W}_1(- n^{-1}\X\trans\Y + \wh{\bSigma}\wt{\u}_2\wt{\v}_2\trans)(\I_q - \v_1^*\v_1\strans){\bxi}_1|
	&= |\a\trans \wt{\W}_1 \wh{\bDelta}(\I_q - \v_1^*\v_1\strans){\bxi}_1|.
\end{align}

Recall that $\wt{\W}_1 = \widehat{\bTheta}\{\I_p + (\wt{z}_{11} - \wt{z}_{22})^{-1}\wh{\bSigma}\wt{\u}_2\wt{\u}_2\trans\}$, where $\wt{z}_{11} = \wt{\u}_1\trans\wh{\bSigma}\wt{\u}_1$ and $\wt{z}_{22} = \wt{\u}_2\trans\wh{\bSigma}\wt{\u}_2$. Denote by $\wt{\w}_i\trans = \widehat{\btheta}_i\trans\{\I_p + (\wt{z}_{11} - \wt{z}_{22})^{-1}\wh{\bSigma}\wt{\u}_2\wt{\u}_2\trans\}$ the $i$th row of $\wt{\W}_1$ for $i = 1, \ldots, p$.
In light of Lemma \ref{lemma:1rk4} in Section \ref{new.Sec.B.5}, it holds that 
\begin{align}\label{wwww}
	\max_{1 \leq i \leq p}\norm{\wt{\w}_i}_0 \leq 2\max\{s_{\max}, 3(r^*+s_u+s_v)\} \ \text{ and } \  \max_{1 \leq i \leq p}\norm{\wt{\w}_i}_2 \leq c.
\end{align}
Then we have that 
\begin{align}
	| \wt{\w}_i\trans \wh{\bDelta} (\I_q - \v_1^*\v_1\strans){\bxi}_1| \nonumber
	& \leq    |\wt{\w}_i\trans n^{-1}\X\trans\E(\I_q - \v_1^*\v_1\strans){\bxi}_1| \nonumber\\
 &\quad+ |\wt{\w}_i\trans \wh{\bSigma}(\wt{\u}_2\wt{\v}_2\trans -\u_2^*\v_2\strans)(\I_q - \v_1^*\v_1\strans){\bxi}_1|. \label{sadafvzxc}
\end{align}
For the first term on the right-hand side of (\ref{sadafvzxc}) above, it follows from the sparsity of $\v_1^*$, and ${\bxi}_1$ that 
	\begin{align}\label{eqta1233}
		& |\wt{\w}_i\trans n^{-1}\X\trans\E(\I_q - \v_1^*\v_1\strans){\bxi}_1|
			\leq    |\wt{\w}_i\trans n^{-1}\X\trans\E {\bxi}_1| +   |\wt{\w}_i\trans n^{-1}\X\trans\E \v_1^*\v_1\strans{\bxi}_1| \nonumber\\[5pt]
		& \leq  \|\wt{\w}_i\trans n^{-1}\X\trans\E \|_{2,s} \norm{{\bxi}_1}_2 + \|\wt{\w}_i\trans n^{-1}\X\trans\E \|_{2,s} \norm{\v_1^*\v_1\strans{\bxi}_1}_2 \nonumber\\[5pt]
		&\leq 2\|\wt{\w}_i\trans n^{-1}\X\trans\E \|_{2,s} \norm{\bxi_1}_2,
	\end{align}
	where $s = c(r^* + s_u + s_v)$ and the last inequality above is due to $\norm{\v_1^*\v_1\strans{\bxi}_1}_2 \leq \norm{{\v_1^*}}_2\abs{{\v}_1\strans\bxi_1} \leq  \norm{\bxi_1}_2 $ for $\norm{{\v}_1^*}_2 =1$.  Note that here, for an arbitrary vector $\x$, $\norm{\x}_{2,s}^2=\max_{\abs{S}\leq s}\sum_{i\in S}x_i^2$ with $S$ standing for an index set.
	
	From \eqref{wwww} and the fact that $n^{-1}\norm{\X\trans\E}_{\max} \leq c\{n^{-1}\log(pq)\}^{1/2}$, it holds that
	\begin{align*}
		\|\wt{\w}_i\trans n^{-1}\X\trans\E \|_{\max} &\leq  \norm{\wt{\w}_i}_1 \| n^{-1}\X\trans\E \|_{\max} \\
  &\leq c\max\{s_{\max}, (r^*+s_u+s_v)\}^{1/2}\{n^{-1}\log(pq)\}^{1/2}.
	\end{align*}
	Then it follows that
	\begin{align*}
		\|\wt{\w}_i\trans n^{-1}\X\trans\E \|_{2,s}  \leq c\max\{s_{\max}, (r^*+s_u+s_v)\}^{1/2}(r^*+s_u+s_v)^{1/2}\{n^{-1}\log(pq)\}^{1/2},
	\end{align*}
	which together with \eqref{eqr11} and \eqref{eqta1233} entails that 
	\begin{align}\label{eqta2233}
		&|\wt{\w}_i\trans n^{-1}\X\trans\E(\I_q - \v_1^*\v_1\strans){\bxi}_1| \nonumber \\
		&~ \leq  c\max\{s_{\max}, (r^*+s_u+s_v)\}^{1/2}(r^*+s_u+s_v)\eta_n^2\{n^{-1}\log(pq)\}/d_1^*.
	\end{align}

	We next bound term $|\wt{\w}_i\trans \wh{\bSigma}(\wt{\u}_2\wt{\v}_2\trans -\u_2^*\v_2\strans)(\I_q - \v_1^*\v_1\strans){\bxi}_1|$ on the right-hand side of (\ref{sadafvzxc}) above.
	Observe that
	\begin{align}
		\| \wh{\bSigma}(\wt{\u}_2\wt{\v}_2\trans -\u_2^*\v_2\strans)\|_2 &\leq 	 \| \wh{\bSigma}(\wt{\u}_2- \u_2^*)\v_2\strans \|_2+\| \wh{\bSigma}\wt{\u}_2(\wt{\v}_2- \v_2^*)\trans \|_2 \nonumber\\	
		&\leq  \|\wh{\bSigma}(\wt{\u}_2- \u_2^*)\|_2 \|\v_2^* \|_2 +  \|\wh{\bSigma}\wt{\u}_2\|_2 \|\wt{\v}_2- \v_2^*\|_2 \nonumber\\&\leq c(r^*+s_u+s_v)^{1/2}\eta_n^2\left\{n^{-1}\log(pq)\right\}^{1/2}, \nonumber
	\end{align}
	where the last inequality above uses $\|\v_1^* \|_2 =1$ and Lemma \ref{lemmauv}. It follows from \eqref{eqr11} and $\|\v_1^* \|_2 =1$ that
	\begin{align*}
		\norm{(\I_q - \v_1^*\v_1\strans){\bxi}_1}_2  &\leq \norm{{\bxi}_1}_2 + \norm{ \v_1^*\v_1\strans{\bxi}_1}_2 \leq 2 \norm{{\bxi}_1}_2\\
		&\leq    c(r^*+s_u+s_v)^{1/2}\eta_n^2\left\{n^{-1}\log(pq)\right\}^{1/2}/d_1^*.
	\end{align*}
	 Together with \eqref{wwww}, it holds that
	\begin{align}\label{atr12212}
		& |\wt{\w}_i\trans \wh{\bSigma}(\wt{\u}_2\wt{\v}_2\trans -\u_2^*\v_2\strans)(\I_q - \v_1^*\v_1\strans){\bxi}_1| \\
  &\leq 	
		\|\wt{\w}_i\|_2 \| \wh{\bSigma}(\wt{\u}_2\wt{\v}_2\trans -\u_2^*\v_2\strans)\|_2 \norm{(\I_q - \v_1^*\v_1\strans){\bxi}_1}_2 \nonumber\\[5pt]
		&\leq c(r^*+s_u+s_v)\eta_n^4\left\{n^{-1}\log(pq)\right\}/d_1^*.
	\end{align}

	Combining \eqref{sadafvzxc}, \eqref{eqta2233}, and \eqref{atr12212}, we can obtain that 
	\begin{align}
		| \wt{\w}_i\trans \wh{\bDelta} (\I_q - \v_1^*\v_1\strans){\bxi}_1| \leq  c\max\{s_{\max}^{1/2}, (r^*+s_u+s_v)^{1/2}, \eta_n^2\}(r^*+s_u+s_v)\eta_n^2\{n^{-1}\log(pq)\}/d_1^*. \nonumber
	\end{align}
Applying \eqref{wwww} again results in  
$$\max_{1 \leq i \leq p}| \wt{\w}_i\trans \wh{\bDelta} (\I_q - \v_1^*\v_1\strans){\bxi}_1| \leq  c\max\{s_{\max}^{1/2}, (r^*+s_u+s_v)^{1/2}, \eta_n^2\}(r^*+s_u+s_v)\eta_n^2\{n^{-1}\log(pq)\}/d_1^*.$$
Thus, for each vector $\a \in \mathbb{R}^p$ we have that 
\begin{align}\label{eqdlq}
	&|\a\trans\wt{\W}_1\wh{\bDelta}(\I_q - \v_1^*\v_1\strans){\bxi}_1| \leq \norm{\a}_1 \norm{\wt{\W}_1\wh{\bDelta}(\I_q - \v_1^*\v_1\strans){\bxi}_1}_{\max}
	\nonumber\\
 &\leq \norm{\a}_0^{1/2} \norm{\a}_2 \max_{1 \leq i \leq p}	| \wt{\w}_i\trans \wh{\bDelta} (\I_q - \v_1^*\v_1\strans){\bxi}_1| \nonumber\\
	&\leq { c \norm{\a}_0^{1/2} \norm{\a}_2 \max\{s_{\max}^{1/2}, (r^*+s_u+s_v)^{1/2}, \eta_n^2\}(r^*+s_u+s_v)\eta_n^2\{n^{-1}\log(pq)\}/d_1^*.}
\end{align}

It remains to bound term  $|\a\trans\wt{\W}_1( {\r}_{\v_1^*}+ {\r}_{\u_2^*}   + {\r}_{\v_2^*}) |$ above. 
Let us recall that the Taylor remainder terms ${\r}_{\v_1^*}, {\r}_{\u_2^*}$, and ${\r}_{\v_2^*}$ satisfy that 
\[ \|\r_{\v_1^*}\|_2 = O(\| \exp^{-1}_{\v_1^*}(\wt{\v}_1) \|_2^2), \ \|\r_{\u_2^*}\|_2  = O(\|\wt{\u}_2 - \u_2^*  \|_2^2), \ \|\r_{\v_2^*}\|_2  = O(\| \exp^{-1}_{\v_2^*}(\wt{\v}_2) \|_2^2). \]
Based on Lemma \ref{lemma:1rk4} that $\norm{\a\trans\wt{\W}_1}_2 \leq c\norm{\a}_0^{1/2}\norm{\a}_2$, from \eqref{eqr11} we have that 
\begin{align*}
    & |\a\trans\wt{\W}_1 {\r}_{\v_1^*}| \leq \norm{\a\trans\wt{\W}_1}_2 \norm{{\r}_{\v_1^*}}_2 \\
    &\leq  c\norm{\a}_0^{1/2}\norm{\a}_2(r^*+s_u+s_v)\eta_n^4\{n^{-1}\log(pq)\}/d_1^{*2}. 
\end{align*}
Then we apply similar arguments to $|\a\trans\wt{\W}_1 {\r}_{\u_2^*} |$ and $|\a\trans\wt{\W}_1{\r}_{\v_2^*} | $. In view of Definition \ref{lemmsofar} and \eqref{eqr22}, it holds that
\begin{align*}
	&\|\wt{\u}_2 - \u_2^*  \|_2^2 \leq  c(r^*+s_u+s_v)\eta_n^4\{n^{-1}\log(pq)\}, \\
	&\|\exp^{-1}_{\v_2^*}(\wt{\v}_2)\|_2^2
	\leq c(r^*+s_u+s_v)\eta_n^4\{n^{-1}\log(pq)\}/d_2^{*2}.
\end{align*}
Similarly, we can show that
\begin{align*}
	&|\a\trans \wt{\W}_1{\r}_{\u_2^*}| \leq \norm{\a\trans\wt{\W}_1}_2 \norm{{\r}_{\u_2^*}}_2 \leq   c\norm{\a}_0^{1/2} \norm{\a}_2  (r^*+s_u+s_v)\eta_n^4\{n^{-1}\log(pq)\}, \\
	&|\a\trans \wt{\W}_1{\r}_{\v_2^*}| \leq \norm{\a\trans\wt{\W}_1}_2 \norm{{\r}_{\v_2 ^*}}_2 \leq  c\norm{\a}_0^{1/2} \norm{\a}_2 (r^*+s_u+s_v)\eta_n^4\{n^{-1}\log(pq)\}/d_2^{*2}.
\end{align*}
Since the nonzero eigenvalues $d^{*2}_{i}$ are at the constant level by Condition \ref{con:nearlyorth}, it follows that
\begin{align}\label{eqawr}
	|\a\trans\wt{\W}_1( {\r}_{\v_1^*}+ {\r}_{\u_2^*}   + {\r}_{\v_2^*}) |
	&\leq 	|\a\trans\wt{\W}_1 {\r}_{\v_1^*}| + 	|\a\trans\wt{\W}_1 {\r}_{\u_2^*}| + 	|\a\trans\wt{\W}_1 {\r}_{\v_2^*}| \nonumber\\
    &	 \leq   c\norm{\a}_0^{1/2} \norm{\a}_2 (r^*+s_u+s_v)\eta_n^4\{n^{-1}\log(pq)\}.
\end{align}

Combining  \eqref{adeee}, \eqref{dassd}, \eqref{eqdlq}, and \eqref{eqawr} yields that 
\begin{align}\label{eq1.2.2}
	&|\a\trans\wt{\W}_1 (\wt{\psi}_1\left(\wt{\u}_1, \wt{\boldeta}_1 \right) - \wt{\psi}_1\left( \wt{\u}_1, \boldeta_1^* \right)) | \nonumber \\
	& \leq { c \norm{\a}_0^{1/2} \norm{\a}_2 \max\{s_{\max}^{1/2}, (r^*+s_u+s_v)^{1/2}, \eta_n^2\}(r^*+s_u+s_v)\eta_n^2\{n^{-1}\log(pq)\}.}
\end{align}
Thus, for $\a\in\mathcal{A}=\{\a\in\R^p:\norm{\a}_0\leq m,\norm{\a}_2=1\}$, we can obtain that 
\begin{align}
	&|\a\trans\wt{\W}_1 (\wt{\psi}_1\left(\wt{\u}_1, \wt{\boldeta}_1 \right) - \wt{\psi}_1\left( \wt{\u}_1, \boldeta_1^* \right)) | \nonumber \\
	& \leq c m^{1/2}\max\{s_{\max}^{1/2}, (r^*+s_u+s_v)^{1/2}, \eta_n^2\}(r^*+s_u+s_v)\eta_n^2\{n^{-1}\log(pq)\}, \label{eq1.2.232}
\end{align}
which completes the proof for the rank-2 case.

\bigskip

\noindent \textbf{Part 2: Extension to the general rank case.} We extend the results using similar arguments as in the first part. 
For the nuisance parameter  $\boldeta_k = [   \v_1\trans, \ldots, \v_{r^*}\trans, \u_1\trans,\ldots, \u_{k-1}\trans, \u_{k+1}\trans, $ $ \ldots, \u_{r^*}\trans,]\trans$,
it follows from the definition of  $\wt{\psi}_k(\u_k,\boldeta_k)$ that 
\begin{align}\label{eqphikk}
	\wt{\psi}_k(\u_k,\boldeta_k)
	& = \der{L}{\u_k} - \M\der{L}{\boldeta_k} \nonumber\\
 &= \der{L}{\u_k} - \left(\M_k^v \der{L}{\v_k}+ \sum_{j \neq k}  \M_j^u \der{L}{\u_j} + \sum_{j \neq k}\M_j^v \der{L}{\v_j}\right).
\end{align}
By Proposition \ref{prop:rankr2},  we see that  $\M_j^u = \0$ and $  \M_j^v = \0$ for $j \in \{ 1, \ldots, r^*\}$ with $j \neq k$, which means that we need only to consider $\v_k$ as the nuisance parameter. In light of the derivatives \eqref{der:1} and \eqref{der:2}, we can deduce that 
\begin{align}
	\wt{\psi}_k(\u_k,\boldeta_k)
	 &= \der{L}{\u_k} - \M_k^v \der{L}{\v_k} \nonumber\\[5pt] 
       & = \wh{\bSigma}\u_k - n^{-1}\X\trans\Y\v_k - \M_k^v ( \v_k\u_k\trans\wh{\bSigma}\u_k - n^{-1}\Y\trans\X\u_k ). \label{t11}
\end{align}

For arbitrary fixed $\M_k^v$, we can see that $\wt{\psi}_k(\u_k, \boldeta_k)$ is only a function of $\u_k$ and $\v_k$, which means that we need only to do the Taylor expansion of $\wt{\psi}_k(\u_k, \boldeta_k)$  with respect to $\v_k$.
Similar to the proof of Lemma \ref{prop:taylor} in Section \ref{sec:proof:taylor}, we can obtain the Taylor expansion of $\wt{\psi}_k(\u_k, \boldeta_k)$ 
\begin{align*}
    \wt{\psi}_k&\left( \u_k,  {\boldeta}_k  \right) = \wt{\psi}_k\left( \u_k, \boldeta_k^* \right)
    +  \der{\wt{\psi}_k( \u_k, {\boldeta}_k  )}{{\v}_k\trans}\Big|_{\v_k^*}(\I_q - \v_k^*\v_k\strans) \exp^{-1}_{\v_k^*}({\v}_k)   + \r_{\v_k^*},
\end{align*}
where  the Taylor remainder term satisfies that
\[ \|\r_{\v_k^*}\|_2 = O(\| \exp^{-1}_{\v_k^*}({\v}_k) \|_2^2). \]
From \eqref{t11}, it holds that
\[ \der{\wt{\psi}_k( \u_k,  {\boldeta}_k   )}{{\v}_k\trans}\Big|_{\v_k^*} = - n^{-1}\X\trans\Y - \u_k\trans\wh{\bSigma}\u_k\M_k^v.\]
Then by Proposition \ref{prop:rankr2} that $\M^v_k =  -z_{kk}^{-1}\wh{\bSigma}\C_{-k}$ and the initial estimates in Definition \ref{lemmsofar}, we have that 
\begin{align}\label{phiuuuu2}
    &\wt{\psi}_k(\wt{\u}_k,\wt{\boldeta}_k) - \wt{\psi}_k(\wt{\u}_k,\boldeta^*_k) \nonumber\\
    & = (- n^{-1}\X\trans\Y + \wh{\bSigma}\wt{\C}_{-k})(\I_q - \v_k^*\v_k\strans) \exp^{-1}_{\v_k^*}(\wt{\v}_k)  + \r_{\v_k^*}  \nonumber\\
	 & =( \wh{\bSigma}(\wt{\C}_{-k} - \C_{-k}^*) -n^{-1}\X\trans\E)(\I_q - \v_k^*\v_k\strans)\exp^{-1}_{\v_k^*}(\wt{\v}_k)  + \r_{\v_k^*},
\end{align}
where we slightly abuse the notation and denote the Taylor remainder term as 
\[\|\r_{\v_k^*}\|_2 = O(\| \exp^{-1}_{\v_k^*}(\wt{\v}_k) \|_2^2). \]

We next bound term $\a\trans \wt{\W}_k (\wt{\psi}_k(\wt{\u}_k,\wt{\boldeta}_k) - \wt{\psi}_k(\wt{\u}_k,\boldeta^*_k) )$ above.
Observe that 
\begin{align*}
	& |\a\trans \wt{\W}_k (\wt{\psi}_k(\wt{\u}_k,\wt{\boldeta}_k) - \wt{\psi}_k(\wt{\u}_k,\boldeta^*_k) )| \\[5pt]
	&\leq |\a\trans\wt{\W}_k( \wh{\bSigma}(\wt{\C}_{-k} - \C_{-k}^*) -n^{-1}\X\trans\E)(\I_q - \v_k^*\v_k\strans)\exp^{-1}_{\v_k^*}(\wt{\v}_k)|
 +	|\a\trans\wt{\W}_k \r_{\v_k^*}|. 
\end{align*}
By Lemma \ref{rankr:boundm0} in Section \ref{new.Sec.B.10}, it can be seen that
\begin{align}
	\norm{ \wh{\bSigma}(\wt{\C}_{-k}  - \C_{-k}^*)}_2 \leq  c (r^*+s_u+s_v)^{1/2}\eta_n^2\left\{n^{-1}\log(pq)\right\}^{1/2}. \label{31ewdasd}
\end{align} 
Denote by $\wt{\w}_{k,i}\trans$ the $i$th row of $\wt{\W}_k$ for $i = 1, \ldots, p$.
By parts (a) and (b) of Lemma \ref{rankr:aw} in Section \ref{new.Sec.B.11}, we have that 
	\begin{align}\label{eqwwwazcs}
		\max_{1 \leq i \leq p}\norm{\wt{\w}_{k,i}}_0 \leq 2\max\{s_{\max}, 3(r^*+s_u+s_v)\} \ \text{ and } \ \max_{1 \leq i \leq p}\norm{\wt{\w}_{k,i}}_2 \leq c.  
	\end{align}
Using similar arguments as for \eqref{eqr21} and \eqref{eqr22}, it holds that
\begin{align}
	&\norm{ \exp^{-1}_{\v_k^*}(\wt{\v}_k)}_0  \leq c(r^*+s_u +s_v), \label{czxdfcaa} \\
	&\|\exp^{-1}_{\v_k^*}(\wt{\v}_k)\|_2
	\leq c(r^*+s_u+s_v)^{1/2}\eta_n^2\{n^{-1}\log(pq)\}^{1/2}/d_k^*.  \label{czxdfcaa2}
\end{align}

Based on results \eqref{31ewdasd}--\eqref{czxdfcaa2} above, we proceed with following the proof for the rank-2 case.
With similar arguments as for 
\eqref{eqta1233} and \eqref{eqta2233}, it follows that
\begin{align}
	&|\wt{\w}_{k,i}\trans n^{-1}\X\trans\E(\I_q - \v_k^*\v_k\strans)\exp^{-1}_{\v_k^*}(\wt{\v}_k)| \nonumber \\
	& \leq  c\max\{s_{\max}, (r^*+s_u+s_v)\}^{1/2}(r^*+s_u+s_v)\eta_n^2\{n^{-1}\log(pq)\}/d_k^*. \nonumber
\end{align}
Further, similar to \eqref{atr12212}, we can deduce that 
\begin{align}
	&|\wt{\w}_{k,i}\trans \wh{\bSigma}(\wt{\C}_{-k} - \C_{-k}^*)(\I_q - \v_k^*\v_k\strans)\exp^{-1}_{\v_k^*}(\wt{\v}_k)| \leq c(r^*+s_u+s_v)\eta_n^4\left\{n^{-1}\log(pq)\right\}/d_k^* \nonumber
\end{align}
and
\begin{align}
	& |\a\trans \wt{\W}_k ( \wh{\bSigma}(\wt{\C}_{-k} - \C_{-k}^*) )(\I_q - \v_k^*\v_k\strans)\exp^{-1}_{\v_k^*}(\wt{\v}_k) | \nonumber\\
	& \leq  c \norm{\a}_0^{1/2} \norm{\a}_2 (r^*+s_u+s_v)\eta_n^4\left\{n^{-1}\log(pq)\right\}/d_k^*.  \label{schkasjhdq}
\end{align} 
Thus, similar to \eqref{eqdlq}, combining the above results gives that
\begin{align}\label{31ewdasd2}
	 & \Big|\a\trans \wt{\W}_k ( \wh{\bSigma}(\wt{\C}_{-k} - \C_{-k}^*) -n^{-1}\X\trans\E)(\I_q - \v_k^*\v_k\strans)\exp^{-1}_{\v_k^*}(\wt{\v}_k) \Big| \\
	& \leq  c \norm{\a}_0^{1/2} \norm{\a}_2 \max\{s_{\max}^{1/2}, (r^*+s_u+s_v)^{1/2}, \eta_n^2 \}(r^*+s_u+s_v)\eta_n^2\{n^{-1}\log(pq)\}
/d_k^*. \nonumber
\end{align}

Moreover, an application of similar arguments as for \eqref{eqawr} shows that
\begin{align}
	&|\a\trans \wt{\W}_k \r_{\v_k^*} | 
	\leq  c\norm{\a}_0^{1/2} \norm{\a}_2  (r^*+s_u+s_v)\eta_n^4\{n^{-1}\log(pq)\}/d_k^{*2}. \label{31ewdasd3}
\end{align}
Under Condition \ref{con:nearlyorth} that the nonzero eigenvalues $d^{*2}_{i}$ are at the constant level, combining the above results yields that 
\begin{align}
	&|\a\trans\wt{\W}_k  (\wt{\psi}_k(\wt{\u}_k,\wt{\boldeta}_k) - \wt{\psi}_k(\wt{\u}_k,\boldeta^*_k) )| \nonumber \\
	& \leq c \norm{\a}_0^{1/2}\norm{\a}_2 \max\{s_{\max}^{1/2}, (r^*+s_u+s_v)^{1/2}, \eta_n^2 \}(r^*+s_u+s_v)\eta_n^2\{n^{-1}\log(pq)\},  \label{saazs12}
\end{align}
which completes the proof of Lemma \ref{prop:taylor12}.

\subsection{Lemma \ref{lemmauv} and its proof} \label{new.Sec.B.3}

\begin{lemma}\label{lemmauv}
	Assume that  Condition \ref{con3} holds and $\wt{\C}$ satisfies Definition \ref{lemmsofar}. Then with probability at least
$1- \theta_{n,p,q}$ with $\theta_{n,p,q}$ given in \eqref{thetapro}, we have that for all sufficiently large $n$ and each $k=1, \ldots, r^*$, 
\begin{enumerate}[label=\rm{(\alph*)}]
	\item $\norm{\wt{\v}_k-\v_k^*}_2 \leq c\gamma_n d_k^{*-1}$,  \ 
 $	\norm{d_k^{*}(\wt{\v}_k - \v_k^*)}_2 \leq c\gamma_n, \ 	\sum_{k=1}^{r^*}\norm{\wt{d}_k(\wt{\v}_k - \v_k^*)}_0 \leq 3(r^*+s_u+s_v)$; 
	\item $\|\wh{\bSigma} \u_k^*\|_2 \leq cd_k^*, \ \|\wh{\bSigma}\wt{\u}_k\|_2 \leq cd_k^*,$ \ 
	$\|\wh{\bSigma}(\wt{\u}_k- \u_k^*)\|_2 \leq c\gamma_n$; 
	\item $\abs{\wt{z}_{kk} - z_{kk}^*} \leq c\gamma_n d_k^*, \ $
 $ \abs{\wt{z}_{kk}^{-1} - z_{kk}^{*-1}} \leq c \gamma_n d_k^{*-3}, \
	| z_{kk}^{*} |^{-1} \leq c d_k^{*-2},
	\  \abs{\wt{z}_{kk} }^{-1} \leq c d_k^{*-2},$
\end{enumerate}
where $\gamma_n = (r^*+s_u+s_v)^{1/2} \eta_n^2\{n^{-1}\log(pq)\}^{1/2}$, $\wt{z}_{kk} =  \wt{\u}_k\trans\wh{\bSigma}\wt{\u}_k, z_{kk}^{*} = \u_k\strans\wh{\bSigma}\u_k^* $, and $c$ is some positive constant.
\end{lemma}

\noindent \textit{Proof}. We first prove part (a). In view of Definition \ref{lemmsofar}, it holds that 
\begin{align*}
	\norm{\wt{d}_k\wt{\v}_k - d_k^*\v^*_k}_2 \leq  c\gamma_n \ \text{ and } \  \abs{d_k^*-\wt{d}_k} \leq c\gamma_n,
\end{align*}
where $ \gamma_n = ({r^*}+s_u+s_v)^{1/2}\eta_n^2\{n^{-1}\log(pq)\}^{1/2}$.
Observe that $d_k^{*}(\wt{\v}_k - \v_k^*) = (\wt{d}_k\wt{\v}_k - d_k^*\v_k^*) + (d_k^*-\wt{d}_k)\wt{\v}_k$. Since $\norm{\wt{\v}_k}_2=1$, we have that
	\begin{align*}
	\norm{d_k^{*}(\wt{\v}_k - \v_k^*)}_2 &\leq \norm{\wt{d}_k\wt{\v}_k - d_k^*\v^*_k}_2 + \abs{d_k^*-\wt{d}_k}\norm{\wt{\v}_k}_2 
		 \leq  c \gamma_n.	
   \end{align*}
Since the true singular values $d_k^* \neq 0$ for each $k=1, \ldots, r^*$, it follows that 
	\begin{align*}
		\norm{\wt{\v}_k - \v_k^*}_2 \leq  c\gamma_n/d_k^{*}.
	\end{align*}
Also, by Definition \ref{lemmsofar} it holds that
\begin{align*}
\sum_{k=1}^{r^*} \norm{\wt{d}_k(\wt{\v}_k - \v_k^*)}_0 &\leq \sum_{k=1}^{r^*}\norm{\wt{d}_k\wt{\v}_k - d_k^*\v^*_k}_0 +  \sum_{k=1}^{r^*}\norm{(d_k^*-\wt{d}_k){\v}_k^*}_0 \\
&\leq \sum_{k=1}^{r^*}\norm{\wt{d}_k\wt{\v}_k - d_k^*\v^*_k}_0 +  \sum_{k=1}^{r^*}\norm{{\v}_k^*}_0 \\
&\leq (r^*+s_u+s_v)[1 + o(1)] + s_v \\
&\leq  3(r^*+s_u+s_v).
\end{align*}

	
	For part (b), let us recall that $\u_k^* = d_k^* \l_k^* $.
	 Since $\norm{\l_k^*}_0 \leq s_u$ and  $\norm{\l_k^*}_2 = 1  $, it follows from Condition \ref{con3} that
	\[ \|\wh{\bSigma} \u_k^*\|_2 = d_k^*\|\wh{\bSigma} \l_k^*\|_2 \leq \rho_u d_k^* \norm{\l_k^*}_2 \leq cd_k^{*}. \]
	From Definition \ref{lemmsofar}, we can show that 
	\[ \norm{\wt{\u}_k - \u_k^* }_0 \leq (r^*+s_u+s_v)[1 + o(1)] \ \text{ and } \ \norm{\wt{\u}_k - \u_k^* }_2 \leq c\gamma_n. \]
	Moreover, by $\norm{\u_k^*}_0 \leq s_u$ and $\norm{\u_k^*}_2 \leq d_k^*$ for sufficiently large $n$, and Condition \ref{con4} that $r^*\gamma_n=o(d^*_{r^*})$, it holds for $\wt{\u}_k$ that
	\begin{align}
		&\norm{\wt{\u}_k}_0  \leq \norm{\wt{\u}_k - \u_k^* }_0 + \norm{\u_k^*}_0 \leq (r^*+s_u+s_v)[1 + o(1)] + s_u \leq 3(r^* + s_u + s_v), \nonumber\\
		& 	\norm{\wt{\u}_k}_2 \leq \norm{\wt{\u}_k - \u_k^* }_2 + \norm{\u_k^*}_2 \leq cd_k^*.\nonumber
	\end{align}
	Then it follows from Condition \ref{con3} that
	\begin{align*}
		 \|\wh{\bSigma} \wt{\u}_k\|_2 \leq \rho_u \norm{\wt{\u}_k}_2 \leq cd_k^* \ \text{ and } \
		 \|\wh{\bSigma}(\wt{\u}_k- \u_k^*)\|_2 \leq   \rho_u \norm{\wt{\u}_k-\u_k^*}_2 \leq  c\gamma_n.
	\end{align*}

	For part (c), using part (b) of this lemma and Definition \ref{lemmsofar}, we can deduce that 
	\begin{align}
		\abs{\wt{z}_{kk} - z_{kk}^*} &= \abs{\wt{\u}_k\trans\wh{\bSigma}\wt{\u}_k-\u_k\strans\wh{\bSigma}\u_k^*} \nonumber\\
	   &\leq \abs{\wt{\u}_k\trans\wh{\bSigma}(\wt{\u}_k-\u_k^*)} + |(\wt{\u}_k-\u_k^*)\trans\wh{\bSigma}\u_k^*| \nonumber\\
	   &\leq \norm{\wt{\u}_k}_2\norm{\wh{\bSigma}(\wt{\u}_k-\u_k^*)}_2 + \norm{\wt{\u}_k-\u_k^*}_2\norm{\wh{\bSigma}\u_k^*}_2 \nonumber\\
	   &\leq c\gamma_n d_k^*. \label{equ1su1}
	\end{align}
Note that
	\begin{align*}
	\abs{\wt{z}_{kk}^{-1} - z_{kk}^{*-1}} = \left| \frac{\wt{z}_{kk} - z_{kk}^*}{\wt{z}_{kk} \cdot z_{kk}^*} \right| =    \left| \frac{\wt{\u}_k\trans\wh{\bSigma}\wt{\u}_k - {\u}_k\strans\wh{\bSigma}{\u}_k^{*}}{\wt{\u}_k\trans\wh{\bSigma}\wt{\u}_k \cdot {\u}_k\strans\wh{\bSigma}{\u}_k^{*}} \right|.
	\end{align*}
	By Condition \ref{con3}, we have that $ {d}^{*2}_k\rho_l \leq {\u}_k\strans\wh{\bSigma}{\u}_k^{*} \leq {d}^{*2}_k\rho_u $.
	Then it follows that
	\begin{align*}
   | z_{kk}^{*-1}| = |{\u}_k\strans\wh{\bSigma}{\u}_k^{*}|^{-1}  \leq  | d_k^{*-2}\rho_l | \leq cd_k^{*-2}.
	\end{align*}
	Together with \eqref{equ1su1}, it yields that 
	\begin{align*}
	\abs{\wt{z}_{kk}^{-1} - z_{kk}^{*-1}} \leq \left| \frac{\wt{\u}_k\trans\wh{\bSigma}\wt{\u}_k - {\u}_k\strans\wh{\bSigma}{\u}_k^{*}}
	{ {\u}_k\strans\wh{\bSigma}{\u}_k^{*}({\u}_k\strans\wh{\bSigma}{\u}_k^{*} + o(1))} \right| \leq c\gamma_n d_k^{*-3}.
	\end{align*}
Furthermore, by $r^*\gamma_n=o(d^*_{r^*})$ in Condition \ref{con4}, we have  for sufficiently large $n$ it holds that 
\[ \abs{\wt{z}_{kk}^{-1}} \leq \abs{z_{kk}^{*-1}} +  \abs{ \wt{z}_{kk}^{-1} - z_{kk}^{*-1} } \leq cd_k^{*-2}. \]
This concludes the proof of Lemma \ref{lemmauv}.


\subsection{Lemma \ref{lemmzzz} and its proof} \label{new.Sec.B.4}

\begin{lemma}\label{lemmzzz}
	Assume that all the conditions of Theorem \ref{theorkr} are satisfied. 
	Then with probability at least
	$1- \theta_{n,p,q}$ with $\theta_{n,p,q}$ given in \eqref{thetapro}, it holds that for all sufficiently large $n$, 
	\begin{align*}
        &|z_{ii}^* - z_{jj}^*| \geq c, \quad \sum_{1 \leq l \leq r^*, \ l \neq k} |z_{kl}^*| =  o(n^{-1/2}) = o(|z_{ii}^* - z_{jj}^*|), \\
        &|\wt{z}_{ii} - \wt{z}_{jj}| \geq c, \quad \sum_{1 \leq l \leq r^*, \ l \neq k}|\wt{z}_{kl}| = O( r^* \gamma_n) = o( |\wt{z}_{ii} - \wt{z}_{jj}| ),
	\end{align*}
	where $c$ is some positive constant and $i, j, k \in \{1, \ldots, r^*\}$ with $i \neq j$. 
\end{lemma}

\noindent \textit{Proof}. 
We will first show that $|z_{ii}^* - z_{jj}^*| \geq c$.
By Condition \ref{con3} and the sparsity of $\u_i^* = d_i^*\l^*_i$, we see that $ {d}^{*2}_i\rho_l \leq z_{ii}^{*} \leq {d}^{*2}_i\rho_u $ and $ {d}^{*2}_j\rho_l \leq z_{jj}^{*} \leq {d}^{*2}_j\rho_u $, which lead to
\begin{align}\label{ediiiac}
	 d_i^{*2} \rho_l - d_j^{*2} \rho_u \leq  z_{ii}^{*}  -z_{jj}^{*}   \leq   d_i^{*2}\rho_u -   d_j^{*2}\rho_l.
\end{align}
In light of Condition \ref{con4}, we have $d^{*2}_{i}  - d^{*2}_{i+1} \geq  \delta_1 d^{*2}_{i}$
for some positive constant $\delta_1 > 1 - (\rho_l/\rho_u)$ with $1 \leq i \leq r^*$. Since $\rho_l, \rho_u$ are positive constants, there exists some positive constant $c_0$ such that $\delta_1 = 1 - (\rho_l/\rho_u) + c_0 $, which further entails that 
\begin{align}\label{ediiiac2}
    d^{*2}_{i}\rho_l  - d^{*2}_{i+1}\rho_u \geq  c_0 \rho_u d^{*2}_{i} \geq c,
\end{align}
where the last inequality above is due to Condition \ref{con:nearlyorth} that $d_i^*$ is at a constant level.

If $i < j$, we have $i + 1 \leq j$ so that $d_{i+1}^{*2} \geq d_{j}^{*2}$. This together with \eqref{ediiiac} and \eqref{ediiiac2} shows that 
\begin{align*}
	z_{ii}^{*}  -z_{jj}^{*}   \geq  d_i^{*2} \rho_l - d_{j}^{*2}  \rho_u \geq d^{*2}_{i}\rho_l  - d^{*2}_{i+1}\rho_u \geq c.
\end{align*}
If $i > j$, using similar arguments we can obtain that 
$z_{jj}^{*}  -z_{ii}^{*}   \geq  c$. Thus, for $i \neq j$ it holds that
\begin{align}\label{zcxzcadas1}
	| z_{ii}^{*}  -z_{jj}^{*}  | \geq c.
\end{align}

We next bound term $\wt{z}_{ii} - \wt{z}_{jj} $ above. By part (c) of  Lemma \ref{lemmauv} and  Condition \ref{con:nearlyorth} that $d_i^*$ is at a constant level,  we can deduce that
\begin{align*}
	|(\wt{z}_{ii} - \wt{z}_{jj}) - (z_{ii}^{*} - z_{jj}^{*})| \leq |\wt{z}_{ii} - z_{ii}^{*}| + |\wt{z}_{jj} - z_{jj}^{*}| \leq  c\gamma_n,
\end{align*}
where $\gamma_n = (r^*+s_u+s_v)^{1/2}\eta_n^2\{n^{-1}\log(pq)\}^{1/2}$. From the assumption of Theorem \ref{theorkr} that $m^{1/2}\kappa_n = o(1)$, we have $\gamma_n = o(1)$.
Together with \eqref{zcxzcadas1}, for all sufficiently large $n$ it follows that
\begin{align}\label{zcxzcadas}
	|\wt{z}_{ii} - \wt{z}_{jj} | \geq 	| z_{ii}^{*}  -z_{jj}^{*}  | - 	|(\wt{z}_{ii} - \wt{z}_{jj}) - (z_{ii}^{*} - z_{jj}^{*})| \geq c.
\end{align}

Now we analyze terms $\sum_{1 \leq l \leq r^*, \ l \neq k}|\wt{z}_{kl}|$ and $\sum_{1 \leq l \leq r^*, \ l \neq k}|z_{kl}^*|$. From Condition \ref{con:nearlyorth}, we have that 
\begin{align}\label{czsqdzzz}
    \sum_{1 \leq l \leq r^*, \ l \neq k}|z_{kl}^*| =  o(n^{-1/2}).
\end{align}
Moreover, it follows that
\begin{align}\label{zijij}
	|\wt{z}_{kl} - z_{kl}^* | & = |  \wt{\u}_k\trans\wh{\bSigma}\wt{\u}_l - {\u}_k\strans\wh{\bSigma}\u_l^*|  \leq |\wt{\u}_k\trans\wh{\bSigma}(\wt{\u}_l - \u_l^*)| + |(\wt{\u}_k - \u_k^*)\trans\wh{\bSigma} \u_l^*| \nonumber \\
	 & \leq  \|\wt{\u}_k\|_2 \|\wh{\bSigma}(\wt{\u}_l - \u_l^*)\|_2 + \|\wt{\u}_k - \u_k^*\|_2 \|\wh{\bSigma} \u_l^*\|_2  \nonumber \\
	 & \leq   c  (r^*+ s_u+s_v)^{1/2}\{n^{-1}\log(pq)\}^{1/2},
\end{align}
where the last inequality above is due to part (a) of Definition \ref{lemmsofar},  part (b) of Lemma \ref{lemmauv}, and Condition \ref{con:nearlyorth} that $d_k^*$ is at a constant level. Then for sufficiently large $n$, it holds that 
\[ |\wt{z}_{kl}| \leq |z_{kl}^*| + |\wt{z}_{kl} - z_{kl}^* | \leq c(r^*+s_u+s_v)^{1/2}\eta_n^2\{n^{-1}\log(pq)\}^{1/2},  \]
which further yields $\sum_{1 \leq l \leq r^*, \ l \neq k}|\wt{z}_{kl}|  = O( r^* \gamma_n)  $.

Let us recall the assumption that  $m^{1/2}\kappa_n = o(1)$ with
$$\kappa_n = \max\{s_{\max}^{1/2} , (r^*+s_u+s_v)^{1/2}, \eta_n^2\} (r^*+s_u+s_v)\eta_n^2\log(pq)/\sqrt{n}.$$
It follows from $\gamma_n = (r^*+s_u+s_v)^{1/2}\eta_n^2\{n^{-1}\log(pq)\}^{1/2}$ and $(r^*+s_u+s_v)^{1/2} \leq \max\{s_{\max}^{1/2} , (r^*+s_u+s_v)^{1/2}, \eta_n^2\}$ that 
\begin{align}\label{czdgggs}
	m^{1/2} (r^*+s_u+s_v) \sqrt{\log(pq)}  \gamma_n = o(1),
\end{align}
which further leads to 
\[\sum_{1 \leq l \leq r^*, \ l \neq k}|\wt{z}_{kl}|  = O( r^* \gamma_n) = o(1). \]
Therefore, along with \eqref{zcxzcadas1}, \eqref{zcxzcadas}, and \eqref{czsqdzzz}, it yields that
$$ \sum_{1 \leq l \leq r^*, \ l \neq k} |z_{kl}^*| = o(|z_{ii}^* - z_{jj}^*|) \ \text{ and } \ \sum_{1 \leq l \leq r^*, \ l \neq k}|\wt{z}_{kl}| = o( |\wt{z}_{ii} - \wt{z}_{jj}| ),$$
which completes the proof of Lemma \ref{lemmzzz}.

\subsection{Lemma \ref{lemma:1rk4} and its proof} \label{new.Sec.B.5}

\begin{lemma}\label{lemma:1rk4}
	Assume that all the conditions of Theorem \ref{theorkr} are satisfied. Let 
	$ \wt{\W}_1 = \widehat{\bTheta}\{\I_p + (\wt{z}_{11} - \wt{z}_{22})^{-1}\wh{\bSigma}\wt{\u}_2\wt{\u}_2\trans\}$ and  ${\W}^{*}_1=   \widehat{\bTheta}\{\I_p + (z_{11}^{*} - z_{22}^{*})^{-1}\wh{\bSigma}{\u}_2^{*}{\u}_2\strans\}$.
	For an arbitrary vector
	$\a = (a_1, \ldots, a_p)\trans \in\R^p$,  with probability at least
	$1- \theta_{n,p,q}$ with $\theta_{n,p,q}$ given in \eqref{thetapro}, we have that for all sufficiently large $n$, 
	\begin{enumerate}[label=\rm{(\alph*)}]
		\item 	$\max_{1 \leq i \leq p}\norm{\w_i^*}_0  \leq 2\max\{s_{\max}, r^*+s_u + s_v \},\\  \max_{1 \leq i \leq p}\norm{\wt{\w}_i}_0 \leq 2\max\{s_{\max}, 3(r^*+s_u + s_v) \}$, \\ 
		$\max_{1 \leq i \leq p}\norm{\wt{\w}_i - \w_i^*}_0 \leq 3 (r^*+s_u + s_v)$;
		\item  $ \max_{1 \leq i \leq p}\norm{\w_i^*}_2 \leq c, \ \max_{1 \leq i \leq p}\norm{\wt{\w}_i}_2 \leq c$,
		\\[5pt] $ \max_{1 \leq i \leq p}\norm{\wt{\w}_i - \w_i^*}_2 \leq  c(r^*+s_u+s_v)^{1/2}\eta_n^2\{n^{-1}\log(pq)\}^{1/2}$;
		\item $ \norm{\a\trans\W^*_1}_2 \leq c \norm{\a}_0^{1/2}\norm{\a}_2$, $ \norm{\a\trans\wt{\W}_1}_2 \leq c \norm{\a}_0^{1/2}\norm{\a}_2,$ \\[5pt] $\norm{\a\trans(\wt{\W}_1-\W_1^*)}_2 \leq c(r^*+s_u+s_v)^{1/2}\eta_n^2\{n^{-1}\log(pq)\}^{1/2} \norm{\a}_0^{1/2}\norm{\a}_2,$
	\end{enumerate}
	where $\wt{\w}_i\trans$ and $ \w_i\strans$ are the $i$th rows of $\wt{\W}_1$ and $\W^*_1$, respectively, with $i = 1, \ldots, p$, and $c$ is some positive constant.
\end{lemma}

\noindent \textit{Proof}. 
It is easy to see that 
\begin{align}\label{eq:aw1}
	\norm{\a\trans\W^*_1}_2 &= \norm{\sum_{i=1}^{p}a_i\w_i\strans}_2 \leq \sum_{i=1}^{p}\abs{a_i}\cdot\norm{\w_i^*}_2 \nonumber \\
	& \leq \norm{\a}_1\max_{1\leq i\leq p}\norm{\w_i^*}_2 \leq  \norm{\a}_0^{1/2}\norm{\a}_2\max_{1\leq i\leq p}\norm{\w_i^*}_2.
\end{align}
Similarly, we also have that 
\begin{align}
	&\norm{\a\trans\wt{\W}_1}_2 \leq  \norm{\a}_0^{1/2}\norm{\a}_2\max_{1\leq i\leq p}\norm{\wt{\w}_i}_2, \label{eq:aw2} \\
	&\norm{\a\trans(\wt{\W}_1-\W_1^*)}_2 \leq \norm{\a}_0^{1/2}\norm{\a}_2\max_{1\leq i\leq p}\norm{\wt{\w}_i-\w_i^*}_2.  \label{eq:aw3}
\end{align}
Then it can be seen that once parts (a) and (b) of this lemma are established, the results in part (c) can be obtained immediately with the aid of \eqref{eq:aw1}--\eqref{eq:aw3}. Thus, it remains to prove parts (a) and (b).

We begin with proving part (a). Since $ \w_i\strans$ is the $i$th row of $\W^*_1$, we have
$ \w_i\strans = \widehat{\btheta}_i\trans \{\I_p + (z_{11}^* - z_{22}^*)^{-1}\wh{\bSigma}{\u}_2^*{\u}_2\strans\}$, where  $\wh{\btheta}_i\trans$ is the $i$th row of $\wh{\bTheta}$.
Noting that $\widehat{\btheta}_i\trans(z_{11}^* - z_{22}^*)^{-1}\wh{\bSigma}{\u}_2^*$ is a scalar and $\norm{ \u_2^*}_0 \leq s_u$, we can deduce that 
$$\max_{1\leq i\leq p}\norm{(\widehat{\btheta}_i\trans(z_{11}^* - z_{22}^*)^{-1}\wh{\bSigma}{\u}_2^*) \cdot {\u}_2\strans}_0 \leq \norm{ \u_2^*}_0 \leq s_u  \leq r^*+s_u+s_v.$$
From Definition \ref{defi2:acceptable}, we see that $\max_{1 \leq i \leq p}\norm{\widehat{\btheta}_i}_0 \leq s_{max}$. 
Hence, it follows that
\begin{align}
	\max_{1\leq i\leq p}\norm{\w_i^*}_0 \leq \max_{1\leq i\leq p}\norm{\widehat{\btheta}_i}_0 + \norm{ \u_2^*}_0 \leq 2\max\{s_{\max}, r^*+s_u + s_v \}.	 \label{sadcqe}
\end{align}
Observe that $\wt{\w}_i\trans = \widehat{\btheta}_i\trans\{\I_p + (\wt{z}_{11} - \wt{z}_{22})^{-1}\wh{\bSigma}\wt{\u}_2 \wt{\u}_2\trans \}$.
By Definition \ref{lemmsofar}, we have that 
 $$\norm{ \wt{\u}_2}_0 \leq \norm{ {\u}_2^*}_0 + \norm{ \wt{\u}_2 -  {\u}_2^*}_0 \leq  3(r^*+s_u + s_v).$$
Then an application of similar arguments as for \eqref{sadcqe} leads to 
\begin{align}
	\max_{1\leq i\leq p}\norm{\wt{\w}_i}_0 \leq \max_{1\leq i\leq p}\norm{\widehat{\btheta}_i}_0 + \norm{ \wt{\u}_2}_0 \leq 2\max\{s_{\max}, 3(r^*+s_u + s_v) \}.	 \nonumber
\end{align}
Further, we can show that
\begin{align*}
	\max_{1\leq i\leq p}\norm{ \wt{\w}_i - \w_i^*}_0
	&\leq \max_{1\leq i\leq p}\norm{(\widehat{\btheta}_i\trans(\wt{z}_{11} - \wt{z}_{22})^{-1}\wh{\bSigma}\wt{\u}_2) \cdot \wt{\u}_2\trans - (\widehat{\btheta}_i\trans (z_{11}^* - z_{22}^*)^{-1}\wh{\bSigma}{\u}_2^*) \cdot {\u}_2\strans}_0 \\
	&\leq 	\max_{1\leq i\leq p}\norm{(\widehat{\btheta}_i\trans(\wt{z}_{11} - \wt{z}_{22})^{-1}\wh{\bSigma}\wt{\u}_2) \cdot (\wt{\u}_2 - \u_2^*)\trans }_0 + \norm{ {\u}_2\strans}_0 \\
	& \leq \norm{ \wt{\u}_2 - \u_2^* }_0 + \norm{{\u}_2^*}_0 \leq  3(r^* + s_u + s_v),
\end{align*}
where the last step above is due to Definition \ref{lemmsofar}.
This completes the proof for part (a).

We next show part (b), which consists of two main steps. Since Condition \ref{con:nearlyorth} is satisfied, the proof below will exploit the fact that the nonzero eigenvalues  $d^{*2}_{i}$ are at the constant level.

\medskip

\noindent\textbf{(1). The upper bound on $\max_{1\leq i\leq p}\norm{\w_i^*}_2$}. Let us recall that 
$$ \w_i\strans = \widehat{\btheta}_i\trans \{\I_p + (z_{11}^{*} - z_{22}^{*})^{-1}\wh{\bSigma}{\u}_2^{*}{\u}_2\strans\}.$$
Under Condition \ref{con3}, it follows from part (b) of  Lemma \ref{lemmauv} and $\norm{{\u}_2^{*}}_2 = d_2^{*} \leq c$ that 
\begin{align}
	\norm{\wh{\bSigma}{\u}_2^{*}{\u}_2\strans}_2 \leq 	\norm{\wh{\bSigma}{\u}_2^{*}}_2 \norm{{\u}_2\strans}_2 \leq c. \nonumber
\end{align}
Also, under Conditions \ref{con3}--\ref{con:nearlyorth}, Lemma \ref{lemmzzz} gives that $|z_{11}^{*} - z_{22}^{*}| \geq c$.
Then we can obtain that
\begin{align}
	\|(z_{11}^{*} - z_{22}^{*})^{-1}\wh{\bSigma}{\u}_2^{*}{\u}_2\strans\|_2 \leq |z_{11}^{*} - z_{22}^{*}|^{-1}\norm{\wh{\bSigma}{\u}_2^{*}{\u}_2\strans}_2 \leq c. \nonumber
\end{align}
Together with Definition \ref{defi2:acceptable} that $\max_{1\leq i\leq p}\|\widehat{\btheta}_i\|_2 \leq c$, it yields that
\begin{align}\label{eq:au3}
	\max_{1\leq i\leq p}\norm{\w_i^*}_2 \leq \max_{1\leq i\leq p}\|\widehat{\btheta}_i\|_2  + \max_{1\leq i\leq p}\|\widehat{\btheta}_i\|_2\| (z_{11}^{*} - z_{22}^{*})^{-1}\wh{\bSigma}{\u}_2^{*}{\u}_2\strans\|_2 \leq c.
\end{align}

\smallskip

\noindent\textbf{(2). The upper bounds on $\max_{1\leq i\leq p}\norm{\wt{\w}_i-\w_i^*}_2$ and $\max_{1\leq i\leq p}\norm{\wt{\w}_i}_2$}. From Definition \ref{defi2:acceptable} that  $\max_{1\leq i\leq p}\|\widehat{\btheta}_i\|_2 \leq c$, we have that 
\begin{align}
	\max_{1\leq i \leq p}\|\wt{\w}_i - \w_i^*\|_2
	& \leq \max_{1\leq i \leq p}\|\wh{\btheta}_i\|_2 \|(\wt{z}_{11} - \wt{z}_{22})^{-1}\wh{\bSigma}\wt{\u}_2\wt{\u}_2\trans - (z_{11}^{*} - z_{22}^{*})^{-1}\wh{\bSigma}{\u}_2^{*}{\u}_2\strans\|_2 \nonumber\\
	&\leq c \|(\wt{z}_{11} - \wt{z}_{22})^{-1}\wh{\bSigma}\wt{\u}_2\wt{\u}_2\trans - (z_{11}^{*} - z_{22}^{*})^{-1}\wh{\bSigma}{\u}_2^{*}{\u}_2\strans\|_2  \nonumber\\
	&\leq c \norm{(\wt{z}_{11} - \wt{z}_{22})^{-1} ( \wh{\bSigma}\wt{\u}_2\wt{\u}_2\trans - \wh{\bSigma}{\u}_2^{*}{\u}_2\strans)}_2  \nonumber\\
 &\quad +  c \norm{[(\wt{z}_{11} - \wt{z}_{22})^{-1} - (z_{11}^{*} - z_{22}^{*})^{-1}]\wh{\bSigma}{\u}_2^{*}{\u}_2\strans  }_2.  \nonumber
\end{align}
We will bound  the two terms introduced above separately. It follows from part (b) of Lemma \ref{lemmauv} and  part (a) of Definition \ref{defi2:acceptable} that 
\begin{align}
	\norm{\wh{\bSigma}{\u}_2^{*}{\u}_2\strans}_2 & \leq 	\norm{\wh{\bSigma}{\u}_2^{*}}_2 \norm{{\u}_2\strans}_2 \leq c, \nonumber\\
	\norm{\wh{\bSigma}\wt{\u}_2\wt{\u}_2\trans - \wh{\bSigma}{\u}_2^{*}{\u}_2\strans}_2 &\leq \norm{\wh{\bSigma} (\wt{\u}_2 - {\u}_2^{*}) {\u}_2\strans }_2 + \norm{\wh{\bSigma}\wt{\u}_2 (\wt{\u}_2 - {\u}_2^{*})\trans  }_2 \nonumber\\
	& \leq  c(r^*+s_u+s_v)^{1/2}\eta_n^2\{n^{-1}\log(pq)\}^{1/2}. \nonumber
\end{align}
Lemma \ref{lemmzzz} implies that $|z_{11}^{*} - z_{22}^{*}| \geq c$ and  $| \wt{z}_{11} - \wt{z}_{22} | \geq c$. Further, it holds that 
\begin{align*}
	& | (\wt{z}_{11} - \wt{z}_{22})^{-1} - (z_{11}^{*} - z_{22}^{*})^{-1} | \\
 & = \left| \frac{1}{z_{11}^{*} - z_{22}^{*}} \cdot \frac{(\wt{z}_{11} - \wt{z}_{22}) - (z_{11}^{*} - z_{22}^{*})}{ (z_{11}^{*} - z_{22}^{*}) + (\wt{z}_{11} - \wt{z}_{22}) - (z_{11}^{*} - z_{22}^{*}) }\right|.
\end{align*}

In view of part (c) of  Lemma \ref{lemmauv},  we have that
\begin{align*}
	|(\wt{z}_{11} - \wt{z}_{22}) - (z_{11}^{*} - z_{22}^{*})| &\leq |\wt{z}_{11} - z_{11}^{*}| + |\wt{z}_{22} - z_{22}^{*}|\\ &\leq  c(r^*+s_u+s_v)^{1/2}\eta_n^2\{n^{-1}\log(pq)\}^{1/2}.
\end{align*}
Together with  $|z_{11}^{*} - z_{22}^{*}| \geq c$, for sufficiently large $n$ it holds that
\begin{align}
	| (\wt{z}_{11} - \wt{z}_{22})^{-1} - (z_{11}^{*} - z_{22}^{*})^{-1} |& = \Big| \frac{(\wt{z}_{11} -\wt{z}_{22}) - (z_{11}^{*} - z_{22}^{*})}{ (z_{11}^{*} - z_{22}^{*})^2 + o((z_{11}^{*} - z_{22}^{*})^2)} \Big| \nonumber\\[5pt]
	& \leq c(r^*+s_u+s_v)^{1/2}\eta_n^2\{n^{-1}\log(pq)\}^{1/2}. \nonumber
\end{align}
Combining the above results gives that
\begin{align}\label{eqwwd}
	\max_{1\leq i \leq p}\|\wt{\w}_i - \w_i^*\|_2
	\leq  c(r^*+s_u+s_v)^{1/2}\eta_n^2\{n^{-1}\log(pq)\}^{1/2}.
\end{align}
Therefore, using \eqref{eq:au3}, \eqref{eqwwd}, and the triangle inequality, we can obtain that for all sufficiently large $n$, 
\begin{align}
	\max_{1\leq i\leq p}\norm{ \wt{\w}_i}_2
	\leq 	\max_{1\leq i\leq p}\norm{ \w_i^* }_2 +	\max_{1\leq i\leq p}\norm{ \wt{\w}_i - \w_i^* }_2  \leq c, \nonumber
\end{align}
which concludes the proof of Lemma \ref{lemma:1rk4}.

\subsection{Lemma \ref{lemma:1rk2} and its proof} \label{new.Sec.B.6}

\begin{lemma}\label{lemma:1rk2}
	Assume that all the conditions of Theorem \ref{theorkr} are satisfied. For $\wt{\bdelta}_1$ defined in \eqref{deltaeq1}  and an arbitrary
	$\a\in\R^p$, with probability at least
	$1- \theta_{n,p,q}$, we have that 
	\begin{align*}
		|\a\trans{\wt{\W}_1} \wt{\bdelta}_1 | \leq c \norm{\a}_0^{1/2}\norm{\a}_2 (r^*+s_u+s_v)\eta_n^4\left\{n^{-1}\log(pq)\right\},
	\end{align*}
	where $\theta_{n,p,q}$ is given in \eqref{thetapro} and $c$ is some positive constant.
\end{lemma}

\noindent \textit{Proof}. 
Notice that
\begin{align}
	|\a\trans{\wt{\W}_1} \wt{\bdelta}_1 | &= | \a\trans {\wt{\W}_1}\wt{z}_{11}^{-1}\wh{\bSigma}\wt{\u}_2\wt{\v}_2\trans\left\{(\wt{\v}_1 - \v_1^*)\wt{\u}_1\trans - (\wt{\v}_2\wt{\u}_2\trans - \v_2^*\u_2\strans)\right\}  \wh{\bSigma}(\wt{\u}_1 - \u_1^*)  | \nonumber\\
	& \leq
	| \a\trans {\wt{\W}_1}\wt{z}_{11}^{-1}\wh{\bSigma}\wt{\u}_2\wt{\v}_2\trans(\wt{\v}_1 - \v_1^*) |
	| \wt{\u}_1\trans  \wh{\bSigma}(\wt{\u}_1 - \u_1^*)  | \nonumber \\
	&\quad + | \a\trans {\wt{\W}_1}\wt{z}_{11}^{-1}\wh{\bSigma}\wt{\u}_2\wt{\v}_2\trans  (\wt{\v}_2\wt{\u}_2\trans - \v_2^*\u_2\strans) \wh{\bSigma}(\wt{\u}_1 - \u_1^*)  |. \nonumber
\end{align}
We aim to bound the two terms introduced above under Condition \ref{con:nearlyorth} that the nonzero eigenvalues $d^{*2}_{i}$ are at the constant level. For the first term above, it follows from Conditions \ref{con3}--\ref{con:nearlyorth} that
\begin{align}
	&| \a\trans {\wt{\W}_1}\wt{z}_{11}^{-1}\wh{\bSigma}\wt{\u}_2\wt{\v}_2\trans(\wt{\v}_1 - \v_1^*) |
	| \wt{\u}_1\trans  \wh{\bSigma}(\wt{\u}_1 - \u_1^*)  | \nonumber \\[5pt]
	&\leq |\wt{z}_{11}^{-1}| \norm{ \a\trans {\wt{\W}_1}}_2\norm{\wh{\bSigma}\wt{\u}_2}_2\norm{\wt{\v}_2}_2\norm{\wt{\v}_1-\v_1^*}_2 \norm{\wt{\u}_1}_2  \norm{\wh{\bSigma}(\wt{\u}_1-\u_1^*)}_2 \nonumber \\[5pt]
	&\leq c \norm{\a}_0^{1/2}\norm{\a}_2  (r^*+s_u+s_v)\eta_n^4 \left\{n^{-1}\log(pq)\right\}, \label{sczaq}
\end{align}
where we have used  Definition \ref{lemmsofar} with  $\norm{\wt{\v}_2}_2 = 1, \norm{\wt{\u}_2}_2 \leq c$, parts (a)--(c) of Lemma \ref{lemmauv}, and part (c) of Lemma \ref{lemma:1rk4}.

For the second term above, let us first bound $	\| \wt{\u}_2\wt{\v}_2\trans -\u_2^*\v_2\strans\|_2$.
In light of part (a) of Definition \ref{lemmsofar} and part (a) of Lemma \ref{lemmauv}, we can deduce that
\begin{align}
	\| \wt{\u}_2\wt{\v}_2\trans -\u_2^*\v_2\strans\|_2 &\leq 	 \| (\wt{\u}_2- \u_2^*)\v_2\strans \|_2+\|\wt{\u}_2(\wt{\v}_2- \v_2^*)\trans \|_2 \nonumber\\[5pt]	
	&\leq  \|\wt{\u}_2- \u_2^*\|_2 \|\v_2^* \|_2 +  \|\wt{\u}_2\|_2 \|\wt{\v}_2- \v_2^*\|_2 \nonumber\\[5pt]
	&\leq c(r^*+s_u+s_v)^{1/2}\eta_n^2\left\{n^{-1}\log(pq)\right\}^{1/2}. \nonumber
\end{align}
Then similar to \eqref{sczaq}, it holds that
\begin{align*}
	&| \a\trans {\wt{\W}_1}\wt{z}_{11}^{-1}\wh{\bSigma}\wt{\u}_2\wt{\v}_2\trans  (\wt{\v}_2\wt{\u}_2\trans - \v_2^*\u_2\strans) \wh{\bSigma}(\wt{\u}_1 - \u_1^*)  | \\[5pt]
	&\leq |\wt{z}_{11}^{-1}| \norm{ \a\trans {\wt{\W}_1}}_2\norm{\wh{\bSigma}\wt{\u}_2}_2\norm{\wt{\v}_2}_2\| \wt{\u}_2\wt{\v}_2\trans -\u_2^*\v_2\strans\|_2  \norm{\wh{\bSigma}(\wt{\u}_1-\u_1^*)}_2 \\[5pt]
	&\leq c   \norm{\a}_0^{1/2}\norm{\a}_2(r^*+s_u+s_v)\eta_n^4\left\{n^{-1}\log(pq)\right\}.
\end{align*}
Thus, combining the above results yields that 
\begin{align}
	| \a\trans {\wt{\W}_1} \wt{\bdelta}_1 | \leq c   \norm{\a}_0^{1/2}\norm{\a}_2(r^*+s_u+s_v)\eta_n^4\left\{n^{-1}\log(pq)\right\}, \nonumber
\end{align}
which completes the proof of Lemma \ref{lemma:1rk2}.

\subsection{Lemma \ref{lemma:gap} and its proof} \label{new.Sec.B.7}

\begin{lemma}\label{lemma:gap}
	Assume that all the conditions of Theorem \ref{theorkr} are satisfied. For $\wt{\M}_1 = -\wt{z}_{11}^{-1}\wh{\bSigma}\widetilde{\u}_2\widetilde{\v}_2\trans$, $\wt{\W}_1 =  \widehat{\bTheta}\{\I_p + (\wt{z}_{11} - \wt{z}_{22})^{-1}\wh{\bSigma}\wt{\u}_2\wt{\u}_2\trans\}$, and an arbitrary
	$\a\in\R^p$, with probability at least
	$	1- \theta_{n,p,q}$ with $\theta_{n,p,q}$ given in \eqref{thetapro}, it holds that
	\begin{align*}
		|\a\trans\wt{\W}_1\wt{\M}_1\v_2^*\u_2\strans\wh{\bSigma}\u_1^*  | = o(\norm{\a}_0^{1/2}\norm{\a}_2 n^{-1/2}).
	\end{align*}
\end{lemma}

\noindent \textit{Proof}. 
	Recall that $\wt{\M}_1 = -\wt{z}_{11}^{-1}\wh{\bSigma}\widetilde{\u}_2\widetilde{\v}_2\trans$. Under Condition \ref{con:nearlyorth} that $d^{*2}_{i}$ are at the constant level, parts (b) and (c) of Lemma \ref{lemmauv} show that  $\norm{\wh{\bSigma}\widetilde{\u}_2 }_2 \leq  c$ and $|\wt{z}_{11}^{-1}| \leq c$. Since $\norm{\widetilde{\v}_2 }_2 = 1$ due to Definition \ref{lemmsofar},
	 we can obtain that 
	\begin{align}
		\norm{\wt{\M}_1}_2 \leq \norm{\wt{z}_{11}^{-1}\wh{\bSigma}\widetilde{\u}_2\widetilde{\v}_2\trans}_2 \leq |\wt{z}_{11}^{-1}| \norm{\wh{\bSigma}\widetilde{\u}_2 }_2 \norm{\widetilde{\v}_2 }_2  \leq c.
	\end{align}
 It further holds that 
	\begin{align}\label{eq4.2.3}
		 |\a\trans\wt{\W}_1 \wt{\M}_1 \v_2^* \u_2\strans\wh{\bSigma}\u_1^*| 
		&\leq \norm{\a\trans\wt{\W}_1}_2 \norm{\wt{\M}_1}_2 \norm{\v_2^*}_2 |\u_2\strans\wh{\bSigma}\u_1^*|   \nonumber\\
		&\leq c \norm{\a}_0^{1/2}\norm{\a}_2  |\l_2\strans\wh{\bSigma}\l_1^*|,
	\end{align}
	where we have used $\norm{\a\trans\wt{\W}_1}_2 \leq c \norm{\a}_0^{1/2}\norm{\a}_2$ in Lemma \ref{lemma:1rk4},
	 $\norm{\v_2^*}_2 =1 $, and $ |\u_2\strans\wh{\bSigma}\u_1^*| \leq c |\l_2\strans\wh{\bSigma}\l_1^*| $. Therefore, under Condition \ref{con:nearlyorth} we have that 
	 \begin{align*}
		|\a\trans\wt{\W}_1\wt{\M}_1\v_2^* \u_2\strans\wh{\bSigma}\u_1^*  | = o(\norm{\a}_0^{1/2}\norm{\a}_2 n^{-1/2}),
	\end{align*}
	which concludes the proof of Lemma \ref{lemma:gap}.

\subsection{Lemma \ref{lemma:1rk3} and its proof} \label{new.Sec.B.8}

\begin{lemma}\label{lemma:1rk3}
	Assume that all the conditions of Theorem \ref{theorkr} are satisfied.  For $\wt{\bepsilon}_1$ defined in \eqref{epeq1}, $h_1$ defined in \eqref{eqh1}, and any $\a\in\mathcal{A}=\{\a\in\R^p:\norm{\a}_0\leq m,\norm{\a}_2=1\}$,
	with probability at least
	$	1- \theta_{n,p,q}$ it holds that
	\begin{align*}
		\abs{-\a\trans{\wt{\W}_1}\wt{\bepsilon}_1 - h_1 / \sqrt{n}} \leq c m^{1/2}(r^*+s_u + s_v)^{3/2}\eta_n^2\{ n^{-1}\log(pq)\},
	\end{align*}
	where $\theta_{n,p,q}$ is given in \eqref{thetapro} and $c$ is some positive constant.
\end{lemma}

\noindent \textit{Proof}. 
Observe that
\begin{align}
	&\abs{-\a\trans\wt{\W}_1\wt{\bepsilon}_1 - h_1 / \sqrt{n}} \nonumber \\
	&\leq n^{-1}|\a\trans \wt{\W}_1 \wt{\M}_1 \E\trans\X\wt{\u}_1 - \a\trans {\W}_1^{*}\M_1^{*} \E\trans\X {\u}_1^{*} |   + n^{-1} \abs{\a\trans(\wt{\W}_1-{\W}_1^*)\X\trans\E\v_1^*} \nonumber\\
	&\leq  n^{-1}|\a\trans \wt{\W}_1 \wt{\M}_1 \E\trans\X (\wt{\u}_1 - {\u}_1^{*} )| +  n^{-1} |(\a\trans \wt{\W}_1 \wt{\M}_1  - \a\trans {\W}_1^{*}\M_1^{*}) \E\trans\X {\u}_1^{*} | \nonumber\\
	&\quad  + n^{-1} \abs{\a\trans(\wt{\W}_1-{\W}_1^*)\X\trans\E\v_1^*}.
	\label{sdwqvcs}
\end{align}
We aim to bound the three terms introduced above separately. Let us first show that 
$\a\trans \wt{\W}_1 \wt{\M}_1$, $\a\trans \wt{\W}_1 \wt{\M}_1  - \a\trans {\W}_1^{*}\M_1^{*}$, and  $\a\trans(\wt{\W}_1-{\W}_1^*)$ are all $s$-sparse with $s=c(r^*+s_u+s_v)$. 
Recall that $\wt{\M}_1 = -\wt{z}_{11}^{-1}\wh{\bSigma}\widetilde{\u}_2\widetilde{\v}_2\trans $ and $\M_1^* = -z_{11}^{*-1}\wh{\bSigma}{\u}_2^{*}{\v}_2\strans$. It follows from part (b) of Lemma \ref{lemmauv} and $\norm{ {\v}_2^* }_0 \leq s_v$ that 
\begin{align}
	\norm{\a\trans \wt{\W}_1 \wt{\M}_1}_0 & = \norm{ (\wt{z}_{11}^{-1}\a\trans\wt{\W}_1\wh{\bSigma}\wt{\l}_2) \cdot \wt{d}_2\wt{\v}_2\trans   }_0 \nonumber\\
	& \leq  \norm{ \wt{d}_2 \wt{\v}_2   }_0 \leq  \norm{ \v_2^* }_0  + \norm{ \wt{d}_2(\wt{\v}_2 - \v_2^* )  }_0 \nonumber\\
 &\leq  c(r^* + s_u + s_v) \label{sawm1}
\end{align}
and
\begin{align} 
 \norm{\a\trans \wt{\W}_1 \wt{\M}_1  - \a\trans {\W}_1^{*}\M_1^{*}}_0 &\leq  \norm{ (\wt{z}_{11}^{-1}\a\trans\wt{\W}_1\wh{\bSigma}\wt{\l}_2) \cdot \wt{d}_2\wt{\v}_2\trans   }_0 + \norm{ ({z}_{11}^{*-1}\a\trans{\W}_1^* \wh{\bSigma}{\u}_2^{*}) \cdot {\v}_2\strans   }_0 \nonumber \\
	& \leq \norm{ \wt{d}_2 \wt{\v}_2   }_0 + \norm{ \v_2^* }_0 \leq  c(r^* + s_u + s_v). \label{sawm2}
\end{align}

Further, from the definitions of $\wt{\W}_1$ and ${\W}_1^*$, we have that 
\begin{align}
	\|\a\trans(\wt{\W}_1-{\W}_1^*)\|_0 & = \|( \a\trans\widehat{\bTheta}(\wt{z}_{11} - \wt{z}_{22})^{-1}\wh{\bSigma}\wt{\u}_2) \cdot \wt{\u}_2\trans  - ( \a\trans\widehat{\bTheta} (z_{11}^{*} - z_{22}^{*})^{-1}\wh{\bSigma}{\u}_2^{*}) \cdot {\u}_2\strans \|_0   \nonumber\\
	&\leq \norm{\wt{\u}_2\trans }_0 + \norm{ {\u}_2\strans }_0 
	\leq  2 \norm{ {\u}_2^* }_0  + \norm{\wt{\u}_2 - {\u}_2^* }_0 \nonumber \\ 
	&\leq c(r^*+s_u+s_v), \label{saaw2}
\end{align}
where the last step above is due to Definition \ref{lemmsofar} and $\norm{ {\u}_2^* }_0 \leq s_u$.
Hence,  combining \eqref{sdwqvcs}--\eqref{saaw2} leads to
\begin{align}
	&\abs{-\a\trans\wt{\W}_1\wt{\bepsilon}_1 - h_1 / \sqrt{n}} \leq  \norm{\a\trans \wt{\W}_1}_2 \norm{\wt{\M}_1}_2 \norm{n^{-1}\E\trans\X (\wt{\u}_1 - {\u}_1^{*} )}_{2,s} \nonumber \\
	&\quad + \norm{\a\trans \wt{\W}_1 \wt{\M}_1  - \a\trans {\W}_1^{*}\M_1^{*}}_2 \norm{ n^{-1}\E\trans\X\u_1^*}_{2,s} \nonumber\\
 & \quad+  \norm{\a\trans(\wt{\W}_1-{\W}_1^*) }_2\norm{ n^{-1}\X\trans\E\v_1^*}_{2,s} \nonumber \\
	&=: A_1 + A_2 + A_3. \label{sdzvcqqqq}
\end{align}
We will provide the upper bounds for the three terms $A_1$, $A_2$, and $A_3$ introduced in (\ref{sdzvcqqqq}) above separately. 

We start with bounding $n^{-1}\norm{\E\trans\X(\wt{\u}_1 - {\u}_1^{*} )}_{2,s}$, $n^{-1}\norm{\E\trans\X\u_1^*}_{2,s}$, and $n^{-1}\norm{\X\trans\E\v_1^*}_{2,s}$. From $n^{-1}\norm{\X\trans\E}_{\max} \leq c\{n^{-1}\log(pq)\}^{1/2} $ and Definition \ref{lemmsofar}, we can deduce that 
\begin{align*}
	&n^{-1}\norm{\E\trans\X(\wt{\u}_1 - {\u}_1^{*} )}_{\max} \leq n^{-1}\norm{\E\trans\X}_{\max}\norm{\wt{\u}_1 - {\u}_1^{*} }_1 \\& \qquad \leq  n^{-1}\norm{\E\trans\X}_{\max}\norm{\wt{\u}_1 - {\u}_1^{*} }_0^{1/2} \norm{\wt{\u}_1 - {\u}_1^{*} }_2 	\leq c(r^*+s_u+s_v)\eta_n^2\{n^{-1}\log(pq)\}, \\
	&n^{-1}\norm{\E\trans\X\u_1^*}_{\max} \leq n^{-1}\norm{\E\trans\X}_{\max}\norm{\u_1^*}_0^{1/2}\norm{\u_1^*}_2 \leq c s_u^{1/2} \{n^{-1}\log(pq)\}^{1/2} d_1^{*}, \\
	&	n^{-1}\norm{\X\trans\E\v_1^*}_{\max}
	\leq n^{-1}\norm{\X\trans\E}_{\max}\norm{\v_1^*}_{0}^{1/2}\norm{\v_1^*}_{2} \leq cs_v^{1/2}\{n^{-1}\log(pq)\}^{1/2}.
\end{align*}
Then it follows that
\begin{align}
	&n^{-1}\norm{\E\trans\X(\wt{\u}_1 - {\u}_1^{*} )}_{2,s} \leq s^{1/2} n^{-1}\norm{\E\trans\X(\wt{\u}_1 - {\u}_1^{*} )}_{\max}
	\leq c s^{3/2}\eta_n^2\{n^{-1}\log(pq)\}, \label{a22ex}\\
	&n^{-1}\norm{\E\trans\X\u_1^*}_{2,s} \leq cs^{1/2}s_u^{1/2}\{n^{-1}\log(pq)\}^{1/2}  d_1^{*}, \label{eqwdaz}\\
	&n^{-1}\norm{\X\trans\E\v_1^*}_{2,s} \leq cs^{1/2}s_v^{1/2}\{n^{-1}\log(pq)\}^{1/2}. \label{eqathe}
\end{align}

Using part (c) of Lemma \ref{lemma:1rk4}, $ \norm{\a}_0 = m $, and $ \norm{\a}_2 = 1 $, we can show that 
\begin{align}
	&\norm{\a\trans \wt{\W}_1}_2 \leq c m^{1/2}, \label{awawaw} \\
	&\norm{\a\trans(\wt{\W}_1-{\W}_1^*) }_2 \leq c m^{1/2} (r^*+s_u+s_v)^{1/2}\eta_n^2\{n^{-1}\log(pq)\}^{1/2}.  \label{awawaw2}
\end{align}
Further, for  $\wt{\M}_1 = -\wt{z}_{11}^{-1}\wh{\bSigma}\widetilde{\u}_2\widetilde{\v}_2\trans $ and $\M_1^* = -z_{11}^{*-1}\wh{\bSigma}{\u}_2^{*}{\v}_2\strans$, it holds that 
\begin{align}
	&\norm{\wt{\M}_1}_2 \leq \norm{\wt{z}_{11}^{-1}\wh{\bSigma}\widetilde{\u}_2\widetilde{\v}_2\trans}_2 \leq |\wt{z}_{11}^{-1}| \norm{\wh{\bSigma}\widetilde{\u}_2 }_2 \norm{\widetilde{\v}_2 }_2  \leq cd_1^{*-2}  d_2^*, \label{eq:mm1} \\
	&\norm{\M_1^*}_2 \leq \norm{z_{11}^{*-1}\wh{\bSigma}{\u}_2^*{\v}_2\strans}_2 \leq |z_{11}^{*-1}| \norm{\wh{\bSigma}{\u}_2^* }_2 \norm{{\v}_2^* }_2  \leq cd_1^{*-2} d_2^*, \label{m1m1r2}
\end{align}
where we have used the results in parts (c) and (d) of Lemma \ref{lemmauv}. Hence,  combining \eqref{a22ex},  \eqref{awawaw}, and \eqref{eq:mm1}, for term $A_1$ above we can obtain that 
\begin{align}
	A_1 &= \norm{\a\trans \wt{\W}_1}_2 \norm{\wt{\M}_1}_2 \norm{n^{-1}\E\trans\X (\wt{\u}_1 - {\u}_1^{*} )}_{2,s} \nonumber \\
	&\leq  c m^{1/2} (r^*+s_u+s_v)^{3/2}\eta_n^2\{n^{-1}\log(pq)\} {d^{*}_2d_1^{*-2}}. \label{daszxc}
\end{align}
With the aid of \eqref{eqathe} and \eqref{awawaw2}, it also holds that 
\begin{align}
	A_3 &= \norm{\a\trans(\wt{\W}_1-{\W}_1^*) }_2\norm{ n^{-1}\X\trans\E\v_1^*}_{2,s} \nonumber\\[5pt]
	&\leq  c m^{1/2} (r^*+s_u+s_v)^{3/2}\eta_n^2\{n^{-1}\log(pq)\}. \label{daszxc3}
\end{align}

It remains to bound term $A_2$ above. From Lemma \ref{lemmauv} and  $\norm{{\v}_2^*}_2 = 1$, we see that 
\begin{align*}
	\norm{\wh{\bSigma}{\u}_2^{*}{\v}_2\strans}_2 & \leq \norm{\wh{\bSigma}{\u}_2^{*}}_2 \norm{{\v}_2\strans}_2   \leq c d_2^{*}, \\[5pt]
	\norm{\wh{\bSigma}\wt{\u}_2\wt{\v}_2\trans -\wh{\bSigma}{\u}_2^{*}{\v}_2\strans}_2 & \leq \|\wh{\bSigma}(\wt{\u}_2- {\u}_2^{*})\|_2 \|{\v}_2\strans \|_2 +\| \wh{\bSigma}\wt{\u}_2\|_2 \|(\wt{\v}_2-{\v}_2^{*})\trans \|_2 \nonumber\\[5pt]
	& \leq  c(r^*+s_u+s_v)^{1/2} \eta_n^2  \{n^{-1}\log(pq)\}^{1/2}.
\end{align*}
Together with the upper bounds for $|\wt{z}_{11}^{-1}|$ and $ |\wt{z}_{11}^{-1}  - z_{11}^{*-1}|$ in  Lemma \ref{lemmauv}, it holds that 
\begin{align}
	\norm{\wt{\M}_1  - \M_1^{*}}_2 &= \norm{ \wt{z}_{11}^{-1}\wh{\bSigma}\widetilde{\u}_2\widetilde{\v}_2\trans  - z_{11}^{*-1}\wh{\bSigma}{\u}_2^*{\v}_2\strans}_2 \nonumber \\
	&\leq  |\wt{z}_{11}^{-1}| \norm{ \wh{\bSigma}\widetilde{\u}_2\widetilde{\v}_2\trans  - \wh{\bSigma}{\u}_2^*{\v}_2\strans}_2 + | \wt{z}_{11}^{-1}  - z_{11}^{*-1}|  \norm{\wh{\bSigma}{\u}_2^*{\v}_2\strans}_2 \nonumber \\
	&\leq c(r^*+s_u+s_v)^{1/2} \eta_n^2  \{n^{-1}\log(pq)\}^{1/2}d_1^{*-2}, \label{m1m2m1m2}
\end{align}
Then a combination of \eqref{awawaw}, \eqref{awawaw2}, \eqref{m1m1r2}, and \eqref{m1m2m1m2} results in 
\begin{align}
	&\norm{\a\trans \wt{\W}_1 \wt{\M}_1  - \a\trans {\W}_1^{*}\M_1^{*}}_2 \leq \norm{\a\trans \wt{\W}_1 (\wt{\M}_1  - \M_1^{*})}_2 + \norm{(\a\trans \wt{\W}_1   - \a\trans {\W}_1^{*})\M_1^{*}}_2 \nonumber\\
	&\leq  \norm{\a\trans \wt{\W}_1 }_2 \norm{\wt{\M}_1  - \M_1^{*}}_2 + \norm{\a\trans( \wt{\W}_1   - {\W}_1^{*})}_2 \norm{\M_1^{*}}_2 \nonumber \\
	&\leq    cm^{1/2} (r^*+s_u+s_v)^{1/2} \eta_n^2  \{n^{-1}\log(pq)\}^{1/2}d_1^{*-2}. \label{awm1m1}
\end{align}

With the aid of \eqref{eqwdaz} and \eqref{awm1m1}, we can deduce that 
\begin{align}
	A_2 &=\norm{\a\trans \wt{\W}_1 \wt{\M}_1  - \a\trans {\W}_1^{*}\M_1^{*}}_2 \norm{ n^{-1}\E\trans\X\u_1^*}_{2,s}\nonumber\\
	& \leq  c m^{1/2} (r^*+s_u+s_v)^{3/2}\eta_n^2\{n^{-1}\log(pq)\} { d_1^{*-1}}. \label{daszxc2}
\end{align}
Therefore, combining \eqref{sdzvcqqqq}, \eqref{daszxc}, \eqref{daszxc2}, and \eqref{daszxc3} yields that
\begin{align}\label{eq4.2.5}
	\abs{-\a\trans\wt{\W}_1\wt{\bepsilon}_1 - h_1 / \sqrt{n}} \leq c m^{1/2}(r^*+s_u + s_v)^{3/2}\eta_n^2\{ n^{-1}\log(pq)\} {d^{*-1}_1}.
\end{align}
Furthermore, under Condition \ref{con:nearlyorth} that $d_1^* $ is at the constant level, we have that 
\begin{align*}
	\abs{-\a\trans\wt{\W}_1\wt{\bepsilon}_1 - h_1 / \sqrt{n}} \leq c m^{1/2}(r^*+s_u + s_v)^{3/2}\eta_n^2\{ n^{-1}\log(pq)\},
\end{align*}
which completes the proof of Lemma \ref{lemma:1rk3}.

\subsection{Lemma \ref{lemm:wexist} and its proof} \label{new.Sec.B.9}

\begin{lemma}\label{lemm:wexist}
	Assume that all the conditions of Theorem \ref{theorkr} are satisfied. Then for each given $k$ with $1 \leq k \leq r^*$, with probability at least
	$	1- \theta_{n,p,q}$ with $\theta_{n,p,q}$ given in \eqref{thetapro}, both $\I_{r^*-1} -\wt{z}_{kk}^{-1}\wt{\U}_{-k}\trans\wh{\bSigma} \wt{\U}_{-k}$ and  $\I_{r^*-1} -z_{kk}^{*-1}\U_{-k}\strans\wh{\bSigma} \U_{-k}^*$ are nonsingular. Moreover, $\wt{\W}_k $ and $\W_k^{*}$ introduced in \eqref{eqwknear1} and \eqref{eqwknear2}, respectively, are well-defined.
\end{lemma}

\noindent \textit{Proof}. We will first analyze matrix $\I_{r^*-1} -\wt{z}_{kk}^{-1}\wt{\U}_{-k}\trans\wh{\bSigma} \wt{\U}_{-k}$,
 which is equivalent to analyzing the nonsingularity of matrix $\A =: \wt{z}_{kk}\I_{r^*-1} -\wt{\U}_{-k}\trans\wh{\bSigma} \wt{\U}_{-k}$.
 For simplicity,
 denote by $\A = (a_{ij}) \in \mathbb{R}^{(r^*-1) \times (r^*-1)}$ with $i, j \in \mathcal{A} = \{1 \leq \ell \leq r^*: \ell \neq k\}$. It can be seen that for each $i, j \in \mathcal{A}$, 
\begin{align}\label{a0aa}
	a_{ij} = \left\{\begin{array}{l}
		\wt{z}_{kk} - \wt{z}_{ii} \quad \text{if} \ i=j, \\
		-\wt{z}_{ij} \quad \text{if} \ i \neq j.
		\end{array}\right.
\end{align} 
From Lemma \ref{lemmzzz}, 
we have that  
$ \sum_{j \in  \mathcal{A}, \, j \neq i} |a_{ij}| = o(|a_{\ell \ell}|)$ for any $i, \ell \in \mathcal{A}$. Then it holds that
\[ |a_{ii}| > \sum_{j \in  \mathcal{A}, \, j \neq i}|a_{ij}| \ \text{ for all } i \in  \mathcal{A},   \]
which shows that $\A$ is strictly diagonally dominant.
Using the Levy--Desplanques Theorem  in \cite{horn2012matrix}, we see that matrix $\A$  is nonsingular, which entails that $\I_{r^*-1} -\wt{z}_{kk}^{-1}\wt{\U}_{-k}\trans\wh{\bSigma} \wt{\U}_{-k}$ is nonsingular. Moreover, with similar arguments we can also show that $\I_{r^*-1} -z_{kk}^{*-1}\U_{-k}\strans\wh{\bSigma} \U_{-k}^*$ is strictly diagonally dominant and thus is nonsingular.
Therefore, we see that both $\wt{\W}_k $ and $\W_k^{*}$ are
well-defined and satisfy the property in Proposition \ref{prop:rankr3},
which concludes the proof of Lemma  \ref{lemm:wexist}.

\subsection{Lemma \ref{rankr:boundm0} and its proof} \label{new.Sec.B.10}

\begin{lemma}\label{rankr:boundm0}
	Assume that all the conditions of Theorem \ref{theorkr} are satisfied. For each given $k$ with $1 \leq k \leq r^*$, 
	with probability at least
	$	1- \theta_{n,p,q}$ with $\theta_{n,p,q}$ given in \eqref{thetapro}, it holds that 
	\begin{align*}
		&\|\U^*_{-k}\|_2 \leq cd_1^*, \ \|\wh{\bSigma} \U^*_{-k}\|_2 \leq cd_1^*, \ \| \wt{\U}_{-k}\|_2 \leq cd_1^*, \ \|\wh{\bSigma} \wt{\U}_{-k}\|_2 \leq cd_1^*, \\[5pt] 
		&\|\wh{\bSigma} (\wt{\U}_{-k} -  \U^*_{-k})\|_2 \leq c\gamma_n, \ \norm{ \wh{\bSigma}(\wt{\C}_{-k}  - \C_{-k}^*)}_2 
		\leq  c \gamma_n, \ \ 	\norm{ \wt{\C}_{-k}  - \C_{-k}^*}_2 
		\leq  c \gamma_n.
	\end{align*}
Moreover, for $\wt{\M}_k = -\wt{z}_{kk}^{-1}\wh{\bSigma}\wt{\C}_{-k}$ and ${\M}_k^* = -{z}_{kk}^{*-1}\wh{\bSigma}{\C}_{-k}^*$, we have that 
	\begin{align*}
		&	\norm{ {\M}_k^* }_2  \leq c d_k^{*-2} d_1^* , \ \norm{ \wt{\M}_k }_2 \leq c d_k^{*-2} d_1^*, \ \ \|\wt{\M}_{k}  - {\M}_{k}^*  \|_2 
		\leq  c \gamma_n d_k^{*-3}d_1^*,
	\end{align*}
	where  $c$ is some positive constant.
\end{lemma}

\noindent \textit{Proof}. 
Let us first bound terms $\|  \wh{\bSigma} {\U}_{-k}^*\|_2$ and $\|  \wh{\bSigma} \wt{\U}_{-k}\|_2$. 
	By definition, it holds that $\norm{{\U}_{-k}^*}_0 \leq   \norm{ \U^*}_0 = s_u $. For any vector $\x \in \mathbb{R}^{r^*-1}$, we see that $\norm{\U^*_{-k}\x}_0 \leq s_u$. It follows from the definition of the induced 2-norm and Condition \ref{con3} that 
\begin{align}\label{uuszfaa22}
	\norm{\wh{\bSigma}\U^*_{-k}}_2 = \sup_{\x\trans\x = 1} \norm{\wh{\bSigma}\U^*_{-k}\x}_2 \leq c \sup_{\x\trans\x = 1}\norm{ \U^*_{-k}\x}_2 \leq c \norm{\U^*_{-k}}_2.
\end{align}
Since $\U_{-k}\strans\U^*_{-k} = \D_{-k}^{*2}$ with $\D_{-k}^{*2} = \diag{d_1^{*2},\ldots, d_{k-1}^{*2}, d_{k+1}^{*2}, \ldots, d_{r^*}^{*2}}$, we can show that 
\begin{align}\label{czasdqz}
	\sup_{\x\trans\x = 1} \norm{ \U^*_{-k}\x}_2^2 = \sup_{\x\trans\x = 1} \x\trans\U_{-k}\strans\U^*_{-k}\x = \sup_{\x\trans\x = 1}\x\trans\D_{-k}^{*2}\x \leq d_1^{*2}, 
\end{align} 
which leads to $\norm{ \U^*_{-k}}_2 \leq cd_1^{*}.$ It also implies that  $\norm{\wh{\bSigma}\U^*_{-k}}_2 \leq cd_1^{*}$. 

In view of Definition \ref{lemmsofar}, we have that 
\begin{align*}
&\norm{\wt{\U}_{-k} - {\U}_{-k}^*}_0 \leq  \norm{\wt{\U} - {\U}^*}_0 \leq 2(r^* + s_u + s_v), \\[5pt]
&\norm{\wt{\U}_{-k} - {\U}_{-k}^*}_2 \leq \norm{\wt{\U}_{-k} - {\U}_{-k}^*}_F \leq \norm{\wt{\U} - {\U}^*}_F \leq  c\gamma_n.
\end{align*}
Then using similar arguments as for \eqref{uuszfaa22}, we can deduce that 
\[ \norm{\wh{\bSigma}(\wt{\U}_{-k} - {\U}_{-k}^*)}_2 \leq c\norm{\wt{\U}_{-k} - {\U}_{-k}^*}_2 \leq c\gamma_n.  \]
Hence, for sufficiently large $n$ it holds that
\begin{align} 
	&\norm{ \wt{\U}_{-k} }_2 \leq \norm{\wt{\U}_{-k} - {\U}_{-k}^*}_2 +\norm{ {\U}^*_{-k}}_2 \leq c d_1^*, \nonumber \\
	&\norm{ \wh{\bSigma}\wt{\U}_{-k}}_2 \leq \norm{\wh{\bSigma}(\wt{\U}_{-k} - {\U}_{-k}^*)}_2 +\norm{ \wh{\bSigma}{\U}^*_{-k}}_2 \leq c d_1^*. \nonumber
\end{align}

For term $	\norm{ \wh{\bSigma}(\wt{\C}_{-k}  - \C_{-k}^*)}_2$ above, it follows that 
\begin{align}\label{czxcaz}
	& \norm{ \wh{\bSigma}(\wt{\C}_{-k}   - \C_{-k}^*)}_2 =	\norm{ \wh{\bSigma}(\wt{\U}_{-k} \wt{\V}_{-k}\trans - \U_{-k}^*\V_{-k}\strans)}_2 \nonumber\\[5pt]
	&\leq 	\norm{ \wh{\bSigma}\wt{\U}_{-k} (\wt{\V}_{-k} -\V_{-k}^*)\trans }_2 +
	\norm{ \wh{\bSigma}(\wt{\U}_{-k} - \U_{-k}^*)\V_{-k}\strans}_2  \nonumber \\[5pt]
	&\leq \norm{ \wh{\bSigma}\wt{\L}_{-k}}_2 \norm{\wt{\D}_{-k}(\wt{\V}_{-k} -\V_{-k}^*)\trans }_2 +
	\norm{ \wh{\bSigma}(\wt{\U}_{-k} - \U_{-k}^*)}_2 \norm{\V_{-k}\strans}_2. 
\end{align} 
Note that $ \wt{\L}_{-k}\trans\wt{\L}_{-k} = \I$. An application of similar arguments as for \eqref{uuszfaa22} and \eqref{czasdqz} leads to  
\[   \norm{ \wh{\bSigma}\wt{\L}_{-k}}_2 \leq c. \]
For term 	$\norm{(\wt{\V}_{-k} -\V_{-k}^*)\wt{\D}_{-k} }_2$ above,  we can deduce that 
\begin{align*}
	\norm{(\wt{\V}_{-k} -\V_{-k}^*)\wt{\D}_{-k} }_2 &\leq \norm{\wt{\V}_{-k}\wt{\D}_{-k} -\V_{-k}^*\D_{-k}^* }_2 + \norm{\wt{\V}_{-k}}_2 \norm{\wt{\D}_{-k} -\D_{-k}^* }_2 \\
	&\leq  \norm{\wt{\V}_{-k}\wt{\D}_{-k} -\V^*_{-k}\D^*_{-k} }_F + \norm{\wt{\D}_{-k} -\D_{-k}^* }_F\\
	&\leq \norm{\wt{\V}\wt{\D} -\V^*\D^*}_F + \norm{\wt{\D} -\D^* }_F
	\leq c\gamma_n,
\end{align*}
where we have used Definition \ref{lemmsofar} and $\norm{\wt{\V}_{-k}}_2  = 1$. Along with $\norm{\V_{-k}^*}_2 = 1$ and $\norm{\wh{\bSigma}(\wt{\U}_{-k} - {\U}_{-k}^*)}_2 \leq c\gamma_n,$ it yields that
\begin{align}\label{czdadqzz}
	\norm{ \wh{\bSigma}(\wt{\C}_{-k}  - \C_{-k}^*)}_2 
	\leq  c \gamma_n.
\end{align} 
Further, using similar arguments we can obtain that
\begin{align}
	\norm{ \wt{\C}_{-k}  - \C_{-k}^*}_2 &=	\norm{ \wt{\U}_{-k} \wt{\V}_{-k}\trans - \U_{-k}^*\V_{-k}\strans}_2 \nonumber\\[5pt]
	&\leq 	\norm{ \wt{\U}_{-k} (\wt{\V}_{-k} -\V_{-k}^*)\trans }_2 +
	\norm{(\wt{\U}_{-k} - \U_{-k}^*)\V_{-k}\strans}_2  \nonumber \\[5pt]
	&\leq \norm{ \wt{\L}_{-k}}_2 \norm{\wt{\D}_{-k}(\wt{\V}_{-k} -\V_{-k}^*)\trans }_2 +
	\norm{\wt{\U}_{-k} - \U_{-k}^*}_2 \norm{\V_{-k}\strans}_2. \nonumber \\
	&\leq  c \gamma_n. \label{dzcvdsff}
\end{align} 

Observe that $\wt{\M}_k = -\wt{z}_{kk}^{-1}\wh{\bSigma}\wt{\U}_{-k}\wt{\V}_{-k}\trans$ and ${\M}_k^* = -{z}_{kk}^{*-1}\wh{\bSigma}{\U}_{-k}^*{\V}_{-k}\strans$.
For $\wt{\M}_k = -\wt{z}_{kk}^{-1}\wh{\bSigma}\wt{\C}_{-k}$, it holds that
\begin{align}
	\norm{ \wt{\M}_k }_2 & \leq |\wt{z}_{kk}^{-1}|
	\|  \wh{\bSigma} \wt{\U}_{-k} \wt{\V}_{-k}\trans \|_2 \leq   |\wt{z}_{kk}^{-1}|
	\|  \wh{\bSigma} \wt{\U}_{-k}\|_2 \| \wt{\V}_{-k}\trans \|_2 \nonumber\\
 &\leq c  d_k^{*-2} d_1^*, \nonumber
\end{align}
where we have used part (c) of Lemma \ref{lemmauv}, $\norm{ \wh{\bSigma}\wt{\U}_{-k}}_2 \leq c d_{1}^*$, and $\| \wt{\V}_k \|_2 = 1$. With the aid of similar arguments, we can show that $$\norm{ {\M}_k^* }_2 \leq c  d_k^{*-2} d_{1}^*.$$ 
For term $\norm{ \wt{\M}_k  - \M_{k}^{*}}_2$,
it follows from part (c) of Lemma \ref{lemmauv}, \eqref{czdadqzz}, $\| \wh{\bSigma} \U^*_{-k} \|_2 \leq c d_{k+1}^*$, and $  \| {\V}^*_{-k} \|_2 = 1$ that 
\begin{align*}
	\norm{ \wt{\M}_k  - \M_{k}^{*}}_2 &\leq  | \wt{z}_{kk}^{-1} - {z}_{kk}^{*-1}| \| \wh{\bSigma} \U^*_{-k} \|_2\| ({\V}^{*}_{-k})\trans \|_2 + |  {z}_{kk}^{*-1}| 	\norm{ \wh{\bSigma}(\wt{\C}_{-k}  - \C_{-k}^*)}_2  \nonumber\\[5pt]
	&\leq  c \gamma_n d_1^*d_k^{*-3}.
\end{align*}
This completes the proof of Lemma \ref{rankr:boundm0}.

\subsection{Lemma \ref{rankr:aw} and its proof} \label{new.Sec.B.11}

\begin{lemma}\label{rankr:aw}
	Assume that all the conditions of Theorem \ref{theorkr} are satisfied. For an arbitrary
	$\a\in\R^p $ and  $\wt{\W}_k $ and $\W_k^{*}$ given in \eqref{eqwknear1} and \eqref{eqwknear2},
	respectively, with probability at least
	$	1- \theta_{n,p,q}$ with $\theta_{n,p,q}$ given in \eqref{thetapro}, we have that 
	\begin{enumerate}[label=\rm{(\alph*)}]
        \item 	$\max_{1 \leq i \leq p}\norm{\w_i^*}_0  \leq 2\max\{s_{\max}, r^*+s_u + s_v \},\\  \max_{1 \leq i \leq p}\norm{\wt{\w}_i}_0 \leq 2\max\{s_{\max}, 3(r^*+s_u + s_v) \}$, \\
		$\max_{1 \leq i \leq p}\norm{\wt{\w}_i - \w_i^*}_0 \leq 3 (r^*+s_u + s_v)$;
		\item  $ \max_{1 \leq i \leq p}\norm{\w_i^*}_2 \leq c, \ \max_{1 \leq i \leq p}\norm{\wt{\w}_i}_2 \leq c$,
		\\[5pt] $ \max_{1 \leq i \leq p}\norm{\wt{\w}_i - \w_i^*}_2 \leq  c(r^*+s_u+s_v)^{1/2}\eta_n^2\{n^{-1}\log(pq)\}^{1/2}$;
		\item $ \norm{\a\trans\W^*_k}_2 \leq c \norm{\a}_0^{1/2}\norm{\a}_2$;
		\item $\norm{\a\trans(\wt{\W}_k-\W^*_k)}_2 \leq c\norm{\a}_0^{1/2}\norm{\a}_2(r^*+s_u+s_v)^{1/2}\eta_n^2\{n^{-1}\log(pq)\}^{1/2}$;
		\item $ \norm{\a\trans\wt{\W}_k}_2 \leq c \norm{\a}_0^{1/2}\norm{\a}_2,$
	\end{enumerate}
    where $\wt{\w}_i\trans$ and $ \w_i\strans$ are the $i$th rows of $\wt{\W}_k$ and $\W^*_k$, respectively, with $i = 1, \ldots, p$, and $c$ is some positive constant.
\end{lemma}

\noindent \textit{Proof}. 
	Similar to the proof of Lemma \ref{lemma:1rk4} in Section \ref{new.Sec.B.5}, with the aid of  \eqref{eq:aw1}--\eqref{eq:aw3} we can obtain immediately the results in parts (c)--(e) once the results in parts (a) and (b) are shown. Hence, it remains to establish parts (a) and (b). We start with proving part (a). Let us recall that 
	\begin{align*}
		&\wt{\W}_k = \widehat{\bTheta} \left\{  \I_p +   \wt{z}_{kk}^{-1}\wh{\bSigma}\wt{\U}_{-k}(\I_{r^*-1} -\wt{z}_{kk}^{-1}\wt{\U}_{-k}\trans\wh{\bSigma} \wt{\U}_{-k})^{-1}\wt{\U}_{-k}\trans\right\}, \\
		&\W^*_k = \widehat{\bTheta} \left\{  \I_p +   z_{kk}^{*-1}\wh{\bSigma} \U_{-k}^*(\I_{r^*-1} -z_{kk}^{*-1}\U_{-k}\strans\wh{\bSigma} \U_{-k}^*)^{-1}\U_{-k}\strans\right\}.
	\end{align*}
    Denote by $\wt{\A} = (\wt{z}_{kk}\I_{r^*-1} -\wt{\U}_{-k}\trans\wh{\bSigma} \wt{\U}_{-k})^{-1}$ and $\A^* = (z_{kk}^{*}\I_{r^*-1} -\U_{-k}\strans\wh{\bSigma} \U_{-k}^*)^{-1}$.
	Then it holds that 
	$$ \w_i\trans = \widehat{\btheta}_i\trans (\I_p  +   \wh{\bSigma}\wt{\U}_{-k}\wt{\A}\wt{\U}_{-k}\trans) \ \text{ and } \ \w_i\strans = \widehat{\btheta}_i\trans (\I_p  +   \wh{\bSigma}\U^*_{-k}\A^*\U_{-k}\strans),$$ where  $\wh{\btheta}_i\trans$ represents the $i$th row of $\wh{\bTheta}$. Let $\bdelta_i\trans = \widehat{\btheta}_i\trans \wh{\bSigma}\wt{\U}_{-k}\wt{\A}$ with $\bdelta_i = (\delta_{ij})$. It follows from $\norm{ {\U}^*}_0 = s_u$ and part (b) of Definition \ref{lemmsofar} that 
\begin{align} \label{esafzq}
	& \max_{1\leq i\leq p}\norm{\widehat{\btheta}_i\trans \wh{\bSigma}\wt{\U}_{-k}\wt{\A}\wt{\U}_{-k}\trans }_0  = \max_{1\leq i\leq p}\norm{\bdelta_i\trans \wt{\U}_{-k}\trans}_0 = \max_{1\leq i\leq p}\norm{\sum_{ \substack{ 1 \leq  j \leq r^* \\ j \neq k}} \delta_{ij} \wt{\u}_j\trans}_0 \nonumber\\
	& \leq  \sum_{\substack{ 1 \leq  j \leq r^* \\ j \neq k}} \norm{ \wt{\u}_j}_0 \leq  \norm{ \wt{\U}}_0 \leq \norm{ \wt{\U} - {\U}^*}_0 +\norm{ {\U}^*}_0 \nonumber\\
 &\leq 2(r^*+s_u+s_v) + s_u \leq 3(r^*+s_u+s_v).
\end{align}
Also, by Definition \ref{defi2:acceptable} we see that $\max_{1 \leq i \leq p}\norm{\widehat{\btheta}_i}_0 \leq s_{max}$. Thus, it holds that
\begin{align*}
\max_{1\leq i\leq p}\norm{\wt{\w}_i}_0 & \leq \max_{1\leq i\leq p}\norm{\widehat{\btheta}_i}_0 + \max_{1\leq i\leq p}\norm{\widehat{\btheta}_i\trans \wh{\bSigma}\wt{\U}_{-k}\wt{\A}\wt{\U}_{-k}\trans }_0  \\
&\leq 2\max\{s_{\max}, 3(r^*+s_u + s_v) \}.
\end{align*}

Similar to \eqref{esafzq}, we can show that 
\begin{equation}
    	\max_{1\leq i\leq p}\norm{\widehat{\btheta}_i\trans \wh{\bSigma}\U^*_{-k}\A^*\U_{-k}\strans }_0  \leq  \norm{ {\U}^*}_0 \leq s_u \leq r^*+s_u + s_v. \label{wonorm0}
\end{equation}
It follows that 
$$
\max_{1\leq i\leq p}\norm{\w_i^*}_0 \leq \max_{1\leq i\leq p}\norm{\widehat{\btheta}_i}_0 +\norm{{\U}^* }_0 \leq 2\max\{s_{\max}, (r^*+s_u + s_v) \}.
$$
Let us further denote by $\bdelta_i\strans = \widehat{\btheta}_i\trans\wh{\bSigma}\U^*_{-k}\A^*$. Then using similar arguments as for \eqref{esafzq} and \eqref{wonorm0},  we can deduce that
\begin{align*}
	\max_{1\leq i\leq p}\norm{ \wt{\w}_i - \w_i^*}_0 &\leq \max_{1\leq i\leq p}\norm{  \widehat{\btheta}_i\trans \wh{\bSigma}\wt{\U}_{-k}\wt{\A}\wt{\U}_{-k}\trans -   \widehat{\btheta}_i\trans\wh{\bSigma}\U^*_{-k}\A^*\U_{-k}\strans   }_0 \\
	&\leq \max_{1\leq i\leq p}\norm{  \bdelta_i\trans(\wt{\U}_{-k} - \U_{-k}^* )\trans +  (\bdelta_i - \bdelta_i^*)\trans \U_{-k}\strans   }_0 \\
	&\leq   \max_{1\leq i\leq p}\norm{  \bdelta_i\trans(\wt{\U}_{-k} - \U_{-k}^* )\trans}_0 +  \max_{1\leq i\leq p}\norm{(\bdelta_i - \bdelta_i^*)\trans \U_{-k}\strans   }_0 \\
	& \leq  \norm{ \wt{\U} - {\U}^*}_0 + \norm{{\U}^* }_0 \leq 3(r^* + s_u + s_v),
\end{align*}
which completes the proof for part (a).

We next proceed with proving part (b), which will consist of two parts.

\medskip

\noindent\textbf{(1). The upper bound on  $ \max_{1 \leq i \leq p}\norm{\w_i^*}_2$}. In light of Definition \ref{defi2:acceptable} that $\max_{1 \leq i \leq p}\norm{\widehat{\btheta}_i}_2 \leq c$, it holds that
	\begin{align}\label{rreq110}
		\max_{1 \leq i \leq p}\norm{\w_i^*}_2 &\leq \max_{1 \leq i \leq p}\norm{\widehat{\btheta}_i}_2 + \max_{1 \leq i \leq p}\norm{\widehat{\btheta}_i \wh{\bSigma}\U^*_{-k}\A^*\U_{-k}\strans}_2 \nonumber \\
		&\leq c(1 + \norm{\wh{\bSigma}\U^*_{-k}}_2 \norm{\A^*}_2 \norm{\U_{-k}\strans}_2).
	\end{align}
We will bound term $\norm{\A^*}_2$. Denote by  $\A_0 = (\A^*)^{-1} = z_{kk}^{*}\I_{r^*-1} - \U_{-k}\strans\wh{\bSigma} \U_{-k}^*$. Then
	using the technical arguments in the proof of Lemma \ref{lemm:wexist} in Section \ref{new.Sec.B.9}, we can see that $\A_0 = z_{kk}^{*}\I_{r^*-1} -\U_{-k}\strans\wh{\bSigma} \U_{-k}^* = (a_{ij})$ is symmetric and strictly diagonally dominant, and $a_{ii} = z_{kk}^* - z_{ii}^*$, $a_{ij} = - z^*_{ij}$ for each $i \neq j$. Moreover, 
	we have that $\sum_{j \neq i}|a_{ij}| = o(|a_{ii}|)$. Let us define
	$$\alpha_1 = \min_i(|a_{ii}| - \sum_{j \neq i}|a_{ij}| ) \ \text{ and } \   \alpha_2 = \min_i(|a_{ii}| - \sum_{j \neq i}|a_{ji}| ).$$
	Then it holds that $\alpha_1 = \alpha_2 \asymp  \min_{i} |a_{ii}| = \min_{i \neq k}|z_{kk}^* - z_{ii}^*|$. It follows from \eqref{ediiiac} and \eqref{ediiiac2} in the proof of  Lemma \ref{lemmzzz} that for each $i \in \{1, \ldots, r^*\}$, 
	\begin{align}
	 z_{ii}^{*}  -z_{i+1, i+1}^{*}  \geq 	d_i^{*2} \rho_l - d_{i+1}^{*2} \rho_u  \geq  c_0 \rho_u d^{*2}_{i} \geq c, \nonumber
	\end{align}
which entails that $ z_{ii}^{*} > z_{i+1, i+1}^{*} $.
We can further see that 
	\begin{align*}
		\min_{i \neq k}|z_{kk}^* - z_{ii}^*| = \min\{|z_{k-1, k-1}^* - z_{kk}^* |,  |z_{kk}^* - z_{k+1, k+1}^*| \} \geq c.
	\end{align*}


	Since  $(\A^*)^{-1} = \A_0$ is symmetric and strictly diagonally dominant, an application of Corollary 2 in \cite{varah1975lower} leads to 
	\begin{align}\label{rreq32}
		\norm{ \A^* }_2  = \norm{ \A_0^{-1} }_2 \leq  \frac{1}{\sqrt{\alpha_1 \alpha_2} }  \leq c ({\min_{i \neq k}|z_{kk}^* - z_{ii}^*|})^{-1} \leq c.
	\end{align}
Also, in light of Lemma \ref{rankr:boundm0} we have that 
\begin{align}
	&\norm{\wh{\bSigma}\U^*_{-k}}_2  \leq cd_1^* \ \text{ and } \
	\norm{\U_{-k}^*}_2  \leq cd_1^*. \label{rreq12}
\end{align}
	Hence, under Condition \ref{con:nearlyorth} that the nonzero eigenvalues $d_i^{*2}$ are at the constant level and Definition \ref{defi2:acceptable} that $\max_{1 \leq i \leq p}\norm{\widehat{\btheta}_i}_2 \leq c$, combining \eqref{rreq32}--\eqref{rreq12} yields  that
	\begin{align}\label{czmla}
	\max_{1 \leq i \leq p}\norm{\widehat{\btheta}_i \wh{\bSigma}\U^*_{-k}\A^*\U_{-k}\strans}_2 \leq  \max_{1 \leq i \leq p}\norm{\widehat{\btheta}_i}_2 \norm{\wh{\bSigma}\U^*_{-k}}_2 \norm{\A^*}_2 \norm{\U_{-k}\strans}_2 \leq c,
	\end{align}
which further results in 
	\begin{align*}
		\max_{1 \leq i \leq p}\norm{\w_i^*}_2 \leq c.
	\end{align*}

\smallskip

	\noindent\textbf{(2). The upper bounds on   $ \max_{1 \leq i \leq p}\norm{\wt{\w}_i - \w_i^*}_2$ and  $ \max_{1 \leq i \leq p}\norm{\wt{\w}_i}_2$}.
	For term  $ \max_{1 \leq i \leq p}\norm{\wt{\w}_i - \w_i^*}_2$ above, in view of Definition \ref{defi2:acceptable} that $\max_{1 \leq i \leq p}\norm{\widehat{\btheta}_i}_2 \leq c$, it holds that
	\begin{align}\label{eqdad}
		\max_{1\leq i \leq p}\|\wt{\w}_i - \w_i^*\|_2
		&\leq \max_{1\leq i \leq p}\|\wh{\btheta}_i\trans\|_2 \norm{  \wh{\bSigma}\wt{\U}_{-k}\wt{\A}\wt{\U}_{-k}\trans -   \wh{\bSigma}\U^*_{-k}\A^*\U_{-k}\strans   }_2 \nonumber \\
		&\leq c\norm{   \wh{\bSigma}\wt{\U}_{-k}\wt{\A}\wt{\U}_{-k}\trans -  \wh{\bSigma}\U^*_{-k}\A^*\U_{-k}\strans   }_2.
	\end{align}
 Some simple calculations give that
	\begin{align}\label{eqttt122}
	 & \wh{\bSigma}\wt{\U}_{-k}\wt{\A}\wt{\U}_{-k}\trans -  \wh{\bSigma}\U^*_{-k}\A^*\U_{-k}\strans    \nonumber\\[5pt]
	 & =  (\wh{\bSigma}\wt{\U}_{-k} - \wh{\bSigma}\U^*_{-k})\wt{\A}\wt{\U}_{-k}\trans + \wh{\bSigma}\U^*_{-k}(\wt{\A} - \A^* )\wt{\U}_{-k}\trans \nonumber\\
  &\quad+  \wh{\bSigma}\U^*_{-k}\A^*(\wt{\U}_{-k}\trans - \U_{-k}\strans ).
	\end{align}
We aim to bound the three terms on the right-hand side of \eqref{eqttt122} above.

Recall that $\wt{\A} = (\wt{z}_{kk}\I_{r^*-1} -\wt{\U}_{-k}\trans\wh{\bSigma} \wt{\U}_{-k})^{-1}$. Observe that $\wt{\A}_0 = \wt{\A}^{-1} = \wt{z}_{kk}\I_{r^*-1} -\wt{\U}_{-k}\trans\wh{\bSigma} \wt{\U}_{-k} $ is also symmetric and strictly diagonally dominant from the proof of Lemma \ref{lemm:wexist}. Using similar arguments as for \eqref{rreq32}, we can deduce that
    \begin{align}
        \norm{\wt{\A}}_2 = \| \wt{\A}_0^{-1}\|_2 \leq c. \label{rreqa1}
    \end{align}
By Definition \ref{lemmsofar}, we have that 
  \begin{align}
  	\norm{\wt{\U}_{-k} - \U^*_{-k}}_2 \leq 	\norm{\wt{\U}_{-k} - \U^*_{-k}}_F \leq \norm{\wt{\U} - \U^*}_F \leq c\gamma_n. \label{equ00kk}
  \end{align}
  An application of Lemma \ref{rankr:boundm0} leads to 
    \begin{align}
        &\norm{\wh{\bSigma}\wt{\U}_{-k}}_2 \leq cd_1^*, \ \
		\norm{\wt{\U}_{-k}}_2  \leq cd_1^*, \label{rreqa2}\\
		&\norm{\wh{\bSigma}(\wt{\U}_{-k} - \U^*_{-k})}_2 \leq c(r^*+s_u+s_v)\eta_n^2\{n^{-1}\log(pq)\}^{1/2}.  \label{eqsu-u}
	\end{align}

	Notice that Condition \ref{con:nearlyorth} implies that the nonzero eigenvalues $d_i^{*2}$ are at the constant level. A combination of  \eqref{rreqa1}--\eqref{eqsu-u} yields that
	\begin{align}\label{eqrg22}
		\norm{  (\wh{\bSigma}\wt{\U}_{-k} - \wh{\bSigma}\U^*_{-k})\wt{\A}\wt{\U}_{-k}\trans}_2 &\leq \norm{\wh{\bSigma}(\wt{\U}_{-k} - \U^*_{-k})}_2\norm{\wt{\A}}_2 \norm{\wt{\U}_{-k}\trans}_2 \nonumber \\
		&\leq  c  (r^*+s_u+s_v)\eta_n^2\{n^{-1}\log(pq)\}^{1/2}.
	\end{align}
	Moreover, it follows from \eqref{rreq32}, \eqref{rreq12}, and \eqref{equ00kk} that
	\begin{align}\label{eqrg32}
		\norm{ \wh{\bSigma}\U^*_{-k}\A^*(\wt{\U}_{-k}\trans - \U_{-k}\strans )}_2 &\leq  \norm{\wh{\bSigma} \U^*_{-k}}_2\norm{\A^*}_2 \norm{\wt{\U}_{-k}- \U_{-k}^* }_2  \nonumber\\
		&\leq  c (r^*+s_u+s_v)\eta_n^2\{n^{-1}\log(pq)\}^{1/2}.
	\end{align}

	We proceed with bounding term $\norm{\wt{\A} - \A^*}_2$ above. From \eqref{rreq32} and \eqref{rreqa1}, we see that $\norm{ {\A}_0^{-1}}_2 = \|\A^*\|_2 \leq c$ and $\norm{ \wt{\A}_0^{-1}}_2 = \|\wt{\A}\|_2 \leq c$. Then it holds that
	\begin{align}
		\norm{\wt{\A} - \A^*}_2 &=  \norm{ \wt{\A}_0^{-1} - \A_0^{-1} }_2 = \norm{ \wt{\A}_0^{-1}( \A_0 - \wt{\A}_0  )\A_0^{-1}  }_2 \nonumber\\
		&\leq \norm{ \wt{\A}_0^{-1}}_2 \norm{  \A_0 - \wt{\A}_0   }_2 \norm{ \A_0^{-1}  }_2 \leq c \norm{  \A_0 - \wt{\A}_0   }_2. \label{eqaa123}
	\end{align}
	It remains to bound term $\norm{  \A_0 - \wt{\A}_0   }_2$ above. Note that
	\begin{align*}
		\norm{ \wt{\A}_0 - \A_0   }_2 &= \norm{ (\wt{z}_{kk}\I_{r^*-1} -\wt{\U}_{-k}\trans\wh{\bSigma} \wt{\U}_{-k}) - (z_{kk}^{*}\I_{r^*-1} -\U_{-k}\strans\wh{\bSigma} \U_{-k}^*)}_2 \\
		& \leq |\wt{z}_{kk} - z_{kk}^{*}| + \norm{\wt{\U}_{-k}\trans\wh{\bSigma} \wt{\U}_{-k} -  \U_{-k}\strans\wh{\bSigma} \U_{-k}^*}_2.
	\end{align*}
    In light of \eqref{rreq12}, \eqref{equ00kk}, and \eqref{rreqa2}, we have that
    \begin{align*}
        & \norm{\wt{\U}_{-k}\trans\wh{\bSigma} \wt{\U}_{-k} -  \U_{-k}\strans\wh{\bSigma} \U_{-k}^*}_2 \leq
        \norm{\wt{\U}_{-k}\trans(\wh{\bSigma} \wt{\U}_{-k} - \wh{\bSigma} \U_{-k}^*)}_2 \\
        &\quad + \norm{(\wt{\U}_{-k}\trans -  \U_{-k}\strans)\wh{\bSigma} \U_{-k}^*}_2 \\
        &\leq      \norm{\wt{\U}_{-k}\trans}_2\norm{\wh{\bSigma} \wt{\U}_{-k} - \wh{\bSigma} \U_{-k}^*}_2 + \norm{\wt{\U}_{-k}\trans -  \U_{-k}\strans}_2 \norm{\wh{\bSigma} \U_{-k}^*}_2 \\
        &\leq  c  (r^*+s_u+s_v)\eta_n^2\{n^{-1}\log(pq)\}^{1/2}.
    \end{align*}
Together with the upper bound of $|\wt{z}_{kk} - z_{kk}^{*}|$ in Lemma \ref{lemmauv},  it yields that
    \begin{align}
        &\norm{ \wt{\A}_0 - \A_0   }_2 \leq  c(r^*+s_u+s_v)\eta_n^2\{n^{-1}\log(pq)\}^{1/2}, \nonumber
    \end{align}
which further entails that
\begin{align}
	 &\norm{\wt{\A} - \A^*}_2 \leq c  (r^*+s_u+s_v)\eta_n^2\{n^{-1}\log(pq)\}^{1/2}. \label{eqa009a}
\end{align}

For term $\wh{\bSigma}\U^*_{-k}(\wt{\A} - \A^* )\wt{\U}_{-k}\trans$ above, it follows from \eqref{rreq12}, \eqref{rreqa2}, and \eqref{eqa009a} that 
    \begin{align}
        \|\wh{\bSigma}\U^*_{-k}(\wt{\A} - \A^* )\wt{\U}_{-k}\trans\|_2 &\leq \|\wh{\bSigma}\U^*_{-k}\|_2 \|\wt{\A} - \A^* \|_2 \|\wt{\U}_{-k}\trans\|_2 \nonumber\\
        &\leq  c (r^*+s_u+s_v)\eta_n^2\{n^{-1}\log(pq)\}^{1/2}. \label{equaua1}
    \end{align}
	Therefore, combining \eqref{eqdad}, \eqref{eqttt122}, \eqref{eqrg22}, \eqref{eqrg32}, and \eqref{equaua1} gives that
	\begin{align*}
		\max_{1\leq i \leq p}\|\wt{\w}_i - \w_i^*\|_2 \leq  c(r^*+s_u+s_v)\{n^{-1}\log(pq)\}^{1/2}.
	\end{align*}
Moreover, from the triangle inequality we have that for sufficiently large $n$, 
\begin{align}
		\max_{1\leq i\leq p}\norm{ \wt{\w}_i}_2
	\leq 	\max_{1\leq i\leq p}\norm{ \wt{\w}_i - \w_i^* }_2 + 	\max_{1\leq i\leq p}\norm{ \w_i^* }_2 \leq c, \nonumber
\end{align}
which completes the proof of part (b). This concludes the proof of
Lemma \ref{rankr:aw}.

\subsection{Lemma \ref{lemma:1rkr} and its proof} \label{new.Sec.B.12}

\begin{lemma}\label{lemma:1rkr}
	Assume that all the conditions of Theorem \ref{theorkr} are satisfied. For $\wt{\bdelta}_k$ defined in \eqref{delrankr} and an arbitrary
	$\a\in\R^p$, with probability at least
	$	1- \theta_{n,p,q}$ it holds that
	\begin{align*}
		|\a\trans \wt{\W}_k \wt{\bdelta}_k | \leq c \norm{\a}_0^{1/2}\norm{\a}_2 (r^*+s_u+s_v)\eta_n^4\left\{n^{-1}\log(pq)\right\}, 
	\end{align*}
	where $\theta_{n,p,q}$ is given in \eqref{thetapro} and $c$ is some positive constant.
\end{lemma}

\noindent \textit{Proof}. 
	Observe that
	\begin{align*}
		|\a\trans \wt{\W}_k \wt{\bdelta}_k | &= | \a\trans \wt{\W}_k \wt{\M}_k \left\{(\wt{\v}_k - \v_k^*)\wt{\u}_k\trans- ( \wt{\C}_{-k}\trans - \C_{-k}\strans)\right\}\wh{\bSigma} (\wt{\u}_k - \u_k^*)  | \\
		& \leq
		| \a\trans \wt{\W}_k \wt{\M}_k (\wt{\v}_k - \v_k^*) |
		| \wt{\u}_k\trans  \wh{\bSigma}(\wt{\u}_k - \u_k^*)  |  \\
  &\quad+ | \a\trans \wt{\W}_k \wt{\M}_k ( \wt{\C}_{-k}\trans - \C_{-k}\strans) \wh{\bSigma}(\wt{\u}_k - \u_k^*)  |.
	\end{align*}
From Lemma \ref{rankr:boundm0}, we have that
	$	\norm{ \wt{\M}_k }_2 \leq c d_k^{*-2} d_1^*. $
It follows from Lemma \ref{rankr:aw} that 
$   	\norm{ \a\trans \wt{\W}_k}_2 \leq c\norm{\a}_0^{1/2}\norm{\a}_2. $
Together with parts (a) and (b) of Lemma \ref{lemmauv}, it holds that 
	\begin{align*}
		& | \a\trans\wt{\W}_k\wt{\M}_k  (\wt{\v}_k - \v_k^*) || \wt{\u}_k\trans  \wh{\bSigma}(\wt{\u}_k - \u_k^*)  | \\
		 &\leq  \norm{ \a\trans \wt{\W}_k}_2\norm{\wt{\M}_k }_2\norm{\wt{\v}_k-\v_k^*}_2 \norm{\wt{\u}_k}_2  \norm{\wh{\bSigma}(\wt{\u}_k-\u_k^*)}_2 \\
		&\leq c \norm{\a}_0^{1/2}\norm{\a}_2 (r^*+s_u+s_v)\eta_n^4 \left\{n^{-1}\log(pq)\right\} d_k^{*-2} d_1^*.
	\end{align*}

Further, by Lemma \ref{rankr:boundm0} we can obtain that
	\begin{align}
		\norm{\wt{\C}_{-k}  - \C_{-k}^*}_2 \leq c (r^*+s_u+s_v)^{1/2}\eta_n^2\left\{n^{-1}\log(pq)\right\}^{1/2}. \nonumber
	\end{align}
	Then it follows that
	\begin{align*}
		& | \a\trans \wt{\W}_k \wt{\M}_k ( \wt{\C}_{-k}\trans - \C_{-k}\strans) \wh{\bSigma}(\wt{\u}_k - \u_k^*)  | \\
		 &\leq  \norm{ \a\trans \wt{\W}_k}_2\norm{\wt{\M}_k }_2\| \wt{\C}_{-k}\trans - \C_{-k}\strans\|_2  \norm{\wh{\bSigma}(\wt{\u}_k-\u_k^*)}_2 \\
		&\leq c \norm{\a}_0^{1/2}\norm{\a}_2  (r^*+s_u+s_v)\eta_n^4 \left\{n^{-1}\log(pq)\right\}d_k^{*-2} d_1^*.
	\end{align*}
Combining the above results leads to 
	\begin{align}
		| \a\trans \wt{\W}_k \wt{\bdelta}_k | \leq c \norm{\a}_0^{1/2}\norm{\a}_2  (r^*+s_u+s_v)\eta_n^4 \left\{n^{-1}\log(pq)\right\}d_k^{*-2}d_1^*. \nonumber
	\end{align}
	Thus, under Condition \ref{con:nearlyorth} that the nonzero eigenvalues $d^{*2}_{i}$ are at the constant level, we can deduce that 
		\begin{align}
		| \a\trans \wt{\W}_k \wt{\bdelta}_k | \leq c \norm{\a}_0^{1/2}\norm{\a}_2  (r^*+s_u+s_v)\eta_n^4 \left\{n^{-1}\log(pq)\right\}, \nonumber
	\end{align}
	which completes the proof of Lemma \ref{lemma:1rkr}.

\subsection{Lemma \ref{lemma:gapr} and its proof} \label{new.Sec.B.13}

\begin{lemma}\label{lemma:gapr}
	Assume that all the conditions of  Theorem \ref{theorkr} are satisfied.
	For $\wt{\M}_k =  -\wt{z}_{kk}^{-1}\wh{\bSigma}\wt{\C}_{-k}$ and $\wt{\W}_k$ defined in \eqref{eqwknear1} and an arbitrary
	$\a\in\R^p$, with probability at least
	$	1- \theta_{n,p,q}$ it holds that
	\begin{align*}
		|\a\trans\wt{\W}_k \wt{\M}_k \C_{-k}\strans \wh{\bSigma} \u_k^*  | = o(\norm{\a}_0^{1/2}\norm{\a}_2  n^{-1/2}),
	\end{align*}
	where $\theta_{n,p,q}$ is given in \eqref{thetapro} and $c$ is some positive constant.
\end{lemma}

\noindent \textit{Proof}. 
	According to the construction that $\wt{\M}_k =  -\wt{z}_{kk}^{-1}\wh{\bSigma}\wt{\C}_{-k}$, it holds that
\begin{align}\label{caeqajhka}
	& \wt{\M}_k \C_{-k}\strans \wh{\bSigma} \u_k^* = \sum_{j \neq k} \wt{\M}_k \v_j^*   \u_j\strans\wh{\bSigma} \u_k^* = -\sum_{j \neq k}\wt{z}_{kk}^{-1}\wh{\bSigma}\sum_{i \neq k}\wt{\u}_i \wt{\v}_i\trans\v_j^*   \u_j\strans\wh{\bSigma} \u_k^* \nonumber\\[5pt]
	&= -\sum_{j \neq k}\wt{z}_{kk}^{-1}\wh{\bSigma}\sum_{i \neq k}\wt{\u}_i {\v}_i\strans\v_j^*   \u_j\strans\wh{\bSigma} \u_k^* -   \sum_{j \neq k}\wt{z}_{kk}^{-1}\wh{\bSigma}\sum_{i \neq k}\wt{\u}_i (\wt{\v}_i\trans - {\v}_i\strans)\v_j^*   \u_j\strans\wh{\bSigma} \u_k^* \nonumber \\[5pt]
	&= -\sum_{j \neq k}\wt{z}_{kk}^{-1}(\wh{\bSigma}\wt{\u}_j + \sum_{i \neq k} \wh{\bSigma}\wt{\u}_i (\wt{\v}_i - {\v}_i^*)\trans\v_j^* )\u_j\strans\wh{\bSigma} \u_k^*,
\end{align}
where the last step above has used ${\v}_i\strans\v_j^* = 0 $ for each $i \neq j$.
For term $ \sum_{i \neq k} \wh{\bSigma}\wt{\u}_i (\wt{\v}_i - {\v}_i^*)\trans\v_j^*$ above, we can deduce that 
\begin{align*}
	\norm{\sum_{i \neq k} \wh{\bSigma}\wt{\u}_i (\wt{\v}_i - {\v}_i^*)\trans\v_j^*}_2 &\leq \sum_{i \neq k}  \| \wh{\bSigma}\wt{\l}_i \|_2 \|\wt{d}_i(\wt{\v}_i - {\v}_i^*)\|_2 \|\v_j^*\|_2 \\
	&\leq c \sum_{i \neq k}\|\wt{d}_i(\wt{\v}_i - {\v}_i^*)\|_2 \leq cr^* \gamma_n ,
\end{align*}
where we have used $\| \wh{\bSigma}\wt{\l}_i \|_2 \leq c$ due to Condition \ref{con3}, $\|\v_j^*\|_2 = 1$, and parts (a) and (b) of Lemma \ref{lemmauv}. 
Recall the condition that  $m^{1/2}\kappa_n = o(1)$ with
$$\kappa_n = \max\{s_{\max}^{1/2} , (r^*+s_u+s_v)^{1/2}, \eta_n^2\} (r^*+s_u+s_v)\eta_n^2\log(pq)/\sqrt{n}.$$
Then an application of similar arguments as for  \eqref{czdgggs} leads to 
$r^*  \gamma_n = o(1).$

Part (b) of Lemma \ref{lemmauv} shows that  $\norm{\wh{\bSigma}\wt{\u}_j}_2 \leq cd_j^* $. Since the nonzero eigenvalues $d_i^{*2}$ are at the constant level by Condition \ref{con:nearlyorth}, for sufficiently large $n$ we have that 
\begin{align*}
	\norm{\wh{\bSigma}\wt{\u}_j + \sum_{i \neq k} \wh{\bSigma}\wt{\u}_i (\wt{\v}_i - {\v}_i^*)\trans\v_j^*}_2 & \leq 
		\norm{\wh{\bSigma}\wt{\u}_j}_2 + \norm{\sum_{i \neq k} \wh{\bSigma}\wt{\u}_i (\wt{\v}_i - {\v}_i^*)\trans\v_j^*}_2 \\
  &\leq  cd_j^*.
\end{align*}
Together with $|\wt{z}_{kk}^{-1}| \leq c d_k^{*-2}$ in part (c) of Lemma \ref{lemmauv}, it follows that
\begin{align}\label{caeqajhka1}
	\norm{\wt{\M}_k \C_{-k}\strans \wh{\bSigma} \u_k^* }_2  
	&\leq \sum_{j \neq k}|\wt{z}_{kk}^{-1}|  \big\|\wh{\bSigma}\wt{\u}_j  + \sum_{i \neq k} \wh{\bSigma}\wt{\u}_i (\wt{\v}_i - {\v}_i^*)\trans\v_j^* \big\|_2  |\u_j\strans\wh{\bSigma} \u_k^*| \nonumber\\[5pt]
	&\leq c   \sum_{j \neq k}   (d_j^{*}/d_k^{*2}) |\u_j\strans\wh{\bSigma} \u_k^*| = c  \sum_{j \neq k}  (d_j^{*2}/d_k^*) |\l_j\strans\wh{\bSigma}\l_k^*|.
\end{align}
Using Condition \ref{con:nearlyorth} that $\sum_{i \neq j} |\l_i\strans\wh{\bSigma}\l_j^*| = o(n^{-1/2})$  and the nonzero eigenvalues $d_i^{*2}$ are at the constant level, and Lemma \ref{rankr:aw} that $\norm{ \a\trans \wt{\W}_k}_2 \leq c\norm{\a}_0^{1/2}\norm{\a}_2$, we can obtain that 
	\begin{align*}
		|\a\trans\wt{\W}_k \wt{\M}_k\C_{-k}\strans \wh{\bSigma} \u_k^*  | 
		&\leq  \norm{\a\trans\wt{\W}_k }_2 \norm{\wt{\M}_k \C_{-k}\strans \wh{\bSigma} \u_k^* }_2 
		\\
  &= o(\norm{\a}_0^{1/2}\norm{\a}_2  n^{-1/2}).
	\end{align*}	
This concludes the proof of Lemma \ref{lemma:gapr}.

\subsection{Lemma \ref{lemma:1rk3r} and its proof} \label{new.Sec.B.14}

\begin{lemma}\label{lemma:1rk3r}
	Assume that all the conditions of Theorem \ref{theorkr} are satisfied. For $\wt{\bepsilon}_k$ defined in \eqref{eprankr}, $h_k$ defined in \eqref{eqhknear}, and any $\a\in\mathcal{A}=\{\a\in\R^p:\norm{\a}_0\leq m,\norm{\a}_2=1\}$,
	with probability at least
	$	1- \theta_{n,p,q}$ we have that 
	\begin{align*}
		\abs{-\a\trans\wt{\W}_k\wt{\bepsilon}_k - h_k / \sqrt{n}} \leq c m^{1/2}  (r^*+s_u + s_v)^{3/2}\eta_n^2\{ n^{-1}\log(pq)\},
	\end{align*}
	where $\theta_{n,p,q}$ is given in \eqref{thetapro} and $c$ is some positive constant.
\end{lemma}

\noindent \textit{Proof}. 
	The proof of Lemma \ref{lemma:1rk3r} follows similar technical arguments as in the proof of Lemma \ref{lemma:1rk3} in Section \ref{new.Sec.B.8}. We will first show that 
$\a\trans \wt{\W}_k \wt{\M}_k$, $\a\trans \wt{\W}_k \wt{\M}_k  - \a\trans {\W}_k^{*}\M_k^{*}$, and  $\a\trans(\wt{\W}_k-{\W}_k^*)$ are $s$-sparse with $s=c(r^*+s_u+s_v)$. It follows from the sparsity of $\U^*$ and $\V^*$ and \eqref{esafzq} that 
\begin{align*}
	\sum_{1 \leq i \leq r^*}\norm{ \u_i^* }_0 \leq s_u, \ \sum_{1 \leq i \leq r^*}\norm{ \v_i^* }_0 \leq s_v, \sum_{1 \leq i \leq r^*}\norm{ \wt{\u}_i }_0 \leq c(r^* + s_u + s_v). 
\end{align*}
Also, in view of part (b) of Lemma \ref{lemmauv}, we have that 
\begin{align*}
	\sum_{1 \leq i \leq r^*}\norm{\wt{d}_i \wt{\v}_i }_0 \leq  \sum_{1 \leq i \leq r^*}\norm{\wt{d}_i (\wt{\v}_i - \v_i^*) }_0 + \sum_{1 \leq i \leq r^*}\norm{ {\v}_i^* }_0 \leq c(r^* + s_u + s_v).  
\end{align*}	
Note that $\wt{\M}_k =  -\wt{z}_{kk}^{-1}\wh{\bSigma} \sum_{i \neq k} \wt{\u}_i\wt{\v}_i\trans$.
    With similar arguments as for  \eqref{sawm1}--\eqref{saaw2}, we can deduce that 
	\begin{align*}
		&\norm{\a\trans \wt{\W}_k \wt{\M}_k}_0 = \norm{ \sum_{i \neq k} ( \a\trans \wt{\W}_k \wt{z}_{ii}^{-1}\wh{\bSigma} \wt{\l}_i) \cdot \wt{d}_i\wt{\v}_i\trans }_0 \leq  \sum_{i \neq k}  \norm{  \wt{d}_i \wt{\v}_i   }_0   \leq c(r^* + s_u + s_v), \\
		&\norm{\a\trans \wt{\W}_k \wt{\M}_k  - \a\trans \W^{*}_k\M_k^{*}}_0 \leq \sum_{i \neq k}  \norm{  \wt{d}_i\wt{\v}_i   }_0 +  \sum_{i \neq k}\norm{ \v_i^* }_0 \leq  c(r^* + s_u + s_v), \\
		&\|\a\trans(\wt{\W}_k -\W^*_k)\|_0 
	 \leq \sum_{i \neq k}  \norm{ \wt{\u}_i   }_0 +  \sum_{i \neq k}\norm{ \u_i^* }_0 \leq c(r^*+s_u+s_v).
\end{align*}

Then similar to \eqref{sdzvcqqqq}, it holds that 
	\begin{align}
		\abs{-\a\trans\wt{\W}_k \wt{\bepsilon}_k - h_k / \sqrt{n}} & \leq  \norm{\a\trans \wt{\W}_k}_2 \norm{\wt{\M}_k}_2 \norm{n^{-1}\E\trans\X (\wt{\u}_k - {\u}_k^{*} )}_{2,s} \nonumber \\[5pt]
		&\quad + \norm{\a\trans \wt{\W}_k \wt{\M}_k  - \a\trans \W^{*}_k \M_k^{*}}_2 \norm{ n^{-1}\E\trans\X\u_k^*}_{2,s}  \nonumber\\
  &\quad+  \norm{\a\trans(\wt{\W}_k-\W^*_k) }_2\norm{ n^{-1}\X\trans\E\v_k^*}_{2,s}.
		\label{sdwqvcs2}
	\end{align}
An application of similar arguments as for \eqref{a22ex}--\eqref{eqathe} gives that
\begin{align}
	&n^{-1}\norm{\E\trans\X(\wt{\u}_k - {\u}_k^{*} )}_{2,s} 
	\leq c s^{3/2}\eta_n^2\{n^{-1}\log(pq)\}, \label{a22ex22}\\[5pt]
	&n^{-1}\norm{\E\trans\X\u_k^*}_{2,s} \leq cs^{1/2}s_u^{1/2}\{n^{-1}\log(pq)\}^{1/2}  d_k^{*}, \label{eqwdaz22}\\[5pt]
	&n^{-1}\norm{\X\trans\E\v_k^*}_{2,s} \leq cs^{1/2}s_v^{1/2}\{n^{-1}\log(pq)\}^{1/2}. \label{eqathe22}
\end{align}

From Lemma \ref{rankr:boundm0}, we have that 
\begin{align}
&\norm{ \M_{k}^{*}}_2 \leq c d_k^{*-2}d_1^*, \ \ \norm{  \wt{\M}_k}_2 \leq c  d_k^{*-2} d_1^*, \label{boundmm} \\
&\norm{ \wt{\M}_k  - \M_{k}^{*}}_2 \leq  c (r^*+s_u+s_v)^{1/2} \eta_n^2\{n^{-1}\log(pq)\}^{1/2}d_k^{*-3} d_1^*. \label{boundmmm}
\end{align}
Along with parts (d) and (e) of Lemma \ref{rankr:aw},  it follows that
\begin{align}
	&\norm{\a\trans (\wt{\W}_k  - \W^{*}_k)}_2 \leq 
	cm^{1/2} (r^*+s_u+s_v)^{1/2} \eta_n^2  \{n^{-1}\log(pq)\}^{1/2},
	\nonumber \\[5pt]
	&\norm{\a\trans \wt{\W}_k \wt{\M}_k  - \a\trans \W^{*}_k \M_k^{*}}_2 \leq  \norm{\a\trans \wt{\W}_k (\wt{\M}_k  -\M_k^{*}) }_2 + \norm{\a\trans (\wt{\W}_k  - \W^{*}_k) \M_k^{*}}_2 \nonumber\\[5pt]
	&~ \leq  \norm{\a\trans \wt{\W}_k}_2 \norm{ \wt{\M}_k  -\M_k^{*} }_2 + \norm{\a\trans (\wt{\W}_k  - \W^{*}_k)}_2 \norm{ \M_k^{*}}_2 \nonumber \\[5pt]
	&~ \leq    cm^{1/2} (r^*+s_u+s_v)^{1/2} \eta_n^2  \{n^{-1}\log(pq)\}^{1/2}d_k^{*-2} d_1^*.  \label{zxcfdfaqe}
\end{align}
Therefore, by   \eqref{sdwqvcs2}--\eqref{zxcfdfaqe}, Lemma \ref{rankr:aw}, and Condition  \ref{con:nearlyorth} that the nonzero eigenvalues $d_i^{*2}$ are at the constant level, we can obtain that
	\begin{align*}
		\abs{-\a\trans\wt{\W}_k\wt{\bepsilon}_k - h_k  / \sqrt{n}} \leq c m^{1/2}  (r^* +s_u + s_v)^{3/2}\eta_n^2\{ n^{-1}\log(pq)\}.
	\end{align*}
This completes the proof of Lemma \ref{lemma:1rk3r}.

\subsection{Lemma \ref{lemmzzzz} and its proof} \label{new.Sec.B.15}

\begin{lemma}\label{lemmzzzz}
	Assume that all the conditions of Theorem \ref{theorkapor} are satisfied. For each given $k$ with $1 \leq k < r^*$,
	with probability at least
	$	1- \theta_{n,p,q}$ with $\theta_{n,p,q}$ given in \eqref{thetapro}, we have that 
	\[  \sum_{k+1 \leq j \leq r^*,\, j \neq i} |z_{ij}^*| =  o( |z_{kk}^* - z_{ii}^*| ) \ \text{ and } \ \sum_{ k+1 \leq j \leq r^*,\, j \neq i} |\wt{z}_{ij}| =  o( |\wt{z}_{kk} - \wt{z}_{ii}| ) \]
	for each $i \in \{k+1, \ldots ,r^*\}$.
\end{lemma}

\noindent \textit{Proof}. 
Let us first analyze terms $|z_{kk}^* - z_{ii}^*| $ and $|\wt{z}_{kk} - \wt{z}_{ii}|$.
For each $ i \in \{k+1, \ldots ,r^*\}$, 
in view of Condition \ref{con3} we have that $ {d}^{*2}_k\rho_l \leq z_{kk}^{*} \leq {d}^{*2}_k\rho_u $ and $ {d}^{*2}_i\rho_l \leq z_{ii}^{*} \leq {d}^{*2}_i\rho_u $, which lead to  
\begin{align}
	& d_k^{*2} \rho_l - d_i^{*2} \rho_u \leq  z_{kk}^{*}  -z_{ii}^{*}   \leq   d_k^{*2}\rho_u -   d_i^{*2}\rho_l. \nonumber
\end{align}
Since $\max_i d_i^* = d_{k+1}^*$, using similar arguments as for \eqref{ediiiac} and \eqref{ediiiac2}, it holds that
\begin{align}\label{zkkjja}
	\min_i  (z_{kk}^{*}  -z_{ii}^{*} ) \geq \min_i (d_k^{*2} \rho_l - d_i^{*2} \rho_u) =  d_k^{*2} \rho_l - d_{k+1}^{*2} \rho_u \geq c d_{k}^{*2}.
\end{align}
Further, from part (c) of  Lemma \ref{lemmauv} and $d_i^* < d_k^*$,  we can obtain that
\begin{align*}
	|(\wt{z}_{kk} - \wt{z}_{ii}) - (z_{kk}^{*} - z_{ii}^{*})| &\leq |\wt{z}_{kk} - z_{kk}^{*}| + |\wt{z}_{ii} - z_{ii}^{*}|\\ &\leq  c(r^*+s_u+s_v)^{1/2}\eta_n^2\{n^{-1}\log(pq)\}^{1/2} d_k^*.
\end{align*}
By the assumption in Theorem \ref{theorkapor} that $m^{1/2}\kappa_n^{(k)} = o(1)$, we have $\gamma_n = o(1)$.
Then for sufficiently large $n$, it follows that
\begin{align}
	|\wt{z}_{kk} - \wt{z}_{ii} | \geq 	| z_{kk}^{*}  -z_{ii}^{*}  | - 	|(\wt{z}_{kk} - \wt{z}_{ii}) - (z_{kk}^{*} - z_{ii}^{*})| \geq c d_k^{*2}. \nonumber 
\end{align}

We next prove that $\sum_{k+1 \leq j \leq r^*, j \neq i} |z_{ij}^*| =  o( |z_{kk}^* - z_{ii}^*| )$. Observe that $|z_{ij}^*| = d_i^* d_j^* |\l_i\strans \wh{\bSigma}\l_j^*|$.
For each $i, j \in \{k+1,\ldots, r^*\}$, when $j < i$, from Condition \ref{con:orth:rankr} we have that 
$$ (d_i^{*2}/d_j^*) |\l_j\strans \wh{\bSigma} \l_i^*| = o(n^{-1/2}).$$
When $j > i$, similarly we can obtain that  $(d_j^{*2}/d_i^*) |\l_j\strans \wh{\bSigma} \l_i^*| = o(n^{-1/2}).$ Then for each $i \in \{k+1,\ldots, r^*\}$, we can deduce that 
\begin{align*}
    \sum_{k+1 \leq j \leq r^*, \, j \neq i}\frac{ |z_{ij}^*|}{|z_{kk}^* - z_{ii}^*|} &\leq c \sum_{k+1 \leq j \leq r^*,\,  j \neq i} \frac{ d_i^* d_j^* |\l_i\strans \wh{\bSigma}\l_j^*|}{ d_k^{*2}} \\[5pt]
    &\leq c \sum_{k+1 \leq j \leq r^*,\,  j  < i} \frac{d_j^{*2}}{d_i^* d_k^{*2} \sqrt{n}} + c \sum_{k+1 \leq j \leq r^*,\,  j  > i} \frac{d_i^{*2}}{d_j^* d_k^{*2} \sqrt{n}}\\[5pt]
    &\leq   \frac{c r^*}{d_{r^*} \sqrt{n}}.
\end{align*}
From Condition \ref{con4} that $r^*\gamma_n = o(d^{*}_{r^*})$ and  $\gamma_n = (r^*+s_u+s_v)^{1/2}\eta_n^2\{n^{-1}\log(pq)\}^{1/2} \geq n^{-1/2}$, it holds that  $r^*n^{-1/2} = o(d^{*}_{r^*})$, which further leads to
\begin{align}\label{czhkasa}
	 \sum_{k+1 \leq j \leq r^*,\,  j \neq i}\frac{ |z_{ij}^*|}{|z_{kk}^* - z_{ii}^*|} = o(1).
\end{align} 

It remains to show that $\sum_{ k+1 \leq j \leq r^*, j \neq i} |\wt{z}_{ij}| =  o( |\wt{z}_{kk} - \wt{z}_{ii}| )$.
For term $|\wt{z}_{ij}|$, an application of similar arguments as for \eqref{zijij} gives that 
\begin{align*}
	|\wt{z}_{ij} - z_{ij}^* | = |  \wt{\u}_i\trans\wh{\bSigma}\wt{\u}_j - {\u}_i\strans\wh{\bSigma}\u_j^*| \leq  c \gamma_n \max\{d_i^*, d_j^*\}.
\end{align*}
It follows that
\[ |\wt{z}_{ij}| \leq |z_{ij}^*| + |\wt{z}_{ij} - z_{ij}^* | = |z_{ij}^*| + O( \gamma_n \max\{d_i^*, d_j^*\}).  \]
Since $|{z}_{kk}^* - {z}_{ii}^* | \geq c d_k^{*2}$ and 
$|\wt{z}_{kk} - \wt{z}_{ii} | \geq c d_k^{*2},$ in light of \eqref{czhkasa}  it suffices to show that $\sum_{k+1 \leq j \leq r^*, \,j \neq i}\gamma_n \max\{d_i^*, d_j^*\} = o(|\wt{z}_{kk} - \wt{z}_{ii} |)$. Since $i, j \geq k+1$ such that $\max\{d_i^*, d_j^*\} \leq d_{k+1}^*$, 
we can show that 
\begin{align*}
	\sum_{k+1 \leq j \leq r^*, \, j \neq i}\frac{  \max\{d_i^*, d_j^*\}\gamma_n }{|\wt{z}_{kk} - \wt{z}_{ii} |} \leq  \frac{c  d_{k+1}^* r^* \gamma_n}{ d_k^{*2}} 
	\leq c  \frac{r^* \gamma_n}{d_{k^*}} \leq c  \frac{r^* \gamma_n}{d_{r^*}} = o(1),
\end{align*}
where we have used Condition \ref{con4} that $r^*\gamma_n = o(d^{*}_{r^*})$.
Therefore, it holds that
\begin{align}
	\sum_{k+1 \leq j \leq r^*, \, j \neq i}\frac{ |\wt{z}_{ij}|}{|\wt{z}_{kk} - \wt{z}_{ii} |} = o(1), \nonumber
\end{align} 
which concludes the proof of Lemma \ref{lemmzzzz}.

\subsection{Lemma \ref{lemm:wexist2} and its proof} \label{new.Sec.B.16}

\begin{lemma}\label{lemm:wexist2}
	Assume that all the conditions of Theorem \ref{theorkapor} are satisfied. For each given $k$ with $1 \leq k \leq r^*$, with probability at least
	$	1- \theta_{n,p,q}$ with $\theta_{n,p,q}$ given in \eqref{thetapro}, both $\I_{r^*-k} - \wt{z}_{kk}^{-1}(\wt{\U}^{(2)})\trans\wh{\bSigma}\wt{\U}^{(2)}$ and  $\I_{r^*-k} -z_{kk}^{*-1}(\U^{*(2)})\trans\wh{\bSigma} \U^{*(2)}$ are nonsingular. Moreover, $\wt{\W}_k $ and $\W_k^{*}$ introduced in \eqref{eqwkge1} and \eqref{eqwkge2}, respectively, are well-defined.
\end{lemma}

\noindent \textit{Proof}. 
Let us first analyze term $\I_{r^*-k} - \wt{z}_{kk}^{-1}(\wt{\U}^{(2)})\trans\wh{\bSigma}\wt{\U}^{(2)}$, which is equivalent to analyzing the nonsingularity of matrix $\A =: \wt{z}_{kk}\I_{r^*-k} - (\wt{\U}^{(2)})\trans\wh{\bSigma}\wt{\U}^{(2)} \in \mathbb{R}^{(r^*-k) \times (r^*-k)}$.
Denote by $\A = (a_{ij})$. Then we can see that for each $i, j \in \{k+1, \ldots, r^*\}$, 
\begin{align}
	a_{ij} = \left\{\begin{array}{l}
		\wt{z}_{kk} - \wt{z}_{ii} \quad \text{ if }  i=j, \\
		-\wt{z}_{ij} \quad \text{ if }  i \neq j.
		\end{array}\right.
\end{align}
From Lemma \ref{lemmzzzz}, it holds that $ \sum_{j \neq i} |a_{ij}| = o(|a_{ii}|)$ for each $i \in \{k+1, \ldots, r^*\}$. Hence, it follows that
\[ |a_{ii}| > \sum_{j \neq i}|a_{ij}|  \]
for all $ i \in  \{k+1, \ldots, r^*\}$,  which entails that matrix $\A$ is strictly diagonally dominant.
With the aid of the Levy--Desplanques Theorem  in \cite{horn2012matrix}, we see that matrix $\A$  is nonsingular and thus matrix $\I_{r^*-k} - \wt{z}_{kk}^{-1}(\wt{\U}^{(2)})\trans\wh{\bSigma}\wt{\U}^{(2)}$ is also nonsingular. Moreover, using similar arguments we can also show that matrix $\I_{r^*-k} -z_{kk}^{*-1}(\U^{*(2)})\trans\wh{\bSigma} \U^{*(2)}$ is nonsingular. Therefore, both $\wt{\W}_k $ and $\W_k^{*}$ are
well-defined and satisfy the property in Proposition \ref{prop:rankapo3}, which completes the proof of Lemma \ref{lemm:wexist2}.

\subsection{Lemma \ref{rankr:boundm} and its proof} \label{new.Sec.B.17}

\begin{lemma}\label{rankr:boundm}
		Assume that all the conditions of Theorem \ref{theorkapor} are satisfied. For each given $k$ with $1 \leq k \leq r^*$, $\wt{\M}_k = -\wt{z}_{kk}^{-1}\wh{\bSigma}\wt{\C}^{(2)}$, and ${\M}_k^* = -{z}_{kk}^{*-1}\wh{\bSigma}{\C}^{*(2)}$,
		with probability at least
		$	1- \theta_{n,p,q}$ we have that 
		\begin{align*}
			&\norm{ {\M}_k^* }_2  \leq c  d_k^{*-2} d_{k+1}^*, \ \norm{ \wt{\M}_k }_2 \leq c  d_k^{*-2}d_{k+1}^*,\\[5pt]
			&\|\wt{\M}_{k}  - {\M}_{k}^*  \|_2 
			\leq  c (r^*+s_u+s_v)^{1/2} \eta_n^2\{n^{-1}\log(pq)\}^{1/2}d_k^{*-2},
		\end{align*}
		where $\theta_{n,p,q}$ is given in \eqref{thetapro} and $c$ is some positive constant.
	\end{lemma}
	
	\noindent \textit{Proof}. 
	The proof of Lemma \ref{rankr:boundm} is similar to that of Lemma \ref{rankr:boundm0} in Section \ref{new.Sec.B.10}. Notice that $ \norm{{\U}^{*(2)}}_0 \leq  \norm{ \U^*}_0 = s_u $ and $(\U^{*(2)})\trans\U^{*(2)} =  \diag{d_{k+1}^{*2},\ldots, d_{r^*}^{*2}} $. 
	Using similar arguments as for \eqref{uuszfaa22} and \eqref{czasdqz}, we can obtain that 
	\begin{align}
		\norm{ \U^{*(2)}}_2 \leq cd_{k+1}^{*} \ \text{ and } \ \norm{\wh{\bSigma}\U^{*(2)}}_2 \leq cd_{k+1}^{*}. \label{dsazzbb}
	\end{align}
From Definition \ref{lemmsofar}, we have that 
\begin{align}
	&\norm{\wt{\U}^{(2)} - {\U}^{*(2)}}_0 \leq  \norm{\wt{\U} - {\U}^*}_0 \leq 2(r^* + s_u + s_v), \nonumber  \\[5pt]
	&\norm{\wt{\U}^{(2)} - {\U}^{*(2)}}_2 \leq \norm{\wt{\U}^{(2)} - {\U}^{*(2)}}_F \leq \norm{\wt{\U} - {\U}^*}_F \leq  c\gamma_n.  \label{dszcczz}
\end{align}
An application of similar arguments as for  \eqref{uuszfaa22} leads to 
\begin{align}
	\norm{\wh{\bSigma}(\wt{\U}^{(2)} - {\U}^{*(2)})}_2 \leq c\norm{\wt{\U}^{(2)} - {\U}^{*(2)}}_2 \leq c\gamma_n. \label{ccjkoi}
\end{align} 
Then for sufficiently large $n$, it follows that
\begin{align} 
	&\norm{ \wt{\U}^{(2)}}_2 \leq \norm{\wt{\U}^{(2)} - {\U}^{*(2)}}_2 +\norm{ {\U}^{*(2)}}_2 \leq c d_{k+1}^*, \label{adzbbb}\\
	&\norm{ \wh{\bSigma}\wt{\U}^{(2)}}_2 \leq \norm{\wh{\bSigma}(\wt{\U}^{(2)} - {\U}^{*(2)})}_2 +\norm{ \wh{\bSigma}{\U}^{*(2)}}_2 \leq c d_{k+1}^*. \label{adzbbb2}
\end{align}

For $\wt{\M}_k = -\wt{z}_{kk}^{-1}\wh{\bSigma}\wt{\C}^{(2)}$, we can deduce that 
	\begin{align*}
		\norm{ \wt{\M}_k }_2 & \leq |\wt{z}_{kk}^{-1}|
		\|  \wh{\bSigma} \wt{\U}^{(2)} (\wt{\V}^{(2)})\trans \|_2 \leq   |\wt{z}_{kk}^{-1}|
		\|  \wh{\bSigma} \wt{\U}^{(2)}\|_2 \| (\wt{\V}^{(2)})\trans \|_2 \\
  &\leq c  d_k^{*-2}d_{k+1}^*,
	\end{align*}
	where we have used part (c) of Lemma \ref{lemmauv}, $\norm{ \wh{\bSigma}\wt{\U}^{(2)}}_2 \leq c d_{k+1}^*$, and $\| \wt{\V}_k^{(2)} \|_2 = 1$.
	With similar arguments, we can show that $$\norm{ {\M}_k^* }_2 \leq c  d_k^{*-2} d_{k+1}^*.$$ 
 For term $\norm{ \wt{\M}_k  - \M_{k}^{*}}_2$, with the aid of similar arguments as for \eqref{czxcaz}--\eqref{czdadqzz}, it holds that 
\begin{align}
	\norm{ \wh{\bSigma}(\wt{\C}^{(2)} - \C^{*(2)})}_2 \leq  c (r^*+s_u+s_v)^{1/2}\eta_n^2\left\{n^{-1}\log(pq)\right\}^{1/2}. \nonumber
\end{align}
Together with part (c) of Lemma \ref{lemmauv}, $\| \wh{\bSigma} \U^{*(2)} \|_2 \leq c d_{k+1}^*$, and $  \| ({\V}^{*(2)})\trans \|_2 = 1$, we can obtain that 
	\begin{align*}
		\norm{ \wt{\M}_k  - \M_{k}^{*}}_2 & \leq  | \wt{z}_{kk}^{-1} - {z}_{kk}^{*-1}| \| \wh{\bSigma} \U^{*(2)} \|_2\| ({\V}^{*(2)})\trans \|_2 + |  {z}_{kk}^{*-1}| \| \wh{\bSigma}(\wt{\C}^{(2)} - \C^{*(2)}  )  \|_2  \nonumber\\[5pt]
		& \leq  c (r^*+s_u+s_v)^{1/2} \eta_n^2\{n^{-1}\log(pq)\}^{1/2} d_k^{*-2}.
	\end{align*}
This concludes the proof of Lemma \ref{rankr:boundm}.

\subsection{Lemma \ref{rankr:aww2} and its proof} \label{new.Sec.B.18}

\begin{lemma}\label{rankr:aww2}
		Assume that all the conditions of Theorem \ref{theorkapor} are satisfied. For each given $k$ with $1 \leq k \leq r^*$, an arbitrary
		$\a\in\R^p $, and  $\wt{\W}_k $ and $\W_k^{*}$ defined in \eqref{eqwkge1} and \eqref{eqwkge2},
		respectively, with probability at least
		$	1- \theta_{n,p,q}$ with $\theta_{n,p,q}$ given in \eqref{thetapro}, we have that 
		\begin{enumerate}[label=\rm{(\alph*)}]
			\item 	$\max_{1 \leq i \leq p}\norm{\w_i^*}_0  \leq 2\max\{s_{\max}, r^*+s_u + s_v \},\\  \max_{1 \leq i \leq p}\norm{\wt{\w}_i}_0 \leq 2\max\{s_{\max}, 3(r^*+s_u + s_v) \}$, \\
			$\max_{1 \leq i \leq p}\norm{\wt{\w}_i - \w_i^*}_0 \leq 3 (r^*+s_u + s_v)$;
			\item  $ \max_{1 \leq i \leq p}\norm{\w_i^*}_2 \leq c, \ \max_{1 \leq i \leq p}\norm{\wt{\w}_i}_2 \leq c$,
			\\[5pt] $ \max_{1 \leq i \leq p}\norm{\wt{\w}_i - \w_i^*}_2 \leq  c(r^*+s_u+s_v)^{1/2}\eta_n^2\{n^{-1}\log(pq)\}^{1/2}d_{k+1}^{*}d_k^{*-2}$;
			\item $ \norm{\a\trans\W_k^*}_2 \leq c \norm{\a}_0^{1/2}\norm{\a}_2$;
			\item $\norm{\a\trans(\wt{\W}_k-\W_k^*)}_2 \leq c \norm{\a}_0^{1/2}\norm{\a}_2 (r^*+s_u+s_v)^{1/2}\eta_n^2\{n^{-1}\log(pq)\}^{1/2}d_{k+1}^{*}d_k^{*-2}$;
			\item $ \norm{\a\trans\wt{\W}_k}_2 \leq c  \norm{\a}_0^{1/2}\norm{\a}_2$,
		\end{enumerate}
		where we denote $d_{r*+1}^* = 0$,  $\wt{\w}_i\trans$ and $ \w_i\strans$ are the $i$th rows of $\wt{\W}_k$ and $\W^*_k$, respectively, with $i = 1, \ldots, p$, and $c$ is some positive constant.
	\end{lemma}

\noindent \textit{Proof}. 
	Similar to the proof of Lemma \ref{lemma:1rk4} in Section \ref{new.Sec.B.5}, an application of similar arguments as for \eqref{eq:aw1}--\eqref{eq:aw3} yields that	once the results in parts (a) and (b) are established, we can obtain immediately the results in parts (c)--(e). Hence, we need only to prove parts (a) and (b), which will be based on the proof of Lemma \ref{rankr:aw} in Section \ref{new.Sec.B.11}. Compared to Lemma \ref{rankr:aw}, we see that the only difference is that matrices $\wt{\U}_{-k}$ and $\U^*_{-k}$ in  $\wt{\W}_k $ and $\W_k^{*}$ of Lemma \ref{rankr:aw}  are now replaced with their submatrices $\wt{\U}^{(2)}$ and $\U^{*(2)}$ in this lemma. As a result, $\wt{\U}^{(2)}$ and $\U^{*(2)}$ will only be more sparse than  $\wt{\U}_{-k}$ and $\U^*_{-k}$, respectively.
	Then applying similar arguments as in the proof of part (a) of Lemma \ref{rankr:aw}, we can obtain the results in part (a) of the current lemma.

	It remains to show part (b), which also follows similar technical arguments as in the proof of part (b) of Lemma \ref{rankr:aw}. In view of  \eqref{dsazzbb}, \eqref{adzbbb}, and \eqref{adzbbb2}, we can deduce that
		\begin{align}\label{rreq1}
			\{\norm{\wh{\bSigma}\U^{*(2)}}_2,  \norm{\wh{\bSigma}\wt{\U}^{(2)}}_2, \
			\norm{\wt{\U}^{(2)}}_2, \norm{\U^{*(2)}}_2  \} \leq c d_{k+1}^*.
		\end{align}				
		Denote by $  \A^* = (z_{kk}^{*}\I_{r^*-k} - (\U^{(2)*})\trans\wh{\bSigma} \U^{*(2)})^{-1}$, $\A_{0}^{-1} = \A^*$,  $\wt{\A} = (\wt{z}_{kk}\I_{r^*-k} -(\wt{\U}^{(2)})\trans\wh{\bSigma} \wt{\U}^{(2)})^{-1}$, and $\wt{\A}_{0}^{-1} = \wt{\A}$. 
		Using the technical arguments in the proof of Lemma \ref{lemm:wexist2} in Section \ref{new.Sec.B.16}, we see that both $\A_{0}$ and $\wt{\A}_{0}$ are strictly diagonally dominant. Similar to \eqref{rreq32}, it holds that 
		\begin{align}\label{rreq321}
			\norm{ \A^* }_2 & = \norm{ \A_{0}^{-1} }_2 \leq c({\min_{k < i  \leq r^*}|z_{kk}^* - z_{ii}^*|})^{-1} \nonumber\\
   &\leq c{|z_{kk}^* - z_{k+1, k+1}^*|}^{-1} \leq cd_k^{*-2},
		\end{align}
		where the last inequality above is due to \eqref{zkkjja}. Moreover, with similar arguments we have that 
		\begin{align}\label{dzxvcewhfkas}
			\norm{\wt{\A} }_2 = \norm{ \A_{0}^{-1} }_2 \leq cd_k^{*-2}.
		\end{align}

		Observe that $\w_i\strans = \widehat{\btheta}_i\trans (\I_p  +   \wh{\bSigma} \U^{*(2)} \A^*(\U^{*(2)})\trans),$ where  $\wh{\btheta}_i\trans$ is the $i$th row of $\wh{\bTheta}$. It follows from Definition \ref{defi2:acceptable} that $\max_{1 \leq i \leq p}\norm{\widehat{\btheta}_i}_2 \leq c$, \eqref{rreq1}, and \eqref{rreq321} that
		\begin{align*}
			\max_{1 \leq i \leq p}\norm{\w_i^*}_2 & \leq \max_{1 \leq i \leq p}\norm{\widehat{\btheta}_i}_2  \norm{\I_p  +     \wh{\bSigma} \U^{*(2)} \A^*(\U^{*(2)})\trans }_2 \\[5pt]
			&\leq \max_{1 \leq i \leq p}\norm{\widehat{\btheta}_i}_2(1 +  \norm{\wh{\bSigma}\U^{*(2)}}_2 \norm{\A^*}_2 \norm{(\U^{*(2)})\trans}_2)
			\\
   &\leq  c \max\{1, d_{k+1}^{*2}/d_{k}^{*2}\} \leq c.
		\end{align*}
  Since $\wt{\w}_i\trans = \widehat{\btheta}_i\trans (\I_p  + \wh{\bSigma}\wt{\U}^{(2)}\wt{\A} (\wt{\U}^{(2)})\trans)$, we can show that 
		\begin{align}\label{eqdad211}
			\max_{1\leq i \leq p}\|\wt{\w}_i - \w_i^*\|_2
			&\leq \max_{1\leq i \leq p}\|\wh{\btheta}_i\trans\|_2 \norm{ \wh{\bSigma}\wt{\U}^{(2)}\wt{\A} (\wt{\U}^{(2)})\trans -    \wh{\bSigma} \U^{*(2)} \A^*(\U^{*(2)})\trans  }_2 \nonumber \\[5pt]
			&\leq c\norm{   \wh{\bSigma}\wt{\U}^{(2)}\wt{\A} (\wt{\U}^{(2)})\trans -    \wh{\bSigma} \U^{*(2)} \A^*(\U^{*(2)})\trans   }_2.
		\end{align}
		By some simple calculations, the term above can be decomposed as
	\begin{align}\label{eqttt122222}
	  & \wh{\bSigma}\wt{\U}^{(2)}\wt{\A} (\wt{\U}^{(2)})\trans -    \wh{\bSigma} \U^{*(2)} \A^*(\U^{*(2)})\trans  =  (\wh{\bSigma}\wt{\U}^{(2)} - \wh{\bSigma} \U^{*(2)})\wt{\A} (\wt{\U}^{(2)})\trans    \nonumber\\
	 & \quad
	 + \wh{\bSigma} \U^{*(2)}(\wt{\A} - \A^* )(\wt{\U}^{(2)})\trans+  \wh{\bSigma} \U^{*(2)}\A^*(\wt{\U}^{(2)} - \U^{*(2)}  )\trans .
	\end{align}

		From \eqref{dszcczz} and \eqref{ccjkoi}, we see that 
		\begin{align}
			&\norm{\wh{\bSigma}(\wt{\U}^{(2)} - \U^{*(2)})}_2 \leq  c    (r^*+s_u+s_v)\eta_n^2\{n^{-1}\log(pq)\}^{1/2},  \label{eqsu-u2}\\
			&\norm{\wt{\U}^{(2)} - \U^{*(2)}}_2 \leq c(r^*+s_u+s_v)\eta_n^2\{n^{-1}\log(pq)\}^{1/2} . \label{equ00kk2}
		\end{align}	
		Also, applying similar arguments as for  \eqref{eqaa123}--\eqref{equaua1}, it holds that 
		\begin{align}\label{aaczewqd}
			\norm{\wt{\A} - \A^*}_2 \leq c (r^*+s_u+s_v)\eta_n^2\{n^{-1}\log(pq)\}^{1/2}d_k^{*-3}.
		\end{align}
	Combining \eqref{rreq1} and  \eqref{eqdad211}--\eqref{aaczewqd} yields that 
		\begin{align}
			\max_{1\leq i \leq p}\|\wt{\w}_i - \w_i^*\|_2
			\leq c(r^*+s_u+s_v)\eta_n^2\{n^{-1}\log(pq)\}^{1/2}d_{k+1}^{*}d_k^{*-2}.  \nonumber
		\end{align}
		Further, by the triangle inequality we have that 
\begin{align}
		\max_{1\leq i\leq p}\norm{ \wt{\w}_i}_2
	\leq 	\max_{1\leq i\leq p}\norm{ \wt{\w}_i - \w_i^* }_2 + 	\max_{1\leq i\leq p}\norm{ \w_i^* }_2 \leq c. \nonumber
\end{align}
For $k = r^*$, it holds that $\wt{\W}_k = \W^*_k = \wh{\bTheta}$. Therefore, based on Definition \ref{defi2:acceptable},  all conclusions of this lemma still hold using similar analysis as above.
This completes the proof of Lemma \ref{rankr:aww2}.

\subsection{Lemma \ref{prop:tay2lor1222r} and its proof} \label{new.Sec.B.19}

\begin{lemma}\label{prop:tay2lor1222r}
Assume that all the conditions of Theorem \ref{theorkapor} are satisfied. Then for $\wt{\W}_k$ given in \eqref{eqwkge1} and  an arbitrary $\a \in \mathbb{R}$, with probability at least
$1- \theta_{n,p,q}$ we have that 
\begin{align*}
	& |\a\trans\wt{\W}_k (\wt{\psi}_k(\wt{\u}_k,\wt{\boldeta}_k) - \wt{\psi}_k(\wt{\u}_k,\boldeta_k^*) )| \\[5pt]
	&\leq  c\norm{\a}_0^{1/2}\norm{\a}_2 \max\{s_{\max}^{1/2} , (r^*+s_u+s_v)^{1/2}, \eta_n^2\} (r^*+s_u+s_v)\eta_n^2\{n^{-1}\log(pq)\}\\
 &\quad\times\max\{d_k^{*-1}, d_k^{*-2}\},
\end{align*}
where $\theta_{n,p,q}$ is given in \eqref{thetapro} and $c$ is some positive constant.
\end{lemma}

\noindent \textit{Proof}. 
The proof of Lemma \ref{prop:tay2lor1222r} is similar to that of Lemma \ref{prop:taylor12} for the general rank case in Section \ref{new.Sec.B.2}. Notice that the nuisance parameter is $\boldeta_k = \left[\v_{k}\trans, \v_{k+1}\trans  \ldots, \v_{r^*}\trans, \u_{k+1}\trans, \ldots, \u_{r^*}\trans \right]\trans$.
By the definition of  $\wt{\psi}_k(\u_k,\boldeta_k)$, we have that 
\begin{align}
	\wt{\psi}_k(\u_k,\boldeta_k)
	&= \der{L}{\u_k} - \M\der{L}{\boldeta_k} \nonumber\\
 &= \der{L}{\u_k} - \left(\M_k^v \der{L}{\v_k}+ \sum_{j = k+1}^{r^*}  \M_j^u \der{L}{\u_j} + \sum_{j = k+1}^{r^*}\M_j^v \der{L}{\v_j}\right). \nonumber
\end{align}
From Proposition \ref{prop:rankapo2},  we see that  $\M_j^u = \0$ and $\M_j^v = \0$ for each $j \in \{ k+1, \ldots, r^*\}$, which means that we need only to consider $\v_k$ as the nuisance parameter. In light of the derivatives \eqref{der:r1} and \eqref{der:r2}, it holds that
\begin{align*}
	\wt{\psi}_k(\u_k,\boldeta_k)
	&= \der{L}{\u_k} - \M_k^v \der{L}{\v_k}  \\[5pt]
	&= \wh{\bSigma}\u_k - n^{-1}\X\trans\Y\v_k + \wh{\bSigma}\wh{\C}^{(1)} \boldsymbol{v}_{k} \\
 &\quad- \M_k^v ( \v_k\u_k\trans\wh{\bSigma}\u_k - n^{-1}\Y\trans\X\u_k + (\wh{\C}^{(1)})\trans\wh{\bSigma}\boldsymbol{u}_{k} ). \label{t11das}
\end{align*}

For each arbitrary fixed $\M_k^v$, we can see from the representation above that $\wt{\psi}_k(\u_k, \boldeta_k)$ is only a function of $\u_k$ and $\v_k$, which entails that we need only to do the Taylor expansion of $\wt{\psi}_k(\u_k, \boldeta_k)$  with respect to $\v_k$.
Similar to the proof of Lemma \ref{prop:taylor} in Section \ref{sec:proof:taylor}, we can obtain the Taylor expansion of $\wt{\psi}_k(\u_k, \boldeta_k)$ 
\begin{align*}
	\wt{\psi}_k \left( \u_k,  {\boldeta}_k  \right) & = \wt{\psi}_k\left( \u_k, \boldeta_k^* \right)
	+  \der{\wt{\psi}_k( \u_k, {\boldeta}_k  )}{{\v}_k\trans}\Big|_{\v_k^*}(\I_q - \v_k^*\v_k\strans) \exp^{-1}_{\v_k^*}({\v}_k) \\
 & \quad+ \r_{\v_k^*},
\end{align*}
where  the Taylor remainder term satisfies that
\[ \|\r_{\v_k^*}\|_2 = O(\| \exp^{-1}_{\v_k^*}({\v}_k) \|_2^2). \]
Moreover, it follows that
\[ \der{\wt{\psi}_k( \u_k,  {\boldeta}_k   )}{{\v}_k\trans}\Big|_{\v_k^*} = - n^{-1}\X\trans\Y  + \wh{\bSigma}\wh{\C}^{(1)} - \u_k\trans\wh{\bSigma}\u_k\M_k^v.\]
Then from Proposition \ref{prop:rankapo2} that $\M^v_k =  -z_{kk}^{-1}\wh{\bSigma}\C^{(2)}$ and the initial estimates in Definition \ref{lemmsofar}, we can deduce that 
\begin{align}
    &\wt{\psi}_k(\wt{\u}_k,\wt{\boldeta}_k) - \wt{\psi}_k(\wt{\u}_k,\boldeta^*_k) \nonumber\\
    &= (- n^{-1}\X\trans\Y  + \wh{\bSigma}\wh{\C}^{(1)}  + \wh{\bSigma}\wt{\C}^{(2)})(\I_q - \v_k^*\v_k\strans) \exp^{-1}_{\v_k^*}(\wt{\v}_k)  + \r_{\v_k^*} \nonumber\\
	 &=( \wh{\bSigma}(\wt{\C}_{-k} - \C_{-k}) -n^{-1}\X\trans\E)(\I_q - \v_k^*\v_k\strans)\exp^{-1}_{\v_k^*}(\wt{\v}_k)  + \r_{\v_k^*}, \nonumber
\end{align}
where we slightly abuse the notation and denote the Taylor remainder term as 
\[\r_{\v_k^*} = O(\| \exp^{-1}_{\v_k^*}(\wt{\v}_k) \|_2^2). \]

We next bound term $\a\trans \wt{\W}_k (\wt{\psi}_k(\wt{\u}_k,\wt{\boldeta}_k) - \wt{\psi}_k(\wt{\u}_k,\boldeta^*_k) )$ above, which will follow similar arguments as for  \eqref{31ewdasd}--\eqref{saazs12}.
Observe that
\begin{align*}
	& |\a\trans \wt{\W}_k (\wt{\psi}_k(\wt{\u}_k,\wt{\boldeta}_k) - \wt{\psi}_k(\wt{\u}_k,\boldeta^*_k) )| \\[5pt]
	&\leq |\a\trans\wt{\W}_k( \wh{\bSigma}(\wt{\C}_{-k} - \C_{-k}^*) -n^{-1}\X\trans\E)(\I_q - \v_k^*\v_k\strans)\exp^{-1}_{\v_k^*}(\wt{\v}_k)|\\
 &\quad
 +	|\a\trans\wt{\W}_k \r_{\v_k^*}|. 
\end{align*}
Denote by $\wt{\w}_{k,i}\trans$ the $i$th row of $\wt{\W}_k$ with $i = 1, \ldots, p$.
From Lemma \ref{rankr:aww2}, we have that 
	\begin{align}
		\max_{1 \leq i \leq p}\norm{\wt{\w}_{k,i}}_0 \leq 2\max\{s_{\max}, 3(r^*+s_u+s_v)\} \ \text{ and } \ \max_{1 \leq i \leq p}\norm{\wt{\w}_{k,i}}_2 \leq c,  \nonumber
	\end{align}
which has the same upper bound as in \eqref{eqwwwazcs}. An application of similar arguments as for \eqref{eqr21} and \eqref{eqr22} leads to 
\begin{align*}
	&\norm{ \exp^{-1}_{\v_k^*}(\wt{\v}_k)}_0  \leq c(r^*+s_u +s_v),  \\
	&\|\exp^{-1}_{\v_k^*}(\wt{\v}_k)\|_2
	\leq c(r^*+s_u+s_v)^{1/2}\eta_n^2\{n^{-1}\log(pq)\}^{1/2}/d_k^*. 
\end{align*}
Similar to \eqref{31ewdasd}, it also holds that
\begin{align}
	\norm{ \wh{\bSigma}(\wt{\C}_{-k}  - \C_{-k}^*)}_2 \leq  c (r^*+s_u+s_v)^{1/2}\eta_n^2\left\{n^{-1}\log(pq)\right\}^{1/2}. \nonumber
\end{align}

In view of the above results, we can see that the upper bound for $|\a\trans\wt{\W}_k  (\wt{\psi}_k(\wt{\u}_k,\wt{\boldeta}_k) - \wt{\psi}_k(\wt{\u}_k,\boldeta^*_k) )|$ is similar to that for the general rank case of Lemma \ref{prop:taylor12} in Section \ref{new.Sec.B.2}. Similar to
\eqref{phiuuuu2}--\eqref{31ewdasd2}, it follows that
\begin{align*}
	& \Big|\a\trans \wt{\W}_k ( \wh{\bSigma}(\wt{\C}_{-k} - \C_{-k}^*) -n^{-1}\X\trans\E)(\I_q - \v_k^*\v_k\strans)\exp^{-1}_{\v_k^*}(\wt{\v}_k) \Big| \\
	& \leq  c \norm{\a}_0^{1/2} \norm{\a}_2 \max\{s_{\max}^{1/2}, (r^*+s_u+s_v)^{1/2}, \eta_n^2 \}(r^*+s_u+s_v)\eta_n^2\{n^{-1}\log(pq)\}
	/d_k^*. \nonumber
\end{align*}
Further, from \eqref{31ewdasd3} we can show that
\begin{align}
	&|\a\trans \wt{\W}_k \r_{\v_k^*} | 
	\leq  c\norm{\a}_0^{1/2} \norm{\a}_2  (r^*+s_u+s_v)\eta_n^4\{n^{-1}\log(pq)\}/d_k^{*2}. \nonumber
\end{align}
 Therefore, it holds that 
\begin{align}
	&|\a\trans\wt{\W}_k  (\wt{\psi}_k(\wt{\u}_k,\wt{\boldeta}_k) - \wt{\psi}_k(\wt{\u}_k,\boldeta^*_k) )| \nonumber \\
	& \leq c \norm{\a}_0^{1/2}\norm{\a}_2 \max\{s_{\max}^{1/2} , (r^*+s_u+s_v)^{1/2}, \eta_n^2\} (r^*+s_u+s_v)\eta_n^2\{n^{-1}\log(pq)\} \nonumber\\
 &\quad \times \max\{d_k^{*-1}, d_k^{*-2}\},  \nonumber
\end{align}
which concludes the proof of Lemma \ref{prop:tay2lor1222r}.

\subsection{Lemma \ref{lemma:1rk3rr} and its proof} \label{new.Sec.B.20}

\begin{lemma}\label{lemma:1rk3rr}
	Assume that all the conditions of Theorem \ref{theorkapor} are satisfied. For $\wt{\bepsilon}_{k}$ defined in \eqref{eqepweak}, $h_{k}$ defined in \eqref{eqhkweak}, and any $\a\in\mathcal{A}=\{\a\in\R^p:\norm{\a}_0\leq m,\norm{\a}_2=1\}$,
	with probability at least
	$	1- \theta_{n,p,q}$ it holds that
	\begin{align*}
		\abs{-\a\trans\wt{\W}_k\wt{\bepsilon}_{k} - h_{k}  / \sqrt{n}} \leq c m^{1/2} (r^*+s_u + s_v)^{3/2}\eta_n^2\{ n^{-1}\log(pq)\}d_k^{*-1},
	\end{align*}
	where $\theta_{n,p,q}$ is given in \eqref{thetapro} and $c$ is some positive constant.
\end{lemma}

\noindent \textit{Proof}. 
This proof follows similar technical arguments as in the proof of Lemma \ref{lemma:1rk3r} in Section \ref{new.Sec.B.14}.
Note that  $ \wt{\M}_{k} = -\wt{z}_{kk}^{-1}\wh{\bSigma}\wt{\C}^{(2)}$ and $\M_{k}^{*} = -z_{kk}^{*-1}\wh{\bSigma}\C^{*(2)}$.
Using similar arguments, we can show that 
$\a\trans \wt{\W}_k \wt{\M}_k$, $\a\trans \wt{\W}_k \wt{\M}_k  - \a\trans {\W}_k^{*}\M_k^{*}$, and  $\a\trans(\wt{\W}_k-{\W}_k^*)$ are all $s$-sparse with $s=c(r^*+s_u+s_v)$. An application of similar arguments as for \eqref{sdwqvcs2} and \eqref{a22ex22}--\eqref{eqathe22} gives that 
\begin{align*}
	&\abs{-\a\trans\wt{\W}_k \wt{\bepsilon}_k - h_k / \sqrt{n}} \leq  \norm{\a\trans \wt{\W}_k}_2 \norm{\wt{\M}_k}_2 \norm{n^{-1}\E\trans\X (\wt{\u}_k - {\u}_k^{*} )}_{2,s} \nonumber \\[5pt]
	&\quad + \norm{\a\trans \wt{\W}_k \wt{\M}_k  - \a\trans \W^{*}_k \M_k^{*}}_2 \norm{ n^{-1}\E\trans\X\u_k^*}_{2,s}  \\
 &\quad+  \norm{\a\trans(\wt{\W}_k-\W^*_k) }_2\norm{ n^{-1}\X\trans\E\v_k^*}_{2,s} \\[5pt]
	&\leq c s^{3/2}\eta_n^2\{n^{-1}\log(pq)\} \norm{\a\trans \wt{\W}_k}_2 \norm{\wt{\M}_k}_2 \\
 &\quad+ cs^{1/2}s_v^{1/2}\{n^{-1}\log(pq)\}^{1/2}\norm{\a\trans(\wt{\W}_k-\W^*_k) }_2 \\[5pt]
	&\quad  + cs^{1/2}s_u^{1/2}\{n^{-1}\log(pq)\}^{1/2}  d_k^{*} \norm{\a\trans \wt{\W}_k \wt{\M}_k  - \a\trans \W^{*}_k \M_k^{*}}_2.
\end{align*}
For the terms above, 
from Lemma \ref{rankr:boundm} we have that 
\begin{align*}
	&\norm{ {\M}_k^* }_2  \leq c  d_k^{*-2} d_{k+1}^*, \ \norm{ \wt{\M}_k }_2 \leq c  d_k^{*-2}d_{k+1}^*,\\[5pt]
	&\|\wt{\M}_{k}  - {\M}_{k}^*  \|_2 
	\leq  c (r^*+s_u+s_v)^{1/2} \eta_n^2\{n^{-1}\log(pq)\}^{1/2}d_k^{*-2},
\end{align*}

Moreover, it follows from parts (d) and (e) of Lemma \ref{rankr:aww2} that
\begin{align*}
	 &\norm{\a\trans(\wt{\W}_k-\W_k^*)}_2 \leq cm^{1/2} (r^*+s_u+s_v)^{1/2}\eta_n^2\{n^{-1}\log(pq)\}^{1/2}d_{k+1}^{*}d_k^{*-2}, \\
&\norm{\a\trans\wt{\W}_k}_2 \leq c  m^{1/2}.
\end{align*}
Then it holds that
\begin{align}
	&\norm{\a\trans \wt{\W}_k \wt{\M}_k  - \a\trans \W^{*}_k \M_k^{*}}_2 \leq  \norm{\a\trans \wt{\W}_k (\wt{\M}_k  -\M_k^{*}) }_2 + \norm{\a\trans (\wt{\W}_k  - \W^{*}_k) \M_k^{*}}_2 \nonumber\\
	&\leq  \norm{\a\trans \wt{\W}_k}_2 \norm{ \wt{\M}_k  -\M_k^{*} }_2 + \norm{\a\trans (\wt{\W}_k  - \W^{*}_k)}_2 \norm{ \M_k^{*}}_2 \nonumber \\
	&\leq    cm^{1/2} (r^*+s_u+s_v)^{1/2} \eta_n^2  \{n^{-1}\log(pq)\}^{1/2}d_k^{*-2}.  \nonumber
\end{align}
Thus, combining the above terms yields that
\begin{align*}
	\abs{-\a\trans\wt{\W}_k\wt{\bepsilon}_{k} - h_{k}  / \sqrt{n}} \leq c m^{1/2}  (r^*+s_u + s_v)^{3/2}\eta_n^2\{ n^{-1}\log(pq)\} d_k^{*-1}.
\end{align*}
This completes the proof of Lemma \ref{lemma:1rk3rr}.

\subsection{Lemma \ref{lemma:gaprk} and its proof} \label{new.Sec.B.21}

\begin{lemma}\label{lemma:gaprk}
	Assume that all the conditions of Theorem \ref{theorkapor} are satisfied.
	For $\wt{\M}_k =  - \wt{z}_{kk}^{-1}\wh{\bSigma}\wt{\C}^{(2)}$ and $\wt{\W}_k$ given in \eqref{eqwkge1} and an arbitrary
	$\a\in\R^p$, with probability at least
	$	1- \theta_{n,p,q}$ it holds that
	\begin{align*}
		|\a\trans\wt{\W}_k \wt{\M}_k (\C^{*(2)})\trans \wh{\bSigma} \u_k^*  | = o(\norm{\a}_0^{1/2}\norm{\a}_2   n^{-1/2}),
	\end{align*}
	where $\theta_{n,p,q}$ is given in \eqref{thetapro} and $c$ is some positive constant.
\end{lemma}

\noindent \textit{Proof}. 
Observe that $\wt{\M}_k =  - \wt{z}_{kk}^{-1}\wh{\bSigma}\wt{\C}^{(2)}$. With the aid of similar arguments as for  \eqref{caeqajhka}--\eqref{caeqajhka1}, it holds that
\begin{align*}
	 \norm{\wt{\M}_k (\C^{*(2)})\trans \wh{\bSigma} \u_k^* }_2  
	\leq  c  \sum_{j = k+1}^{r^*}  (d_j^{*2}/d_k^*) |\l_j\strans\wh{\bSigma}\l_k^*|.
\end{align*}
Together with Condition \ref{con:orth:rankr} that $\sum_{j = k+1}^{r^*}  (d_j^{*2}/d_k^*) |\l_j\strans\wh{\bSigma}\l_k^*| = o(n^{-1/2})$ and Lemma \ref{rankr:aww2} that $\norm{ \a\trans \wt{\W}_k}_2 \leq c\norm{\a}_0^{1/2}\norm{\a}_2$, it follows that
	\begin{align*}
		|\a\trans\wt{\W}_k \wt{\M}_k\C_{-k}\strans \wh{\bSigma} \u_k^*  | 
		& \leq  \norm{\a\trans\wt{\W}_k }_2 \norm{\wt{\M}_k \C_{-k}\strans \wh{\bSigma} \u_k^* }_2 
		\\
  &= o(\norm{\a}_0^{1/2}\norm{\a}_2  n^{-1/2}),
	\end{align*}	
which concludes the proof of Lemma \ref{lemma:gaprk}.

\subsection{Lemma \ref{lemma:k21rk2} and its proof} \label{new.Sec.B.22}

\begin{lemma}\label{lemma:k21rk2}
	 Assume that all the conditions of Theorem \ref{theorkapor} are satisfied.  For $\wt{\bdelta}_{k}$ defined in \eqref{eqdeweak} and an arbitrary	$\a\in\R^p$,
	with probability at least
	$	1- \theta_{n,p,q}$ it holds that
\begin{align}
	|\a\trans\wt{\W}_k\wt{\bdelta}_{k}|
	& \leq c \norm{\a}_0^{1/2}\norm{\a}_2  (r^*+s_u+s_v)^{1/2}\eta_{n}^2\{\log(pq)\}^{1/2} d_{k+1}^{*}d_k^{*-3}(\sum_{i=1}^{k-1}d_i^*)/n  \nonumber  \\
	& \quad +c \norm{\a}_0^{1/2}\norm{\a}_2(r^*+s_u+s_v)\eta_{n}^4\{n^{-1}\log(pq)\}  d_k^{*-2} d_{k+1}^{*}, \nonumber
\end{align}
	where $\theta_{n,p,q}$ is given in \eqref{thetapro} and $c$ is some positive constant.
	\end{lemma}

\noindent \textit{Proof}. 
Denote by $\wt{\bdelta}_k = \wt{\bdelta}_{0,k} +  \wt{\bdelta}_{1,k}$ with 
\begin{align*}
&\wt{\bdelta}_{0,k}  = \wt{\M}_k ((\wt{\v}_k - \v_k^*)\wt{\u}_k\trans - (\wt{\C}^{(2)} - \C^{*(2)})\trans)  \wh{\bSigma}(\wt{\u}_k - \u_k^*),\\[5pt]
&\wt{\bdelta}_{1,k} = - \wt{\M}_k  (\wh{\C}^{(1)} - \C^{*(1)})\trans  \wh{\bSigma} \wt{\u}_k.
\end{align*}
The derivation for the upper bound on $|\a\trans\wt{\W}_k\wt{\bdelta}_{0,k}|$ is similar to that in the proof of Lemma \ref{lemma:1rkr} in Section \ref{new.Sec.B.12}.
From Lemma \ref{rankr:boundm}, we see that $\norm{ \wt{\M}_k }_2 \leq c  d_k^{*-2}d_{k+1}^*$. 
Part (e) of Lemma \ref{rankr:aww2} entails that $ \norm{\a\trans\wt{\W}_k}_2 \leq c  \norm{\a}_0^{1/2}\norm{\a}_2.$ 
Further, observe that $\wt{\C}^{(2)}$  and $ \C^{*(2)}$ are submatrices of $\wt{\C}_{-k}$  and $ \C^{*}_{-k}$, respectively. It follows from similar arguments as for  \eqref{czxcaz}--\eqref{dzcvdsff} that 
\begin{align*}
	\norm{ \wt{\C}^{(2)}  - \C^{*(2)}}_2 
	\leq  c \gamma_n,
\end{align*} 
where $ \gamma_n = ({r^*}+s_u+s_v)^{1/2}\eta_n^2\{n^{-1}\log(pq)\}^{1/2}$. From the above results and Lemma \ref{lemmauv}, we can deduce that 
\begin{align}
|\a\trans\wt{\W}_k\wt{\bdelta}_{0,k}|  &\leq 
\norm{\a\trans \wt{\W}_k }_2 \norm{\wt{\M}_k }_2\norm{(\wt{\v}_k - \v_k^*)\wt{\u}_k\trans - (\wt{\C}^{(2)} - \C^{*(2)})\trans}_2 \norm{ \wh{\bSigma}(\wt{\u}_k - \u_k^*) }_2 \nonumber \\[5pt]
&\leq c \norm{\a}_0^{1/2}\norm{\a}_2   \gamma_n d_k^{*-2}d_{k+1}^* (\norm{\wt{\v}_k - \v_k^*}_2 \norm{\wt{\u}_k\trans }_2 + \norm{(\wt{\C}^{(2)} - \C^{*(2)})\trans}_2) \nonumber\\[5pt]
&\leq c \norm{\a}_0^{1/2}\norm{\a}_2  (r^*+s_u+s_v)\eta_{n}^4\{n^{-1}\log(pq)\}   d_k^{*-2}d_{k+1}^*. \label{czdjkqdaf}
\end{align}

We next bound term $|\a\trans\wt{\W}_k\wt{\bdelta}_{1,k}|$ above.
It can be seen that
\begin{align*}
\a\trans\wt{\W}_k\wt{\bdelta}_{1,k} & = - \a\trans\wt{\W}_k\wt{\M}_k (\wh{\C}^{(1)} - \C^{*(1)})\trans  \wh{\bSigma} {\u}_k^*  \\
&\quad+ \a\trans\wt{\W}_k\wt{\M}_k (\wh{\C}^{(1)} - \C^{*(1)})\trans  \wh{\bSigma} ({\u}_k^* -\wt{\u}_k ). 
\end{align*}
Notice that  $\wt{\v}_j\trans \wt{\v}_i = 0$ and $\v_j\strans\v_i^* = 0$ for each $ 1 \leq i \leq k- 1 $ and $k+1 \leq j \leq r^*$. It holds that
\begin{align}
    \wt{\M}_k \wt{\v}_i = -\wt{z}_{kk}^{-1} \widehat{\bSigma} \sum_{j = k+1}^{r^*} \wt{\u}_j \wt{\v}_j\trans \wt{\v}_i = \0 \ \text{ and } \ \M^*_k {\v}_i^* = - z_{kk}^{*-1} \wh{\bSigma}\sum_{j= k+1 }^{r^*} \u_i^*\v_j\strans\v_i^* =  \0. \nonumber
\end{align}
Then we can show that 
\begin{align}
& - \a\trans \wt{\W}_k\wt{\M}_k (\wh{\C}^{(1)} - \C^{*(1)})\trans  \wh{\bSigma} {\u}_k^* = - \a\trans\wt{\W}_k\wt{\M}_k \sum_{i=1}^{k-1} (\wt{\v}_i\wt{\u}_i^{ T} - \v_i^*\u_i\strans)  \wh{\bSigma} {\u}_k^*  \nonumber  \\[5pt]
	& =  \a\trans\wt{\W}_k\wt{\M}_k \sum_{i=1}^{k-1}   \v_i^*\u_i\strans  \wh{\bSigma} {\u}_k^* =  \a\trans\wt{\W}_k\wt{\M}_k \sum_{i=1}^{k-1} (\wt{\v}_i -  \v_i^* - \wt{\v}_i )  \u_i\strans  \wh{\bSigma} {\u}_k^* \nonumber  \\[5pt]
	& =  \a\trans\wt{\W}_k\wt{\M}_k \sum_{i=1}^{k-1} (\wt{\v}_i -  \v_i^*)  \u_i\strans  \wh{\bSigma} {\u}_k^*.
\end{align}
Hence, it follows that 
\begin{align*}
	 \a\trans\wt{\W}_k\wt{\bdelta}_{1,k} &= - \a\trans\wt{\W}_k\wt{\M}_k \sum_{i=1}^{k-1} (\wt{\v}_i -  \v_i^*)  \u_i\strans  \wh{\bSigma} {\u}_k^* + \a\trans\wt{\W}_k\wt{\M}_k (\wh{\C}^{(1)} - \C^{*(1)})\trans  \wh{\bSigma} ({\u}_k^* -\wt{\u}_k ) \\
	 &=: B_1 + B_2.
\end{align*}
We will bound the two terms $B_1$ and $B_2$ introduced above separately.

For the first term $B_1$ above, we have that 
\begin{align*}
	&| \sum_{i=1}^{k-1}\a\trans\wt{\W}_k\wt{\M}_k ( \wt{\v}_i- \v_i^*) {\u}_i\strans \wh{\bSigma}{\u}_k^*| \\
	&\leq  \norm{\a\trans\wt{\W}_k}_2 \norm{\wt{\M}_k}_2  \sum_{i=1}^{k-1}\norm{ d_i^* (\wt{\v}_i- \v_i^*)}_2 |d_k^*| |{\l}_i\strans \wh{\bSigma}{\l}_k^*| \\
	&\leq c  \norm{\a}_0^{1/2}\norm{\a}_2  d_k^{*-1} d_{k+1}^{*} \sum_{i=1}^{k-1}\norm{ d_i^* (\wt{\v}_i- \v_i^*)}_2 |{\l}_i\strans \wh{\bSigma}{\l}_k^*|,
\end{align*}
where the last step above has used Lemma \ref{rankr:boundm} and part (c) of Lemma \ref{rankr:aww2}. 
For term $|{\l}_i\strans \wh{\bSigma}{\l}_k^*|$ with $i = 1, \ldots, k$,
in view of Condition \ref{con:orth:rankr} it holds that 
\[  (d_k^{*2}/d_i^*) |\l_i\strans \wh{\bSigma} \l_k^*| = o(n^{-1/2}).  \]
Together with part (a) of Lemma \ref{lemmauv}, we can obtain that  
\begin{align*}
	\sum_{i=1}^{k-1}\norm{ d_i^* (\wt{\v}_i- \v_i^*)}_2 |{\l}_i\strans \wh{\bSigma}{\l}_k^*| \leq c \gamma_n \sum_{i=1}^{k-1} \frac{d_i^*}{d_k^{*2} \sqrt{n}},
\end{align*}
where $ \gamma_n = ({r^*}+s_u+s_v)^{1/2}\eta_n^2\{n^{-1}\log(pq)\}^{1/2}$.
This further leads to 
\begin{align}\label{sczcqqq0}
	| \sum_{i=1}^{k-1}\a\trans\wt{\W}_k\wt{\M}_k ( \wt{\v}_i- \v_i^*) {\u}_i\strans \wh{\bSigma}{\u}_k^*| 
\leq c  \norm{\a}_0^{1/2}\norm{\a}_2    \gamma_n {\sum_{i=1}^{k-1}d_i^*d_{k+1}^{*}}{d_k^{*-3} n^{-1/2}}. 
\end{align}

For term $B_2$ above, let us first bound term $\norm{ \wh{\C}^{(1)} - \C^{*(1)}}_2$. Observe that $\wh{\C}^{(1)} = \wt{\U}^{(1)} (\wt{\V}^{(1)})\trans$ and $\C^{*(1)} = \U^{*(1)}(\V^{*(1)})\trans$, where $\wt{\U}^{(1)} = (\wt{\u}_1, \ldots, \wt{\u}_{k-1})$, $\wt{\V}^{(1)} = (\wt{\v}_1, \ldots, \wt{\v}_{k-1})$, ${\U}^{*(1)} = ({\u}^*_1, \ldots, {\u}^*_{k-1})$, and ${\V}^{*(1)} = ({\v}^*_1, \ldots, {\v}^*_{k-1})$. Then it follows that
\begin{align*}
	\norm{ \wh{\C}^{(1)} - \C^{*(1)}}_2 &=	\norm{ \wt{\U}^{(1)} (\wt{\V}^{(1)})\trans - \U^{*(1)}(\V^{*(1)})\trans}_2 \nonumber\\[5pt]
	&\leq 	\norm{\wt{\U}^{(1)} (\wt{\V}^{(1)} -\V^{*(1)})\trans }_2 +
	\norm{ (\wt{\U}^{(1)} - \U^{*(1)})(\V^{*(1)})\trans}_2  \nonumber \\[5pt]
	&\leq \norm{ \wt{\L}^{(1)}}_2 \norm{\wt{\D}^{(1)}(\wt{\V}^{(1)} -\V^{*(1)})\trans }_2 +
	\norm{ \wt{\U}^{(1)} - \U^{*(1)}}_2 \norm{(\V^{*(1)})\trans}_2. \nonumber
\end{align*} 
It can be seen that $\norm{ \wt{\L}^{(1)}}_2 = \norm{(\V^{*(1)})\trans}_2 = 1$.
For term 	$\norm{(\wt{\V}^{(1)} -\V^{*(1)})\wt{\D}^{(1)}}_2,$  we have that 
\begin{align*}
	\norm{(\wt{\V}^{(1)} -\V^{*(1)})\wt{\D}^{(1)}}_2 &\leq \norm{\wt{\V}^{(1)}\wt{\D}^{(1)} -\V^{*(1)}\D^{*(1)} }_2 + \norm{\wt{\V}^{(1)}}_2 \norm{\wt{\D}^{(1)} -\D^{*(1)} }_2 \\
	&\leq  \norm{\wt{\V}^{(1)}\wt{\D}^{(1)} -\V^{*(1)}\D^{*(1)} }_F +  \norm{\wt{\D}^{(1)} -\D^{*(1)} }_F\\
	&\leq \norm{\wt{\V}\wt{\D} -\V^*\D^*}_F + \norm{\wt{\D} -\D^* }_F \\
	&\leq  c(r^*+s_u+s_v)^{1/2}\eta_{n}^2\{n^{-1}\log(pq)\}^{1/2},
\end{align*}
where we have used Definition \ref{lemmsofar} and $\norm{\wt{\V}_{-k}}_2  = 1$. Moreover, from Definition \ref{lemmsofar} it holds that
\begin{align*}
		\norm{ \wt{\U}^{(1)} - \U^{*(1)}}_2 & \leq 	\norm{ \wt{\U}^{(1)} - \U^{*(1)}}_F \leq 	\norm{ \wt{\U} - \U^{*}}_F \\
  &\leq c(r^*+s_u+s_v)^{1/2}\eta_{n}^2\{n^{-1}\log(pq)\}^{1/2}.
\end{align*}

Combining the above results gives that
\begin{align*}
	\norm{ \wh{\C}^{(1)} - \C^{*(1)}}_2 \leq  c (r^*+s_u+s_v)^{1/2}\eta_{n}^2\{n^{-1}\log(pq)\}^{1/2}.
\end{align*} 
It follows from part (c) of Lemma \ref{lemmauv}, Lemma \ref{rankr:boundm}, and Lemma \ref{rankr:aww2} that
\begin{align}
	& |\a\trans \wt{\W}_k\wt{\M}_k (\wh{\C}^{(1)} - \C^{*(1)})\trans  \wh{\bSigma} ({\u}_k^* -\wt{\u}_k )| \nonumber\\
	&\leq  \norm{\a\trans\wt{\W}_k}_2 \norm{\wt{\M}_k}_2    \norm{\wh{\C}^{(1)} - \C^{*(1)}}_2 \norm{ \wh{\bSigma}(\wt{\u}_k -{\u}_k^* )}_2 \nonumber \\
	&\leq c \norm{\a}_0^{1/2}\norm{\a}_2 (r^*+s_u+s_v)\eta_{n}^4\{n^{-1}\log(pq)\}  d_k^{*-2} d_{k+1}^{*}. \label{sczcqqq}
\end{align}
Thus, a combination of \eqref{sczcqqq0} and \eqref{sczcqqq} yields that
\begin{align}
	|\a\trans\wt{\W}_k\wt{\bdelta}_{1,k}|
	& \leq c \norm{\a}_0^{1/2}\norm{\a}_2  (r^*+s_u+s_v)^{1/2}\eta_{n}^2\{n^{-1}\log(pq)\}^{1/2} {\sum_{i=1}^{k-1}d_i^*d_{k+1}^{*}}{d_k^{*-3} n^{-1/2}} \nonumber  \\
	& \quad + c\norm{\a}_0^{1/2}\norm{\a}_2 (r^*+s_u+s_v)\eta_{n}^4\{n^{-1}\log(pq)\}  d_k^{*-2} d_{k+1}^{*}. \nonumber
\end{align}
Along with \eqref{czdjkqdaf}, it follows that
\begin{align}
	|\a\trans\wt{\W}_k\wt{\bdelta}_{k}|
	& \leq c \norm{\a}_0^{1/2}\norm{\a}_2  (r^*+s_u+s_v)^{1/2}\eta_{n}^2\{\log(pq)\}^{1/2} d_{k+1}^{*}d_k^{*-3}(\sum_{i=1}^{k-1}d_i^*)/n  \nonumber  \\
	& \quad +c \norm{\a}_0^{1/2}\norm{\a}_2(r^*+s_u+s_v)\eta_{n}^4\{n^{-1}\log(pq)\}  d_k^{*-2} d_{k+1}^{*}. \nonumber
\end{align}
This completes the proof of Lemma \ref{lemma:k21rk2}.

\blue{
\subsection{Lemma \ref{lemma:1rk3r1} and its proof} \label{new.Sec.B.260}

\begin{lemma}\label{lemma:1rk3r1}
	Assume that all the conditions of Theorem \ref{corollary1} are satisfied. For $\wt{\bepsilon}_k$ defined in \eqref{eprankr22}, $h_k = -\a\trans \W^{*}_k \M_{k}^{*} \E\trans\X {\u}_k^{*}/\sqrt{n}
+  \a\trans\W^*_k\X\trans\E\v_k^*/\sqrt{n}$, and any $\a\in\mathcal{A}=\{\a\in\R^p:\norm{\a}_0\leq m,\norm{\a}_2=1\}$,
	with probability at least
	$	1- \theta_{n,p,q}^{\prime \prime}$ we have that 
	\begin{align*}
		\abs{-\a\trans\wt{\W}_k\wt{\bepsilon}_k - h_k / \sqrt{n}} \leq c m^{1/2}  (r^*+s_u + s_v)^{2}\eta_n^2\{ n^{-1}\log(pq)\},
	\end{align*}
	where $\theta_{n,p,q}^{\prime \prime}$ is given in \eqref{thetapro77} and $c$ is some positive constant.
\end{lemma}

\noindent \textit{Proof}. 
	The proof of this lemma  follows similar technical arguments as in the proof of Lemma \ref{lemma:1rk3r}.
	By the proof of Lemma \ref{lemma:1rk3r}, it holds that 
$\a\trans \wt{\W}_k \wt{\M}_k$, $\a\trans \wt{\W}_k \wt{\M}_k  - \a\trans {\W}_k^{*}\M_k^{*}$, and  $\a\trans(\wt{\W}_k-{\W}_k^*)$ are $s$-sparse with $s=c(r^*+s_u+s_v)$. Then similar to \eqref{sdzvcqqqq}, we have that 
	\begin{align}
		&\abs{-\a\trans\wt{\W}_k \wt{\bepsilon}_k - h_k / \sqrt{n}}  \leq  |\a\trans \wt{\W}_k \wt{\M}_k n^{-1}\E\trans\X (\wt{\u}_k - {\u}_k^{*} )| \nonumber \\[5pt]
		&\quad + |\a\trans (\wt{\W}_k \wt{\M}_k  -  \W^{*}_k \M_k^{*}) n^{-1}\E\trans\X\u_k^*|  
 +  |\a\trans(\wt{\W}_k-\W^*_k)  n^{-1}\X\trans\E\v_k^*|.
		\label{sdwqvcs21}
	\end{align}

We first analyze term $|\a\trans \wt{\W}_k \wt{\M}_k n^{-1}\E\trans\X (\wt{\u}_k - {\u}_k^{*} )|$. Note that 
\begin{align*}
	\norm{\a\trans \wt{\W}_k\wt{\M}_k}_0 \leq s, \  \ \norm{\a\trans \wt{\W}_k\wt{\M}_k}_2 \leq c m^{1/2}.
\end{align*}
In addition, it holds that
\begin{align*}
	\norm{\wt{\u}_k - {\u}_k^{*}}_0 \leq s, \ \ \norm{\wt{\u}_k - {\u}_k^{*}}_2 \leq c \gamma_n.
\end{align*}
Thus, we can show that $ \a\trans \wt{\W}_k\wt{\M}_k/\norm{\a\trans \wt{\W}_k\wt{\M}_k}_2 \in K_q(s)$ and $  (\wt{\u}_k - {\u}_k^{*})/\norm{\wt{\u}_k - {\u}_k^{*}}_2 \in K_p(s)$, where $K_p(s):= \{ \mathbf{b} \in \mathbb{R}^p: \|\mathbf{b}\|_0 \leq s, \|\mathbf{b}\|_2 = 1 \}$. Conditional on the second fold of data, an application of Lemma \ref{lemm:ds} yields that 
\begin{align}\label{eqsaca}
	&|\a\trans \wt{\W}_k \wt{\M}_k n^{-1}\E\trans\X (\wt{\u}_k - {\u}_k^{*} )| \nonumber \\
	&\leq  \left|\frac{\a\trans \wt{\W}_k \wt{\M}_k }{\norm{\a\trans \wt{\W}_k\wt{\M}_k}_2}n^{-1}\E\trans\X \frac{\wt{\u}_k - {\u}_k^{*} }{\norm{\wt{\u}_k - {\u}_k^{*}}_2}\right| \norm{\a\trans \wt{\W}_k\wt{\M}_k}_2 \norm{\wt{\u}_k - {\u}_k^{*}}_2 \nonumber\\
	&\leq \sup_{\boldsymbol{h}_1 \in K_p(s), \boldsymbol{h}_2 \in K_q(s)} \left|\boldsymbol{h}_2\trans n^{-1}\E\trans\X \boldsymbol{h}_1\right| \norm{\a\trans \wt{\W}_k\wt{\M}_k}_2 \norm{\wt{\u}_k - {\u}_k^{*}}_2 \nonumber\\ 
	&\leq c m^{1/2} s^{1/2} \{n^{-1}\log(pq)\}^{1/2}  (r^*+s_u+s_v)^{1/2} \eta_n^2\{n^{-1}\log(pq)\}^{1/2} \nonumber\\
	&= c m^{1/2} (r^*+s_u+s_v)\eta_n^2\{n^{-1}\log(pq)\}.
\end{align} 

For the second and third terms in \eqref{sdwqvcs21}, 
observe that 
\begin{align*}
	&\|\a\trans (\wt{\W}_k \wt{\M}_k  -  \W^{*}_k \M_k^{*})\|_0 \leq s,   \|\a\trans (\wt{\W}_k \wt{\M}_k  -  \W^{*}_k \M_k^{*})\|_2 \leq c m^{1/2} \gamma_n, \norm{\u_k^*}_0 \leq s, \norm{\u_k^*}_2 \leq c, \\
	&\|\a\trans (\wt{\W}_k \wt{\M}_k  -  \W^{*}_k \M_k^{*})\|_0 \leq s, \ \  \|\a\trans (\wt{\W}_k \wt{\M}_k  -  \W^{*}_k \M_k^{*})\|_2 \leq c \gamma_n, \norm{\v_k^*}_0 \leq s, \norm{\v_k^*}_2 \leq c.
\end{align*}
Based on the above results, similar to \eqref{eqsaca}, applying Lemma \ref{lemm:ds} leads to 
\begin{align}
	&|\a\trans (\wt{\W}_k \wt{\M}_k  -  \W^{*}_k \M_k^{*}) n^{-1}\E\trans\X\u_k^*| \leq  c m^{1/2} (r^*+s_u+s_v)\eta_n^2\{n^{-1}\log(pq)\}, \\
   &|\a\trans(\wt{\W}_k-\W^*_k)  n^{-1}\X\trans\E\v_k^*| \leq  c m^{1/2} (r^*+s_u+s_v)\eta_n^2\{n^{-1}\log(pq)\}. \label{szcasdz}
\end{align}

Therefore, combining \eqref{sdwqvcs21}--\eqref{szcasdz} gives that 
	\begin{align*}
		\abs{-\a\trans\wt{\W}_k\wt{\bepsilon}_k - h_k  / \sqrt{n}} \leq c m^{1/2}  (r^* +s_u + s_v)^{2}\eta_n^2\{ n^{-1}\log(pq)\}.
	\end{align*}
This completes the proof of Lemma \ref{lemma:1rk3r1}.

\subsection{Lemma \ref{lemm:tayds} and its proof} \label{new.Sec.B.270}

\begin{lemma}\label{lemm:tayds}
Assume that all the conditions of Theorem \ref{corollary1} are satisfied. Then for any $\a\in\mathcal{A}=\{\a\in\R^p:\norm{\a}_0\leq m,\norm{\a}_2=1\}$ and $\a\trans\wt{\W}_k (\wt{\psi}_k(\wt{\u}_k,\wt{\boldeta}_k) - \wt{\psi}_k(\wt{\u}_k,\boldeta_k^*) )$ given in \eqref{ukrankeq22}, with probability at least
$1- \theta_{n,p,q}^{\prime \prime}$ we have
\begin{align*}
	& |\a\trans\wt{\W}_k (\wt{\psi}_k(\wt{\u}_k,\wt{\boldeta}_k) - \wt{\psi}_k(\wt{\u}_k,\boldeta_k^*) )| \\[5pt]
	& \leq  c m^{1/2} \max\{s_{\max}^{1/2}, (r^*+s_u+s_v)^{1/2}\eta_n^2  \}(r^*+s_u+s_v)^{1/2}\eta_n^2\{n^{-1}\log(pq)\},
\end{align*}
where $\theta_{n,p,q}^{\prime \prime}$ is given in \eqref{thetapro77} and $c$ is some positive constant.
\end{lemma}

\noindent \textit{Proof}. The proof of Lemma \ref{lemm:tayds} is similar to the proof of Lemma \ref{prop:taylor12}.
By \eqref{phiuuuu2}, we have that 
\begin{align}
    \wt{\psi}_k(\wt{\u}_k,\wt{\boldeta}_k) - \wt{\psi}_k(\wt{\u}_k,\boldeta^*_k) =( \wh{\bSigma}(\wt{\C}_{-k} - \C_{-k}^*) -n^{-1}\X\trans\E)(\I_q - \v_k^*\v_k\strans)\exp^{-1}_{\v_k^*}(\wt{\v}_k)  + \r_{\v_k^*}, \nonumber
\end{align}
where the Taylor remainder term is 
$\|\r_{\v_k^*}\|_2 = O(\| \exp^{-1}_{\v_k^*}(\wt{\v}_k) \|_2^2). $
Then we bound term $\a\trans \wt{\W}_k (\wt{\psi}_k(\wt{\u}_k,\wt{\boldeta}_k) - \wt{\psi}_k(\wt{\u}_k,\boldeta^*_k) )$.

First, the upper bounds on $\a\trans \wt{\W}_k( \wh{\bSigma}(\wt{\C}_{-k} - \C_{-k}^*))(\I_q - \v_k^*\v_k\strans)\exp^{-1}_{\v_k^*}(\wt{\v}_k)$ and $\a\trans \wt{\W}_k\r_{\v_k^*}$ follow the same argument as in \eqref{schkasjhdq} and \eqref{31ewdasd3} and are given by 
\begin{align}
	& \Big|\a\trans \wt{\W}_k  \wh{\bSigma}(\wt{\C}_{-k} - \C_{-k}^*) (\I_q - \v_k^*\v_k\strans)\exp^{-1}_{\v_k^*}(\wt{\v}_k) \Big| \nonumber \\
   &\quad \quad \leq  c m^{1/2} (r^*+s_u+s_v)\eta_n^4\{n^{-1}\log(pq)\}, \label{dsaczxc} \\
   &|\a\trans \wt{\W}_k \r_{\v_k^*} | 
   \leq  c m^{1/2}  (r^*+s_u+s_v)\eta_n^4\{n^{-1}\log(pq)\}. \label{dsaczxcaa}
\end{align}

Next, we bound term $\a\trans \wt{\W}_k n^{-1}\X\trans\E(\I_q - \v_k^*\v_k\strans)\exp^{-1}_{\v_k^*}(\wt{\v}_k)$.
Denote by $\wt{\w}_{k,i}\trans$ the $i$th row of $\wt{\W}_k$ for $i = 1, \ldots, p$.
By parts (a) and (b) of Lemma \ref{rankr:aw} in Section \ref{new.Sec.B.11}, we have that 
	\begin{align}
		\max_{1 \leq i \leq p}\norm{\wt{\w}_{k,i}}_0 \leq 2\max\{s_{\max}, 3(r^*+s_u+s_v)\} \ \text{ and } \ \max_{1 \leq i \leq p}\norm{\wt{\w}_{k,i}}_2 \leq c.  \nonumber
	\end{align}
Similar to \eqref{eqr21} and \eqref{eqr22}, it holds that
\begin{align*}
	&\norm{ \exp^{-1}_{\v_k^*}(\wt{\v}_k)}_0  \leq c(r^*+s_u +s_v),  \ \|\exp^{-1}_{\v_k^*}(\wt{\v}_k)\|_2
	\leq c \gamma_n.
\end{align*}
Then it follows that
\begin{align}
	&|\wt{\w}_{k,i}\trans n^{-1}\X\trans\E(\I_q - \v_k^*\v_k\strans)\exp^{-1}_{\v_k^*}(\wt{\v}_k)| \nonumber \\
	& \leq |\wt{\w}_{k,i}\trans n^{-1}\X\trans\E \exp^{-1}_{\v_k^*}(\wt{\v}_k)| + |\wt{\w}_{k,i}\trans n^{-1}\X\trans\E \v_k^*| |\v_k\strans\exp^{-1}_{\v_k^*}(\wt{\v}_k)| \nonumber \\
	&\leq |\wt{\w}_{k,i}\trans n^{-1}\X\trans\E \exp^{-1}_{\v_k^*}(\wt{\v}_k)| + c \gamma_n \left|\wt{\w}_{k,i}\trans n^{-1}\X\trans\E \v_k^*\right|, \nonumber
\end{align}
where the last step above is due to $|\v_k\strans\exp^{-1}_{\v_k^*}(\wt{\v}_k)| \leq  \norm{\v_k^*}_2 \norm{\exp^{-1}_{\v_k^*}(\wt{\v}_k)}_2 \leq c \gamma_n $.

Denote by $s_1 = c \max\{s_{\max}, (r^*+s_u+s_v)\}$ and $s_2 = c(r^*+s_u+s_v)$.
For the first term above, by Lemma \ref{lemm:ds} we have that 
\begin{align*}
	|\wt{\w}_{k,i}\trans n^{-1}\X\trans\E \exp^{-1}_{\v_k^*}(\wt{\v}_k)| &\leq \left|\frac{\wt{\w}_{k,i}\trans}{\norm{\wt{\w}_{k,i}}_2}  n^{-1}\X\trans\E \frac{\exp^{-1}_{\v_k^*}(\wt{\v}_k)}{\norm{\exp^{-1}_{\v_k^*}(\wt{\v}_k)}_2} \right| \norm{\wt{\w}_{k,i}}_2 \norm{\exp^{-1}_{\v_k^*}(\wt{\v}_k)}_2 \\ 
	&\leq \sup_{\boldsymbol{h}_1 \in K_p(s_1), \boldsymbol{h}_2 \in K_q(s_2)} \left|\boldsymbol{h}_1\trans n^{-1}\X\trans\E \boldsymbol{h}_2\right|  \norm{\wt{\w}_{k,i}}_2 \norm{\exp^{-1}_{\v_k^*}(\wt{\v}_k)}_2  \\
	&\leq c \max(s_1^{1/2},s_2^{1/2})\left\{n^{-1}\log(pq)\right\}^{1/2} \gamma_n \\
	&= c \max\{s_{\max}^{1/2}, (r^*+s_u+s_v)^{1/2}\} (r^*+s_u+s_v)^{1/2}\eta_n^2 \left\{n^{-1}\log(pq)\right\}.
\end{align*}
Hence, it follows that 
\begin{align*}
	|\a\trans&\wt{\W}_{k}\trans n^{-1}\X\trans\E \exp^{-1}_{\v_k^*}(\wt{\v}_k)| \\ 
	& \leq c m^{1/2} \max\{s_{\max}^{1/2}, (r^*+s_u+s_v)^{1/2}\} (r^*+s_u+s_v)^{1/2}\eta_n^2 \left\{n^{-1}\log(pq)\right\}.
\end{align*}

Therefore, combining the above result with \eqref{dsaczxc} and \eqref{dsaczxcaa} yields that
\begin{align}
	&|\a\trans\wt{\W}_k  (\wt{\psi}_k(\wt{\u}_k,\wt{\boldeta}_k) - \wt{\psi}_k(\wt{\u}_k,\boldeta^*_k) )| \nonumber \\
	& \leq c m^{1/2} \max\{s_{\max}^{1/2}, (r^*+s_u+s_v)^{1/2}\eta_n^2  \}(r^*+s_u+s_v)^{1/2}\eta_n^2\{n^{-1}\log(pq)\},  \nonumber
\end{align}
which completes the proof of Lemma \ref{lemm:tayds}.
}

\blue{
\subsection{Lemma \ref{lemm:ds} and its proof}
\begin{lemma}\label{lemm:ds}
	Assume that Conditions \ref{cone}--\ref{con3} hold and let $K_p(s):= \{ \mathbf{b} \in \mathbb{R}^p: \|\mathbf{b}\|_0 \leq s, \|\mathbf{b}\|_2 = 1 \}$. Then for some $s_1 \leq p, s_2 \leq q$, there exist some constants $C$ and $c$ such that with probability at least $1- 2 (pq)^{-c \left(s_1 \vee s_2\right)}$,
$$
\sup _{\u \in K_p(s_1), \v \in K_q\left(s_2\right)}\left|\u\trans n^{-1}\X\trans\E \v\right| \leq C \sqrt{\frac{\left(s_1 \vee s_2\right) \log (pq)}{n}}.
$$
\end{lemma}

\noindent \textit{Proof.} 
Let $\mathbf{x}_i$ and $\mathbf{e}_j$ represent the $i$th and $j$th columns of $\mathbf{X} \in \mathbb{R}^{n \times p}$ and $\mathbf{E} \in \mathbb{R}^{n \times q}$, respectively. Note that we consider the fixed design setting that $\X$ is given. The following calculations will be carried out in the case where $\X$ is given.  Under Condition \ref{cone} that $\E\sim\N(\0,\I_n\otimes\bSigma_e)$, it holds that 
\[
\mathbb{E}\left[ {n^{-1}\u\trans {\X}\trans {\E} \v}\right] = \mathbb{E}[ n^{-1}\sum_{j=1}^q \sum_{i=1}^p u_i \mathbf{x}_i^T \mathbf{e}_j v_j] =   0.
\]
Furthermore, we have that
\begin{align*}
\text{var}(n^{-1}\u\trans {\X}\trans {\E} \v)
& = ((n^{-1} {\X}\u)\otimes\v)\trans(\I_n\otimes\bSigma_e)((n^{-1} {\X}\u)\otimes\v) \\
& = n^{-1} \u\trans {\X}\trans \I_n  n^{-1} {\X}\u   \v\trans \bSigma_e \v   \\
&\leq n^{-1} \u\trans \wh{\bSigma} \u \cdot \v\trans \bSigma_e \v \\ &\leq n^{-1} \norm{\u}_2 \norm{\wh{\bSigma} \u}_2 \norm{\v}_2 \norm{\bSigma_e}_2\norm{\v}_2  \leq c n^{-1} \sigma_{\max}^2,
\end{align*}
where the last inequality above has used Condition \ref{con3} and the fact that $\norm{\u}_2 = 1$ and $\norm{\v}_2 = 1$. Hence, under Condition \ref{cone}, it follows that 
\begin{align}\label{eqwazcgg}
	\mathbb{P}(|n^{-1}\u\trans {\X}\trans {\E} \v| > t) \leq 2\exp( - \frac{nt^2}{2 c \sigma_{{\max }}^2} ).
\end{align}

Denote by $$
K_1\left(s_1\right)=\bigcup_{U \subseteq\{1, \ldots, p\}:|U| \leq s_1} A(U),   \ \  K_2\left(s_2\right)=\bigcup_{V \subseteq\{1, \ldots, q\}:|V| \leq s_2} B(V) 
$$
with
\begin{align*}
	&A(U):=\left\{\u \in \mathbb{R}^{p}:\|\u\|_{2}=1, \operatorname{supp}(\u) \subseteq U\right\}, \\
	&B(V):=\left\{\v \in \mathbb{R}^{q}:\|\v\|_{2}=1, \operatorname{supp}(\v) \subseteq V\right\}.
\end{align*}
Let us define $\Delta_{AB}:= \sup _{\u \in A(U), \v \in B(V)}\Delta_{AB}^{\u, \v}$ with $\Delta_{AB}^{\u, \v} =\left|\u\trans n^{-1}\X\trans\E \v\right|$ and fixed $\u \in A(U), \v \in B(V)$. Let $\left\{\u_{1}, \ldots, \u_{M_{1}}\right\}$ be a $1 / 3$-net of $A(U)$; that is, for any $\u \in A(U)$, there exists some $\u_{i}$ such that $\left\|\u-\u_{i}\right\|_{2} \leq 1 / 3$. It is well-known from \cite{ledoux2013probability} that we can construct a $1 / 3$-net of $A(U)$ with $M_{1} \leq 9^{2 s_1}$, which implies that $\log M_{1} \leq 2 s_1 \log 9$. By the same argument, we can construct a $1 / 3$-net $\left\{\v_{1}, \ldots, \v_{M_{2}}\right\}$ of $B(V)$ with $M_{2} \leq 9^{2 s_{2}}$, which entails that $\log M_{net} \leq 2 s_2 \log 9$. Then we can bound
$$
\begin{aligned}
	\Delta_{AB} & \leq \Delta_{AB}^{\u_{i}, \v_{j}}+\left|\Delta_{AB}^{\u_{i}, \v_{j}}-\Delta_{AB}^{\u_{i}, \v}\right|+\left|\Delta_{AB}^{\u_{i}, \v}-\Delta_{AB}^{\u, \v}\right| \\
& \leq \Delta_{AB}^{\u_{i}, \v_{j}}+\Delta_{AB}\left\|\u-\u_{i}\right\|_{2}+\Delta_{AB}\left\|\v-\v_{j}\right\|_{2} \\
& \leq \max _{i \in\left\{1, \ldots, M_{1}\right\}, \ j \in\left\{1, \ldots, M_{2}\right\}} \Delta_{AB}^{\u_{i}, \v_{j}}+1 / 3 \Delta_{AB}+1 / 3 \Delta_{AB}.
\end{aligned}
$$
This leads to 
\begin{equation}
	\Delta_{AB} \leq 3 \max _{i \in\left\{1, \ldots, M_{1}\right\}, \ j \in\left\{1, \ldots, M_{2}\right\}} \Delta_{AB}^{\u_{i}, \v_{j}}. \label{zzadq}
\end{equation}

By the definitions of $U$ and $V$, the numbers of choices of $U$ and $V$ are upper bounded by $\binom{p}{s_1} \leq p^{s_1}$ and $\binom{p}{s_2} \leq p^{s_2}$, respectively. Denote by $s_0 = \max(s_1, s_2)$.
Then by applying the union bound and \eqref{zzadq}, we can deduce that 
$$
\begin{aligned}
& \mathbb{P}\left(\sup _{\u \in K_p\left(s_1\right), \v \in K_q\left(s_2\right)}\left|\u\trans n^{-1}\X\trans\E\v\right| \geq   3c\sqrt{\frac{s_0 \log (pq)}{n}}\right) \\
& =\mathbb{P}\left(\sup_{\u \in \bigcup_{U :|U| \leq s_1} A(U), \v \in \bigcup_{V :|V| \leq s_2} B(V)}\left|\u\trans n^{-1}\X\trans\E\v\right| \geq  3c\sqrt{\frac{s_0 \log (pq)}{n}}\right) \\
& \leq p^{s_1}q^{s_2} \max _{U:|U| \leq s_1, V:|V| \leq s_2} \mathbb{P}\left(\sup_{\u \in A(U), \v \in B(V)}\left|\u\trans n^{-1}\X\trans\E\v\right|\geq 3 c \sqrt{\frac{s_0 \log (pq)}{n}}\right) \\
& \leq (pq)^{s_0} \max _{U:|U| \leq s_1, V:|V| \leq s_2} \mathbb{P}\left(\Delta_{AB} \geq 3 c \sqrt{\frac{s_0 \log (pq)}{n}}\right) \\
& \leq (pq)^{s_0} \mathbb{P}\left(\max _{i \in\left\{1, \ldots, M_{1}\right\}, \ j \in\left\{1, \ldots, M_{2}\right\}} \Delta_{AB}^{\u_{i}, \u_{j}} \geq c \sqrt{\frac{s_0 \log (pq)}{n}}\right) \\
& \leq (pq)^{s_0} M_{1}M_{2} \max _{i, j} \mathbb{P}\left(\left|\u_i\trans n^{-1}\X\trans\E\v_j\right| \geq c \sqrt{\frac{s_0 \log (pq)}{n}}\right) := A_0.
\end{aligned}
$$

Therefore, from $\log M_{1} \leq 2 s_1 \log 9$, $\log M_{2} \leq 2 s_2 \log 9$, and \eqref{eqwazcgg}, we can obtain that 
$$
\begin{aligned}
A_0 \leq & 2 (pq)^{s_0} M_{1}M_{2} \exp(-c  s_0 \log (pq) )\\
\leq & 2\exp \left(s_0 \log (pq)+ \log M_{1} + \log M_{2}-c s_0 \log (pq)\right) \\
\leq & 2\exp \left(s_0 \log (pq)+4 s_0 \log 9-c s_0 \log (pq)\right) \\
\leq & 2\exp \left(-c s_0 \log (pq)\right) \leq  2 (pq)^{-c s_0}.
\end{aligned}
$$
This completes the proof of Lemma \ref{lemm:ds}.
}

\section{Additional technical details} \label{new.Sec.C}

\subsection{The Taylor expansion on the Riemannian manifold} \label{sec1.1.1}

To facilitate our technical analysis, let us first introduce briefly some necessary background on the Riemannian manifold. For more detailed and rigorous introduction to the Riemannian manifold, see, e.g., \cite{do1992riemannian}.
Let $\mathcal{M}$ be a $p$-dimensional compact Riemannian manifold. For a given $\X \in \mathcal{M}$,  the tangent space to $\mathcal{M}$ at $\X$ is a $p$-dimensional linear space and will be denoted as $T_{X}\mathcal{M}$.
A Riemannian metric $g_X$ is defined at each point $\X \in \mathcal{M}$ by the map $g_X: T_X\mathcal{M} \times T_X\mathcal{M} \rightarrow \mathbb{R}$ and is an inner product on the tangent space $T_X\mathcal{M}$. For $\bxi_1, \bxi_2 \in T_X\mathcal{M}$, denote the inner product as  $\langle\bxi_1,\bxi_2\rangle = g_X(\bxi_1,\bxi_2)$.
Then the inner product induces a norm $\| \cdot \|$, which is denoted as $\| \bxi \| =\sqrt{\langle\bxi, \bxi\rangle  }$ for each $\bxi \in T_{X}\mathcal{M}$. 
Given $\X \in \mathcal{M}$ and its tangent vector $\bxi \in T_X\mathcal{M}$, let  $\gamma(t ; \X, \bxi)$ be the  geodesic (the  locally length-minimizing curve) satisfying that $\gamma(0; \X, \bxi)=\X$ and $\dot{\gamma}(0; \X, \bxi)=\bxi$.
The exponential map is defined through the geodesic. Specifically, the exponential map at a point $\X$ is defined as
\begin{equation}\label{expon}
	\operatorname{exp}_{X}: T_{X} \mathcal{M} \rightarrow \mathcal{M},\, \bxi \mapsto \operatorname{exp}_{X} \bxi=\gamma(1 ; \X, \bxi).
\end{equation}

By Theorem 3.7 and Remark 3.8 in \cite{do1992riemannian}, for each point  $\X \in \mathcal{M}$ a normal neighborhood $S$ of $\X \in \mathcal{M}$ is the one that satisfies (1) each point $\Y$ in $S$ can be joined to $\X$ by a unique geodesic $\gamma(t ; \X, \bxi)$, $0 \leq t \leq 1$, with $\gamma(0; \X, \bxi)=\X$ and ${\gamma}(1; \X, \bxi)=\Y$; and (2) the  exponential map $\exp_{X}$  is a local diffeomorphism between a neighborhood of $\0 \in T_X\mathcal{M}$ and a  neighborhood $S$ of $\X \in \mathcal{M}$. Since $\exp_{X}$  is a local diffeomorphism,  the exponential map is defined only locally in that it maps a small neighborhood of $\0 \in T_X\mathcal{M}$ to a neighborhood $S$ of $\X \in \mathcal{M}$. Denote by $\exp^{-1}_{X}$ the inverse of the exponential map. For each point $\Y$ in $S$, we can connect two points $\X$ and $\Y$ by the exponential map that $\exp_{X}\bxi = \Y$, or equivalently, $\bxi = \exp_{X}^{-1}\Y $ for $\bxi \in T_X\mathcal{M}$.

\begin{lemma}\label{lemm:taylormanifold}(\citep[Lemma A.3]{ mukherjee2010learning})
	Let $\mathcal{M}$ be a  compact Riemannian manifold. Assume that $f $ is a  twice differentiable function on $\mathcal{M}$. Denote by $\nabla_{\mathcal{M}} f(\X)$ the gradient of $f$.
	Then there exists a constant $C>0$ such that for all $\X \in \mathcal{M}$ and $\bxi \in T_X \mathcal{M}$, $\norm{\bxi} \leq \epsilon_0$ with some $\epsilon_0 > 0$, the first-order Taylor expansion below satisfies that 
	\begin{equation}\label{tay0}
		\| f\left(\exp _{X}(\bxi)\right) - f(\X) -\left\langle\nabla_{\mathcal{M}} f(\X), \bxi\right\rangle \| \leq C\| \bxi \|^2.
	\end{equation}
\end{lemma}

From \eqref{tay0} in Lemma \ref{lemm:taylormanifold} above,	 we can write the first-order Taylor expansion on the Riemannian manifold as 
\begin{equation}\label{taylor:eq}
	f\left(\exp _{X}(\bxi)\right) = f(\X)+\left\langle\nabla_{\mathcal{M}} f(\X), \bxi\right\rangle + O(\| \bxi \|^2)
\end{equation}
or
\begin{equation}\label{taylor:eq2}
	f(\Y) = f(\X)+\left\langle\nabla_{\mathcal{M}} f(\X), \exp_{X}^{-1}\Y\right\rangle + O(\| \exp_{X}^{-1}\Y \|^2)
\end{equation}
for  $\bxi = \exp_{X}^{-1}\Y $ with $\bxi \in T_X\mathcal{M}$.

\subsection{The geometry of the Stiefel manifold} \label{sec:stiefel}

We now focus on a special manifold, the so-called Stiefel manifold. We will briefly
introduce some  necessary background on the Stiefel manifold. The Stiefel manifold $\text{St}(p,n)$ denotes the set of all orthonormal $p$-frames in the Euclidean space $\mathbb{R}^n$, where the $p$-frame is a set of $p$ orthonormal vectors in  $\mathbb{R}^n$. Specifically, the Stiefel manifold is given by  $\text{St}(p,n) = \{\X \in \R^{n \times p}: \X\trans\X = \I_p \}$. For $p = 1$, the Stiefel manifold $\text{St}(p,n)$ reduces to the unit sphere $\mathcal{S}^{n-1}$ in $\R^{n}$, where  $\mathcal{S}^{n-1} := \{\x \in \mathbb{R}^{n}: \x\trans\x = 1  \}$. For $p = n$, the Stiefel manifold $\text{St}(p,n)$ becomes the orthogonal group $O(n)$, where $O(j) := \{ \mathbf{O}_j \in \mathbb{R}^{j \times j} : \mathbf{O}_j\trans\mathbf{O}_j = \I_j \}$ for each positive integer $j$. We can also represent the Stiefel manifold  as  $\text{St}(p,n) = O(n) / O(n-p)$. 
 See, e.g.,  \cite{edelman1998geometry,lv2013impacts} for more details on these representations.

Denote by $T_{X}\operatorname{St}(p,n)$ the tangent space of the Stiefel manifold. The tangent space $T_{X}\operatorname{St}(p,n)$ admits the form
\begin{align}\label{sti:tangent}
	T_{X}\operatorname{St}(p,n) = \left\{ \X\A + \B: \A \in \R^{p \times p}, \A = - \A\trans, \B \in \R^{n \times p}, \X\trans\B = \0   \right\}.
\end{align}
According to the Stiefel manifold representation of orthonormal matrices in \cite{edelman1998geometry} and \cite{chen2012sparse}, for each given $\X^* \in \operatorname{St}(p,n)$, matrices on the Stiefel manifold $\operatorname{St}(p,n)$ can be represented as
\begin{align}\label{exponv2}
	\{\X = \exp_{X^*}\bxi^*: \bxi^* = \X^*\A + \B \in  T_{X^*}\operatorname{St}(p,n)\},
\end{align}
where $\exp_{X^*}$ is the exponential map defined in \eqref{expon} above. 
For the Stiefel manifold, it is common to use the canonical metric as suggested in \cite{edelman1998geometry}. For the tangent vector $\bxi = \X\A + \B \in T_{X}\operatorname{St}(p,n)$, the canonical metric $\left\langle \cdot, \cdot \right\rangle_c $ is given by 
\begin{align}\label{metric1}
	\left\langle \bxi, \bxi \right\rangle_c &= \operatorname{tr}({\bxi\trans (\I_n - \frac{1}{2}\X\X\trans) \bxi })\nonumber\\
	&= \frac{1}{2} \operatorname{tr}({\A\trans\A}) + \operatorname{tr}({\B\trans\B}),
\end{align}
where the second equality above can be derived easily using $\X\trans\X = \I_p$ and $\X\trans\B = 0 $. With such canonical metric, we can obtain the gradient of a function on the Stiefel manifold below.
\begin{lemma}(\citep[Section 2.4.4]{edelman1998geometry})\label{lemmgradst}
    For a real-valued function $f$ defined on the Stiefel manifold $\operatorname{St}(p,n)$, let $\nabla_X f$ be the gradient of $f$ at $\mathbf{X} \in \operatorname{St}(p,n)$. Then it holds that
    \begin{align*}
        \nabla_X f =   \der{f}{\X} - \X\der{f}{\X\trans}\X .
    \end{align*}
\end{lemma}

Let us now consider a special case that $p=1$ for the Stiefel manifold  $\operatorname{St}(p,n)$. For such case,  the Stiefel manifold $\text{St}(p,n)$ reduces to the unit sphere $\mathcal{S}^{n-1}$. For $\mathbf{x} \in \operatorname{St}(1,n)$ and its tangent space $T_{\mathrm{x}}\operatorname{St}(1,n)$, from  \eqref{sti:tangent} we see that $\A = 0$. Then we can write $T_{\mathrm{x}}\operatorname{St}(1,n)$ as
\begin{align}\label{tagvec}
	T_{\mathrm{x}}\operatorname{St}(1,n) = \{\B: \B \in \R^{n}, \mathbf{x}\trans\B = \0  \}.
\end{align}
In addition, similar to \eqref{exponv2}, for given $\mathbf{x}^* \in \operatorname{St}(1,n)$, vectors on the Stiefel manifold $\operatorname{St}(1,n)$ can be represented as
\begin{align}\label{exponv222}
	\{\mathbf{x} = \exp_{\mathrm{x}^*}\bxi^*:   \mathbf{x}\strans\bxi^* = \0\}.
\end{align}
Then for the tangent vector $\bxi = \B \in T_{\mathrm{x}}\operatorname{St}(1,n)$, the canonical metric is given by 
\begin{align}
	\left\langle \bxi, \bxi \right\rangle_c &= \operatorname{tr}({\bxi\trans (\I_n - \frac{1}{2}\x\x\trans) \bxi })\nonumber\\
	&= \operatorname{tr}({\B\trans\B}) =  \operatorname{tr}({\bxi\trans\bxi}) \nonumber\\
 &:= 	\left\langle \bxi, \bxi \right\rangle_e, \nonumber
\end{align}
where $\left\langle \bxi, \bxi \right\rangle_e =  \operatorname{tr}({\bxi\trans\bxi})$ represents the (usual) Euclidean metric. Thus,
for the case of $p =1$ the canonical metric is in fact  equivalent to the Euclidean metric. To simplify the notation, denote the metric $\left\langle \cdot , \cdot \right\rangle$ on $\operatorname{St}(1,n)$ as
\begin{align}\label{metric2}
	\left\langle \bxi, \bxi \right\rangle =   \operatorname{tr}({\bxi\trans\bxi}).
\end{align}
Moreover, since $\bxi \in T_{\mathrm{x}}\operatorname{St}(1,n)$ is an $n$-dimensional vector and  $\left\langle \bxi, \bxi \right\rangle =  \operatorname{tr}({\bxi\trans\bxi}) = \bxi\trans\bxi$, such metric induces the norm $\norm{\bxi}_2^2 =  {\bxi\trans\bxi}$.

For the Stiefel manifold $\operatorname{St}(1,n)$,  the gradient given in Lemma \ref{lemmgradst} above can be written as
\begin{align}\label{grad:sti:eq}
	\nabla_{\mathrm{x}} f & =   \der{f}{\x} - \x\der{f}{\x\trans}\x =   \der{f}{\x} - \x{\x\trans}\der{f}{\x} \nonumber\\
 & = (\I_n - \x{\x\trans})  \der{f}{\x} .
\end{align}
Furthermore, we can characterize the geodesic on  $\operatorname{St}(1,n)$, i.e., on the unit sphere $\mathcal{S}^{n-1}$ below.
\begin{lemma}(\citep[Example 5.4.1]{AbsMahSep2008})\label{lemmgeodsti}
	For the unit sphere $\mathcal{S}^{n-1}$  with  metric \eqref{metric2}, let $\mathbf{x} \in \mathcal{S}^{n-1}$ and $\bxi$ be the tangent vector in the tangent space of $\mathcal{S}^{n-1}$ at $\mathbf{x}$.
	Let $t \mapsto  \gamma(t ; \mathbf{x}, \bxi)$ be the geodesic on $\mathcal{S}^{n-1}$ with $\gamma(0; \mathbf{x}, \bxi)=\mathbf{x}$ and $\dot{\gamma}(0; \mathbf{x}, \bxi)=\bxi$. Then the geodesic admits the representation 
    \begin{align*}
        \gamma(t ; \mathbf{x}, \bxi) = \mathbf{x}  \cos( \norm{\bxi}_2 t) + \frac{\bxi}{\norm{\bxi}_2} \sin(\norm{\bxi}_2 t).
    \end{align*}
\end{lemma}

{
\section{Additional insights into constructions of \texorpdfstring{$\M$}{M} and \texorpdfstring{$\W$}{W}}\label{appsec:issue}

Here, we show that if one directly considers the SVD constraints in $\der{\wt{\psi}_k}{\boldeta_k}$ and calculates the regular derivatives in the Euclidean space, it will result in deficiency in the degrees of freedom and there would not exist a valid $\W$ matrix.

For simplicity, let us consider the rank-$2$ case. The following derivations are similar to those in the proofs of Lemma \ref{3-7:prop:2} in Section \ref{near2:sec1}, and Propositions \ref{prop:rankr2} and \ref{prop:rankr3} in Sections \ref{new.Sec.A.11} and \ref{new.Sec.A.12}, respectively. 
Note that we have 
\begin{align*}
    L
     & = (2n)^{-1}\Big\{\norm{\Y}_F^2 + \u_1\trans\X\trans\X\u_1\v_1\trans\v_1 + \u_2\trans\X\trans\X\u_2\v_2\trans\v_2 + 2\u_1\trans\X\trans\X\u_2\v_1\trans\v_2 \nonumber \\
     & \quad~\ - 2\u_1\trans\X\trans\Y\v_1 - 2\u_2\trans\X\trans\Y\v_2\Big\}.
\end{align*}
After some calculations, we can deduce that 
\begin{align}
     & \der{L}{\u_1} = \wh{\bSigma}\u_1 \v_1\trans\v_1 - n^{-1}\X\trans\Y\v_1,                 \label{3-7:der55:1}                                      \\
     & \der{L}{\v_1} = \v_1\u_1\trans\wh{\bSigma}\u_1 + \v_2\u_1\trans\wh{\bSigma}\u_2 - n^{-1}\Y\trans\X\u_1, \label{3-7:der55:2} \\
     & \der{L}{\u_2} = \wh{\bSigma}\u_2 \v_2\trans\v_2 - n^{-1}\X\trans\Y\v_2,                                          \label{3-7:der55:3}             \\
     & \der{L}{\v_2} = \v_2\u_2\trans\wh{\bSigma}\u_2 + \v_1\u_1\trans\wh{\bSigma}\u_2 - n^{-1}\Y\trans\X\u_2. \label{3-7:der55:4}
\end{align}

Denote by $\boldeta_{1} = (\v_1\trans, \v_2\trans, \u_2\trans )\trans,$ and $\boldeta_{1}^* = (\v_1\strans, \v_2\strans, \u_2\strans )\trans.$
Utilizing the derivatives given in \eqref{3-7:der55:2}--\eqref{3-7:der55:4} and the constraints $\norm{\v_1}_2=\norm{\v_2}_2=1$ and $\v_1\trans\v_2=0$, we can obtain that 
\begin{align*}
    \M\der{L}{\boldeta_{1}}\Big|_{\boldeta_{1}^*}
     & = \M_1\left\{\v_1^*\u_1\trans\wh{\bSigma}(\u_1 - \u_1^*) - n^{-1}\E\trans\X\u_1\right\} - n^{-1}\M_3\X\trans\E\v_2^* \\
     & \quad~ + \M_2\left\{\v_1^*\u_2\strans\wh{\bSigma}(\u_1 - \u_1^*) - n^{-1}\E\trans\X\u_2^*\right\}                  \\
     & = (\M_1\v_1^*\u_1\trans + \M_2\v_1^*\u_2\strans)\wh{\bSigma}(\u_1 - \u_1^*) - n^{-1}\Big\{\M_1\E\trans\X\u_1 \\
     & \quad~  + \M_2\E\trans\X\u_2^* + \M_3\X\trans\E\v_2^*\Big\}                                                                 \\
     & = (\M_1\v_1\u_1\trans + \M_2\v_1\u_2\trans)\wh{\bSigma}(\u_1 - \u_1^*) + \bdelta_1^{\prime},
\end{align*}
where we define 
\begin{align}\label{3-7:eq:p22rop:2:1}
    \bdelta_1^{\prime}
     & = - n^{-1}\left\{\M_1\E\trans\X\u_1 + \M_2\E\trans\X\u_2^* + \M_3\X\trans\E\v_2^*\right\} - \nonumber                                 \\
     & \quad~ \left\{\M_1(\v_1 - \v_1^*)\u_1\trans - \M_2(\v_1\u_2\trans - \v_1^*\u_2\strans)\right\}\wh{\bSigma}(\u_1 - \u_1^*).
\end{align}

Then with the derivative in \eqref{3-7:der55:1}, we can represent $\wt{\psi}(\u_1,\boldeta_{1}^*)$ as 
\begin{align}\label{3-7:eq:p22rop:2:2}
    \wt{\psi}(\u_1,\boldeta_{1}^*)
     & = \der{L}{\u_1}\Big|_{\boldeta_{1}^*} - \M\der{L}{\boldeta_{1}}\Big|_{\boldeta_{1}^*} \nonumber                 \\
     & = (\I_p - \M_1\v_1\u_1\trans - \M_2\v_1\u_2\trans)\wh{\bSigma}(\u_1 - \u_1^*) + \bdelta_1,
\end{align}
where $\bdelta_1$ is equal to $-\bdelta_1^{\prime} - n^{-1}\X\trans\E\v_1^*$. Thus, combining \eqref{3-7:eq:p22rop:2:1} and \eqref{3-7:eq:p22rop:2:2} yields that 
\begin{align}\label{rank2etatest}
    \wt{\psi}(\u_1,\boldeta_{1}^*)
    = (\I_p - \M_1\v_1\u_1\trans - \M_2\v_1\u_2\trans)\wh{\bSigma}(\u_1 - \u_1^*) + \bepsilon + \bdelta,
\end{align}
where $\bepsilon = n^{-1}\left\{\M_1\E\trans\X\u_1  + \M_2\E\trans\X\u_2^* + \M_3\X\trans\E\v_2^*\right\} - n^{-1}\X\trans\E\v_1^*$ and
\begin{align*}
    \bdelta = \left\{\M_1(\v_1 - \v_1^*)\u_1\trans - \M_2(\v_1\u_2\trans - \v_1^*\u_2\strans)\right\}\wh{\bSigma}(\u_1 - \u_1^*).
\end{align*}

Next we proceed with the construction of matrix $\M = [\M_1, \M_2, \M_3]$.
Utilizing the derivatives in \eqref{3-7:der55:1}--\eqref{3-7:der55:4} and after some calculations, it holds that 
\begin{align*}
     & \derr{L}{\boldeta_{1}}{\boldeta_{1}\trans} =
    \left[
        \begin{array}{ccc}
            \u_1\trans\wh{\bSigma}\u_1\I_q  & \u_1\trans\wh{\bSigma}\u_2\I_q                                                     & \v_2\u_1\trans\wh{\bSigma} \\
                                                 &                                                                                   &                                      \\
             \u_1\trans\wh{\bSigma}\u_2\I_q                     &      \u_2\trans\wh{\bSigma}\u_2\I_q                                                                 & 2\v_2\u_2\trans\wh{\bSigma} + \v_1\u_1\trans\wh{\bSigma} - n^{-1}\Y\trans\X                   \\
                                                 &                                                                                   &                                      \\
            \0  & 2\wh{\bSigma}\u_2 \v_2\trans -n^{-1}\X\trans\Y  & \wh{\bSigma} \v_2\trans\v_2
        \end{array}
    \right],                                                                                                 \\
     & \derr{L}{\u_1}{\boldeta_{1}\trans} = \left[2  \wh{\bSigma} \u_1\v_1\trans -n^{-1}\X\trans\Y, \0, \0 \right].
\end{align*}
Note that $n^{-1}\X\trans\Y = \wh{\bSigma}\C + \wh{\bSigma}(\C^*-\C) + n^{-1}\X\trans\E$, where $\C = \u_1\v_1\trans + \u_2\v_2\trans$. Plugging it into the above derivatives, we have that 
$$\derr{L}{\boldeta_{1}}{\boldeta_{1}\trans} = \A + \bDelta_{a} \ \text{ and } \ \derr{L}{\u_1}{\boldeta_{1}\trans}=\B + \bDelta_{b},$$
where
\begin{align*}
     & \A = \left[
        \begin{array}{ccc}
            \u_1\trans\wh{\bSigma}\u_1\I_q  & \u_1\trans\wh{\bSigma}\u_2\I_q  &   \v_2\u_1\trans\wh{\bSigma}                        \\
                                                 &                               &                                                                \\
              \u_1\trans\wh{\bSigma}\u_2\I_q                      & \u_2\trans\wh{\bSigma}\u_2\I_q                  &  \v_2\u_2\trans\wh{\bSigma} \\
                                                 &                               &                                                                \\
          \0 & -\wh{\bSigma}\u_1\v_1\trans + \wh{\bSigma}\u_2\v_2\trans& \wh{\bSigma} \v_2\trans\v_2
        \end{array}
    \right],                                                                                     \\
     & \bDelta_{a} = \left[
        \begin{array}{ccc}
            \0 & \0                                             & \0                                       \\
            \0 & \0                                             &  (\C-\C^*)\trans\wh{\bSigma} - n^{-1}\E\trans\X\\
            \0 & \wh{\bSigma}(\C-\C^*) - n^{-1}\X\trans\E & \0
        \end{array}
    \right],                                                                                     \\
     & \B = \left[\wh{\bSigma}\u_1\v_1\trans - \wh{\bSigma}\u_2\v_2\trans,\0,\0\right], ~
    \bDelta_{b} = \left[\wh{\bSigma}(\C - \C^*) - n^{-1}\X\trans\E,\0,\0\right].
\end{align*}

Then we aim to find matrix $\M\in\R^{p\times(p+2q)}$ satisfying that $\B - \M\A = \0$. It is equivalent to solving the following equations
\begin{align*}
    &\u_1\trans\wh{\bSigma}\u_1\M_1 + \u_1\trans\wh{\bSigma}\u_2\M_2 - \wh{\bSigma}\u_1\v_1\trans + \wh{\bSigma}\u_2\v_2\trans = \0, \\
    &\u_1\trans\wh{\bSigma}\u_2\M_1 + \u_2\trans\wh{\bSigma}\u_2\M_2 - \M_3 \wh{\bSigma}\u_1\v_1\trans + \M_3 \wh{\bSigma}\u_2\v_2\trans  = \0,\\
    &\M_1\v_2\u_1\trans\wh{\bSigma} + \M_3\wh{\bSigma} \v_2\trans\v_2 + \M_2\v_2\u_2\trans\wh{\bSigma} = \0.
\end{align*}
Observe that the the key terms in the modified score function \eqref{rank2etatest} related to the construction of matrix $\W$ is $\I_p - \M_1\v_1\u_1\trans - \M_2\v_1\u_2\trans$. For simplicity, here we provide the explicit expressions of matrices $\M_1$ and $\M_2$, respectively. It is worth mentioning that matrix $\M_3$ can be derived following the the above equations and matrices $\M_1$ and $\M_2$ easily. 

Recall that $z_{11}=\u_1\trans\wh{\bSigma}\u_1$, $z_{22}=\u_2\trans\wh{\bSigma}\u_2$, and $z_{12}=\u_1\trans\wh{\bSigma}\u_2$. 
It holds that 
\begin{align}
    &z_{11} \M_1 + z_{12}\M_2 - \wh{\bSigma}\u_1\v_1\trans + \wh{\bSigma}\u_2\v_2\trans = \0, \label{cv1} \\
    &z_{12} \M_1 + z_{22} \M_2 - \M_3 \wh{\bSigma}\u_1\v_1\trans + \M_3 \wh{\bSigma}\u_2\v_2\trans  = \0, \label{cv2} \\
    &\M_1\v_2\u_1\trans\wh{\bSigma} + \M_3\wh{\bSigma} \v_2\trans\v_2 + \M_2\v_2\u_2\trans\wh{\bSigma} = \0. \label{cv3}
\end{align}
Then we solve the above equations under the SVD constraints $\norm{\v_1}_2=\norm{\v_2}_2=1$ and $\v_1\trans\v_2=0$.
From \eqref{cv3}, it follows that 
$  \M_3\wh{\bSigma} = - \M_1\v_2\u_1\trans\wh{\bSigma} - \M_2\v_2\u_2\trans\wh{\bSigma}$. Combining it with \eqref{cv2} leads to 
\begin{align*}
\left(z_{12} \mathbf{M}_1+z_{22} \mathbf{M}_2\right)\left(\mathbf{I}_q-\boldsymbol{v}_2 \boldsymbol{v}_2\trans\right)+\left(z_{11} \mathbf{M}_1+z_{12} \mathbf{M}_2\right) \boldsymbol{v}_2 \boldsymbol{v}_1\trans=\mathbf{0}.
\end{align*}
Then using $z_{11} \M_1 + z_{12}\M_2 =  \wh{\bSigma}\u_1\v_1\trans - \wh{\bSigma}\u_2\v_2\trans$ in \eqref{cv1}, we can further show that 
\begin{align}\label{sadwq}
\left(z_{12} \mathbf{M}_1+z_{22} \mathbf{M}_2\right)\left(\mathbf{I}_q-\boldsymbol{v}_2 \boldsymbol{v}_2\trans\right) =   \wh{\bSigma}\u_2\v_1\trans.
\end{align}

Moreover, \eqref{cv1} also implies that 
$\mathbf{M}_1= z_{11}^{-1}(  \wh{\bSigma}\u_1\v_1\trans - \wh{\bSigma}\u_2\v_2\trans - z_{12} \mathbf{M}_2)$. Combining it with  \eqref{sadwq}, we can obtain the solutions for matrices $\M_1$ and $\M_2$. For simplicity, let us denote by 
\begin{align*}
\begin{aligned}
& \alpha_1=-z_{11}^{-1} z_{12}\left(z_{11} z_{22}-z_{12}^2\right)^{-1}, \quad \alpha_2=z_{11}^{-1}\left[1-z_{11}^{-1}z_{12}^2\left(z_{11} z_{22}-z_{12}^2\right)^{-1}\right], \\
& \beta_1=\left(z_{11} z_{22}-z_{12}^2\right)^{-1}, \quad \beta_2= - z_{11}^{-1}z_{12}\left(z_{11} z_{22}-z_{12}^2\right)^{-1}.
\end{aligned}
\end{align*}
Then matrices $\M_1$ and $\M_2$ are given by 
\begin{align}
&\mathbf{M}_1 =\alpha_1 \widehat{\boldsymbol{\Sigma}} \boldsymbol{u}_2 \boldsymbol{v}_1\trans +\alpha_2 \widehat{\boldsymbol{\Sigma}} \boldsymbol{u}_1 \boldsymbol{v}_1\trans -z_{11}^{-1} \widehat{\boldsymbol{\Sigma}} \boldsymbol{u}_2 \boldsymbol{v}_2\trans, \label{m13qew} \\
&\mathbf{M}_2  =\beta_1 \widehat{\boldsymbol{\Sigma}} \boldsymbol{u}_2 \boldsymbol{v}_1\trans+\beta_2 \widehat{\boldsymbol{\Sigma}} \boldsymbol{u}_1 \boldsymbol{v}_1\trans. \label{m13qew3}
\end{align}

Since $\B - \M\A = \0$, it holds that 
\begin{align*}
    \derr{L}{\u_1}{\boldeta_{1}\trans} - \M\derr{L}{\boldeta_{1}}{\boldeta_{1}\trans}
    =  \bDelta_{b} - \M \bDelta_{a}  = \bDelta,
\end{align*}
where $\bDelta = \left[\bDelta_1,\bDelta_2,\bDelta_3\right]$ with 
\begin{align*}
    \bDelta_1 &= \wh{\bSigma}(\C-\C^*) - n^{-1}\X\trans\E,\\
        \bDelta_2 &= \M_3 \left\{n^{-1}\X\trans\E - \wh{\bSigma}(\C-\C^*)\right\}, 
   \\
    \bDelta_3 &= \M_2\left\{n^{-1}\E\trans\X - (\C-\C^*)\trans\wh{\bSigma}\right\}.
\end{align*}

Next we move on to the construction of matrix $\W$. In view of the modified score function \eqref{rank2etatest}, we can see that there is no intrinsic term similar to terms \eqref{laybias} and \eqref{laybias2}. This means that if we can find an appropriate matrix $\W$ to control the bias in the first term of \eqref{rank2etatest}, we may be able to use the construction of matrices $\M$ and $\W$ to directly make inference without imposing the strong and weak orthogonality conditions (Conditions \ref{con:nearlyorth} and \ref{con:orth:rankr}). However, we will show that such matrix $\W$ simply does \textit{not} exist.

In view of matrices $\mathbf{M}_1$ and $\mathbf{M}_3$ given in \eqref{m13qew} and \eqref{m13qew3}, we have that 
\begin{align*}
\mathbf{M}_1 \boldsymbol{v}_1 \boldsymbol{u}_1^{\mathrm{T}}=\alpha_1 \widehat{\boldsymbol{\Sigma}} \boldsymbol{u}_2 \boldsymbol{u}_1^{\mathrm{T}}+\alpha_2 \widehat{\boldsymbol{\Sigma}} \boldsymbol{u}_1 \boldsymbol{u}_1^{\mathrm{T}}, \quad \mathbf{M}_3 \boldsymbol{v}_1 \boldsymbol{u}_2^{\mathrm{T}}=\beta_1 \widehat{\boldsymbol{\Sigma}} \boldsymbol{u}_2 \boldsymbol{u}_2^{\mathrm{T}}+\beta_2 \widehat{\boldsymbol{\Sigma}} \boldsymbol{u}_1 \boldsymbol{u}_2^{\mathrm{T}}.
\end{align*}
Then some simplifications give that 
\begin{align*}
\mathbf{I}_p-\mathbf{M}_1 \boldsymbol{v}_1 \boldsymbol{u}_1^{\mathrm{T}}-\mathbf{M}_3 \boldsymbol{v}_1 \boldsymbol{u}_2^{\mathrm{T}}=\mathbf{I}_p-\widehat{\boldsymbol{\Sigma}} \boldsymbol{u}_2\left(\alpha_1 \boldsymbol{u}_1+\beta_1 \boldsymbol{u}_2\right)^{\mathrm{T}}-\widehat{\boldsymbol{\Sigma}} \boldsymbol{u}_1\left(\alpha_2 \boldsymbol{u}_1+\beta_2 \boldsymbol{u}_2\right)^{\mathrm{T}}.
\end{align*}

Denote by $\mathbf{L}_2=\left[-\widehat{\boldsymbol{\Sigma}} \boldsymbol{u}_1,-\widehat{\boldsymbol{\Sigma}} \boldsymbol{u}_2\right] \in \mathbb{R}^{p \times 2}$ and $\mathbf{R}_2=\left[\alpha_2 \boldsymbol{u}_1+\beta_2 \boldsymbol{u}_2, \alpha_1 \boldsymbol{u}_1+\beta_1 \boldsymbol{u}_2\right]^{\mathrm{T}} \in \mathbb{R}^{2 \times p}$. Then we have 
\begin{align*}
\mathbf{I}_p-\mathbf{M}_1 \boldsymbol{v}_1 \boldsymbol{u}_1^{\mathrm{T}}-\mathbf{M}_3 \boldsymbol{v}_1 \boldsymbol{u}_2^{\mathrm{T}}=\mathbf{I}_p+\mathbf{L}_2 \mathbf{R}_2 .
\end{align*}
By the Sherman--Morrison--Woodbury formula, $\mathbf{I}_p+\mathbf{L}_2 \mathbf{R}_2$ is nonsingular if and only if $\mathbf{I}_2+\mathbf{R}_2 \mathbf{L}_2$ is nonsingular. However, it is easy to see that 
\begin{align*}
\begin{aligned}
\mathbf{I}_2+\mathbf{R}_2 \mathbf{L}_2 & =\left[\begin{array}{cc}
1-\left(\alpha_2 z_{11}+\beta_2 z_{12}\right) & -\left(\alpha_2 z_{12}+\beta_2 z_{22}\right) \\
-\left(\alpha_1 z_{11}+\beta_1 z_{12}\right) & 1-\left(\alpha_1 z_{12}+\beta_1 z_{22}\right)
\end{array}\right] \\[5pt]
& =\left[\begin{array}{cc}
1-\left(\alpha_2 z_{11}+\beta_2 z_{12}\right) & -\left(\alpha_2 z_{12}+\beta_2 z_{22}\right) \\
0 & 0
\end{array}\right],
\end{aligned}
\end{align*}
which shows that matrix $\mathbf{I}_p-\mathbf{M}_1 \boldsymbol{v}_1 \boldsymbol{u}_1^{\mathrm{T}}-\mathbf{M}_3 \boldsymbol{v}_1 \boldsymbol{u}_2^{\mathrm{T}}$ is in fact singular. Therefore, we see that such matrix $\W$ does \textit{not} exist \textit{without} incorporating the Stiefel manifold structure imposed by the SVD constraints. 
}

\end{document}

%% file: command.tex
\newcommand{\norm}[1]{\|#1\|}
\newcommand{\vect}[1]{\text{vec}(#1)}
\newcommand{\wh}[1]{\widehat{#1}}
\newcommand{\wt}[1]{\widetilde{#1}}

\newcommand{\diag}[1]{\text{diag}\{#1\}}

\newcommand{\abs}[1]{\vert #1 \vert}

\newcommand{\der}[2]{\frac{\partial{#1}}{\partial{#2}}}

\newcommand{\derr}[3]{\frac{\partial^2{#1}}{\partial{#2}\partial{#3}}}

\newcommand{\bDelta}{\boldsymbol{\Delta}}
\newcommand{\bdelta}{\boldsymbol{\delta}}
\newcommand{\bSigma}{\boldsymbol{\Sigma}}

\newcommand{\bbeta}{\boldsymbol{\beta}}
\newcommand{\balpha}{\boldsymbol{\alpha}}
\newcommand{\bTheta}{\boldsymbol{\Theta}}
\newcommand{\btheta}{\boldsymbol{\theta}}
\newcommand{\bepsilon}{\boldsymbol{\epsilon}}
\newcommand{\bmu}{\boldsymbol{\mu}}

\newcommand{\boldeta}{\boldsymbol{\eta}}
\newcommand{\bxi}{\boldsymbol{\xi}}
\newcommand{\N}{\mathcal{N}}

\newtheorem{theorem}{Theorem}
\newtheorem{lemma}{Lemma}
\newtheorem{condition}{Condition}

\newtheorem{proposition}{Proposition}
\newtheorem{definition}{Definition}
\newtheorem{corollary}{Corollary}
\def\trans{^{\rm T}}
\def\strans{^{* T}}

\def\R{{\mathbb R}}

\def\C{{\bf C}}
\def\X{{\bf X}}
\def\Y{{\bf Y}}
\def\U{{\bf U}}
\def\V{{\bf V}}
\def\D{{\bf D}}
\def\I{{\bf I}}
\def\E{{\bf E}}

\def\A{{\bf A}}
\def\B{{\bf B}}
\def\W{{\bf W}}
\def\L{{\bf L}}

\def\Q{{\bf Q}}
\def\T{{\bf T}}
\def\M{{\bf M}}

\def\x{\mathbf{x}}

\def\e{\boldsymbol{e}}

\def\u{\boldsymbol{u}}
\def\v{\boldsymbol{v}}
\def\z{\boldsymbol{z}}
\def\a{\boldsymbol{a}}
\def\b{\boldsymbol{b}}

\def\w{\boldsymbol{w}}
\def\h{\boldsymbol{h}}

\def\r{\boldsymbol{r}}

\def\l{\boldsymbol{l}}

\def\0{\boldsymbol{0}}
\def\1{\boldsymbol{1}}

\newcommand{\brho}{\boldsymbol{\rho}}
\renewcommand{\ldots}{\cdots}
\renewcommand{\trans}{^T}
\renewcommand{\N}{N}